%% file: SSSthesis.tex
\documentstyle[12pt,epsfig,uthesis]{report}

\newcommand{\be}{\begin{equation}}
\newcommand{\ee}{\end{equation}}
\newcommand{\ba}{\begin{eqnarray}}
\newcommand{\ea}{\end{eqnarray}}

\catcode`\@=11

\@addtoreset{equation}{chapter}
\@addtoreset{table}{chapter}
\renewcommand{\theequation}{\thechapter.\arabic{equation}}

\renewcommand\thetable{\thechapter.\@arabic\c@table}
\@addtoreset{figure}{chapter}

\catcode`@=11
\def\seceqaa{\@addtoreset{equation}{section}
           \def\theequation{A\arabic{equation}}}
\def\seceqbb{\@addtoreset{equation}{section}
           \def\theequation{B\arabic{equation}}}
\def\seceqcc{\@addtoreset{equation}{section}
           \def\theequation{C\arabic{equation}}}
\def\seceqdd{\@addtoreset{equation}{section}
           \def\theequation{D\arabic{equation}}}
\def\seceqee{\@addtoreset{equation}{section}
           \def\theequation{E\arabic{equation}}}
\def\seceqff{\@addtoreset{equation}{section}
           \def\theequation{F\arabic{equation}}}
\def\seceqgg{\@addtoreset{equation}{section}
           \def\theequation{G\arabic{equation}}}
\def\seceqhh{\@addtoreset{equation}{section}
           \def\theequation{H\arabic{equation}}}
\def\seceqjj{\@addtoreset{equation}{section}
           \def\theequation{J\arabic{equation}}}
\def\seceqll{\@addtoreset{equation}{section}
           \def\theequation{L\arabic{equation}}}
\catcode`@=11

\begin{document}

\begin{titlepage}
\begin{center}
{\large\bf Topics In Large Volume Swiss-Cheese Compactification Geometries\footnote{This review article is based on author's Ph.D.~thesis,
           defended on March 28, 2011.}}

\vskip 0.3cm
{\large\bf Pramod Shukla\footnote{E-mail: {\tt pramodmaths@gmail.com}}} \\
\vskip 0.3cm
{Department of Physics\footnote{The address from October 1, 2011 will be Max-Planck-Institut f\"{u}r Physik (Werner-Heisenberg-Institut),
F\"{o}hringer Ring 6, 80805 M\"{u}nchen, Germany.}\\
 Indian Institute of Technology Roorkee\\
 Roorkee, 247667  Uttarakhand, India}\\

\vskip 1cm
\pagenumbering{roman}
{\bf\Large Abstract}
\end{center}
\noindent
{In this review article, we present a systematic study of large volume type IIB string compactifications that addresses several interesting issues in string cosmology and string phenomenology within a single string compactification scheme. In the context of string cosmology, with the inclusion of perturbative and non-perturbative $\alpha^\prime$ corrections to the K\"{a}hler potential and instanton generated superpotential (without the inclusion of D3/D7-branes), we discuss the issues like obtaining a metastable non-supersymmetric $dS$ minimum without adding anti-$D3$ brane and obtaining slow roll inflation with the required number of 60 e-foldings along with non-trivial non-Gaussianities and gravitational waves. For studying cosmology and phenomenology within a single string compactification scheme, we provide a geometric resolution to a long-standing tension between LVS cosmology and LVS phenomenology after incorporating the effect of a single mobile spacetime-filling $D3$-brane and stacks of fluxed $D7$-branes wrapping the ``big" divisor of a Swiss-Cheese CY. Using GLSM techniques and the toric data for the Swiss-Cheese CY, we calculate geometric K\"{a}hler potential in LVS limit which are subdominant as compared to the tree level and (non-)perturbative contributions. We propose an alternate possibility of supporting some (MS)SM-like model in a framework with $D7$-branes wrapping the big divisor (unlike the previous LVS models) after realizing $g_{YM}\sim{\cal O}(1)$ in our setup. A detailed study of addressing several interesting issues in supersymmetry breaking scenarios in the context of D3/D7 Swiss-Cheese phenomenology, like realizing ${\cal O}(TeV)$ gravitino and explicit calculation of various soft masses and couplings is presented. Further, we show the possibility of realizing fermions masss scales ($\sim(MeV-GeV)$) of first two generations and order $eV$ neutrino mass scale along with an estimate for proton lifetime ($\tau_P\le 10^{61}$ years) from a SUSY GUT-type dimension-six operator. Apart from the issues related to (string) cosmology/phenomenology, we also discuss some other interesting issues on implications of moduli stabilization via inclusion of fluxes in type IIB compactification scenarios. These issues include the existence of area codes, `inverse-problem' related to non-supersymmetric black hole attractors and existence of fake superpotentials.}
\end{titlepage}

%
\vskip 1cm
\prefacesection{Acknowledgment}
\bigskip
\noindent
{\small This review article is based on my Ph.D.~thesis work which is a culmination of a pleasant journey of my academic life at Indian Institute of Technology Roorkee, India. First of all I am deeply grateful to my supervisor Dr. Aalok Misra whose unlimited continuous support and guidance made the thesis possible. I would like to thank (Department of Physics,) Indian Institute of Technology Roorkee, Roorkee, India for providing the required research facilities during the research period. I gratefully acknowledge the financial support provided by Council of scientific and Industrial Research (CSIR), New Delhi in the form junior/senior research fellowships during the period of research. I thank Abdus Salam's International Center for Theoretical Physics (ICTP), Italy for providing me financial support (twice in March, 2009 and March 2010) to attend spring schools there. I would like to thank String Theory Group (especially Prof. M. Bianchi) Tor vergata, University of Rome; The Institut de Physique Théorique (IPhT) Saclay (especially Prof. I. Bena), Saclay for their nice hospitality during my (short) visits to these institutions. I would also like to thank {\it The Excellence Cluster ``Origin and Structure of the Universe"} TUM, Munich for providing me financial support for attending the ``International school on strings and Fundamental Physics" at Garching, Munich. Finally, I would like to thank the various funding agencies and research centers in India for their indirect support (through the organizers) during the coferences/workshops/schools I attended (in India) for providing me opportunity to interact with experts of my field and for nice hospitality at respective places.}
%






\afterpreface
\include{chap1}
\include{chap2}

\include{chap3}

\include{chap4}

\include{chap5}

\include{chap6}
\include{chap7}

\renewcommand{\chaptermark}[1]{         
      \markboth{ \thechapter.\ #1}{}} %
\appendix
\include{Appendix}

\include{bib}

\end{document}

%% file: chap1.tex
\chapter{Introduction And General Motivations}
\markboth{nothing}{\bf 1. Introduction And General Motivations}

{\hskip1.4in{\it ``If we do discover a complete theory (of the Universe) it would be the ultimate triumph of human reason-for then we would know the mind of God".}}

\hskip4.2in - Stephen Hawking.

\section{Introduction}
General relativity and quantum mechanics are two of the most exciting achievements of Physics which provided a completely different understanding of the laws governing the nature. There are four fundamental interactions, namely Electromagnetic, Weak, Strong and Gravitational which dictate the rules for each and every physical processes around us. Based on the classification of four interactions, different theories are formulated and provide satisfactory results in particular regimes of validity. However, as there are many processes which involve more than one interaction, one requires a unified structure of all these four interactions in a single theory in order to answer which theory is to be used for explaining the physics in particular processes involving the interplay of more than one interaction. For example, early time cosmology or physics of black holes are two (sub)areas with the interplay of quantum and gravitational effects.

Presently, the Standard Model of particle physics, which has been experimentally tested in various aspects of observations involving the interaction of elementary particles (see \cite{SMcheck}), is believed to be quite satisfactory for short distance physics and involves three of the four fundamental interactions except the effects of gravity. Although the Standard Model provides a realistic description of renormalizable gauge theories, however it has several loop holes, like it has many free parameters, the Higgs mass is not protected against quantum corrections, it does not explain the presence of dark matter and most importantly it does not include gravity. These, along with the possible motivations for new physics (beyond SM) in explaining issues of non-zero neutrino masses, unification of gauge couplings, proton stability etc., show that an ultimate theory which could answer all the questions in physics is still missing and hence motivates a quest for unification to search the most fundamental theory.

The search for an ``ultimate theory" starts with a natural extension of the Standard Model with a beautiful symmetry identifying bosons with fermions- supersymmetry\footnote{See \cite{susyIntro} and references therein for a review.} - which puts several crucial constraints and  simplifies the theory. Supersymmetry doubles the spectrum of the theory with the inclusion of superpartners to each and every particle of the theory. However, as any direct observation of supersymmetry has not been found yet, this implies that if such a symmetry exists at all, it should be broken at some energy scale. The supersymmetric Standard Model provides resolutions to some of the problems of the Standard Model \cite{susyExtensions}, e.g. Higgs mass is protected by supersymmetry against quantum corrections as large quantum corrections are forbidden. Also within the framework of minimally supersymmetric Standard Model (MSSM), the renormalization group flow predicts a unification of all three gauge-couplings at a high energy scale ($\sim 10^{16} {\rm GeV}$) in a magical way making MSSM a possible candidate for some new physics beyond Standard Model. Moreover, supersymmetry in its local form naturally includes gravity as supersymmetry is a spacetime symmetry. The models with supersymmetric extensions have been very attractive on gravity as well as gauge theory sides and enormous amount of work has been done in this regard (see \cite{susyExtensions,susyinitials,sugra}).

\section{General Motivations For String Theory: A Brief (Historic) Review}

\hskip2.5in{{\it ``String theory is an attempt at a deeper description of nature by thinking of an elementary particle not as a little point but as a little loop of vibrating string".}}

\hskip4.2in - Edward Witten.\\

All the theories of real world Physics are valid up to a particular energy scale and physical observations require specific probe energy, e.g. the energy scales for observing a mountain and its atomic constituents are different and if one is interested in observing the sub-atomic/nuclear constituents of the mountain, one has to probe with more energy and accordingly, one has to increase the regime of validity of the particular theory depending on the energy scales involved. The notion of unification is something which makes a theory more interesting and at some stage, this appears to be a natural demand for extending the boundaries of validity of the theory compelling the same to be a realistic one. There have been several path breaking unifications with a completely new understanding of various fundamental questions and the examples are: unification of the Physics of ``Apples" with the Physics of ``Moon" in Newton's gravitational law implying the unification of rules governing the dynamics of celestial and terrestrial bodies, unification of electricity and magnetism are encoded in a set of Maxwell's equations in a single framework of ``electromagnetic theory", unification of notions of ``space" and ``time" into ``spacetime" and mass energy equivalence. Presently, we have a well settled and experimentally tested theory of short distance physics, namely Quantum Mechanics and a very beautiful theory of massive objects, namely General Relativity \cite{GR_wald,GRtests} and to unify these two is among the most outstanding challenges in physics.

The gravitational interaction shows quite different behavior as compared to the other three interactions in several aspects, as unlike the other three it can not be switched off as it is inherent in the spacetime itself; it is the weakest etc. Two massive objects are supposed to interact gravitationally via exchange of a mediating spin-two boson, namely graviton (yet to be observed experimentally) and in a perturbative regime of a conventional quantum field theory, quantization of gravity is not possible due to its non-renormalizable nature. There have been several attempts to overcome this problem \cite{GR_nonrenormal,GR_QMcorrections,LoopQG}, however the most promising candidate for a unified theory of all interactions, is String Theory.

String theory had been initially proposed for explaining the spectrum of hadrons and their interactions almost five decades ago in the 1960's and at that time it could not receive much attention as the same was formulated in higher dimensions (contrary to the fact that we live in four spacetime dimensions) and appearance of a massless spin-2 object in its spectrum (contrary to requirements in the hadronic regime). After that the second among the aforementioned ``uninvited guests" which caused string theory for getting ruled out in the first episode, has been realized to be a boon and in 1974, the massless spin-2 particle appearing in string spectrum was interpreted to be the graviton \cite{Scherk_Schwarz1974} and the identifying arguments have been boosted with the investigations that the low energy limits of (super)string theory is supergravity in higher dimensions. Actually in order to make string theory stable and tachyon free, supersymmetric extensions are required and consistent formulation demands 10-spacetime dimensions. Further, there has been five consistent superstring theories, namely Type IIA, Type IIB, Type I and two Heterotic superstring theories with gauge groups $E_8\times E_8$ and $SO(32)$. All five superstring theories are related in a single framework of ``M-Theory" the low energy limit of which is an 11-dimensional supergravity. Some nice reviews of these are available in \cite{GSW,STbook,stringtheoryReview,book}. Further, as brane world scenarios can be naturally embedded in string theory and interactions (except gravity) can be localized on the brane providing a completely different understanding of gauge interactions in a geometric way, and various distinguished aspects of gravity can be attributed to extra hidden dimensions as gravity can propagate through extra dimensions. This way, string theory has been elevated to be a possible candidate having the required prospects to unify all interactions.

Apart from being a promising candidate for unifying all interactions, string theory also provides a consistent regulator as a resolution to the divergences appearing due to point interactions in usual Quantum field theories, as in string theory framework, the most fundamental ingredient for interactions is a one-dimensional extended object -the string. The basic idea in string theory is that all elementary particles appear to be various excitation modes of a fundamental vibrating string. The past and ongoing investigations have proven string theory to be most promising after connecting a 10-dimensional superstring theory with four dimensional real world via a process called ``compactification". The concept of compactification has been introduced in 1920's in the context of realizing four-dimensional Einstein's general relativity and Maxwell's electromagnetic theory in a single framework of a five-dimensional theory of gravity \cite{KK}. In the process of compactification, the extra six-dimensions of a 10-dimensional superstring theory are supposed to be curled up as suitable class(es) of manifold(s), and the most studied ones are Calabi Yau manifolds. In a generic string compactification, the 10-dimensional spacetime splits (apart from a warped factor) in a product of two spaces,
\begin{equation}
{\cal M}_{10} \equiv {\cal M}_4\otimes {\cal M}_6
\end{equation}
where ${\cal M}_4$ is the maximally symmetric four-dimensional space and ${\cal M}_6$ is the compactification
manifold. Here, the choice of suitable class(es) of compactification manifolds is crucial and the possible manifolds are large in number. The required mathematical notions for Calabi Yau compactifications can be found in \cite{Mathematicalreview} (and reference therein). Also an alternative to compactification was proposed (in \cite{Randall_Sundaram}), but could not receive much attention. The process of compactification results in a plethora of scalars called moduli. These moduli play very exciting role in real world physics, as these could be possible candidates for inflaton, dark energy scalar field, a possible quintessence field and also could be a possible dark matter candidate, and thus have many possible interesting physical implications. Inspite of their various useful implications in explaining the real world, on the flip side, moduli have been troubling the researchers from the very beginning of there appearance starting from moduli stabilization and also cause several cosmological problems yet to be resolved in a satisfactory way. These moduli are dictated by shape and size of internal manifold geometries along with the fluxes. The complex structure moduli and axion-dilaton were stabilized with the inclusion of fluxes \cite{fluxesGiddindsetal,Grana} while the K\"{a}hler moduli could be stabilized only after including non-perturbative effects in \cite{KKLT}.

\section{String Cosmology and String Phenomenology}

\hskip1.2in{\it{``Nature composes some of her loveliest poems for the microscope and the telescope".}}
- Theodore Roszak, (from ``Where the Wasteland Ends, 1972.")\\

String theory as a quantum theory of gravity must have explanations for all questions in particle physics as well as cosmology as an effective description. In this context, it has several interesting connections with the four dimensional real world forming a mesh of string cosmology, string phenomenology, $AdS/QCD$ and recently emerging $AdS/CMT$ within a single framework of a fundamental theory. String theory has a very rich mathematical structure and has contributed to several fields such as (Mirror Symmetry, Topological Strings etc. in) Algebraic Geometry, Topology as well as Number Theory making string theory useful beyond the physical implications as well by utilizing it as a mathematical tool.

In the context of string compactifications, obtaining de-Sitter vacua, embedding inflation scenarios fulfilling cosmological/astrophysical requirements and realizing the Standard Model (along with its matter content) have been some of the major issues and for the past decade or so, a lot of progress has been made in realistic model building in string theory and obtaining numbers which could be matched (directly or indirectly) with some experimental data thereby serving as a testing laboratory of string theory. A very interesting class of models for realistic model building of cosmology as well as phenomenology has been the L(arge) V(olume) S(cenarios) class of models which has been developed in the context of Type IIB orientifold compactifications \cite{Balaetal2}.

For the first time, de-Sitter solution in string theory framework was realized in an ${\cal N} = 1$ type IIB orientifold compactification after inclusion of non-perturbative effects in the superpotential \cite{KKLT}. Although the KKLT(-type) models could realize $dS$-minimum fixing all moduli, but they did so by using some uplifting mechanism such as adding anti-$D3$ brane, introduced by hand, for uplifting $AdS$ minimum to $dS$-minimum. Followed by the same another interesting class of models has been proposed (see \cite{Balaetal2}) in  the context of Type IIB orientifold compactifications with inclusion of ${\alpha^\prime}^3$-corrections (of \cite{BBHL}) to the K\"{a}hler potential. In such scenarios a non-supersymmetric $AdS$-minimum has been realized in the LVS limit and then the realized $AdS$-minimum could be uplifted to a $dS$-minimum with any of the uplifting mechanisms \cite{otherupliftings,westphal}. After achieving the prerequisites (moduli stabilization and de-Sitter realizations) of string cosmology, the interesting issue is to embed inflationary scenarios in string theoretic setup and this has been one of the attractive areas of research in recent few years (see \cite{KKLMMT,kallosh}). There have been inflationary models in KKLT-type as well as in LVS-type with the inflaton field getting identified with brane separation (in brane inflation \cite{kesav,Dasguptaetal,braneinflation}), K\"{a}hler moduli, axion as well as the Wilson line moduli (in modular inflation (see \cite{AxionInflation,gravwavesKallosh,axionicswisscheese,kahlerinflation} and references therein)).

{\it Motivated by studies in string cosmology, we start with a systematic study of large volume type IIB string compactification that addresses issues in string cosmology like obtaining a metastable non-supersymmetric $dS$-minimum without adding anti-$D3$ brane \cite{dSetal} and obtaining slow-roll axionic inflation with the required number of $60$ e-foldings \cite{axionicswisscheese}. In addition to the complex-structure moduli-dependent and the nonperturbative instanton generated superpotential (as in KKLT \cite{KKLT}) as well as the inclusion of perturbative $\alpha^\prime$-corrections (as in LVS \cite{Balaetal2}), we also include ``non-perturbative" $\alpha^\prime$-corrections (coming from the world-sheet instantons) in the K\"{a}hler potential written out utilizing the (subgroup of ) $SL(2,{\bf Z})$ symmetry of the underlying type IIB theory (see \cite{Grimm}).}

Although the idea of inflation was initially introduced to explain the homogeneous and isotropic nature of the universe at large scale structure \cite{FirstInflation,cosmoproblem,linde}, its best advantage is reflected while studying inhomogeneities and anisotropries of the universe, which are encoded in non-linear effects (parameterized by ``$f_{NL},\, \tau_{NL}$") and the ``tensor-to-scalar ratio" $r$ seeding the non-Gaussianities of the primordial curvature perturbation as well as signature of gravity waves, which are expected to be observed by PLANCK if $f_{NL}\sim {\cal O} (1)$ and $r \sim {\cal O} (10^{-2})$ \cite{Planck}. As these parameters give a lot of information about the dynamics inside the universe, the theoretical prediction of large/finite (detectable) values of the non-linear parameters $f_{NL},\, \tau_{NL}$ as well as ``tensor-to-scalar ratio" $r$ has received a lot of attention in the recent few years \cite{Shiu,Yokoyamafnl,Yokoyama,Bartolotheoryobservations,Bartolocmbaniso,Bartolocmbaniso,Hajiancmb_data,Kogo,CMb,Alabidi,Byrnes_Wands,Byrnestrispectrum,alishahiha,beyondslowroll,Linde_Book}.

Motivated by the ongoing PLANCK satellite experiments in cosmology, we turn towards studies related to issues in cosmology such as realizing ${\cal O}(1)$ non-linearity parameter $f_{NL}$ as a signature of non-Gaussianities and finite/detectable tensor-to-scalar ratio as some signals of gravitational waves. {\it We realized non-Gaussianities parameter $f_{NL}\sim {\cal O}(10^{-2})$ for slow roll, and $f_{NL}\sim {\cal O} (1)$ for beyond slow-roll scenarios as well as $r\sim {\cal O}(10^{-3})$ with loss of scale invariance well within experimental bound $|n_R - 1| < 0.05$ in our LVS Swiss-Cheese orientifold setup \cite{largefNL_r}.}

Further, string phenomenology has been an active area of research for a long time resulting in an enormous number of attempts available in the literature and a few can be found in \cite{susybreakingKKLT,moresusybreaking,susykklt2,SM1,SMreview,Pheno1,Pheno2}. In this context, realizing the (MS)SM spectrum, study of SUSY-breaking phenomena along with realizing its low energy matter content, obtaining non-zero neutrino masses and their mixing (as signatures of physics beyond the Standard Model) have attracted a lot of attention. These phenomenological aspects are very challenging from the point of view of looking for their string theoretic origin and have been of quite an interest which will span some of the salient features of the review article. On the way of embedding (MS)SM in and realizing its matter content from string phenomenology, the questions of supersymmetry breaking and its transmission to the visible sector are among the most challenging issues - the first being mainly controlled by the moduli potentials while the second one by the coupling of supersymmetry breaking fields to the visible sector matter fields. The breaking of supersymmetry which is encoded in the soft terms, is supposed to occur in a hidden sector and then communicated to the visible sector (MS)SM via different mediation processes (e.g. gravity mediation, anomaly mediation, gauge mediation) among which although none is clearly preferred, gravity mediation is the most studied one due to its efficient computability. The study of supersymmetry breaking in string theory context was initiated long back \cite{susyinitials} and a lot of work has been done in this direction (see \cite{Banksetal,Quevedosusy2,granagrimm,ibanezuranga,Ibanez,Lustetal,conlonLVSsusy} and references therein).
A more controlled study of supersymmetry-breaking has been possible only after all moduli could be stabilized with the inclusion of fluxes along with non-perturbative effects. Since it is possible to embed the chiral gauge sectors (like that of the (MS)SM) as well as inflationary scenarios in D-brane Models with fluxes, the study of $D$-branes Models have been fascinating since the discovery of $D$-branes (see \cite{SM2,Uranga,QuevedoMSSM,SM3} and references therein). In the context of $dS$ realized in the KKLT setup, the uplifting term from the $D3$-brane causes the soft supersymmetry breaking. In a generic sense, the presence of fluxes generate the soft supersymmetry breaking terms. The soft terms in various models in the context of gauge sectors realized on fluxed D-branes have been calculated (see \cite{Lustetal,ibanezfont} and references therein).

Similar to the context of $dS$ realization and its cosmological implications, the LVS class of models has been found to be exciting steps towards realistic supersymmetry breaking with some natural advantages such as the large volume not only suppresses the string scale but also the gravitino mass and results in the hierarchically small scale of supersymmetry breaking. Moreover the study of LVS models in the context of ${\cal N} = 1$ type IIB orientifold compactification in the presence of $D7$-branes, has also been quite attractive and promising for the phenomenological purposes because in such models, $D7$-brane wrapping the smaller cycle produces the qualitatively similar gauge coupling as that of the Standard Model and also with the magnetized $D7$- branes, the Standard Model chiral matter can be realized from strings stretching between stacks of $D7$-branes \cite{Lustetal,ibanezfont,Jockers_thesis,conloncal}. In one of such models, RG evolutions of soft-terms to the weak scale have been studied to have a low energy spectra by using the RG equations of MSSM (assuming that only charged matter content below the string scale is the MSSM) and it was found that with $D7$ chiral matter fields, low energy supersymmetry-breaking could be realized with a small hierarchy between the gravitino mass and soft supersymmetry-breaking terms \cite{conloncal}. A much detailed study with fluxed $D3/D7$ branes has been done in the context of ${\cal N} = 1$ type IIB orientifold compactification (see e.g. \cite{Jockers_thesis,conloncal}) and it has been found that the ${\cal N} = 1$ coordinates get modified with the inclusion of $D3$ and $D7$-branes. The gauge coupling
realized on $D7$-branes wrapping a four-cycle depends mainly on the size modulus of the wrapped four-cycle and also on the complex structure as well as axion-dilaton modulus after including the loop-corrections, which in the dilute flux limit (without loop-corrections) is found to be dominated by the size modulus of the wrapped four-cycle \cite{conloncal}.

In the models having branes at singularities, it has been argued that at the leading order, the soft terms vanish for the no-scale structure which gets broken at higher orders with the inclusion of (non-)perturbative $\alpha^\prime$-corrections to the K\"{a}hler potential\footnote{However, it has been observed (in \cite{loops}) that the inclusin of mere perturbative string loop-corrections to the K\"{a}hler potential do not violate the no-scale structure and is referered as ``extended no-scale structure".} resulting in the non-zero soft-terms at higher orders. In the context of LVS phenomenology in such models with $D$-branes at singularities \cite{SM1}, it has been argued that all the leading order contributions to the soft supersymmetry-breaking (with gravity as well as anomaly mediation processes) still vanish and the non-zero soft terms have been calculated in the context of gravity mediation with inclusion of loop-corrections.

{\it In order to support (MS)SM-like models and framing string cosmology as well as string phenomenology in a single large volume Swiss-Cheese setup, we extend our LVS cosmology setup with inclusion of a single spacetime filling mobile $D3$-brane along with stack(s) of fluxed $D7$-brane(s) wrapping the ``big" divisor (unlike the previously studied LVS models in which ``small" divisor wrapping has been done in order to realize ${\cal O}(1)$ gauge coupling) of a Swiss-Cheese Calabi Yau. After constructing appropriate local involutively-odd harmonic one-forms on the ``big" divisor and considering the Wilson line moduli contributions, we realize ${\cal O}(1)$ $g_{YM}$ with $D7$-branes wrapping the big divisor $\Sigma_B$ in the rigid limit of wrapping.}

There has been a long-standing tension between cosmology and phenomenology - the former demands the scale of inflation to be only a couple of orders less than the GUT scale while the latter demands the supersymmetry breaking at TeV scale resulting in a hierarchy in the energy scales involved on two sides and hence in order to incorporate cosmology and phenomenology in a single string theoretic setup, one needs to reconcile the scales involved as a resolution to the tension. We propose a possible geometric resolution (in \cite{D3_D7_Misra_Shukla}) which is translated into figuring out a way of obtaining a TeV gravitino when dealing with LVS phenemenology and a $10^{12}$ GeV gravitino when dealing with LVS cosmology in the early inflationary epoch of the universe, within the same setup. The holomorphic pre-factor in the superpotential coming from the space-time filling mobile $D3$-brane position moduli - section of (the appropriate) divisor bundle \cite{Ganor1_2,Maldaetal_Wnp_pref} - plays a crucial role. We show that as the mobile space-time filling $D3$-brane moves from a particular non-singular elliptic curve embedded in the Swiss-Cheese Calabi-Yau to another non-singular elliptic curve, it is possible to obtain $10^{12} $ GeV gravitino during the primordial inflationary era supporting the cosmological/astrophysical data as well as a TeV gravitino in the present era supporting the required SUSY breaking at TeV scale within the same set up, for the same volume of the Calabi-Yau stabilized at around $10^6 \, l_s^6$.

Finally, the compelling evidence of non-zero neutrino masses and their mixing has attracted a lot of attention as it supports the idea why one should think about physics beyond the Standard Model. Models with seesaw mechanism giving small Majorana neutrino masses have been among the most popular ones (see \cite{Mohapatra_group}). In the usual seesaw mechanisms, a high intermediate scale of right handed neutrino (where some new physics starts) lying between TeV and GUT scale, is involved, which has a natural geometric origin in the class of large volume models, as suggested in \cite{conlon_neutrino}, and as will be explicitly shown in this chapter {\bf 5} of this review article. The issue of proton stability which is a generic prediction of Grand unified theories, has been a dramatic outcome of Grand unified theories beyond SM. Although proton decay has not been experimentally observed, usually in Grand unified theories which provide an elegant explanation of various issues of real wold physics, the various decay channels are open due to higher dimensional baryon (B) numbers violating operators. However the life time of the proton (in decay channels) studied in various models has been estimated to be quite large (as $\tau_p \sim M_X$ with $M_X$ being some high scale) \cite{Prot_Decay_review}. Further, studies of dimension-five and dimension-six operators relevant to proton decay in SUSY GUT as well as String/M theoretic setups, have been of great importance in obtaining estimates on the lifetime of the proton (See \cite{Prot_Decay_review} and references therein). In our $D3/D7$ Swiss-Cheese LVS setup, we explored the possibility of realizing fermion (the first two-generation leptons/quarks) mass scales of ${\cal O}(MeV-GeV)$ and (first two-generation neutrino-like) $\le {\cal O}({\rm eV})$ masses, the latter via lepton number violating non-renormalizable dimension-five operators \cite{MS_LVSfermions}. Also, we showed that there are no SUSY GUT-type dimension-five operators pertaining to proton decay and estimate the proton lifetime from a SUSY GUT-type four-fermion dimension-six operator \cite{MS_LVSfermions}. We will elaborate on these issues in the string cosmology and string phenomenology portions in the upcoming chapters of the review article.

\section{Issues in (Fluxed) Swiss-Cheese Compactification Geometries}
{\hskip0.5in {\it ``The hidden harmony is better than the obvious."}}

\hskip 4in - Alexander Pope.

In the context of type IIB compactification, flux compactifications have been extensively studied from the point of view of moduli stabilization (See \cite{fluxesGiddindsetal,Grana} and references therein). Though, generically only the complex structure moduli get stabilized by turning on fluxes and one needs to consider non-perturbative moduli stabilization for the K\"{a}hler moduli \cite{KKLT}. In the context of type II compactifications, it is naturally interesting to look for examples wherein it may be possible to stabilize the complex structure moduli (and the axion-dilaton modulus) at different points of the moduli space that are finitely separated, for the same value of the fluxes. This phenomenon is referred to as ``area codes" that leads to formation of domain walls. Further, there is a close connection between flux vacua and black-hole attractors. It has been shown that extremal black holes exhibit an interesting phenomenon - the attractor mechanism \cite{attractor}. In the same, the moduli are ``attracted" to some fixed values determined by the charges of the black hole, independent of the asymptotic values of the moduli. Supersymmetric black holes at the attractor point, correspond to minimizing the central charge and the effective black hole potential, whereas nonsupersymmetric attractors \cite{nonsusybh1,nonsusybh}, at the attractor point, correspond to minimizing only the potential and not the central charge \cite{nonsusybh2}. Another interesting aspect of testing string theory is studies related to black hole physics, as black holes are like theoretical laboratories for stringy models. We discuss the realizations of the aforementioned aspects (in the context of flux compactification and black hole attractors along with existence of fake superpotential) in our Swiss-Cheese setup in chapter {\bf 6} of the review article.

\section{Overview of the review article}

The review article is implicitly divided into three parts: LVS Cosmology, LVS phenomenology and some other interesting implications on flux compactification side of a type IIB Swiss-Cheese (orientifold) compactification. After discussing prime motivations for my research work and reviewing the relevant literature with earlier studies in (this) chapter {\bf 1}, we start with building our LVS cosmology setup providing sufficient relevant pieces of information about various (non-)perturbative $\alpha^\prime$-corrections as well as loop-corrections to the K\"{a}hler potential (which we show to be subdominant), and non-perturbative instanton correction to superpotential along with their modular completions in chapter {\bf 2}. We also provide some relevant geometric information (see \cite{Candelasetal,DDF,denef_LesHouches}) about the Swiss-Cheese class of Calabi Yaus among which, we are using the one, expressed as an projective variety in ${\bf WCP}^4[1, 1, 1, 6, 9]$ throughout the review article to compactify a ten-dimension type IIB string theory of which we consider the resulting low energy effective theory for addressing some interesting aspects of our four-dimensional real world.

Chapter {\bf 3} of the review article is devoted to several interesting cosmological implications of our setup on LVS cosmology side \cite{axionicswisscheese,dSetal,largefNL_r} with the prime result of realizing a metastable non-supersymmetric de-Sitter minimum in a more natural way (without the addition of $D3$-branes). Using (non-)perturbative $\alpha^\prime$-corrections to the K\"{a}hler potential and non-perturbative instanton corrections to the superpotential, we discuss the possibility of realizing a non-supersymmetric $dS$-minimum in the LVS limit of the internal manifold and there was no need for adding $D3$-brane (\`{a} la KKLT) by hand \cite{KKLT}. Moving one more step towards realistic stringy cosmological model building as a test of string theory, after realizing non-supersymmetric $dS$-minimum, we address the issue of embedding (axionic) slow roll inflationary scenarios in \cite{axionicswisscheese} without the ``$\eta$-problem" and show that a linear combination of NS-NS axions provides a flat direction for the inflaton field to inflate from a saddle point to some nearest $dS$-minimum. Moreover, in the context of studies related to structure formations in string cosmology, we realize non-Gaussianities parameter $f_{NL}\sim {\cal O}(10^{-2})$ for slow roll, and $f_{NL}\sim {\cal O}(1)$ for beyond slow-roll scenarios in our LVS Swiss-Cheese orientifold setup \cite{largefNL_r}. Further, using general (not specific to string theory) considerations of Hamilton-Jacobi formalism and some algebraic geometric inputs, after imposing the freezing out of the curvature perturbations at super horizon scales we show the possibility to realize tensor-to-scalar ratio $r\sim {\cal O}(10^{-3})$ along with loss of scale invariance lying within the experimental bounds $|n_R - 1| < 0.05$. We close the chapter {\bf 3} by making some interesting observations pertaining to the possibility of inflaton field being a Cold Dark Matter (CDM) candidate as well as a quintessence field in some corner(s) of the moduli space, given that axions are stabilized at sub-Planckian Vevs.

In chapter {\bf 4}, we start with the extension of our LVS Swiss-Cheese cosmology setup of chapter {\bf 2} with the inclusion of a mobile spacetime-filling $D3$-brane and stacks of $D7$-branes wrapping $\Sigma_B$ along with supporting $D7$-brane fluxes and discuss several phenomenological issues. On the geometric side to enable us to work out the complete K\"{a}hler potential, we estimate the geometric K\"{a}hler potential (of the two divisors $\Sigma_B$ and $\Sigma_S$) for Swiss-Cheese Calabi-Yau ${\bf WCP}^4[1, 1, 1, 6, 9]$ using its toric data and GLSM techniques in the large volume limit. The geometric K\"{a}hler potential is first expressed, using a general theorem due to Umemura \cite{Umemura}, in terms of genus-five Siegel Theta functions or in the LVS limit genus-four Siegel Theta functions. Later, using a result due to Zhivkov, for purposes of calculations, we express the same in terms of derivatives of genus-two Siegel Theta functions \cite{Kimura,Zhivkov}. We also provide a ``geometric resolution" to a long standing tension between LVS cosmology and LVS phenomenology pertaining to realizing TeV gravitino for phenomenology and $10^{12}$ GeV gravitino for cosmology. Finally, we close the chapter {\bf 4} with a discussion on the possibility of realizing ${\cal O}(1)$ gauge coupling in our setup with the inclusion of possible competing contribution coming from Wilson line moduli \cite{D3_D7_Misra_Shukla}.

In chapter {\bf 5}, we estimate various soft supersymmetry breaking masses/parameters in the context of $D3/D7$ LVS Swiss-Cheese setup framed in the last chapter and realize order TeV gravitino and gaugino masses in the context of gravity mediated supersymmetry breaking. We observe that anomaly mediated gaugino mass contribution is suppressed by the standard loop factor as compared to gravity mediated contribution. The $D3$-brane position moduli and the $D7$-brane Wilson line moduli are found to be heavier than gravitino. Further, we find a (near) universality in the masses, $\mu$-parameters, physical Yukawa couplings and the $\mu B$-terms for the $D3$-brane position moduli - the two Higgses in our construction - and a hierarchy in the same set and a universality in the A terms on inclusion of the $D7$-brane Wilson line moduli. Based on phenomenological intuitions we further argue that the Wilson line moduli is to be identified with the squarks (sleptons) (at least the first two families) of MSSM as the Yukawa couplings for the same are negligible \cite{Sparticles_MS}. Building up on some more phenomenological aspects of our setup, we discuss the RG flow of the slepton and squark masses to the EW scale and in the process show that related integrals are close to the mSUGRA point on the ``SPS1a slope".

For realizing more realistic implications on phenomenology side, we show the possibility of generating fermion mass scales of $MeV-GeV$ range which can possibly be related to first two generations of quarks/leptons and realize neutrino mass scales of $\le {eV}$ via lepton number violating non-renormalizable dimension-five operators which could possibly be related to first two generations of neutrinos \cite{MS_LVSfermions}. We also show that there are no SUSY GUT-type dimension-five operators corresponding to proton decay and close the chapter {\bf 5} with an estimate of the proton lifetime from a SUSY GUT-type four-fermion dimension-six operator to be $10^{61}$ years.

Apart from issues related to string cosmology and string phenomenology, chapter {\bf 6} of the review article includes some other interesting issues in type IIB in the context of flux compactifications and related black hole attractors \cite{dSetal}. In this context, we discuss the issues of existence of ``area codes", ``inverse-problem" related to non-supersymmetric black hole attractors and existence of fake superpotentials. We argue the existence of extended ``area codes" where in complex structure moduli and the axion-dilaton modulus can be stabilized at points near as well as away from the two singular conifold loci of the Swiss-Cheese CY for the same values of the NS-NS and RR fluxes. As regards supersymmetric and non-supersymmetric black-hole attractors in ${\cal N} =2$  Type II compactifications on the same CY, we explicitly solve the inverse problem which is to calculate the electric and magnetic charges of the extremal black hole potential, given the extremum values of the moduli. In the same context, we also show explicitly the existence of ``fake superpotentials" as a consequence of non-unique superpotentials for the same black-hole potential corresponding to reversal of signs of some of the electric and magnetic charges which we explicitly show by constructing a constant symplectic matrix for our two-paramater Swiss-Cheese Calabi-Yau.

Chapter {\bf 7} contains an overall summary and conclusions along with interesting
future directions. Finally, we provide appendices and a bibliography to close the review article.

%% file: chap2.tex
\chapter{Large Volume Swiss-Cheese Orientifold Setup}
\markboth{nothing}{\bf 2. Large Volume Swiss-Cheese Orientifold Setup}

\hskip1in{\it{`` To those who do not know mathematics it is difficult to get across a real feeling as to the beauty, the deepest beauty, of nature ... If you want to learn about nature, to appreciate nature, it is necessary to understand the language that she speaks in".}}
\hskip3.7in - Richard Feynman.

\section{Introduction}
In the context of realistic model building in string theory KKLT and LVS class of models are among the most popular ones. Both of these models are developed in the context of IIB orientifold compactifications. In the KKLT class of models, one considers the tree level contributions in the K\"{a}hler potential coming from complex structure, axion dilaton as well as K\"{a}hler structure deformations, and non-perturbative effects (from gaugino-condensation or $ED3$-instantons) along with the flux contributions in the superpotential. These effects are sufficient to fix all moduli and one realizes a supersymmetric $AdS$ minimum which is then uplifted to non-supersymmetric metastable $dS$ minimum with the inclusion of $\overline{D3}$-brane in the setup\cite{KKLT}. A drawback of KKLT models is the control over corrections remain marginal and situation becomes more problematic with inclusion of more K\"{a}hler moduli in the setup. Followed by the KKLT model, in addition to the ingredients of the KKLT setup, it has been observed that if one includes perturbative ${\alpha^\prime}^3$-corrections (of \cite{BBHL}) in the K\"{a}hler potential in the context of Type IIB compactification on an orientifold of a two-parameter Calabi Yau, consistency requires that the divisor volumes of the two divisor moduli ($\tau_b$ and $\tau_s$) are stabilized at hierarchically separated values ($\tau_b\sim {\cal V}^{2/3}$ and $\tau_s\sim ln{\cal V}$, ${\cal V}$ being the Calabi Yau volume), developing a new and extremely interesting class of models in ``large volume scenarios" \cite{Balaetal2}. In these models also, all moduli are stabilized and one realizes a non-supersymmetric $AdS$-minima in the large volume limit which could be uplifted \`{a} la KKLT. The most basic idea behind LVS models, is to balance a non-perturbative correction depending exponentially on the smaller divisor volume against a perturbative correction depending inversely on the larger divisor volume, thus potentially giving rise to exponentially large overall volume.

In this chapter, we provide a detailed information of our large volume setup, which (or its D3/D7 extension) we will be using throughout the review article. We consider Type IIB compactified  on an orientifold of a Swiss-Cheese Calabi Yau with the inclusion of (non-)perturbative $\alpha^{\prime}$-corrections to the K\"{a}hler potential and non-perturbative instanton corrections to the superpotential along with the flux superpotential. We also include modular completions of the K\"{a}hler potential and the superpotential.

The chapter is structured as follows: We will be starting with details of the Swiss-Cheese class of Calabi-Yaus (and in particular a projective variety in $\bf{WCP}^4 [1,1,1,6,9]$) with interesting geometric information regarding the same in section {\bf 2}. In section {\bf 3}, we discuss the choice of involution for orientifolding the Swiss-Cheese Calabi Yau in the context of Type IIB compactifications and summarize the spectrum of resulting four-dimensional ${\cal N}=1$ effective theory after orientifold truncation. Section {\bf 4} contains some brief reviews on inclusion of (non-)perturbative $\alpha^\prime$-corrections as well as string loop-corrections to the K\"{a}hler potential and in section {\bf 5}, we provide relevant pieces of information on non-perturbative effects in the superpotential along with flux generated Gukov-Vafa-Witten contribution. The following section {\bf 6} contains the modular completions of the K\"{a}hler potential and the superpotential and the final section {\bf 7}, subsequently summaries our large volume Swiss-Cheese (cosmology) setup.

\section{Swiss-Cheese Calabi Yau}
The ``Swiss Cheese" class of Calabi Yau is used to denote those Calabi-Yau's whose volume can be written as \cite{DDF,denef_LesHouches,Conlonthesis}:
\begin{equation}
\label{eq:Swiss-Cheese_classEQ}
{\cal V}=(\tau^B + \sum_{i\neq B} a_i\tau^S_i)^{\frac{3}{2}} - (\sum_{j\neq B}b_j\tau^S_j)^{\frac{3}{2}} - ...,
\end{equation}
where $\tau^B$ is the volume of the big divisor and $\tau^S_i$ are the volumes of the $h^{1,1}-1$ (corresponding to the (1,$h^{1,1}-1$)-signature of the Hessian) small divisors. The big divisor governs the size of the Swiss-Cheese and the small divisors control the size of the holes of the same Swiss- Cheese.

The Swiss Cheese Calabi-Yau we have been using, is a two-parameter Calabi-Yau obtained as a resolution of the degree-18 hypersurface in $\bf{WCP}^4[1,1,1,6,9]$:
\begin{equation}
\label{eq:hypersurface}
x_1^{18} + x_2^{18} + x_3^{18} + x_4^3 + x_5^2 - 18\psi \prod_{i=1}^5x_i - 3\phi x_1^6x_2^6x_3^6 = 0,
\end{equation}
which has one ``big" $\Sigma_B(x_5=0)$ and one ``small" $\Sigma_S(x_4=0)$ divisors \cite{DDF,denef_LesHouches}. The aforementioned Calabi-Yau has $h^{1,1}=2$ and $h^{2,1}=272$ with a large discrete symmetry group given by $\Gamma={\bf Z}_6\times{\bf Z}_{18}$ (${\bf Z}_6:(0,1,3,2,0,0); {\bf Z}_{18}:(1,-1,0,0,0)$ (See \cite{Candelasetal}) relevant to construction of the mirror \`{a} la Greene-Plesser prescription. As in \cite{DDF,Kachru_et_al}, if one assumes that one is working with a subset of periods of $\Gamma$-invariant cycles - the six periods corresponding to the two complex structure deformations in (\ref{eq:hypersurface}) will coincide with the six periods of the mirror - the complex structure moduli absent in (\ref{eq:hypersurface}) will appear only at a higher order in the superpotential because of $\Gamma$-invariance and can be consistently set to zero \cite{Kachru_et_al} and thus, effectively only two complex structure moduli are activated. Further, defining $\rho\equiv (3^4.2)^{\frac{1}{3}}\psi$, the singular loci of the Swiss-Cheese Calabi Yau are in $\bf{ WCP}^2[3,1,1]$ with homogenous coordinates $[1,\rho^6,\phi]$ and are given as under:
\begin{enumerate}
\item
${\rm Conifold\ Locus 1}: \{(\rho,\phi)|(\rho^6+\phi)^3=1\}$
\item
${\rm Conifold\ Locus 2}: \{(\rho,\phi)|\phi^3=1\}$
\item
${\rm Boundary}: (\rho,\phi)\rightarrow\infty$
\item
${\rm Fixed\ point\ of\ quotienting}$: The fixed point $\rho=0$ of ${\cal A}^3$ where
${\cal A}:(\rho,\phi)\rightarrow(\alpha\rho,\alpha^6\phi)$, where $\alpha\equiv e^{\frac{2\pi i}{18}}$.\end{enumerate}
The aforementioned information about the singular loci will be important while discussing the possibility of ``area codes" in the context of (complex structure) moduli stabilization via flux compactifications in chapter {\bf 6} of the review article.

Let us now elaborate of the relevance of the Swiss-Cheese hypersurface (\ref{eq:hypersurface}) in the context of Type IIB compactifications elucidated by the F-theoretic description \cite{denef_LesHouches}. F-theory proposed by Vafa, provides a non-perturbative completion of Type IIB in a geometric way uplifting the same to 12-dimensional space. From the point of view of orientifold limit of F-theory \cite{Sen}, F-theory compactified on an elliptically fibred  Calabi-Yau four-fold $X_4$ (with projection $\pi$) is equivalent to Type IIB compactified on the base $B_3$ (of the same $X_4$), which is a Calabi-Yau three fold $Z$-orientifold with $O3/O7$ planes. Here, the base $B_3$ could either be a Fano three-fold or an $n$-twisted ${\bf {CP}}^1$-fibration over ${\bf{ CP}}^2$ such that pull-back of the divisors in $CY_3$ automatically satisfy Witten's unit-arithmetic genus condition \cite{Witten}; the Euler characteristics $\chi(D)\equiv \sum_j (-)^j h^{0,j}=1$ for a divisor $D$ \cite{DDF,denef_LesHouches}.
In the latter case, the base $B_3$ (which is to be an $n$-twisted ${\bf {CP}}^1$-fibration over ${\bf{ CP}}^2$), is given by the following toric data:
$$
\begin{array}{c|ccccc}
&D_1&D_2&D_3&D_4&D_5 \\ \hline
{\bf C}^*&1&1&1&-n&0\\
{\bf C}^*&0&0&0&1&1
\end{array}$$
where the divisors $D_{1,2,3}$ are pullbacks of three lines in ${\bf{CP}}^2$ and the divisors $D_{4,5}$ are two sections of the fibration.

From the point of view of M-theory compactified on $X_4$, the non-perturbative superpotential receives non-zero contributions from $M5$-brane instantons involving wrapping around uplifts {\bf V} to $X_4$ of ``vertical" divisors in $B_3$. These vertical divisors are defined such that $\pi({\rm\bf V})$ is a proper subset of $B_3$ and are either components of singular fibers or are pull-backs of smooth divisors in $B_3$. There exists a Weierstrass model $\pi_0:{\cal W}\rightarrow B_3$ and its resolution $\mu: X_4\rightarrow {\cal W}$. For the vertical divisors being components of singular fibers, $B_3$ can be taken to be a ${\bf{CP}}^1$-bundle over $B_2$ with ADE singularity of the Weierstrass model along $B_2$. From the Type IIB point of view, this corresponds to pure Yang-Mills with ADE gauge groups on $D7$-branes wrapping $B_2$; the vertical divisors are hence referred to as ``gauge-type" divisors. The pullbacks of smooth divisors in $B_3$ need not have a gauge theory interpretation - they are hence referred to as ``instanton-type" divisors.

Now, utilizing Witten's prescription of ``gauged linear sigma model" description of ``toric varieties", overall volume of the Calabi Yau three-fold base $B_3$ can be computed in terms of two-cycle volumes and hence in terms of divisor volumes. Writing the K\"{a}hler class $J=\xi^1D_1+\xi^2D_5$, where $D_1$ and $D_5=D_4+nD_1$ are divisors dual to the holomorphic curves $C^1=D_1\cdot D_4$ and $C^2=D_1\cdot D_2$ in the Mori Cone (for which $\int_{C_i}J>0$) such that $\int_{C_{1,2}}J=\xi^{1,2}$, the volume of the Calabi Yau three-fold base $B_3$ is given by:
\begin{equation}
{\cal V}_{B_3}=\frac{1}{6}\left(\xi^1D_1+\xi^2D_5\right)^3.
\end{equation}
Further, using the divisor intersection numbers; $D_1^3=0, D_1^2D_5=1, D_1D_5^2=n,D_5^3=n^2$, the volumes of divisor $D_{4,5}$ are given by: ${\cal V }_{D_4=D_5-nD_1}=\left(\frac{\partial}{\partial\xi^2}-n\frac{\partial}{\partial\xi^1}\right) {\cal V}_{B_3}$
and ${\cal V}_{D_5}=\frac{\partial}{\partial\xi^2} {\cal V}_{B_3}$. One hence obtains:
\begin{eqnarray}
& & {\cal V}_{D_5}=\frac{\left(\xi^1+n\xi^2\right)^2}{2}; \, \, \, {\cal V}_{D_4}=\frac{{\xi^1}^2}{2}\nonumber\\
& & {\cal V}_{B_3}=\frac{\sqrt{2}}{3n}\left({\cal V}_{D_5}^{3/2}-{\cal V}_{D_4}^{3/2}\right),
\end{eqnarray}
implying that $B_3$ is of the ``Swiss Cheese" type wherein the ``big" divisor $D_5$ contributes positively and the ``small" divisor $D_4$ contributes negatively. Also,
\begin{equation}
{\cal V}_{D_4\cap D_5}= \left(\frac{\partial}{\partial\xi^2}-n\frac{\partial}{\partial\xi^1}\right)\frac{\partial}{\partial\xi^2}
{\cal V}_{B_3}=0
\end{equation}indicating that $D_4(\equiv\Sigma_S)$ and $D_5(\equiv\Sigma_B)$ do not intersect implying that there is no contribution to the one-loop contribution to the K\"{a}hler potential from winding modes corresponding to strings winding non-contractible 1-cycles in the intersection locus corresponding to stacks of intersecting $D7$-branes wrapped around $D_{4,5}$ (See \cite{BHP,loops} and section {\bf 4}). Finally, for $n=6$ \cite{DDF}, the $CY_4$ will be the resolution of a Weierstrass model with $D_4$ singularity along the first section and an $E_6$ singularity along the second section. The $CY_3$ turns out to be a unique Swiss-Cheese Calabi Yau - an elliptic fibration over ${\bf{CP}}^2$ - in ${\bf{WCP}}^4[1,1,1,6,9]$ given by (\ref{eq:hypersurface}) and the overall volume (${\cal V}$) of this Swiss-Cheese Calabi Yau is given as\footnote{In the later notations throughout the review article, we will drop the subscript from ${\cal V}_{B_3}$ and will denote the overall Swiss-Cheese Calabi Yau volume as ${\cal V}$ and the two complexified divisors volumes by $\tau_b$ and $\tau_s$, which are the standard large volume notations used.},
\begin{equation}
\label{eq:CYvolume}
{\cal V}=\frac{1}{9\sqrt{2}}\left(\tau_b^{3/2}-\tau_s^{3/2}\right)
\end{equation}
In large volume scenarios, the divisor volumes are scaled as: $\tau_b\sim {\cal V}^{\frac{2}{3}}, \, \, \tau_s\sim ln{\cal V}$ and thus two divisor volumes are hierarchically separated in the large volume limit.
\section{Orientifolding the Swiss-Cheese Calabi Yau: Choice of Involution}
The  effective low energy theory derived from an orientifold compactification and analyzing its subsequent spectrum have been among most fascinating studies in string theory. The standard Calabi-Yau compactifications of type IIB lead to an effective ${\cal N} = 2$ four-dimensional supergravity in low energy limits and spectrum is further truncated after orientifolding and only half of the supersymmetries survive resulting in a ${\cal N}=1$ supersymmetric theory \cite{Jockers_thesis}. An orientifold is an orbifold which is modded out by a group structure: ${\cal O}\equiv(-)^{F_L}\Omega_p \, \, \sigma$, where $\Omega_p$ is the world-sheet parity and $F_L$ is the space-time fermion number in the left-moving sector and  $\sigma$ is a holomorphic, isometric involution under which the K\"{a}hler form $J$ is even while the complex structure $\Omega$ has two possibilities:
\begin{equation}
\label{eq:involution}
\sigma^*(J)=J,\ \sigma^*(\Omega)=\pm\Omega.
\end{equation}
The choice of the above two possible involutions necessitates the inclusion of D-branes in order to cancel the tadpoles after supporting the two sets of orientifold planes: $O5/O9$ (for choice $\sigma^*(\Omega)=\Omega$) and $O3/O7$ (for choice $\sigma^*(\Omega)=-\Omega$) at the fixed point of these involutions.
Let us start with reviewing the massless bosonic spectrum of the ten-dimensional type IIB which includes fields the dilaton $\phi$ with an axion $l=C_0$, the metric $g$, an NS-NS two-form $B_2$, RR forms $C_2$ and $C_4$ which has a self-dual field strength in the R-R sector. The orientifolds projections of these fields are encoded in the following set of equations:
\begin{eqnarray}
\label{eq:field_involutions}
& & \sigma^{*}\left(\begin{array}{c}
                      \phi \\
                      l \\
                      g \\
                                          \end{array}
\right)= \left(\begin{array}{c}
                      \phi \\
                      l \\
                      g \\
                                         \end{array}
\right)\, \, ; \, \, \, \, \, \sigma^{*}\left(\begin{array}{c}
                      B_2 \\
                      C_2 \\
                      C_4
                                          \end{array}
\right)=  \left(\begin{array}{c}
                      -B_2 \\
                      -C_2\\
                      C_4
                                          \end{array}
\right)
\end{eqnarray}
In the four-dimension effective compactified theory, these ten-dimensional fields are expanded out in terms of harmonic forms and only the invariant states of the full orientifold projection survive. The harmonic forms are in one-to-one correspondence with the elements of the cohomology groups $H^{(p,q)}$ which split into direct sum of even and odd eigenspaces under the orientifold action as $H^{(p,q)}\equiv H_+^{(p,q)}\oplus H_{-}^{(p,q)}$.

With the choice of involution $\sigma^*(J)=J,\ \sigma^*(\Omega)=-\Omega$, the K\"{a}hler form $J$ can be expanded in even basis as $J=v^\alpha\omega_\alpha $, while the two-forms $B_2$ and $C_2$ in the odd basis of $H^{(1,1)}(CY_3,{\bf Z})$ as $B_2=b^a\omega_a$ and $C_2=c^a\omega_a$. Further, these are complexified as \begin{eqnarray}
& & -B_2+iJ=t^A\omega_A=-b^a\omega_a+iv^\alpha\omega_\alpha; \, \, \, G^a = c^a - \tau b^a\nonumber\\
& & C_4 = D_2^{\alpha}\wedge \omega_\alpha + V^{\tilde{\alpha}}\wedge\alpha_{\tilde{\alpha}}+ U_{\tilde{\alpha}}\wedge\beta^{\tilde{\alpha}} + {\tilde\rho}_{\alpha} {\tilde{\omega^\alpha}}
\end{eqnarray} where $(\omega_a,\omega_\alpha)$ form canonical bases for ($H^2_-(CY_3,{\bf Z}), H^2_+(CY_3,{\bf Z})$) and $(\alpha_{\tilde{\alpha}},\beta^{\tilde{\alpha}})$ is a real symplectic basis of $H^3_+(CY_3,{\bf Z})$ while ${\tilde{\omega^\alpha}}$ is a basis of $H^4_+(CY_3,{\bf Z})$, and $\tau=C_0+i e^{-\phi}$. Further $b^a, \, c^a \, {\rm and}\,  \rho^\alpha$ are spacetime scalars, $(V^{\tilde{\alpha}}, U_{\tilde{\alpha}})$ are spacetime one forms and $D_2^{\alpha}$ is a spacetime two-form. Further, as field strength of RR four-form is self dual, it projects out half of the fields from $C_4$ and the resulting ${\cal N}=1$ massless bosonic spectrum of type IIB after orientifolding is summarized in {\bf Table 2.1}.

Here it is understood that $a$ indexes the real subspace of real dimensionality $h^{1,1}_-=2$; the  {complexified} K\"{a}hler moduli correspond to $H^{1,1}(CY_3)$ with {complex} dimensionality $h^{1,1}=2$ or equivalently real dimensionality equal to 4. So, even though $G^a=c^a-\tau b^a$ (for real $c^a$ and $b^a$ and complex $\tau$) is complex, the number of $G^a$'s is indexed by $a$ which runs over the real subspace $h^{1,1}_-(CY_3)$ and the divisor-volume moduli are complexified by RR four-form axions. Further, it is important to note here that we have $h^{1,1}_-(CY_3)=2$\footnote{For explicit construction of a basis: $Dim_{\bf {R}}H^{1,1}_-(CY_3, {\bf Z})=2$, see Appendix {\bf A.1}.}.

\begin{table}[htbp]
\centering
\begin{tabular}{|l|l|l|}
\hline
 &  & \\
  & $h_-^{(2,1)}$ & $z^{\tilde a}$ \\
Chiral multiplet & $h_+^{(1,1)}$ & $(v^\alpha, \, \, \rho_{\alpha})$ \\
  & $h_-^{(1,1)}$ & $(b^a, \, \, c^a)$\\
   & 1 & $(\phi, C_0)$\\
    &  & \\
    \hline
    &  & \\
Vector multiplet & $h_+^{(2,1)}$ & $V^{\tilde\alpha}$ \\
&  & \\
\hline
 &  & \\
Gravity multiplet & 1 & $g_{\mu\nu}$ \\
   &  & \\
\hline
\end{tabular}
\caption{${\cal N}=1$ massless bosonic spectrum of Type IIB Calabi Yau orientifold}
\end{table}
\section{(Non-) perturbative $\alpha^{\prime}$ and String Loop-Corrections to the K\"{a}hler Potential}
In this section, we provide a brief discussion on the higher derivative corrections to the K\"{a}hler potential coming from perturbative $\alpha^{\prime}$-corrections, and the non-perturbative world sheet instanton corrections. In the context of Type IIB orientifold compactifications, the ${\cal N}=1$ moduli space is locally factorizable into a Special K\"{a}hler manifold and a K\"{a}hler manifold derivable from the parent ${\cal N}=2$ special K\"{a}ler and quaternionic manifolds respectively \cite{Grimm,Jockers_thesis}:
$${\cal M}_{{\cal N}=1}={\cal M}^{{\cal N}=1}_{sk}\left(\subset{\cal M}^{{\cal N}=2}_{sk}\right)\times{\cal M}^{{\cal N}=1}_k \left(\subset{\cal M}^{{\cal N}=2}_q\right)$$ where:
\begin{equation}
\label{eq:Ksk}
K_{sk}=-ln\Biggl[i\int_{CY_3}\Omega(z^{\tilde{a}})\wedge{\bar \Omega}({\bar z}^{\tilde{a}})\Biggr], \ \ \ \ \ \tilde{a}=1,...,h^{2,1}_-(CY_3).
\end{equation}
Thus the special K\"{a}hler sector induces the K\"{a}hler potential due to complex structure deformations. Further, defining $$\rho\equiv 1 + t^A\omega_A(\in H^2(CY_3)) - F_A\tilde{\omega}^A(\in H^4(CY_3)) + (2F - t^AF_A){\rm vol}(CY_3),$$ where $A\equiv(\alpha=1,...,h^{1,1}_+,a=1,...,h^{1,1}_-)$. Further, with this $\rho$, one can define appropriate ${\cal N}=1$ coordinates $\tau, G^a$ and $T_\alpha$ (in which the metric is manifestly K\"{a}hler) via a new coordinate $\rho_c$ as below.
\begin{eqnarray}
\label{eq:N=1 coordinates}
& & \rho_c\equiv e^{-B_2}\wedge C^{RR}(=C_0+C_2+C_4) + i Re(e^{-\phi} \rho)\equiv \tau + G^a \omega_a + T_\alpha {\tilde\omega}^\alpha\nonumber\\
& & \tau = C_0 + i e^{-\phi}; \, \, \, \, \, \, \, G^a = c^a - \tau b^a \nonumber\\
& & T_\alpha = \frac{1}{2} i e^{-\phi} \kappa_{\alpha\beta\gamma} v^\beta v^\gamma - {\tilde\rho}_\alpha +\frac{1}{2} \kappa_{\alpha a b} c^a c^b -\frac{1}{2(\tau-\bar\tau)} \kappa_{\alpha a b } G^a (G^b - \bar{G}^b)\nonumber\\
\end{eqnarray}
$\tilde{\rho}_\alpha$ being defined via $C_4$(the RR four-form potential)$=\tilde{\rho}_\alpha\tilde{\omega}_\alpha, \tilde{\omega}_\alpha\in H^4_+(CY_3,{\bf Z})$ and the complexified divisor volumes are defined as: $\rho_s=\tilde{\rho}_s-i\tau_s$ and $\rho_b=\tilde{\rho}_b-i\tau_b$. Now, the K\"{a}hler potential for the quaternion sector is given as,
\begin{equation}
\label{eq:Kq}
K_q = -2 ln\Bigl[\int_{CY_3}e^{-2\phi}\langle\rho,\rho\rangle_{\rm Mukai}\Bigr]=-2ln\Biggl[ie^{-2\phi}\biggl(2(F-{\bar F})-(F_\alpha+{\bar F}_{\alpha})(t^\alpha-{\bar t}^{\alpha})\biggr)\Biggr]\nonumber\\
\end{equation}
The prepotential up to tree level contributions is,
\begin{equation}
\label{eq:Ftree}
F=-\frac{1}{3!} \, \, \kappa_{ABC} \, \, t^At^Bt^C
\end{equation}
where $t^A$'s are sizes of the two-cycle and only $\kappa_{\alpha\beta\gamma}, \kappa_{\alpha bc}$ intersection numbers are non-zero. In the up coming subsections, we discuss (non-)perturbative $\alpha^{\prime}$-corrections to the prepotential (\ref{eq:Ftree}) and hence to the K\"{a}hler potential.

\subsection{Inclusion of Perturbative $\alpha^\prime$-Corrections to the K\"{a}hler Potential}
Let us provide a brief summary of the inclusion of perturbative $\alpha^\prime$-corrections to the K\"{a}hler potential in type IIB string theory compactified on Calabi-Yau three-folds with NS-NS and RR fluxes turned on, as discussed in \cite{BBHL}. As the most dominant perturbative $\alpha^{\prime}$ contribution in Type IIB appears at $(\alpha^\prime)^3-$ level and the same contributing to the K\"{a}hler moduli space metric are contained in
\begin{equation}
\label{eq:nonpert1}
\int d^{10}x\sqrt{g}e^{-2\phi}\left(R + (\partial\phi)^2 + (\alpha^\prime)^3\frac{\zeta(3) J_0}{3.2^{11}}
+ (\alpha^\prime)^3(\bigtriangledown^2\phi) Q\right),
\end{equation}
where
\begin{eqnarray}
\label{eq:nonpert2}
& &  J_0\equiv t^{M_1N_1...M_4N_4}t_{M_1^\prime N_1^\prime....M_4^\prime N_4^\prime}
R^{M_1N_1}_{\ \ \ \ \ M_1^\prime N_1^\prime}...R^{M_4N_4}_{\ \ \ \ \ M_4^\prime N_4^\prime}\nonumber\\
& & + \frac{1}{4}\epsilon^{ABM_1N_1...M_4N_4}\epsilon_{ABM_1^\prime N_1^\prime...M_4^\prime N_4^\prime}
R^{M_1^\prime N_1^\prime}_{\ \ \ \ \ M_1 N_1}...R^{M_4^\prime N_4^\prime}_{\ \ \ \ \ M_4 N_4},
\end{eqnarray}
the second term in (\ref{eq:nonpert2}) being the ten-dimensional generalization of the eight-dimensional
Euler density, and
\begin{eqnarray}
\label{eq:nonpert3}
& &  t^{IJKLMNPQ}\equiv-\frac{1}{2}\epsilon^{IJKLMNPQ} -\frac{1}{2}\biggl[(\delta^{IK}\delta^{JL}
-\delta^{IL}\delta^{JK})(\delta^{MP}\delta^{NQ}-\delta^{MQ}\delta^{NP})\nonumber\\
& & +(\delta^{KM}\delta^{LN}-\delta^{KN}\delta^{LM})(\delta^{PI}\delta^{QJ}-\delta^{PJ}\delta^{QI})
+(\delta^{IM}\delta^{JN}-\delta^{IN}\delta^{JM})(\delta^{KP}\delta^{LQ}-\delta^{KQ}\delta^{LP})
\biggr]\nonumber\\
& & +\frac{1}{2}\left[\delta^{JK}\delta^{KM}\delta^{NP}\delta^{QI}+\delta^{JM}\delta^{NK}\delta^{LP}\delta^{QI}
+\delta^{JM}\delta^{NP}\delta^{QK}\delta^{KI}\right]\nonumber\\
& & {\rm +\ 45\ terms\ obtained\ by\ antisymmetrization\ w.r.t.}\ (ij),(kl),(mn),(pq),
\end{eqnarray}
and
\begin{equation}
\label{eq:nonpert31}
Q\equiv\frac{1}{12(2\pi)^3}\biggl(R_{IJ}^{\ \ }R_{KL}^{\ \ MN}R_{MN}^{\ \ IJ}
-2 R_{I\ K}^{\ K\ L}R_{K\ L}^{\ M\ N}R_{M\ N}^{\ I\ J}\biggr).
\end{equation}
The perturbative world-sheet corrections to the hypermultiplet moduli space of Calabi-Yau three-fold compactifications of
type II theories are captured by the prepotential:
\begin{equation}
\label{eq:Falpha}
F=-\frac{1}{3!} \, \, \kappa_{ABC} \, \, t^At^Bt^C - \frac{i}{2} \, \, \zeta(3) \, \, \chi(CY_3)
\end{equation}
Substituting (\ref{eq:Falpha}) in (\ref{eq:Kq}) we have the following expression of the K\"{a}hler potential, which includes perturbative ${\alpha^\prime}^3$-correction, 

\begin{equation}
\label{eq:Kalpha}
K = K_{sk} - ln[-i(\tau-{\bar\tau})] - 2 ln \Biggl[{\cal V}+\left(\frac{(\tau-{\bar\tau})}{(2i)}\right)^{\frac{3}{2}}\biggl(2\zeta(3)\chi(CY_3))\Biggr]
\end{equation}
substituting which into the ${\cal N}=1$ potential
$V = e^K\left(g^{i{\bar j}}D_iW{\bar D}_{\bar j}{\bar W} - 3 |W|^2\right)$ (one sums over all the moduli),
one gets:
\begin{eqnarray}
\label{eq:nonpert7}
& & V = e^K\Biggl[(G^{-1})^{\alpha{\bar\beta}}D_\alpha W D_{\bar\beta}{\bar W} + (G^{-1})^{\tau{\bar\tau}}D_\tau W
D_{\bar\tau}{\bar W}
- \frac{9\hat{\xi}\hat{{\cal V}}e^{-\phi_0}}{(\hat{\xi}-\hat{{\cal V}})(\hat{\xi} + 2\hat{{\cal V}})}(W{\bar D}_{\bar\tau}{\bar W}\nonumber\\
& & + {\bar W}D_\tau W) -3\hat{\xi}\frac{((\hat{\xi}) ^2 + 7\hat{\xi}\hat{{\cal V}} + (\hat{{\cal V}})^2)}{(\hat{\xi}-\hat{{\cal V}})
(\hat{\xi} + 2\hat{{\cal V}})^2}|W|^2\Biggr],
\end{eqnarray}
the hats being indicative of the Einstein frame - in our subsequent discussion, we will drop the hats for
notational convenience. The structure of the $\alpha^\prime$-corrected potential shows that the no-scale structure is no longer preserved due to explicit dependence of $V$ on $\hat{{\cal V}}$ and the $|W|^2$ term is not canceled. Also it is interesting to observe from ${\cal N}=1$ scalar potential above, that these effects are volume-suppressed implying that models in large volume scenarios have naturally better control against perturbative $\alpha^\prime$-corrections.

\subsection{Inclusion of Non-Perturbative $\alpha^\prime$-Corrections to the K\"{a}hler Potential}
In the context of Type IIB Calabi Yau orientifold compactifications, the non-perturbative $\alpha^\prime$-corrections come from world sheet instantons and the same have been assumed to be volume suppressed in LVS studies done so far. However it has been shown in \cite{Grimm} that using the holomorphic, isometric involution $\sigma$ of (\ref{eq:field_involutions}), these contributions are not volume-suppressed and inclusion of the same makes our LVS Swiss-Cheese setup quite different. These world sheet instanton contributions inherited from the underlying ${\cal N} = 2$ theory are encoded as a shift $F_{ws}$ in prepotential in the large volume limit as the last term in the expression below \cite{Grimm}.
\begin{equation}
\label{eq:Fws}
F=-\frac{1}{3!} \, \, \kappa_{ABC} \, \, t^At^Bt^C - \frac{i}{2} \, \, \zeta(3) \, \, \chi(CY_3) + i\sum_{\beta\in H_2^-(CY_3,{\bf Z})}n^0_\beta \, \, Li_3(e^{ik_at^a}),
\end{equation}
where the degree of holomorphic curves $\beta\in H^2(CY_3, \bf Z)$ is $k_a=\int_{\beta}\omega_a$. The above expression is one more generalization of tree level prepotential (\ref{eq:Ftree}) after including perturbative ${\alpha^\prime}^3$-corrections in (\ref{eq:Falpha}). In the Einstein frame, the K\"{a}hler potential with inclusion of perturbative ${\alpha^\prime}^3$-correction as  well as non-perturbative (world sheet instanton) correction takes the form as below.
\begin{equation}
\label{eq:Kbefore_modular_completion}
K = K_{sk} - ln[-i(\tau-{\bar\tau})] - 2 ln \Biggl[{\cal V}+\left(\frac{(\tau-{\bar\tau})}{(2i)}\right)^{\frac{3}{2}}\biggl(2\zeta(3)\chi(CY_3) - 4 Im F_{ws}\biggr)\Biggr]
\end{equation}
Using ({\ref{eq:Fws}}), the world sheet instanton contributions involve $e^{i k_A t^A}$, where $t^A$ $(=-B_2 + i J)$ is the complexified two-cycle size and $k_A$'s are degrees of holomorphic curves $\beta \in H^2_-(CY_3, \bf Z)$. The choice of involution $\sigma$ implies a direct sum splitting of $H^{(1,1)}(CY_3, \bf Z)$ in even and odd eigenspaces and hence $A=(\alpha (=1,2,..h_+^{(1,1)}), a (=1,2,..h_-^{(1,1)}))$. With the choice of involution we have been using, in the odd sector we have $v^a=0$ and so even in the large volume limit: $e^{i k_A t^A}= e^{-i k_a b^a}$, which is not volume suppressed. Next, as the world sheet instanton contributions get induced by all holomorphic curves $\beta$'s (by world sheet wrapping the holomorphic curves), one has to sum over all such (involutively appropriate) curves with genus-zero Gopakumar-Vafa invariants appearing as coefficients. It has been shown (in \cite{Klemm_GV}) that these coefficient could be very large (even $\sim {10}^{20}$) with appropriate choice of holomorphic, isometric involution $\sigma$. Subsequently, we find in our studies related to model building in string cosmology as well as in string phenomenology that these corrections are large enough to even compete with tree level contribution due to the very large genus-zero Gopakumar-Vafa invariants $n^0_\beta$ and play extremely crucial roles, e.g. in string cosmology for solving the $\eta$-problem in slow roll inflationary scenarios. We will see these interesting implication as we proceed in this review article.

\subsection{Inclusion of Perturbative String Loop-Corrections to the K\"{a}hler Potential}
In this subsection, we briefly review the (relevant) results of \cite{loops,BHP} in the context of the inclusion of perturbative string loop-corrections to the K\"{a}hler potential. Stabilization of the K\"{a}hler moduli with perturbative string loop-corrections (without incorporating non-perturbative effects in the supurpotential) has been initiated in \cite{BergHackKor1} in the context of ${\bf T}^6/{{\bf Z}\times {\bf Z}}$ orientifold compactifications and then have been studied to the Calabi Yau's case in \cite{loops}. Several subsequent implications of the inclusion of string loop-corrections have been studied in \cite{loops,BHP,BergHackKor1}. The string loop-corrections arise from two sources; via the exchange of Kaluza-Klein (KK) modes between D7-branes (or O7-planes) and D3-branes (or O3-planes, both localized in the internal space), or via the exchange of winding strings between intersecting stacks of D7-branes (or between intersecting D7-branes and O7-planes), at 1-loop level is given as \cite{loops}.
\begin{equation}
\label{eq:Kloops}
K_{1-loop} = \sum_{i=1}^{h^{(1,1)}} \frac{{\cal C}_i^{KK}(U_\alpha,{\bar U}_{\tilde\alpha}) \, \, \left(a_{i_k}t^k\right)}{\frac{(\tau-\bar\tau)}{2i}{\cal V}} + \sum_{i=1}^{h^{(1,1)}} \frac{{\cal C}_i^{W}(U_\alpha,{\bar U}_{\tilde\alpha})  }{\left(a_{i_k}t^k\right){\cal V}}
\end{equation}
where $\left(a_{i_k}t^k\right)$ is a linear combination of two-cycle (complexified) volumes $t^A$ while ${\cal C}_i^{KK}(U_\alpha,{\bar U}_{\tilde\alpha})$ and ${\cal C}_i^{W}(U_\alpha,{\bar U}_{\tilde\alpha})$ are some unknown complex structure dependent functions (arising due to KK and winding modes exchange respectively) which could be assumed to be fixed by flux stabilizations and hence one can pick out the dependence on the K\"{a}hler moduli in terms of Calabi Yau volume scalings and hence utilize large volume scenarios. {\it Here it is important to point out that for our Swiss-Cheese Calabi Yau, the two divisor $\Sigma_B$ and $\Sigma_S$ do not intersect (See \cite{Curio+Spillner}) implying that there is no contribution from winding modes corresponding to strings winding non-contractible 1-cycles in the intersection locus corresponding to stacks of intersecting $D7$-branes wrapped around the two divisors.} Further, for our setup the loop-contributions can arise only from KK modes corresponding to closed string or 1-loop open-string exchange between $D3$- and $D7$-(or $O7$-planes)branes wrapped around the two divisors.

After inclusion of perturbative ${\alpha^\prime}^3$-corrections and world sheet instanton correction along with the string one-loop corrections to the K\"{a}hler potential, one can make an interesting observation that the K\"{a}hler potential without (non-) perturbative $\alpha^\prime$-corrections shows ``no scale property" even after including the string loop effects- ``the extended no scale structure". This also supports the existence of large volume minima because the string loop-corrections are subdominant as compared to the perturbative ${\alpha^\prime}^3$-corrections and hence makes the LVS class of models more robust.

\section{Flux and Non-Perturbative Corrections to the Superpotential}
The superpotential, being a holomorphic function of chiral superfields is not renormalized in perturbation theory. However, it receives crucial non-perturbative corrections either through D-instantons or gaugino condensation. In the context of Type IIB Calabi Yau compactifications in the presence of fluxes, $F_3$ and $H_3$ (the field strengths of RR and NS-NS two-forms $C_2$ and $B_2$ respectively) induces a geometro-fluxed holomorphic contribution to the superpotential given as \cite{GVW,Grana}
\begin{equation}
\label{eq:fluxW}
W_{flux} = \int_{CY_3}G_3\wedge\Omega ; \, \, \, \, G_3 = F_3 + \tau H_3
\end{equation}
Based on the action for the Euclidean $D3$-brane world volume (denoted by $\Sigma_S$) action \\
$iT_{D3}\int_{\Sigma_4} e^{-\phi}\sqrt{g-B_2+F}+T_{D3}\int_{\Sigma_4}e^C\wedge e^{-B_2+F}$, the nonperturbative
superpotential coming from a $D3$-brane wrapping a divisor $\Sigma_S\in H^4(CY_3/\sigma,{\bf Z})$ such that the unit arithmetic genus condition of Witten \cite{Witten} is satisfied, will be proportional to
(See \cite{Grimm})
\begin{equation}
\label{eq:nonpertW1}
W_{n.p.}\sim e^{\frac{1}{2}\int_\Sigma e^{-\phi}(-B_2+iJ)^2-i\int_\Sigma(C_4-C_2\wedge B_2+\frac{1}{2}C_0B_2^2)}
= e^{iT_\alpha\int_\Sigma\tilde{\omega}_\alpha}\equiv e^{in_\Sigma^\alpha T_\alpha},
\end{equation}
where $C_{0,2,4}$ are the RR potentials. The pre-factor multiplying (\ref{eq:nonpertW1}) is assumed to factorize into a function of the ${\cal N}=1$ coordinates $\tau,G^a$ and a function of the other moduli. With the above mentioned brief information, the general form of superpotential can be given as,
\begin{equation}
\label{eq:Wbefore_modular_completion}
W = \int_{CY_3}G_3\wedge\Omega + \sum_{\Sigma} f_{\Sigma}(\tau, G^a, ...) e^{i n^\alpha_\Sigma T_\alpha},\, \, n^\alpha_\Sigma = \int_{\Sigma} \tilde{\omega}^\alpha
\end{equation}

\section{Modular Completions of the K\"{a}hler Potential and Superpotential}
The Type IIB superstring theory has two beautiful symmetries: $SL(2,\bf Z)$ and  an axionic shift symmetry. Assuming that the resulting ${\cal N}=1$ theory after considering the orientifold projection, has some discrete subgroup $\Gamma_S\subset SL(2,{\bf Z})$ of underlying ${\cal N} =2 $ Type IIB superstring theory still surviving. Let us investigate the effects of imposing the above mentioned two symmetries, which can be equivalently translated in the transformations below \cite{Grimm}.\\

{\bf Under $\Gamma_S\subset SL(2,{\bf Z})$ transformations:}
\begin{eqnarray}
\label{eq:sl2z}
& & \tau\rightarrow\frac{a\tau + b}{c\tau + d}; \, \, \{a,b,c,d\}\in {\bf Z}: ad-bc=1\nonumber\\
& & \left(\begin{array}{c} C_2\\B_2\end{array}\right)\rightarrow
\left(\begin{array}{cc} a & b \\ c & d \end{array}\right)
\left(\begin{array}{c} C_2\\B_2\end{array}\right),
\nonumber\\
& & G^a\rightarrow\frac{G^a}{(c\tau + d)},\nonumber\\
& & T^\alpha\rightarrow T_\alpha + \frac{c}{2}\frac{\kappa_{\alpha ab}G^a G^b}{(c\tau + d)};
\end{eqnarray}
{\bf Under axionic shift symmetry, $b^a\rightarrow b^a + 2\pi n^a$:}
\begin{eqnarray}
\label{eq:shiftsymmetry}
& & G^a\rightarrow G^a - 2\pi\tau n^a,\nonumber\\
& & T_\alpha\rightarrow T_\alpha - 2\pi\kappa_{\alpha ab}n^aG^b + 2\pi^2\tau\kappa_{\alpha ab}n^an^b,
\end{eqnarray}
From (\ref{eq:Kbefore_modular_completion}), we observe that, the K\"{a}hler potential induced by complex structure deformations $K_{sk}$ coming from special K\"{a}hler sector, are invariant under $\Gamma_S\subset SL(2,{\bf Z})$ as well as under axionic shift symmetry. Also, in the quaternionic sector, ${\cal V}$ being Einstein frame Calabi Yau volume and $\chi{(CY)}$ being Euler characteristic of the (Swiss-Cheese) Calabi Yau, are geometric quantities and hence are invariant under the aforementioned symmetries. However, $(\tau-\bar\tau)$ transforms non-trivially under $\Gamma_S$ as,
\begin{equation}
(\tau-\bar\tau)\longrightarrow \frac{(\tau-\bar\tau)}{|c \tau + d |^2}
\end{equation}
and \begin{eqnarray}
& & Im F_{ws}(\tau, G) = \frac{1}{2}  \sum_{\beta\in H_2^-(CY_3,{\bf Z})} n^0_\beta \Biggl[L_{i_3}\left(e^{i\frac{k_a (G^a-\bar{G}^a)}{\tau-\bar\tau}}\right)+L_{i_3}\left(e^{-i\frac{k_a (G^a-\bar{G}^a)}{\tau-\bar\tau}}\right)\Biggr]\nonumber\\
& & \hskip 0.95in = \sum_{\beta\in H_2^-(CY_3,{\bf Z})} \sum_{n=1}^{\infty} \frac{n^0_\beta}{n^3} \, \, Cos\left(n\frac{k_a (G^a-\bar{G}^a)}{\tau-\bar\tau}\right)
\end{eqnarray}
which imply that the perturbative corrections to the K\"{a}hler potential are independent of axions making the same invariant under axionic shift symmetry, and the non-perturbative world-sheet instanton corrections show invariance under axionic shift symmetry, as component of NS-NS two form $B_2$ appears in $Cos\left(n\frac{k_a (G^a-\bar{G}^a)}{\tau-\bar\tau}\right)$ through $G^a$'s. Further, perturbative ${\alpha^{\prime}}^3$ correction and non-perturbative worldsheet instanton contributions do not respect $\Gamma_S$ symmetry in a generic sense which might be due to the fact that all relevant corrections (in large volume limit) have not been included and corrections due to $D(-1)$ branes as well as the reduction of $D1$ instantons have not trivial effects even in large volume limits \cite{Llanaetal,Green+Gutperle}. After the inclusion of such contributions, the modular completion of the K\"{a}hler potential (\ref{eq:Kbefore_modular_completion}) has been conjectured (in \cite{Llanaetal,Green+Gutperle,Grimm}) as below.
\begin{equation}
\label{eq:Kafter_modular_completion}
K = K_{sk} - ln[-i(\tau-{\bar\tau})] - 2 ln \Biggl[{\cal V}+\frac{1}{2}\chi(CY_3) f(\tau,\bar\tau) - 4 g\bigl(\tau,\bar\tau,G^a,\bar{G}^a\bigr)\Biggr]
\end{equation}
such that $e^K\longrightarrow |c\tau + d|^2 e^K$, where the candidate modular completion functions are conjectured as below.
\begin{eqnarray}
& & f(\tau,\bar\tau) = \sum_{m,n\in{\bf Z}^2/(0,0)}
\frac{({\bar\tau}-\tau)^{\frac{3}{2}}}{(2i)^{\frac{3}{2}}|m+n\tau|^3}; \, \, \, g\bigl(\tau,\bar\tau,G^a,\bar{G}^a)\bigr) = \sum_{\beta\in H_2^-(CY_3,{\bf Z})} n^0_\beta \nonumber\\
& & \times \sum_{m,n\in{\bf Z}^2/(0,0)}
\frac{({\bar\tau}-\tau)^{\frac{3}{2}}}{(2i)^{\frac{3}{2}}|m+n\tau|^3} \, \, \, Cos\left((n+m\tau)k_a\frac{(G^a-{\bar G}^a)}{\tau - {\bar\tau}}
 - mk_aG^a\right)\nonumber\\
\end{eqnarray}
In the above mentioned equations, the Eisentein Series $f(\tau,\bar\tau)$ reproduces the known results of the inclusion of perturbative ${\alpha^{\prime}}^3$-corrections (of \cite{BBHL}) for $n=0$. Also, a summation over a two-dimensional lattice without origin, is incorporated to make modular completion manifest. Further, in the context of modular completion of non-perturbative world sheet instanton corrections, one can observe that all the $SL(2,\bf Z)$ images of world sheet instantons are summed over and the result of non-perturbative correction in (\ref{eq:Kbefore_modular_completion}) corresponds to a particular case; $m=0$. \cite{Llanaetal}

Finally, before discussing the modular completion of the superpotential, we close the modular completion of K\"{a}hler potential portion with an intuitive modular completion for string loop-corrections in (\ref{eq:Kloops}) based on transformation $(\tau-\bar\tau)\longrightarrow \frac{(\tau-\bar\tau)}{|c\tau + d|^2}$ under $\Gamma_S\subset SL(2,\bf Z)$ and summing over all points of two-dimensional lattice without origin as proposed in the Eisentein series earlier with the following result.
\begin{equation}
\label{eq:Kloopmodular_completed}
K_{1-loop}\sim\frac{C^{KK\ (1)}_s(U_\alpha,{\bar U}_{\bar\alpha})\sqrt{\tau_s}}{{\cal V}\left(\sum_{(m,n)\in{\bf Z}^2/(0,0)}\frac{\frac{(\tau-{\bar\tau})}{2i}}{|m+n\tau|^2}\right)} + \frac{C^{KK\ (1)}_b(U_\alpha,{\bar U}_{\bar\alpha})\sqrt{\tau_b}}{{\cal V}\left(\sum_{(m,n)\in{\bf Z}^2/(0,0)}\frac{\frac{(\tau-{\bar\tau})}{2i}}{|m+n\tau|^2}\right)}
\end{equation}
Now, we discuss the modular completion of the superpotential (\ref{eq:Wbefore_modular_completion}). As $e^K$ is a modular function of weight $+2$ and physical gravitino mass-squared is given by the combination  $e^K|W|^2$, this requires the superpotential, W to be a modular form of weight $-1$ (apart from a phase factor) i.e. $W\longrightarrow (c\tau +d)^{-1} W$. By using the transformations properties of $\tau$ and $B_2,C_2$ under modular subgroup $\Gamma_S$, one can easily see that flux-generated contribution to superpotential is indeed of modular weight $-1$. To estimate the same for non-perturbative superpotential contribution is extremely non-trivial, as modular parameter $\tau$ appears through the holomorphic pre-factor, which involves computation of determinants of instanton fluctuations. However, one can estimate a possible modular invariant candidate for holomorphic pre-factor of $e^{in^\alpha T_\alpha}$ with the following obeservations.
\begin{itemize}
\item {The prefactor should compensate the $\Gamma_S$ transformations of $T_{\alpha}$ appearing through $e^{i n^\alpha T_\alpha}$ which (using (\ref{eq:sl2z})) is, $e^{\frac{i n^\alpha}{2}\frac{c \, \kappa_{sab}G^a G^b}{c\tau +d}}$ }
\item{The prefactor should compensate the axionic shift symmetry transformations of $T_{\alpha}$ appearing through $e^{i n^\alpha T_\alpha}$ which (using (\ref{eq:shiftsymmetry})) is, $e^{i n^\alpha (- 2\pi\kappa_{\alpha ab}n^aG^b + 2\pi^2\tau\kappa_{\alpha ab}n^an^b)}$.}
\item{The prefactor should also have a factor $(c\tau +d)^{-1}$ in order to make full W of modular weight $-1$.}
\end{itemize}
Fortunately, all the above mentioned requirements are satisfied in transformation of Jacobi-forms of index $n$, which are given as a sum of theta-functions and modular forms; $\Theta_n(\tau, G^a)=\sum \frac{\theta_{n^\alpha}(\tau,G)}{f(\tau)}$ and finally, one arrives at the following modular completed form of the superpotential (W).
\begin{equation}
\label{eq:nonpert9}
W = \int_{CY_3}G_3\wedge\Omega + \sum_{n^\alpha}\frac{\theta_{n^\alpha}(\tau,G)}{f(\tau)}e^{in^\alpha T_\alpha},
\end{equation}
where the theta function is given as:
\begin{equation}
\label{eq:nonpert11}
\theta_{n^\alpha}(\tau,G)=\sum_{m_a}e^{\frac{i\tau m^2}{2}}e^{in^\alpha G^am_a}.
\end{equation}
In (\ref{eq:nonpert11}), $m^2=C^{ab}m_am_b, C_{ab}=-\kappa_{\alpha^\prime ab}$, $\alpha=\alpha^\prime$. More details of modular completions can be found in \cite{Grimm,Eichleretal_thetafunction}.

\section{Our LVS Setup: In a Nutshell}

In the mid of 2007 we started a systematic study of issues in string cosmology in the context of type IIB ``Swiss-Cheese" orientifold compactification in the LVS limit. What makes our setup different from the earlier LVS setups studied is the inclusion of non-perturbative world-sheet instanton corrections to the K\"{a}hler potential and the modular completed expressions of K\"{a}hler potential and superpotential.

With the inclusion of perturbative (using \cite{BBHL}) and non-perturbative (using \cite{Grimm}) $\alpha^\prime$-corrections as well as the loop corrections (using \cite{BHP,loops}), the (closed string moduli dependent)  K\"{a}hler potential for the two-parameter ``Swiss-Cheese" Calabi-Yau expressed as a projective variety in ${\bf WCP}^4[1,1,1,6,9]$, can be shown to be given by:
\begin{eqnarray*}
\label{eq:nonpert81}
& & K = - ln\left(-i(\tau-{\bar\tau})\right) -ln\left(-i\int_{CY_3}\Omega\wedge{\bar\Omega}\right)\nonumber\\
 & & - 2\ ln\Biggl[{\cal V} + \frac{\chi(CY_3)}{2}\sum_{m,n\in{\bf Z}^2/(0,0)}
\frac{({\bar\tau}-\tau)^{\frac{3}{2}}}{(2i)^{\frac{3}{2}}|m+n\tau|^3}\nonumber\\
& & - 4\sum_{\beta\in H_2^-(CY_3,{\bf Z})} n^0_\beta\sum_{m,n\in{\bf Z}^2/(0,0)}
\frac{({\bar\tau}-\tau)^{\frac{3}{2}}}{(2i)^{\frac{3}{2}}|m+n\tau|^3}cos\left((n+m\tau)k_a\frac{(G^a-{\bar G}^a)}{\tau - {\bar\tau}}
 - mk_aG^a\right)\Biggr]\nonumber\\
 & & +\frac{C^{KK\ (1)}_s(U_\alpha,{\bar U}_{\bar\alpha})\sqrt{\tau_s}}{{\cal V}\left(\sum_{(m,n)\in{\bf Z}^2/(0,0)}\frac{\frac{(\tau-{\bar\tau})}{2i}}{|m+n\tau|^2}\right)} + \frac{C^{KK\ (1)}_b(U_\alpha,{\bf U}_{\bar\alpha})\sqrt{\tau_b}}{{\cal V}\left(\sum_{(m,n)\in{\bf Z}^2/(0,0)}\frac{\frac{(\tau-{\bar\tau})}{2i}}{|m+n\tau|^2}\right)}.
\end{eqnarray*}
In the aforementioned equation, the first line and $-2\ ln({\cal V})$ are the tree-level contributions. The second (excluding the volume factor in the argument of the logarithm) and third lines are the perturbative and non-perturbative $\alpha^\prime$ corrections; $\{n^0_\beta\}$ are the genus-zero Gopakumar-Vafa invariants that count the number of genus-zero rational curves. The fourth line is the 1-loop contribution; $\tau_s$ is the volume of the ``small" divisor and $\tau_b$ is the volume of the ``big" divisor. One sees from (\ref{eq:nonpert81}) that in the LVS limit, loop corrections are sub-dominant as compared to the perturbative and non-perturbative $\alpha^\prime$ corrections.

Further, modular completed form of the superpotential including the non-perturbative instantons along with flux induced contribution is given as:
\begin{equation}
\label{eq:nonpert9}
W = \int_{CY_3}G_3\wedge\Omega + \sum_{n^\alpha}\frac{\theta_{n^\alpha}(\tau,G)}{f(\tau)}e^{in^\alpha T_\alpha}.
\end{equation}

We will be using our LVS Swiss-Cheese (cosmology) setup build up in this chapter for addressing various interesting issues in string cosmology in the next chapter and for addressing issues in string phenomenology, we will be augmenting this setup with the inclusion  of a single spacetime filling mobile $D3$-brane along with stack(s) of fluxed $D7$-brane(s) wrapping the ``big" divisor of the Swiss-Cheese Calabi Yau in chapter {\bf 4}.


%% file: chap3.tex
\chapter{LVS Swiss-Cheese Cosmology}
\markboth{nothing}{\bf 3. LVS Swiss-Cheese Cosmology}

\hskip1in{\it{`` There is at least one philosophical problem in which all thinking men are interested. It is the problem of cosmology: the problem of understanding the world including ourselves, and our knowledge, as part of the world...."}}

\hskip3.7in -Popper, Sir Karl Raimund\footnote{From ``The Logic of Scientific Discovery (1934)", preface to 1959 edition.}

\section{Introduction}

In String Cosmology, obtaining $dS$ vacua and embedding of inflationary scenarios have been two major issues for a long time. In the context of realizing $dS$ vacua,  the complex structure moduli and the axion-dilaton modulus were stabilized with the inclusion of fluxes \cite{fluxesGiddindsetal,Grana} and the K\"{a}hler moduli could be stabilized only with inclusion of non-perturbative effects. A supersymmetric $AdS$ minimum was obtained in Type IIB orientifold compactification which was uplifted to a non-supersymmetric metastable $dS$ by adding $\overline D3$-brane, in \cite{KKLT}. Subsequently, several other uplifting mechanisms were proposed \cite{otherupliftings}. In a different approach with more than one K\"{a}hler modulus  in the context of the Type IIB orientifold compactification in the large volume scenarios, a non-supersymmetric $AdS$ was realized with the inclusion of perturbative ${\alpha^{\prime}}^3$ correction to the K\"{a}hler potential which was then uplifted to $dS$ vacuum \cite{Balaetal2}. Followed by this, again in the context of Type IIB orientifold compactification in large volume scenarios, we showed in \cite{dSetal} that with the inclusion of (non-)perturbative $\alpha^{\prime}$ corrections to the K\"{a}hler potential and instanton corrections to the superpotential, one can realize {\it non}-supersymmetric metastable $dS$ solution in a more natural way without having to add an uplifting term (via inclusion of $\overline D3$-brane).

Once the de-Sitter solution is realized, the next issue to look at is embedding of inflation in string theory  that has been a field of recent interest because of several attempts to construct inflationary models in the context of string theory to reproduce CMB and WMAP observations \cite{kallosh1,Krause,wmap,KKLMMT}. These Inflationary models are also supposed to be good candidates for ``testing"  string theory \cite{kallosh1,Krause}. Initially, the idea of inflation was introduced to explain some cosmological problems like the horizon problem, homogeneity problem, monopole problem etc.\cite{FirstInflation,cosmoproblem,linde}. Some ``slow roll" conditions were defined (with ``$\epsilon$" and ``$\eta$" parameters) as sufficient conditions for inflation to take place for a given potential. In string theory it was a big puzzle to construct inflationary models due to the problem of stability of compactification of internal manifold, which is required for getting a potential which could drive the inflation and it was possible to {\it rethink} about the same only after the volume modulus (as well as complex structure and axion-dilaton) could be stabilized by introducing non-perturbative effects (resulting in a meta-stable dS also) \cite{KKLT}. Subsequently, several models have been constructed with different approaches such as ``brane inflation" (for example $ D3/\overline{D3}$ branes in a warped geometry, with the brane separation as the inflaton field, D3/D7 brane inflation model \cite{KKLMMT,kesav,Dasguptaetal,braneinflation}) and ``modular inflation" \cite{ book,alphacorrection,kahlerinflation}, but all these models were having the so called $\eta$- problem which was argued to be solved by fine tuning some parameters of these models. The models with multi scalar fields (inflatons) have also been proposed  to solve the $\eta$ problem \cite{Assisted}.

Meanwhile in the context of type IIB string compactifications, the idea of ``racetrack inflation" was proposed by adding an extra exponential term with the same  K\"{a}hler modulus but with a different weight in the expression for the superpotential \cite{pillado2}. This was followed by ``Inflating in a better racetrack"  proposed by Pillado et al \cite{pillado1} considering two K\"{a}hler moduli in superpotential; it was also suggested that inflation may be easier to achieve if one considers more (than one) K\"{a}hler moduli. The potential needs to have a flat direction which provides a direction for the inflaton to inflate. For the multi-K\"{a}hler moduli, the idea of treating the ``smaller" K\"{a}hler modulus as inflaton field was also proposed \cite{kahlerinflation,westphal}. The idea of  ``axionic inflation" in the context of type IIB compactifications shown by Grimm and Kallosh et al \cite{Grimm,AxionInflation,gravwavesKallosh}, seemed to be of great interest for stringy inflationary scenarios \cite{AxionInflation,gravwavesKallosh}.

Although inflationary scenario has been initially introduced to explain the homogeneous and isotropic nature of the universe at large scale structure \cite{FirstInflation,cosmoproblem,linde}, it also gives extremely interesting results while studying inhomogeneities and anisotropries of the universe, which is a consequence of the vacuum fluctuations of the inflaton as well as the metric fluctuations. These fluctuations result in non-linear effects (parametrized by ``$f_{NL},\tau_{NL}$") seeding the non-Gaussianity of the primordial curvature perturbation, which are expected to be observed by PLANCK, with non-linear parameter $f_{NL} \sim {\cal O}(1)$ \cite{Planck}. Along with the non-linear parameter $f_{NL}$, the ``tensor-to-scalar ratio" $r$ is also one of the key inflationary observables, which measures the anisotropy arising from the gravity-wave(tensor) perturbations and the signature of the same is expected to be predicted by the PLANCK if the tensor-to-scalar ratio $r \sim 10^{-2}$ \cite{Planck}. As these parameters give a lot of information about the dynamics inside the universe, the theoretical prediction of large/finite (detectable) values of the non-linear parameters ``$f_{NL},\tau_{NL}$" as well as ``tensor-to-scalar ratio" $r$ has received a lot of attention for recent few years \cite{Hajiancmb_data,Kogo,Alabidi,Byrnes_Wands,Byrnestrispectrum,alishahiha,beyondslowroll,kinney2,finite_r1,
finite_r,slowrollflow,slowflowturner,kinney3,Maldacenadeltan}.

For estimating the non-linear parameter $f_{NL}$, a very general formalism (called the ``${\delta N}$ -formalism") was developed and applied for some models \cite{deltaN}. Initially, the parameter $f_{NL}$ was found to be suppressed (to undetectable value) by the slow roll parameters in case of the single inflaton model. Followed by this, several models with multi-scalar fields have been proposed but again with the result of the non-linear parameter $f_{NL}$ of the order of the slow-roll parameters as long as the slow-roll conditions are satisfied \cite{Yokoyamafnl,Yokoyama,Bartolotheoryobservations,Sasaki,Rigopoulosnlpertb,Rigopouloslargenongaussianity,Rigopoulosbispectra,Seerylidsey,Battefeldmultifield,Choihall,Wandstwofield,natural}. Recently, considering multi-scalar inflaton models, Yokoyama et al have given a general expression for calculating the non-linear parameter $f_{NL}$ (using ${\delta N}$ -formalism) for non-separable potentials\cite{Yokoyamafnl} and found the same to be suppressed again by the slow-roll parameter $\epsilon$ (with an enhancement by exponential of ${{\cal O}(1)}$ quantities). In the work followed by the same as a generalization to beyond slow-roll cases, the authors have proposed a model for getting finite $f_{NL}$  violating the slow roll conditions temporarily \cite{ Yokoyama}. The observable ``tensor-to-scalar ratio" $r$, characterizing the amount of anisotropy arising from scalar-density perturbations (reflected as the CMB quadrupole anisotropy) as well as the gravity-wave perturbations arising through the tensorial metric fluctuations, is crucial for the study of temperature/angular anisotropy from the CMB observations. The ``tensor-to-scalar ratio" $r$ is defined as the ratio of squares of the amplitudes of the tensor to the scalar perturbations defined through their corresponding power spectra. Several efforts have been made for getting large/finite value of `` $r$ " with different models, some resulting in small undetectable values while some predicting finite bounds for the same \cite{Alabidi,kinney2,finite_r1,finite_r,slowrollflow,slowflowturner,kinney3}.

In this chapter, developing on our LVS Swiss-Cheese Type IIB orientifold setup \cite{axionicswisscheese} framed in the previous chapter, we discuss the possibilities of realizing several interesting cosmological issues like:
\begin{itemize}
\item
getting dS minimum without the need of any uplifting mechanism (e.g. adding $\overline{D3}$-branes \`{a} la KKLT \cite{KKLT}),

\item
realizing multi-axionic Large Volume Swiss-Cheese inflationary scenarios with number of e-foldings $N_e\sim{\cal O}(10)$,

\item
realizing Non-trivial finite non-Gaussienities $f_{NL}\sim{\cal O}({10}^{-2}-{10}^{0})$ and Finite (detectable) tensor-to-scalar ratio $r\sim10^{-3}$, with the loss of scale invariance within experimental bound: $|n_R-1|\le{0.05}$ such that the curvature perturbations are ``frozen" at super horizon scales.
\end{itemize}

We also discuss some peculiar interesting observations like,

\begin{itemize}
\item
the inflaton as a dark matter candidate at least in some corner of the moduli space,
\item
the inflaton as a quintessence to explain dark energy, once again, at least in some corner - the same as above - of the moduli space.
\end{itemize}

\vskip-0.5cm
It is exciting that all the aforementioned cosmological issues are realized in a single string theoretic setup. In realizing these, the crucial input from algebraic geometry that we need is the fact that Gopakumar-Vafa invariants of genus-zero rational curves for compact Calabi-Yau three-folds expressed as projective varieties can be very large for appropriate choice of holomorphic, isometric involution \cite{Klemm_GV} required for orientifolding the Swiss-Cheese Calabi Yau. This is utilized when incorporating the non-perturbative $\alpha^\prime$ contribution to the K\"{a}hler potential.

This chapter is organized as follows: We start with addressing the issues of getting a large-volume non-supersymmetric dS vacuum ({\it without having to add any uplifting term} \cite{dSetal}) with the inclusion of non-perturbative $\alpha^\prime$-corrections to the K\"{a}hler potential that survive orientifolding and instanton contributions to the superpotential in section {\bf 2}. In section {\bf 3}, we discuss the possibility of realizing axionic slow-roll inflation with the required number of e-foldings to be 60, with the NS-NS axions providing the flat direction for slow roll inflation to proceed starting from a saddle point and proceeding towards the nearest dS minimum. In section {\bf 4}, using the techniques of \cite{Yokoyamafnl,Yokoyama}, we discuss the possibility of getting non-trivial/finite ${\cal O}({10}^{-2})$ non-Gaussianities in slow-roll and ${\cal O}(1)$ non-Gaussianities in slow-roll violating scenarios. In section {\bf 5}, based on general arguments not specific to our (string-theory) set-up and using the techniques of \cite{beyondslowroll,kinney2}, we show that ensuring ``freezeout" of curvature perturbations at super horizon scales, one can get a tensor-scalar ratio  $r\sim{\cal O}(10^{-3})$ in the context of slow-roll scenarios with loss of scale invariance within the experimental bound $|n_R-1|\le{0.05}$. Finally, in section {\bf 6}, we summarize the chapter by giving some arguments to show the possibility of identifying the inflaton, responsible for slow-roll inflation, to also be a dark matter candidate as well as a quintessence field for axions with sub-Planckian Vevs.

\section{Getting dS Vacuum Without $\overline{D3}$-Branes}

In this section, we discuss our work pertaining to getting a de Sitter minimum without the addition of anti-$D3$ branes in the context of type IIB Swiss-Cheese orientifold compactifications in the large volume limit. As built up in chapter {\bf 2}, the K\"{a}hler potential inclusive of the perturbative (using \cite{BBHL})and non-perturbative (using \cite{Grimm}) $\alpha^\prime$-corrections and one- and two-loop corrections (using \cite{BHP,loops}) can be shown to be given by:
\begin{eqnarray}
\label{eq:nonpert81}
& & K = - ln\left(-i(\tau-{\bar\tau})\right) -ln\left(-i\int_{CY_3}\Omega\wedge{\bar\Omega}\right)\nonumber\\
& & - 2\ ln\Biggl[{\cal V} + \frac{\chi(CY_3)}{2}\sum_{m,n\in{\bf Z}^2/(0,0)}
\frac{(\frac{{\bar\tau}-\tau}{2i})^{\frac{3}{2}}}{|m+n\tau|^3}\nonumber\\
& & - 4\sum_{\beta\in H_2^-(CY_3,{\bf Z})} n^0_\beta\sum_{m,n\in{\bf Z}^2/(0,0)}
\frac{({\bar\tau}-\tau)^{\frac{3}{2}}}{(2i)^{\frac{3}{2}}|m+n\tau|^3}cos\left((n+m\tau)k_a\frac{(G^a-{\bar G}^a)}{\tau - {\bar\tau}}
 - mk_aG^a\right)\Biggr]\nonumber\\
 & & +\frac{C^{KK\ (1)}_s(U_\alpha,{\bar U}_{\bar\alpha})\sqrt{\tau_s}}{{\cal V}\left(\sum_{(m,n)\in{\bf Z}^2/(0,0)}\frac{\frac{(\tau-{\bar\tau})}{2i}}{|m+n\tau|^2}\right)} + \frac{C^{KK\ (1)}_b(U_\alpha,{\bar U}_{\bar\alpha})\sqrt{\tau_b}}{{\cal V}\left(\sum_{(m,n)\in{\bf Z}^2/(0,0)}\frac{\frac{(\tau-{\bar\tau})}{2i}}{|m+n\tau|^2}\right)}.\nonumber\\
\end{eqnarray}
In (\ref{eq:nonpert81}), the first line and $-2\ ln({\cal V})$ are the tree-level contributions, the remaining part (excluding the volume factor in the argument of the logarithm) and second line are the perturbative and non-perturbative $\alpha^\prime$ corrections; $\tau_s$ is the volume of the ``small" divisor and $\tau_b$ is the volume of the ``big" divisor. The loop-contributions arise from KK modes corresponding to closed string or 1-loop open-string exchange between $D3$- and $D7$-(or $O7$-planes)branes wrapped around the ``s" and ``b" divisors - note that the two divisors do not intersect (See \cite{Curio+Spillner}) implying that there is no contribution from winding modes corresponding to strings winding non-contractible 1-cycles in the intersection locus corresponding to stacks of intersecting $D7$-branes wrapped around the ``small" and ``big" divisors. Also $n^0_\beta$'s are the genus-0 Gopakumar-Vafa invariants for the holomorphic curve $\beta$ and $k_a=\int_\beta\omega_a$, and $G^a=c^a-\tau b^a$, the real RR two-form potential $C_2=c_a\omega^a$ and the real NS-NS two-form potential $B_2=b_a\omega^a$. We denote the complexified divisor volumes as: $\rho_s=\tilde{\rho}_s-i\tau_s$ and $\rho_b=\tilde{\rho}_b-i\tau_b$ where $\tilde{\rho}_\alpha$ being defined via $C_4$(the RR four-form potential)$=\tilde{\rho}_\alpha\tilde{\omega}_\alpha, \tilde{\omega}_\alpha\in H^4_+(CY_3,{\bf Z})$.

The non-perturbative instanton-corrected superpotential \cite{Grimm} as described in chapter {\bf 2} is:
\begin{equation}
\label{eq:nonpert9}
W = \int_{CY_3}G_3\wedge\Omega + \sum_{n^\alpha}\frac{\theta_{n^\alpha}(\tau,G)}{f(\tau)}e^{in^\alpha T_\alpha},\nonumber\\
\end{equation}
Now the metric corresponding to the K\"{a}hler potential in (\ref{eq:nonpert81}), will be given as:
\begin{equation}
\label{eq:nonpert13}
{\cal G}_{A{\bar B}}=\left(\begin{array}{cccc}
\partial_{\rho_s}{\bar\partial}_{{\bar\rho_s}}K & \partial_{\rho_s}{\bar\partial}_{{\bar\rho_b}}K
& \partial_{\rho_s}{\bar\partial}_{{\bar G}^1}K
& \partial_{\rho_s}{\bar\partial}_{{\bar G}^2}K\\
\partial_{\rho_b}{\bar\partial}_{{\bar\rho_s}}K & \partial_{\rho_b}{\bar\partial}_{{\bar\rho_b}}K
& \partial_{\rho_b}{\bar\partial}_{{\bar G}^1}K
& \partial_{\rho_b}{\bar\partial}_{{\bar G}^2}K\\
\partial_{G^1}{\bar\partial}_{{\bar\rho}_s}K & \partial_{G^1}{\bar\partial}_{{\bar\rho}_b}K &
\partial_{G^1}{\bar\partial}_{{\bar G}^1}K & \partial_{G^1}{\bar\partial}_{{\bar G}^2}K \\
\partial_{G^2}{\bar\partial}_{{\bar\rho}_s}K & \partial_{G^2}{\bar\partial}_{{\bar\rho}_b}K &
\partial_{G^2}{\bar\partial}_{{\bar G}^2}K & \partial_{G^2}{\bar\partial}_{{\bar G}^2}K
\end{array}\right),
\end{equation}
where $A,B\equiv\rho^{s,b},G^{1,2}$. Now, in the Large Volume Scenario (LVS) limit: ${\cal V}\rightarrow\infty, \tau_s\sim ln {\cal V},\ \tau_b\sim {\cal V}^{\frac{2}{3}}$. In this limit, the inverse metric is given as:{\footnote{The detailed calculation of the form of K\"{a}hler metric and its inverse are given in Appendix {\bf A.2}, where we also argue that the one-loop corrections are sub-dominant w.r.t. the perturbative and non-perturbative $\alpha^\prime$ corrections.}}
\begin{equation}
\label{eq:nonpert16}
{\cal G}^{-1}\sim\left(\begin{array}{cccc}
-{\cal Y}\sqrt{ln {\cal V}} & {\cal V}^{\frac{2}{3}}ln {\cal V}
& \frac{-i{\cal Z}ln {\cal V}}{{\cal X}_2}& 0 \\
{\cal V}^{\frac{2}{3}}ln {\cal V} & {\cal V}^{\frac{4}{3}} &
\frac{i{\cal Z}{\cal V}^{\frac{2}{3}}}{k_1{\cal X}_2}& 0\\
 \frac{i{\cal Z} ln {\cal V}}{{\cal X}_2}
 & \frac{-i{\cal Z}{\cal V}^{\frac{2}{3}}}{k_1{\cal X}_2}
& \frac{1}{(k_1^2-k_2^2){\cal X}_1}& \frac{k_2}{(k_1k_2^2-k_1^3){\cal X}_1} \\
0 & 0 & \frac{k_2}{(k_1k_2^2-k_1^3){\cal X}_1} & \frac{1}{(k_1^2-k_2^2){\cal X}_1}
\end{array}\right),
\end{equation}
where
\begin{eqnarray}
\label{eq:nonpert14}
& & {\cal Z}(\tau)\equiv \sum_c\sum_{m,n}A_{n,m,n_{k^c}}(\tau) sin(nk.b + mk.c),\ A_{n,m,n_{k^c}}(\tau)\equiv \frac{(n+m\tau)}{|n+m\tau|^3}\nonumber\\
&& {\cal Y}\equiv {\cal V}_E + \frac{\chi}{2}\sum_{m,n\in{\bf Z}^2/(0,0)}
\frac{(\tau - {\bar\tau})^{\frac{3}{2}}}{(2i)^{\frac{3}{2}}|m+n\tau|^3} \nonumber\\
& &
- 4\sum_{\beta\in H_2^-(CY_3,{\bf Z})}n^0_\beta\sum_{m,n\in{\bf Z}^2/(0,0)}
\frac{(\tau - {\bar\tau})^{\frac{3}{2}}}{(2i)^{\frac{3}{2}}|m+n\tau|^3}cos\left((n+m\tau)k_a\frac{(G^a-{\bar G}^a)}{\tau - {\bar\tau}}
 - mk_aG^a\right),\nonumber\\
& & {\cal X}_1\equiv\frac{\sum_{\beta\in H_2^-(CY_3,{\bf Z})}n^0_\beta\sum_{m,n\in{\bf Z}^2/(0,0)}e^{-\frac{3\phi_0}{2}}|n+m\tau|^3|
A_{n,m,n_{k^c}}(\tau)|^2cos(nk.b + mk.c)}{{\cal Y}}
 \nonumber\\
& & + \frac{|\sum_{\beta\in H_2^-(CY_3,{\bf Z})}n^0_\beta\sum_{m,n\in{\bf Z}^2/(0,0)}e^{-\frac{3\phi_0}{2}}|n+m\tau|^3A_{n,m,n_{k^c}}(\tau)sin(nk.b + mk.c)|^2}{{\cal Y}^2},
\nonumber\\
& & {\cal X}_2\equiv \sum_{\beta\in H_2^-(CY_3,{\bf Z})}n^0_\beta\sum_{m,n\in{\bf Z}^2/(0,0)}|n+m\tau|^3|
A_{n,m,n_{k^c}}(\tau)|^2cos(nk.b + mk.c).\nonumber\\
\end{eqnarray}
Having extremized the superpotential w.r.t. the complex structure moduli and the axion-dilaton modulus, the ${\cal N}=1$ potential
will be given by:
\begin{eqnarray}
\label{eq:nonpert17}
& & V = e^K\Biggl[\sum_{A,B=\rho_\alpha,G^a}\Biggl\{({\cal G}^{-1})^{A{\bar B}}\partial_A W_{np}{\bar\partial}_{\bar B}{\bar W_{np}}
+ \left(({\cal G}^{-1})^{A{\bar B}}(\partial_A K){\bar\partial}_{\bar B}{\bar W_{np}}W + c.c.\right)\Biggr\} \nonumber\\
& & + \left(\sum_{A,B=\rho_\alpha,G^a}({\cal G}^{-1})^{A{\bar B}}\partial_A K{\bar\partial}_{\bar B}K - 3\right)|W|^2 + \sum_{\alpha,{\bar\beta}\in{\rm c.s.}}({\cal G}^{-1})^{\alpha{\bar\beta}}\partial_{\alpha} K_{c.s.}{\bar\partial}_{\bar\beta}K_{c.s.}|W_{np}|^2
\Biggr],\nonumber\\
\end{eqnarray}
where the total superpotential $W$ is the sum of the complex structure moduli Gukov-Vafa-Witten superpotential $W_{c.s.}$
and the non-perturbative superpotential $W_{np}$ arising because of instantons (obtained by wrapping of
$ED3$-branes around the divisors with complexified volumes $\rho_s$ and $\rho_b$).
Now, using:

\begin{eqnarray}
\label{eq:nonpert18}
& & \partial_{\rho_\alpha}W=\frac{\theta_{n^\alpha}(\tau,G^a)}{f(\tau)}e^{in^\alpha T_\alpha}in^\alpha
\left(-\frac{e^{-\phi}}{2}\right); \ \ \partial_{G^a}W=\sum_{n^\alpha}\frac{e^{i\frac{\tau m^2}{2}}}{f(\tau)}e^{im_aG^an^\alpha}e^{in^\alpha T_\alpha}\nonumber\\
& & \left(im_an^\alpha - in^\alpha\frac{\kappa_{\alpha ab}}{2|\tau-{\bar\tau}|^2}\left[{\bar\tau}(G^b-{\bar G}^b)
+({\bar\tau}G^b-\tau{\bar G}^b) + \frac{(2G^b - {\bar G}^b)}{(\tau-{\bar\tau})}\right]\right).\nonumber\\
\end{eqnarray}
in the large-volume limit, one forms tables {\bf Table A.1-A.3}. One therefore sees from {\bf Table A.1} that the dominant term in $({\cal G}^{-1})^{A{\bar B}}\partial_A W_{np}{\bar\partial}_{\bar B}{\bar W_{np}}$ is:
\begin{equation}
\label{eq:nonpert181}
({\cal G}^{-1})^{\rho_s{\bar\rho}_s}|\partial_{\rho_s}W_{np}|^2\sim\frac{{\cal Y}\sqrt{ln {\cal V}}}{{\cal V}^{2n^se^{-\phi}}}e^{-2\phi}(n^s)^2\left|\frac{\theta_{n^s}(\tau,G)}{f(\tau)}
\right|^2e^{-2n^s Im(T_1)}.
\end{equation}
From table {\bf Table A.2}, we see that the dominant term in $({\cal G}^{-1})^{A{\bar B}}(\partial_A K){\bar\partial}_{\bar B}{\bar W_{np}}W$ is:
\begin{eqnarray}
\label{eq:nonpert19}
& & ({\cal G}^{-1})^{\rho_s{\bar\rho}_s}\partial_{\rho_s}K{\bar\partial}_{\bar\rho_s}{\bar W}_{np}W+c.c.\sim \frac{W_{c.s.} ln {\cal V}}{{\cal V}^{n^s e^{-\phi}}}\left(\frac{\theta_{n^s}({\bar\tau},{\bar G})}{f(\eta({\bar\tau}))}
\right)\nonumber\\
 & & e^{-in^s(-\tilde{\rho_s}+\frac{1}{2}\kappa_{1ab}
\frac{{\bar\tau}G^a-\tau{\bar G}^a}{({\bar\tau}-\tau)}\frac{(G^b-{\bar G}^b)}{({\bar\tau}-\tau)} -
\frac{1}{2}\kappa_{sab}\frac{G^a(G^b-{\bar G}^b)}{(\tau-{\bar\tau})})}+c.c.
\end{eqnarray}
From table {\bf Table A.3}, the dominant and sub-dominant terms in $({\cal G}^{-1})^{A{\bar B}}\partial_A K{\bar\partial}_{\bar B}K|W|^2$ are:
\begin{eqnarray}
& & \hskip-.1cm ({\cal G}^{-1})^{\rho_b{\bar\rho}_b}|\partial_{\rho_b}K|^2|W|^2\sim \left[\frac{{\cal V}}{{\cal V} + \xi}\approx 1-\frac{\xi}{{\cal V}}+{\cal O}\left(\frac{1}{{\cal V}^2}\right)\right]|W|^2; \nonumber\\
& & \hskip-.1cm \biggl[(G^{-1})^{\rho_s{\bar\rho}_s}|\partial_{\rho_s}K|^2 + ({\cal G}^{-1})^{\rho_s{\bar\rho}_b}\partial_{\rho_s}{\bar\partial}_{\bar\rho_b}K + c.c.\biggr]|W|^2 \sim \left[\frac{(ln {\cal V})^{\frac{3}{2}}}{({\cal V}+\xi)} + \frac{(ln {\cal V})^{\frac{3}{2}}}{({\cal V}+\xi)^2}\right]|W|^2\nonumber\\
& & \hskip-.1cm \approx\biggl[\frac{(ln {\cal V})^{\frac{3}{2}}}{{\cal V}}-\xi\frac{(ln {\cal V})^{\frac{3}{2}}}{{\cal V}^2}
+ \frac{(ln {\cal V})^{\frac{3}{2}}}{{\cal V}^2} -\xi\frac{(ln {\cal V})^{\frac{3}{2}}}{{\cal V}^3}+{\cal O}\left(\frac{1}{{\cal V}^3}\right)\biggr]|W|^2\nonumber\\
\end{eqnarray}
respectively and the $\xi$-independent terms together cancel the ``-3" in (\ref{eq:nonpert16}). One notes that there are additional terms of ${\cal O}\left(\frac{1}{{\cal V}}\right)$ that one gets from
$\Bigl[({\cal G}^{-1})^{G^1{\bar G^1}}|\partial_{G^1}K|^2 + ({\cal G}^{-1})^{G^2{\bar G^2}}|\partial_{G^2}K|^2 +({\cal G}^{-1})^{G^1{\bar G^2}}\partial_{G^1}K{\bar\partial}_{\bar G^2}K\Bigr]|W|^2$, which is given by:

\begin{equation}
\label{eq:nonpert20}
\frac{|W|^2}{{\cal V}}\left(\frac{3k_2^2+k_1^2}{k_1^2-k_2^2}\right)
\frac{\left|\sum_c\sum_{n,m\in{\bf Z}^2/(0,0)}e^{-\frac{3\phi}{2}}A_{n,m,n_{k^c}}(\tau) sin(nk.b+mk.c)\right|^2}
{\sum_{c^\prime}\sum_{m^\prime,n^\prime\in{\bf Z}^2/(0,0)} e^{-\frac{3\phi}{2}}|n+m\tau|^3
|A_{n^\prime,m^\prime,n_{k^{c^{\prime}}}}(\tau)|^2 cos(n^\prime k.b+m^\prime k.c)},
\end{equation}
which one sees {\it can be either positive or negative}.To summarize, from (\ref{eq:nonpert18}) - (\ref{eq:nonpert20}), one gets the following potential:
\begin{eqnarray}
\label{eq:nonpert21}
& & V\sim\frac{{\cal Y}\sqrt{ln {\cal V}}}{{\cal V}^{2n^s+2}}e^{-2\phi}(n^s)^2\frac{\left(\sum_{m^a}e^{-\frac{m^2}{2g_s} + \frac{m_ab^a n^s}{g_s} + \frac{n^s\kappa_{sab}b^ab^b}{2g_s}}\right)^2}{\left|f(\tau)\right|^2}
\nonumber\\
& & +\frac{W \, ln {\cal V}}{{\cal V}^{n^s+2}}\left(\frac{\theta_{n^s}({\bar\tau},{\bar G})}{f(\eta({\bar\tau}))}
\right)e^{-in^s(-\tilde{\rho_1}+\frac{1}{2}\kappa_{1ab}
\frac{{\bar\tau}G^
a-\tau{\bar G}^a}{({\bar\tau}-\tau)}\frac{(G^b-{\bar G}^b)}{({\bar\tau}-\tau)} -
\frac{1}{2}\kappa_{1ab}\frac{G^a(G^b-{\bar G}^b)}{(\tau-{\bar\tau})})}+c.c.\nonumber\\
& & +
\frac{|W|^2}{{\cal V}^3}\left(\frac{3k_2^2+k_1^2}{k_1^2-k_2^2}\right)
\frac{\left|\sum_c\sum_{n,m\in{\bf Z}^2/(0,0)}e^{-\frac{3\phi}{2}}A_{n,m,n_{k^c}}(\tau) sin(nk.b+mk.c)\right|^2}
{\sum_{c^\prime}\sum_{m^\prime,n^\prime\in{\bf Z}^2/(0,0)} e^{-\frac{3\phi}{2}}|n+m\tau|^3
|A_{n^\prime,m^\prime,n_{k^{c^{\prime}}}}(\tau)|^2 cos(n^\prime k.b+m^\prime k.c)}\nonumber\\
& & +\frac{\xi|W|^2}{{\cal V}^3}.
\nonumber\\
\end{eqnarray}
On comparing (\ref{eq:nonpert21}) with the analysis of \cite{Balaetal2}, one sees that for generic values of
the moduli $\rho_\alpha, G^a, k^{1,2}$ and ${\cal O}(1)\ W_{c.s.}$, {\it and $n^s=1$}, analogous to \cite{Balaetal2}, the second term
dominates; the third term is a new term. As argued in \cite{denef_LesHouches,DDF}, a complete set of divisors lying within the K\"{a}hler cone, need be considered so that the complexstructure
moduli-dependent superpotential $W_{c.s.}\sim W_{\rm ED3}$ - the $ED3$-instanton superpotential - therefore only O(1) D3-instanton numbers $n^s$ corresponding to wrapping of the ED3-brane around the small
divisor $\Sigma_S$, contribute. We would hence consider either $W_{c.s.}\sim W_{\rm ED3}(ns = 1)$ for $W\sim W_{\rm ED3}(ns = 1)$ or $ W_{c.s.} \sim -W_{\rm ED3}(ns = 1)$ with $W \sim W_{\rm ED3}(ns = 2)$. Unlike usual LVS (for which $W_{c.s.}\sim{\cal O}(1)$) and similar to KKLT scenarios (for which $W_{c.s.}\ll 1$), in either of the cases for us, we have $W_{c.s.}\ll 1$ in large volume limit; we would henceforth assume that the fluxes and complex structure moduli have been so fine tuned/fixed that $W_{c.s}\sim \pm W_{\rm ED3}(n^s=1)$. Further, from studies related to study of axionic slow-roll inflation in Swiss-Cheese models \cite{axionicswisscheese}, it becomes necessary to take $n^s\geq 2$.  We assume that the fundamental-domain-valued $b^a$'s satisfy: $\frac{|b^a|}{\pi}<<1$. This implies that the first term in (\ref{eq:nonpert21}) - $|\partial_{\rho^s}W_{np}|^2$ - a positive definite term and denoted henceforth by $V_I$, is the most dominant. Hence, if a minimum exists, it will be positive. To evaluate the extremum of the potential, one sees that:
\begin{eqnarray*}
\label{eq:extrV-c_b}
& & \partial_{c^a}V_I\sim- 4\frac{\sqrt{ln {\cal V}}}{{\cal V}^{2n^s+2}}\sum_{\beta\in H_2^-(CY_3,{\bf Z})} n^0_\beta\sum_{m,n\in{\bf Z}^2/(0,0)}mk^a \frac{({\bar\tau}-\tau)^{\frac{3}{2}}}{(2i)^{\frac{3}{2}}|m+n\tau|^3} sin(n k.b + mk.c)\nonumber\\
& &\frac{\left(\sum_{m^a}e^{-\frac{m^2}{2g_s} + \frac{m_ab^a n^s}{g_s} + \frac{n^s\kappa_{sab}b^ab^b}{2g_s}}\right)^2}{\left|f(\tau)\right|^2}=0; \ \  \Leftrightarrow nk.b + mk.c = N\pi;
\end{eqnarray*}
\begin{eqnarray}
\label{eq:extrV-c_b}
& & \partial_{b^a}V_I\bigg|_{nk.b + mk.c = N\pi}\sim \sum_{m^a \in{2\pi{\bf Z}}} \Biggl(\frac{{\cal V}\sqrt{ln {\cal V}}}{{\cal V}^{2n^s+1}}\frac{e^{-\frac{m^2}{2g_s} + \frac{m_{a^\prime}b^{a^\prime} n^s}{g_s} + \frac{n^s\kappa_{sa^\prime b^\prime}b^{a^\prime}b^{b^\prime}}{2g_s}}e^{-\frac{m^2}{2g_s} + \frac{m_ab^a n^s}{g_s} + \frac{n^s\kappa_{sab}b^ab^b}{2g_s}}}{\left|f(\tau)\right|^2}\nonumber\\
& & \hskip 1.5in \times \left(\frac{n^s m_a}{g_s} + \frac{n^s\kappa_{sab}b^b}{g_s}\right)\Biggr)=0.\nonumber\\
\end{eqnarray}
Now, given the ${\cal O}(1)$ triple-intersection numbers and super sub-Planckian NS-NS axions, we see that potential $V_I$ gets automatically extremized for $D1$-instanton numbers $m_a>>1$. However, if the NS-NS axions get stabilized as per $\frac{n^s m^a}{g_s} + \frac{n^s \kappa_{sab}b^b}{g_s}=0$, satisfying $\partial_{b^a}V=0$, this would imply that the NS-NS axions get stabilized at a rational multiple of $\pi$ ($b^a=-\frac{2\pi(N\in{\bf Z})}{\kappa_{sab}}$). It turns out that the locus $nk.b + mk.c = N\pi$ for $|b^a|<<\pi$ and $|c^a|<<\pi$ corresponds to a flat saddle point with the NS-NS axions providing a flat direction - See \cite{axionicswisscheese}. Here, the point is that the extremization of the potential w.r.t.$b^a$'s and $c^a$'s in the large volume limit yields a saddle point at $sin(nk.b+mk.c)=0$ and for those degree-$k^a$ holomorphic curves $\beta$ for which $b^a\sim -m^a/{\kappa}$ (assuming that $\frac{n k^a m_a}{\pi\kappa}\in{\bf Z}$). The latter corresponds to the small values of $m^a$ (as $b^a$'s are sub-Planckian). Large values of $m^a$'s (which are also permitted by induced shift symmetry of $m^a$'s due to that of axions in $W_{D1-instanton}$) although don't satisfy $b^a\sim -m^a/{\kappa}$, are damped because of $\exp({-m^2/2g_s})$, especially in the $g_s<<1$ limit, the weak coupling limit in which the LVS scenarios are applicable.

Analogous to \cite{Balaetal2}, for all directions in the moduli space with ${\cal O}(1)$ $W_{c.s.}$ and away from $D_iW_{cs}=D_\tau W=0=\partial_{c^a}V=\partial_{b^a}V=0$, the ${\cal O}(\frac{1}{{\cal V}^2})$ contribution
of $\sum_{\alpha,{\bar\beta}\in{c.s.}}({\cal G}^{-1})^{\alpha{\bar\beta}}D_\alpha W_{cs}{\bar D}_{\bar\beta}{\bar W}_{cs}$  dominates over (\ref{eq:nonpert21}),
ensuring that that there must exist a minimum, and given the positive definiteness of the potential $V_I$, this will be a dS minimum. There has been no need to add any $\overline{D3}$-branes as in KKLT to generate a dS vacuum. Also, interestingly, one can show that the condition $nk.b + mk.c = N \pi$ gurantees that the slow roll parameters ``$\epsilon$" and ``$\eta$" are much smaller than one for slow roll inflation beginning from the saddle point and proceeding along an NS-NS axionic flat direction towards the nearest dS minimum (See \cite{axionicswisscheese}). The arguments related to the life-time of the dS minimum in the literature estimate the lifetime to be
$\sim e^{\frac{2\pi^2}{V_0}}$ where the minimum value of the potential, $V_0\sim \frac{\sqrt{ln {\cal V}}}{{\cal V}^N}$ for $N\geq5$. The lifetime, hence, can be made arbitrarily large as ${\cal V}$ is increased.

\section{Realizing Axionic Slow Roll Inflation}
In this section, we discuss the possibility of getting slow roll inflation along a flat direction provided by the NS-NS axions starting from a saddle point and proceeding to the nearest dS minimum. In what follows, we will assume that the volume moduli for the small and big divisors and the axion-dilaton modulus have been stabilized. All calculations henceforth will be in the axionic sector - $\partial_a$ will imply $\partial_{G^a}$ in the following.

The slow-roll inflation parameters ($\epsilon,\, \eta$) are defined as (in $M_p=1$ units).
\begin{equation}
\epsilon\equiv\frac{{\cal G}^{ij}\partial_iV\partial_jV}{2V^2},\ N^i_{\ j}\equiv\frac{{\cal G}^{ik}\left(\partial_k\partial_jV - \Gamma^l_{jk}\partial_lV\right)}{V},
\end{equation}
and $\eta\equiv$ is the most negative eigenvalue of the above matrix $N^i_{\ j}$. In terms of the real axions,
\begin{equation}
\label{eq:realN}
N=\left(\begin{array}{cccc}
N^{c^1}_{\ c^1} & N^{c^1}_{\ c^2} & N^{c^1}_{\ b^1} & N^{c^1}_{b^2} \\
N^{c^2}_{\ c^1} & N^{c^2}_{\ c^2} & N^{c^2}_{\ b^1} & N^{c^2}_{b^2} \\
N^{b^1}_{\ c^1} & N^{b^1}_{\ c^2} & N^{b^1}_{\ b^1} & N^{b^1}_{b^2} \\
N^{b^2}_{\ c^1} & N^{b^2}_{\ c^2} & N^{b^2}_{\ b^1} & N^{b^2}_{b^2} \\
\end{array}\right).
\end{equation}
In terms of the complex $G^{1,2}$ and ${\bar G}^{{\bar 1},{\bar 2}}$,
\begin{eqnarray}
\label{eq:compN}
& & N^{c^1}_{\ c^1}=\frac{{\bar\tau}}{{\bar\tau}-\tau}N^{G^1}_{\ G^1}
-\frac{\tau}{{\bar\tau}-\tau}N^{{\bar G}^{\bar 1}}_{\ G^1}
+\frac{{\bar\tau}}{{\bar\tau}-\tau}N^{G^1}_{\ {\bar G}^{\bar 1}}
-\frac{\tau}{{\bar\tau}-\tau}N^{{\bar G}^{\bar 1}}_{\ G^1},\nonumber\\
& & N^{c^1}_{\ c^2}=\frac{{\bar\tau}}{{\bar\tau}-\tau}N^{G^1}_{\ G^2}
+\frac{\tau}{{\bar\tau}-\tau}N^{G^1}_{\ {\bar G}^{\bar 2}}
-\frac{\tau}{{\bar\tau}-\tau}N^{{\bar G}^{\bar 1}}_{\ G^2}
-\frac{\tau}{{\bar\tau}-\tau}N^{{\bar G}^{\bar 1}}_{\ {\bar G}^{\bar 2}},\nonumber\\
& & N^{c^1}_{\ b^1}=-\frac{|\tau|^2}{{\bar\tau}-\tau}N^{G^1}_{\ G^1}
+\frac{\tau^2}{{\bar\tau}-\tau}N^{{\bar G}^{\bar 1}}_{\ G^1}
-\frac{{\bar\tau}^2}{{\bar\tau}-\tau}N^{G^1}_{\ {\bar G}^{\bar 1}}
+\frac{|\tau|^2}{{\bar\tau}-\tau}N^{{\bar G}^{\bar 1}}_{\ {\bar G}^1},\nonumber\\
& & N^{c^1}_{\ b^2}=-\frac{|\tau|^2}{{\bar\tau}-\tau}N^{G^1}_{\ G^2}
-\frac{|\tau|^2}{{\bar\tau}-\tau}N^{{\bar G}^{\bar 1}}_{\ G^2}
+\frac{\tau}{{\bar\tau}-\tau}N^{G^1}_{\ {\bar G}^{\bar 2}}
+\frac{|\tau|^2}{{\bar\tau}-\tau}N^{{\bar G}^{\bar 1}}_{\ {\bar G}^{\bar 2}},\ {\rm etc}.
\end{eqnarray}
The first derivative of the potential is given by:
\begin{eqnarray}
\label{eq:dV}
& & \partial_aV\bigg|_{D_{c.s.}W=D_\tau W=0}=(\partial_a K)V+e^K\biggl[{\cal G}^{\rho_s{\bar\rho}_s}((\partial_a \partial_{\rho_s}W_{np} {\bar\partial}_{\bar\rho_s}){\bar W}_{np}+\partial_{\rho_s}W_{np}\partial_a{\bar\partial_{\bar\rho_s}}{\bar W}_{np})\nonumber\\
& & +\partial_a{\cal G}^{\rho_s{\bar\rho}_s}\partial_{\rho_s}W_{np}{\bar\partial}_{\bar\rho_s}{\bar W}_{np}\biggr].
\end{eqnarray}
The most dominant terms in (\ref{eq:dV}) of ${\cal O}(\frac{\sqrt{ln {\cal V}}}{{\cal V}^{2n^s+1}})$ that could potentially violate the requirement ``$\epsilon<<1$" are of the type:
\begin{itemize}
\item
e.g. $e^K(\partial_a{\cal G}^{\rho_s{\bar\rho}_s})(\partial_bW_{np}){\bar\partial}_{\bar c}{\bar W}_{np}$, is proportional to $\partial_a\chi_2$, which at the locus
$sin(nk.b + mk.c)$, vanishes;
\item
e.g. $e^K {\cal G}^{\rho_s{\bar\rho}_s}\partial_a\partial_b W_{np}{\bar\partial}_{\bar c}{\bar W}_{np}$: the contribution to $\epsilon$ will be $\frac{(n^s)^2e^{-\frac{2\alpha}{g_s}}{\cal V}}{\sum_{\beta\in H_2} n^0_\beta}$. Now, it turns out that the genus-0 Gopakumar-Vafa integer invariants $n^0_\beta$'s for compact Calabi-Yau's of a projective variety in weighted complex projective spaces for appropriate degree of the holomorphic curve, can be as large as $10^{20}$ and even higher \cite{Klemm_GV} thereby guaranteeing that the said contribution to $\epsilon$ respects the slow roll inflation requirement.
\end{itemize}
One can hence show from (\ref{eq:dV}) that along $sin(nk.b + mk.c)$, $\epsilon<<1$ is always satisfied. Next, to evaluate $N^a_{\  b}$ and the Hessian, one needs to evaluate the second derivatives of the potential and components of the affine connection. In this regard, one needs to evaluate, e.g.:
\begin{eqnarray}
\label{eq:ddV}
& & {\bar\partial}_{\bar d}\partial_aV=({\bar\partial}_{\bar d}\partial_aK)V+\partial_aK{\bar\partial}_{\bar d}V +e^K\biggl[{\bar\partial}_{\bar d}\partial_a{\cal G}^{\rho_s{\bar\rho_s}}\partial_{\rho_s}W_{np}{\bar\partial}_{\bar\rho_s}{\bar W}_{np}
+ \partial_a{\cal G}^{\rho_s{\bar\rho_s}}{\bar\partial}_{\bar d} \bigl(\partial_{\rho_s}W_{np}\nonumber\\
& & {\bar\partial}_{\bar\rho_s}{\bar W}_{np}\bigr)
+ {\bar\partial}_{\bar d}{\cal G}^{\rho_s{\bar\rho_s}}\partial_a\left(\partial_{\rho_s}W_{np}{\bar\partial}_{\bar\rho_s}{\bar W}_{np}\right) + {\cal G}^{\rho_s{\bar\rho_s}}\partial_a{\bar\partial}_{\bar d}\left(\partial_{\rho_s}W_{np}{\bar\partial}_{\bar\rho_s}{\bar W}_{np}\right)\biggr].
\end{eqnarray}
One can show that at saddle point locus $sin(nk.b + mk.c)$, the most dominant term in (\ref{eq:ddV}) comes from $e^K{\cal G}^{\rho_s{\bar\rho}_s}\partial_b\partial_{\rho_s}W_{np}{\bar\partial}_{\bar c}{\bar\partial_{\bar\rho_s}}{\bar W}_{np}\sim \frac{{\cal V}n^sg_s\kappa}{\sum_{\beta\in H_2} n^0_\beta}$.
Now, the large values of the genus-0 Gopakumar-Vafa invariants again nullifies this contribution to $\eta$.\footnote{These $\epsilon<<1$ and $\eta<<1$ realizations are based on assuming the choice of holomorphic, isometric involution $\sigma$ such that genus-zero Gopakumar-Vafa invariants $n^\beta_0$'s can be very large ($\sim 10^{20}$ \cite{Klemm_GV}). These estimates have been made more explicit in the next section, where we compare how large these $n^\beta_0$'s are needed in terms of Calabi Yau volume ${\cal V}$.}

Now, the affine connection components, in the LVS limit, are given by:
\begin{equation}
\label{eq:affine}
\Gamma^a_{bc}={\cal G}^{a{\bar d}}\partial_b{\cal G}_{c{\bar d}}\sim\left[\left(\frac{{\bar\tau}}{{\bar\tau}-\tau}\right)\partial_{c^a}
+\left(\frac{1}{{\bar\tau}-\tau}\right)\partial_{b^a}\right]{\cal X}_1\equiv {\cal O}({\cal V}^0),
\end{equation}
implying that
\begin{equation}
\label{eq:affineContrN}
N^{\bar a}_{\ b}\ni\frac{{\cal G}^{c{\bar a}}\Gamma^d_{cb}\partial_dV}{V}\sim\frac{{\cal V}
\sum_{m,n\in{\bf Z}^2/(0,0)}
\frac{({\bar\tau}-\tau)^{\frac{3}{2}}}{(2i)^{\frac{3}{2}}|m+n\tau|^3} sin(nk.b + mk.c)\frac{\sqrt{ln {\cal V}}}{{\cal V}^{1 + 2n^s}}}{\sum_{\beta\in H_2^-(CY_3,{\bf Z})} (n^0_\beta)^2\frac{\sqrt{ln{\cal V}}}{{\cal V}^{1 + 2n^s}}}.
\end{equation}

We thus see that in the LVS limit and because of the large genus-0 Gopakumar-Vafa invariants, this contribution is nullified - note that near the locus $sin(nk.b + mk.c)$, the contribution is further damped. Thus the ``$\eta$ problem" of \cite{KKLMMT} is solved.

We will  show that one gets a saddle point at $\{(b^a,c^a)|nk.b + mk.c=N_{(m,n;k^a)}\pi\}$ and the NS-NS axions provide a flat direction. We will work out the slow-roll inflation direction along which inflation proceeds between the saddle point and the minimum. Now,
the Hessian or the mass matrix ${\cal M}$ of fluctuations is defined as:
\begin{equation}
\label{eq:Hessian}
{\cal M}=\left(\begin{array}{cc}
2{\rm Re}\left(\partial_a{\partial_{\bar b}}V + \partial_a\partial_bV\right) & -2{\rm Im}\left(\partial_a{\bar\partial_{\bar b}}V+\partial_a\partial_bV\right)\\
-2{\rm Im}\left(\partial_a{\bar\partial_{\bar b}}V-\partial_a\partial_bV\right) &
2{\rm Re}\left(\partial_a{\partial_{\bar b}}V - \partial_a\partial_bV\right)\\
\end{array}\right).
\end{equation}
An eigenvector of the Hessian is to be understood to denote the following fluctuation direction:
\begin{equation}
\label{eq:evectordirec}
\left(\begin{array}{c}
\delta c^1 - A \delta b^1 \\
\delta c^2 - A \delta b^2 \\
-\frac{1}{g_s}\delta b^1 \\
-\frac{1}{g_s}\delta b^2
\end{array}\right).
\end{equation}
One can show that near $nk.b+mk.c=N\pi$ and $b^a\sim-\frac{m^a}{\kappa}\sim\frac{N\pi}{nk^a}$, assuming that $\frac{nk.m}{\pi\kappa}\in{\bf Z}$:
\begin{eqnarray}
\label{eq:GVvolmore1}
& & \partial_a\partial_bV=\Lambda_1{\bar\tau}^2n^2k_ak_b + \Lambda_1{\bar\tau}nmk_ak_b +\Lambda_2|\kappa_{1ab}|,\nonumber\\
& & \partial_a{\bar\partial_{\bar b}}V=-\Lambda_1|\tau|^2n^2k_ak_b-\Lambda_1{\bar\tau}nmk_ak_b-\Lambda_2|\kappa_{1ab}|,
\end{eqnarray}
where
\begin{eqnarray}
\label{eq:Lambdas}
& & \Lambda_1\equiv\frac{4}{|\tau-{\bar\tau}|^2}\frac{\sqrt{ln{\cal V}}}{\cal V}\sum_{\beta\in H_2^-}\frac{n^0_\beta}{\cal V}\sum_{(m,n)\in{\bf Z}^2/(0,0)}\frac{\left(\frac{\tau - {\bar\tau}}{2i}\right)^{\frac{3}{2}}}{|m+n\tau|^3}\nonumber\\
& & \frac{\left(\sum_{m^a}e^{-\frac{m^2}{2g_s} + \frac{m_ab^a n^1}{g_s} + \frac{n^1\kappa_{1ab}b^ab^b}{2g_s}}\right)^2}{\left|f(\tau)\right|^2},\nonumber\\
& & \Lambda_2\equiv\frac{2}{|\tau-{\bar\tau}|^2}\frac{\sqrt{ln{\cal V}}}{\cal V}\frac{\left(\sum_{m^a}e^{-\frac{m^2}{2g_s} + \frac{m_ab^a n^1}{g_s} + \frac{n^1\kappa_{1ab}b^ab^b}{2g_s}}\right)^2}{\left|f(\tau)\right|^2}\nonumber\\
& & \sum_{m^a,\ {\rm no\ sum\ over\ }a}e^{-\frac{m^2}{2g_s} + \frac{m_ab^a n^1}{g_s} + \frac{n^1\kappa_{1ab}b^ab^b}{2g_s}}.\nonumber\\
\end{eqnarray}
In the limit $A>>1$, one gets the Hessian:
\begin{equation}
\left(\begin{array}{cccc}
-\frac{2}{g_s^2}\Lambda_1n^2k_1^2 & -\frac{2}{g_s^2}n^2k_1k_2 & \frac{2A}{g_s}\Lambda_1n^2k_1^2 &
\frac{2A}{g_s}\Lambda_1n^2k_1k_2\\
-\frac{2}{g_s^2}\Lambda_1n^2k_1k_2 & -\frac{2}{g_s^2}n^2k_2^2 & \frac{2A}{g_s}\Lambda_1n^2k_1k_2 &
\frac{2A}{g_s}\Lambda_1n^2k_2^2\\
\frac{2A}{g_s^2}\Lambda_1n^2k_1^2 & \frac{2A}{g_s^2}n^2k_1k_2 & 2A^2\Lambda_1n^2k_1^2 - |{\cal X}|&
2A^2\Lambda_1n^2k_1k_2\\
\frac{2A}{g_s^2}\Lambda_1n^2k_1k_2 & \frac{2A}{g_s^2}n^2k_2^2 & 2A^2\Lambda_1n^2k_1k_2 &
2A^2\Lambda_1n^2k_2^2 - |{\cal X}|
\end{array}\right),
\end{equation}
where ${\cal X}\equiv 2\Lambda_2|\kappa_{1ab}|$.
The eigenvalues are given by:
$$\{0,-|{\cal X}|,\frac{2 A^2 {k_1}^2 {\Lambda_1} n^2 g_s^3+2 A^2 {k_2}^2 {\Lambda_1} n^2 g_s^3-|{\cal X}| g_s^3-2 {k_1}^2 {\Lambda_1} n^2 g_s-2 {k_2}^2 {\Lambda_1} n^2 g+\sqrt{\cal Z}}{2 g_s^3},$$\\
$$   -\frac{-2 A^2 {k_1}^2 {\Lambda_1}
   n^2 g_s^3-2 A^2 {k_2}^2 {\Lambda_1} n^2 g_s^3+|{\cal X}| g_s^3+2 {k_1}^2 {\Lambda_1} n^2 g_s+2 {k_2}^2 {\Lambda_1} n^2 g_s+\sqrt{\cal Z}}{2 g_s^3}\},$$where
   $${\cal Z}\equiv g_s^2
   \biggl(8 g_s \left({k_1}^2+{k_2}^2\right) {\Lambda_1} \left(2 A^2 (g_s+1) \left({k_1}^2+{k_2}^2\right) {\Lambda_1} n^2-g_s |{\cal X}|\right) n^2$$
   $$+\left(-|{\cal X}| g_s^2+2
   \left(A^2 g_s^2-1\right) {k_1}^2 {\Lambda_1} n^2+2 \left(A^2 g_s^2-1\right) {k_2}^2 {\Lambda_1} n^2\right)^2\biggr).$$
The four eigenvectors are given by:
\begin{eqnarray*}
\label{eq:evectors}
&& \left(
\begin{array}{c}
 -\frac{{k_2}}{{k_1}} \\ 1 \\ 0 \\ 0 \end{array}\right) ; \ \ \left(\begin{array}{c} 0 \\ 0 \\ -\frac{{k_2}}{{k_1}} \\ 1\end{array}\right); \nonumber\\
 \end{eqnarray*}
 \begin{eqnarray}
\label{eq:evectors}
& & \left(\begin{array}{c}
 -\frac{{k_1} \left(2 A^2 {k_1}^2 {\Lambda_1} n^2 g_s^3+2 A^2 {k_2}^2 {\Lambda_1} n^2 g_s^3-|{\cal X}| g_s^3+4 A^2 {k_1}^2 {\Lambda_1} n^2 g_s^2+4 A^2 {k_2}^2 {\Lambda_1}
   n^2 g_s^2+2 {k_1}^2 {\Lambda_1} n^2 g_s+2 {k_2}^2 {\Lambda_1} n^2 g_s+\sqrt{\cal Z}\right)}{A g_s {k_2} \left(-2 A^2 {k_1}^2 {\Lambda_1} n^2 g_s^3-2 A^2 {k_2}^2 {\Lambda_1} n^2 g_s^3+|{\cal X}| g_s^3+2
   {k_1}^2 {\Lambda_1} n^2 g_s+2 {k_2}^2 {\Lambda_1} n^2 g_s+\sqrt{\cal Z}\right)} \\
   -\frac{2 A^2 {k_1}^2 {\Lambda_1} n^2 g_s^3+2 A^2 {k_2}^2 {\Lambda_1} n^2 g_s^3-|{\cal X}| g_s^3+4 A^2 {k_1}^2
   {\Lambda_1} n^2 g_s^2+4 A^2 {k_2}^2 {\Lambda_1} n^2 g_s^2+2 {k_1}^2 {\Lambda_1} n^2 g_s+2 {k_2}^2 {\Lambda_1} n^2 g_s+\sqrt{\cal Z}}{A g_s \left(-2 A^2 {k_1}^2 {\Lambda_1} n^2 g_s^3-2 A^2
   {k_2}^2 {\Lambda_1} n^2 g_s^3+|{\cal X}| g_s^3+2 {k_1}^2 {\Lambda_1} n^2 g_s+2 {k_2}^2 {\Lambda_1} n^2 g_s+\sqrt{\cal Z}\right)}
   \\ \frac{{k_1}}{{k_2}} \\ 1   \end{array}\right); \nonumber\\
& & \left(\begin{array}{c}
\frac{{k_1} \left(2 A^2 {k_1}^2 {\Lambda_1} n^2 g_s^3+2 A^2 {k_2}^2 {\Lambda_1} n^2 g_s^3-|{\cal X}| g_s^3+4 A^2 {k_1}^2 {\Lambda_1} n^2 g_s^2+4 A^2 {k_2}^2 {\Lambda_1}
n^2 g_s^2+2 {k_1}^2 {\Lambda_1} n^2 g_s+2 {k_2}^2 {\Lambda_1} n^2 g_s-\sqrt{\cal Z}\right)}{A g_s {k_2} \left(2 A^2 {k_1}^2 {\Lambda_1} n^2 g_s^3+2 A^2 {k_2}^2 {\Lambda_1} n^2 g_s^3-|{\cal X}| g_s^3-2 {k_1}^2
{\Lambda_1} n^2 g_s-2 {k_2}^2 {\Lambda_1} n^2 g_s+\sqrt{\cal Z}\right)} \\
   \frac{2 A^2 {k_1}^2 {\Lambda_1} n^2 g_s^3+2 A^2 {k_2}^2 {\Lambda_1} n^2 g_s^3-|{\cal X}| g_s^3+4 A^2 {k_1}^2
   {\Lambda_1} n^2 g_s^2+4 A^2 {k_2}^2 {\Lambda_1} n^2 g_s^2+2 {k_1}^2 {\Lambda_1} n^2 g_s+2 {k_2}^2 {\Lambda_1} n^2 g_s-\sqrt{\cal Z}}{A g_s \left(2 A^2 {k_1}^2 {\Lambda_1} n^2 g_s^3+2 A^2
   {k_2}^2 {\Lambda_1} n^2 g_s^3-|{\cal X}| g_s^3-2 {k_1}^2 {\Lambda_1} n^2 g_s-2 {k_2}^2 {\Lambda_1} n^2
   g_s+\sqrt{\cal Z}\right)} \\  \frac{{k_1}}{{k_2}} \\     1 \end{array}\right)\nonumber\\
\end{eqnarray}
From the second eigenvector in (\ref{eq:evectors}), one sees that the NS-NS axions provide a flat direction. From the set of eigenvalues, one sees that for $g_s<<1$, the fourth eigenvalue is negative and hence the corresponding fourth eigenvector in (\ref{eq:evectors}) provides the unstable direction. One sees that for $g_s<<1$, the eigenvectors are insensitive to $|{\cal X}|$. Further, in the fourth eigenvector in (\ref{eq:evectors}), the top two components are $\sim{\cal O}(g_s)$ and hence negligible as compared to the third and fourth components in the same eigenvector - this justifies taking a linear combination of the NS-NS axions as flat unstable directions for the slow-roll inflation to proceed. The kinetic energy terms for the NS-NS and RR axions can be written as:
\begin{equation}
\label{eq:kinax1}
\left(\begin{array}{cccc}
\partial_\mu c^1 & \partial_\mu c^2 & \partial_\mu b^1 & \partial_\mu b^2
\end{array}\right) {\cal K} \left(\begin{array}{c} \partial^\mu c^1 \\ \partial^\mu c^2 \\ \partial^\mu b^1 \\ \partial^\mu b^2 \end{array}\right),
\end{equation}
where

\begin{equation}
\label{eq:kinax2}
{\cal K}\equiv{\cal X}_1\left(\begin{array}{cccc}
k_1^2 & k_1k_2 & -(\tau+{\bar\tau})k_1^2 & -(\tau+{\bar\tau})k_1k_2 \\
k_1k_2 & k_2^2 & -(\tau+{\bar\tau})k_1k_2 & -(\tau+{\bar\tau})k_1k_2 \\
-(\tau+{\bar\tau})k_1^2 & -(\tau+{\bar\tau})k_1k_2 & |\tau|^2k_1^2 & |\tau|^2k_1k_2 \\
-(\tau+{\bar\tau})k_1k_2 & -(\tau+{\bar\tau})k_2^2 & k_1k_2|\tau|^2 & k_2^2|\tau|^2
\end{array}\right).
\end{equation}
Writing $\tau=A+\frac{i}{g_s}$, the eigenvalues of ${\cal K}$ are given by:
\begin{equation}
\label{eq:diagevs}
{\cal X}_1\left\{ 0,0,\frac{\left( 1 + \left( 1 + A^2 \right) \,g_s^2 +
       {\sqrt{\cal S}} \right) \,
     \left( {{k_1}}^2 + {{k_2}}^2 \right) }{2\,g_s^2},
  \frac{\left( 1 + \left( 1 + A^2 \right) \,g_s^2 - {\sqrt{\cal S}} \right) \,\left( {{k_1}}^2 + {{k_2}}^2 \right) }{2\,g_s^2}\right\}
\end{equation}
where ${\cal S}\equiv 1 + 2\,\left( -1 + A^2 \right) \,g_s^2 + \left( 1 + 14\,A^2 + A^4 \right) \,g_s^4$.
The basis of axionic fields that would diagonalize the kinetic energy terms is given by:
\begin{equation}
\label{eq:diagbasis}
\left(\begin{array}{c} {\frac{{k_1}\,\left( {b^2}\,{k_1} - {b^1}\,{k_2} \right) \,{\sqrt{1 + \frac{{{k_2}}^2}{{{k_1}}^2}}}}{{{k_1}}^2 + {{k_2}}^2}} \\ {\frac{{k_1}\,\left( {c2}\,{k_1} - {c^1}\,{k_2} \right) \,{\sqrt{1 + \frac{{{k_2}}^2}{{{k_1}}^2}}}}{{{k_1}}^2 + {{k_2}}^2}} \\ {\frac{{k_2}\,\Omega_1\,\left( {b^1}\,\left( 1 + \left( -1 + A^2 \right) \,g_s^2 +          {\sqrt{\cal S}} \right) \,{k_1} +       A^2\,{b^2}\,g_s^2\,{k_2} + {b^2}\,        \left( 1 - g_s^2 + {\sqrt{\cal S}} \right) \,        {k_2} - 4\,A\,g_s^2\,\left( {c^1}\,{k_1} + {c2}\,{k_2} \right)  \right) }{4\,     {\sqrt{2}}\,{\sqrt{\cal S}}\,     \left( {{k_1}}^2 + {{k_2}}^2 \right) }} \\ {\frac{{k_2}\,     \Omega_1\,     \left( {b^1}\,\left( -1 - \left( -1 + A^2 \right) \,g_s^2 +          {\sqrt{\cal S}} \right) \,{k_1} -       A^2\,{b^2}\,g_s^2\,{k_2} + {b^2}\,        \left( -1 + g_s^2 + {\sqrt{\cal S}} \right) \,        {k_2} + 4\,A\,g_s^2\,\left( {c^1}\,{k_1} + {c2}\,{k_2} \right)  \right) }{4\,     {\sqrt{2}}\,{\sqrt{\cal S}}\,     \left( {{k_1}}^2 + {{k_2}}^2 \right) } }\end{array}\right),
\end{equation}
where $\Omega_1\equiv {\sqrt{-\left( \frac{\left( -1 - \left( 1 + 14\,A^2 + A^4 \right) \,g_s^4 +{\sqrt{\cal S}} +\left( -1 + A^2 \right) \,g_s^2\,\left( -2 +{\sqrt{\cal S}} \right)\right) \, \left( {{k_1}}^2 + {{k_2}}^2 \right) }{A^2\,g_s^4\,{{k_2}}^2} \right) }}.$ This tells us that in the $g_s<<1$ limit, there are two NS-NS axionic basis fields in terms of which the axionic kinetic terms are diagonal -
${\cal B}^1\equiv\frac{\left( b^2k_1 - b^1k_2 \right)}{\sqrt{k_1^2 + k_2^2}}$, and ${\cal B}^2\equiv\frac{1}{2g_s\sqrt{2k_2^2(k_1^2+k_2^2)}}(b^1k_1 + b^2k_2)$.
By solving for $b^1$ and $b^2$ in terms of
${\cal B}^1$ and ${\cal B}^2$, and plugging into the mass term, one finds that the mass term for $B^2$ and not $B^1$, becomes proportional to $g_s^2(B^2)^2$ - given that the inflaton must be lighter than its non-inflatonic partner, one concludes that $\frac{1}{2g_s\sqrt{2k_2^2(k_1^2+k_2^2)}}(b^1k_1 + b^2k_2)$ must be identified with the inflaton.


\section{Realizing Non-Trivial Non-Gaussianities: Finite $f_{NL}$}
We now proceed to showing the possibility of getting finite values for the non-linearity parameter $f_{NL}$ in two different contexts. First, we show the same for slow-roll inflationary scenarios. Second, we show the same when the slow-roll conditions are violated.
\subsection{Finite $f_{NL}$ in Slow-Roll Inflationary Scenarios}
In \cite{axionicswisscheese}, we discussed the possibility of getting slow roll inflation along a flat direction provided by the NS-NS axions starting from a saddle point and proceeding to the nearest dS minimum. In what follows, we will assume that the volume moduli for the small and big divisors and the axion-dilaton modulus have been stabilized. All calculations henceforth will be in the axionic sector - $\partial_a$ will imply $\partial_{G^a}$ in the following.
On evaluation of the slow-roll inflation parameters (in $M_p=1$ units), we found that ${\epsilon}\sim{\frac{(n^s)^2}{(k^2 g_s^{\frac{3}{2}}\Delta){\cal V}}}$ and ${\eta}\sim{\frac{1}{k^2 g_s^{\frac{3}{2}} \Delta}}[g_s n^s \kappa_{1ab} +{\frac{(n^s)^2}{\sqrt{\cal V}}} \pm{n^s k^2 g_s^{\frac{3}{2}}\Delta}]$\footnote{The $g_s$ and k-dependence of $\epsilon$ and $\eta$ was missed in \cite{axionicswisscheese}. The point is that the extremization of the potential w.r.t.$b^a$'s and $c^a$'s in the large volume limit yields a saddle point at $sin(nk.b+mk.c)=0$ and at those maximum degree-$k^a$ holomorphic curves $\beta$ for which $b^a\sim -m^a/{\kappa}$ (assuming that $\frac{nk.m}{\pi\kappa}\in{\bf Z}$).} where $\Delta\equiv{\frac{\sum_{\beta\in H_2^-(CY_3,{\bf Z})} n^{0}_{\beta}}{{\cal V}}}$ and we have chosen Calabi-Yau volume ${\cal V}$ to be such that ${\cal V}\sim e^{\frac{4\pi^2}{g_s}}$ (similar to \cite{LargeVcons}). Using Castelnuovo's theory of study of moduli spaces that are fibrations of Jacobian of curves over the moduli space of their deformations, for compact Calabi-Yau's expressed as projective varieties in weighted complex projective spaces (See \cite{Klemm_GV}) one sees that for given degrees of the holomorphic curve and appropriate choice of holomorphic, isometric involution, the genus-0 Gopakumar-Vafa invariants can be very large to compensate the volume factor appearing in the expression for $\eta$. Hence the slow-roll conditions can be satisfed, and in particular, there is no ``$\eta$"-problem. By investigating the eigenvalues of the Hessian, we showed (in \cite{axionicswisscheese}) that one could identify a linear combination of the NS-NS axions (``$k_2b^2+k_1b^1$") with the inflaton and the slow-roll inflation starts from the aforementioned saddle-point and ends when the slow-roll conditions were violated, which most probably corresponded to the nearest dS minimum, one can show that (in $M_p=1$ units)
\begin{equation}
\label{eq:Ne def}
N_e=-\int_{{\rm in:\ Saddle\ Point}}^{{\rm fin:\ dS\ Minimum}}\frac{1}{\sqrt{\epsilon}}d{\cal I}\sim
  \frac{k g_s^{3/4}\sqrt{\sum_{\beta\in H_2}n^0_\beta}}{n^s }.
\end{equation}  We will see that one can get $N_e\sim{60}$ e-foldings in the context of slow roll as well as slow roll violating scenarios.
Now before explaining how to get the non-linear parameter ``$f_{NL}$" relevant to studies of non-Gaussianities,
to be ${\cal O}(10^{-2})$ in our slow-roll LVS Swiss-Cheese orientifold setup, let us summarize  the formalism and results of \cite{Yokoyamafnl} in which the authors analyze the primordial non-Gaussianity in multi-scalar field inflation models (without the explicit form of the potential) using the slow-roll conditions and the $\delta N$ formalism of (\cite{Maldacenadeltan}) as the basic inputs.

Assuming that the time derivative of scalar field $\phi^a(t)$ is not independent of $\phi^a(t)$ (as in the case of standard slow-roll inflation) the background $e$-folding number between an initial hypersurface at $t=t_*$ and a final hypersurface at $t=t_{\rm c}$ (which is defined by
$N \equiv \int H dt$) can be regarded as a function of the homogeneous background field configurations $\phi^a(t_*)$ and $\phi^a(t_{\rm c})$ (on the initial and final hypersurface at $t=t_*$ and $t=t_{\rm c}$ respectively). i.e.
\begin{equation}
N\equiv N(\phi^a(t_{\rm c}),\phi^a(t_*))~.
\end{equation}
By considering $t_{\rm c}$ to be a time when the background trajectories have converged, the curvature perturbation $\zeta$ evaluated at $t=t_{\rm c}$ is given by $\delta N(t_{\rm c},\phi^a(t_*))$ (using the $\delta N$ formalism). After writing the $\delta N(t_{\rm c},\phi^a(t_*))$ upto second order in field perturbations $\delta{\phi^a(t_*)}$ (on the initial flat hypersurface at $t=t_*$) the curvature perturbation $\zeta(t_{\rm c})$ becomes
\begin{equation}
\label{deltaNsecond}
\zeta(t_{\rm c}) \simeq \delta N(t_{\rm c},\phi^a_{*}) = \partial_{a}N^*\delta
\phi^a_{*} + {1 \over 2}\partial_{a}\partial_{b}N^*\delta \phi^a_* \delta \phi^b_*~,
\end{equation}
and using the power spectrum correlator equations and $\zeta ({\bf x}) = \zeta_G({\bf x}) -{3 \over 5}f_{NL}\zeta_G^2({\bf x})$, where $\zeta_G({\bf x})$ represents the Gaussian part, one can arrive at
\begin{equation}
- {6 \over 5}f_{NL} \simeq{\partial^{a}N_*\partial^{b}N_*\partial_{a}\partial_{b}N^*\over
   \left(\partial_{c}N^*\partial^{c}N_*\right)^2}~
\label{fNLdeltaN}
\end{equation}
with the assumption that the field perturbation on the initial flat hypersurface, $\delta \phi^a_*$, is Gaussian.

For the generalization of the above in the context of non-Gaussianties, the authors assumed the so called ``relaxed" slow-roll conditions (RSRC) (which is $\epsilon \ll 1$ and $|\eta_{ab}| \ll 1$) for all the scalar fields, and introduce a time $t_{\rm f}$, at which the RSRC are still satisfied. Then for calculating  $\zeta(t_{\rm c})$, they express $\delta\phi^a(t_{\rm f})$ in terms of $\delta\phi^a_*$, with the scalar field expanded as $\phi^a \equiv {\phi_0}^a + \delta \phi^a$ and then evaluate  $N(t_{\rm c}, \phi^a(t_{\rm f}))$ (the $e$-folding number to reach $\phi^{a(0)}(t_{\rm c})$ starting with $\phi^a=\phi^a(t_{\rm f})$) and with the calculation of $\zeta(t_{\rm c})$ in terms of derivatives of field variations of $N^f$ (making use of the background field equations in variable $N$ instead of time variable) and comparing the same with (\ref{deltaNsecond}) and using (\ref{fNLdeltaN}) one arrives at the following general expression for the non-linear parameter $f_{NL}$:
\begin{eqnarray}
\label{eq:fNL_def}
& & -\frac{6}{5}f_{NL} = \nonumber\\
& & \frac{\partial_a\partial_bN^f\Lambda^a_{a^\prime}{\cal G}^{a^\prime a^{\prime\prime}}\partial_{a^{\prime\prime}}N_* \Lambda^b_{b^\prime}{\cal G}^{b^\prime b^{\prime\prime}}\partial_{b^{\prime\prime}}N_* + \int_{N_*}^{N_f}dN \partial_cNQ^c_{df}\Lambda^d_{d^\prime}{\cal G}^{d^\prime d^{\prime\prime}}\partial_{d^{\prime\prime}}N_*
\Lambda^f_{f^\prime}{\cal G}^{f^\prime f^{\prime\prime}}\partial_{f^{\prime\prime}}N_*}{({\cal G}^{kl}\partial_kN^*\partial_lN_*)^2}\nonumber\\
\end{eqnarray}
with the following two constraints (See \cite{Yokoyamafnl}) required:
\begin{equation}
\label{eq:slowrollconstraint}
\biggl|\frac{{\cal G}^{ab}\left({\partial_b} V\right)_{;a}}{V}\biggr|\ll\sqrt{\frac{{\cal G}^{ab}\partial_{a} V\partial_{b}V}{V^2}}\ {\rm and}\
\bigl|Q^a_{bc}\bigr|\ll\sqrt{\frac{{\cal G}^{ab}\partial_{a} V\partial_{b}V}{V^2}}
\end{equation}
where the semicolon implying a covariant derivative involving the affine connection.
In (\ref{eq:fNL_def}) and (\ref{eq:slowrollconstraint}), ${\cal G}^{ab}$'s are the components of the moduli space metric along the axionic directions given as,

\begin{equation}
\label{eq:inversemetric}
{\cal G}^{ab}\sim {\frac{{\cal Y}}{k^2 g_s^{\frac{3}{2}} \sum_{\beta\in H_2^-(CY_3,{\bf Z})} n^{0}_{\beta}}}\equiv{\frac{1}{k^2 g_s^{\frac{3}{2}} \Delta}}
\end{equation} Further,
\begin{eqnarray}
\label{eq:LambdaPQ_defs}
& & \Lambda^a_b\equiv \left(T e^{\int_{N_*}^{N_f}P(N)dN}\right)^a_b,\ P^a_b\equiv\left[-\frac{\partial_{a^\prime}({\cal G}^{aa^\prime}\partial_bV)}{V} + \frac{{\cal G}^{aa^\prime}\partial_{a^\prime}V\partial_bV}{V^2}\right];\nonumber\\
& & Q^a_{bc}\equiv\left[-\frac{\partial_{a^\prime}({\cal G}^{aa^\prime}\partial_b\partial_cV)}{V}
+\frac{\partial_{a^\prime}({\cal G}^{aa^\prime}\partial_{(b}V)\partial_{c)}V}{V^2}
+\frac{{\cal G}^{aa^\prime}\partial_{a^\prime}V\partial_b\partial_cV}{V^2}
-2\frac{{\cal G}^{aa^\prime}\partial_{a^\prime}V\partial_bV\partial_cV}{V^3}\right]\nonumber\\
\end{eqnarray}
where $V$ is the scalar potential and as the number of e-folding is taken as a measure of the period of inflation (and hence as the time variable), the expression for $\Lambda^I_J$ above, has a time ordering $T$ with the initial and final values of number of e-foldings $N_*$ and $N_f$ respectively. From the definition of $P^I_J$ and $\Lambda^I_J$, one sees that during the slow-roll epoch, $\Lambda^I_J=\delta^I_J$.

After using (\ref{eq:nonpert21}) along with:
\begin{eqnarray}
\label{eq:constraints}
& & \sum_{m_a\in 2{\bf Z}\pi}{e^{-\frac{m^2}{2g_s} + \frac{m_ab^a n^s}{g_s} + \frac{n^s\kappa_{1ab}b^ab^b}{2g_s}}}\sim 1 \nonumber\\
& & \sum_{m_a\in 2{\bf Z}\pi} m_a {e^{-\frac{m^2}{2g_s} + \frac{m_ab^a n^s}{g_s} + \frac{n^s\kappa_{1ab}b^ab^b}{2g_s}}}\sim e^{\frac{-2\pi^2}{g_s}}\sim \frac{1}{\sqrt{\cal V}}, \nonumber\\
& & \end{eqnarray}
for sub-planckian $b^a$'s, one arrives at the following results (along the slow-roll direction $sin(n k_ab^a+m k_ac^a)=0$) :
\begin{eqnarray}
\label{eq:derpotential}
& & \frac{\partial_{a} V}{V}\sim \frac{n^s}{\sqrt{\cal V}};
\frac{\partial_{a}\partial_{b} V}{V}\sim g_s n^s \kappa_{1ab} +\frac{(n^s)^2}{\sqrt{\cal V}} \pm n^s k^2 g_s^{\frac{7}{2}}\Delta,\nonumber\\
& & \frac{\partial_a \partial_{b}\partial_{c}V}{V} \sim {\frac{n^s}{\sqrt{\cal V}}}\Biggl[g_s (n^s) \kappa_{1ab} +(n^s)^2  \pm g_s^{\frac{7}{2}} n^s k^2 \Delta\Biggr]
\end{eqnarray}
Further using the above, one sees that the $\epsilon$ and $\eta$ parameters along with $Q^a_{bc}$ (appearing in the expression of $f_{NL}$) are given as under:
\begin{equation}
\label{eq:epeta}
\epsilon\sim {\frac{(n^s)^2}{{\cal V} g_s^{\frac{3}{2}} k^2 \Delta}}; \eta \sim \frac{1}{g_s^{\frac{3}{2}}k^2 \Delta}\Biggl[ g_s n^s \kappa_{1ab} +\frac{(n^s)^2}{\sqrt{\cal V}} \pm {n^s k^2 g_s^{\frac{7}{2}} \Delta}\Biggr],\nonumber\\
\end{equation}
and
\begin{equation}
\label{eq:Q}
Q^a_{bc}\sim \frac{n^s}{{g_s}^{\frac{3}{2}}k^2\Delta{\sqrt{\cal V}}}\biggl[(n^s)^2 +\frac{(n^s)^2}{\sqrt{\cal V}}-\frac{(n^s)^2}{{\cal V}}\pm {n^s k^2 g_s^{\frac{7}{2}}\Delta}\biggr]
\end{equation}
Now in order to use the expression for $f_{NL}$, the first one of required constraints  (\ref{eq:slowrollconstraint}) results in the following inequality:
\begin{eqnarray}
\label{eq:srconstraint1}
|\delta|\equiv\Biggl|g_s n^s \kappa_{1ab} +\frac{(n^s)^2}{\sqrt{\cal V}} - n^s g_s^{\frac{7}{2}} k^2 \Delta\Biggr|\ll{n^s}{\sqrt\frac{g_s^{\frac{3}{2}} k^2 \Delta}{{\cal V}}}
\end{eqnarray}
Now we solve the above inequality for say $|\delta|\sim{\frac{1}{{\cal V}}}$, which is consistent with the constraint requirement along with the following relation
\begin{equation}
\label{eq:Delta}
\Delta\equiv\frac{\sum_{\beta\in H_2^-(CY_3,{\bf Z})} n^{0}_{\beta}}{\cal Y}\sim {\frac{1}{k^2 g_s^{\frac{7}{2}}}\Biggl[g_s \kappa_{1ab}+\frac{n^s}{\sqrt{\cal V}}}\Biggr]\sim{\frac{1}{k^2 g_s^{\frac{5}{2}}}}
\end{equation}
The second constraint is
\begin{equation}
\label{eq:constraint_2}
\Biggl| (n^s)^2+\frac{(n^s)^2}{\sqrt{\cal V}}+\frac{(n^s)^2}{{\cal V}}- g_s^{\frac{7}{2}} k^2 (n^s)\Delta\Biggr|\ll{\sqrt{k^2 \Delta g_s^{\frac{3}{2}}}}
\end{equation}
Given that we are not bothering about precise numerical factors, we will be happy with ``$<$" instead of a strict ``$\ll$" in (\ref{eq:constraint_2}).
Using (\ref{eq:Delta}) in the previous expressions (\ref{eq:epeta},\ref{eq:Q}) for $\epsilon,\eta$ and $Q^a_{bc}$, we arrive at the final expression for the slow-roll parameters $\epsilon,\eta$ and $Q^a_{bc}$ as following:
\begin{eqnarray}
& & \epsilon \sim {G^{ab}}\frac{(n^s)^2}{{\cal V}}\sim \frac{g_s (n^s)^2}{{\cal V}}; |\eta| \sim G^{ab}|\delta|\sim\frac{g_s }{{\cal V}}\nonumber\\
& & (Q^a_{bc})_{max}\sim G^{ab}{\sqrt{k^2 \Delta g_s^{\frac{3}{2}}}}\biggl({\frac{n^s}{{\sqrt{\cal V}}}}\biggr)\sim{n^s{\sqrt\frac{g_s}{{\cal V}}}}
\end{eqnarray}

As the number of e-foldings satisfies $\partial_{I}N=\frac{V}{\partial_I V}\sim\frac{\sqrt{{\cal V}}}{n^s \sqrt{g_s}}$, which is almost constant and hence $\partial_{I}\partial_{J}N\sim0$. Consequently the first term of (\ref{eq:fNL_def}) is negligible and the maximum contribution to the non-Gaussianities parameter $f_{NL}$ coming from the second term is given by:
\begin{equation}
\label{eq:fNLII}
\frac{\int_{N_*}^{N_f}dN \partial_cNQ^a_{bc}\Lambda^b_{b^\prime}{\cal G}^{b^\prime b^{\prime\prime}}\partial_{b^{\prime\prime}}N
\Lambda^c_{c^\prime}{\cal G}^{c^\prime c^{\prime\prime}}\partial_{c^{\prime\prime}}N}{({\cal G}^{df}\partial_dN\partial_fN)^2}\le(Q^a_{bc})_{max}\sim{n^s{\sqrt\frac{g_s}{{\cal V}}}}.
\end{equation}
This way, for Calabi-Yau volume ${\cal V}\sim{10}^{6}$, D3-instanton number $n^s\sim {\cal O}(1)$ with $n^s\sim g_s\sim k^2$ implying the slow-roll parameters\footnote{These values are allowed for the curvature perturbation freeze-out at the superhorizon scales, which is discussed in the section pertaining to finite tensor-to-scalar ratio.} $\epsilon\sim{0.00028}, |\eta|\sim {10}^{-6}$ with the number of e-foldings $N_e\sim{60}$, one obtains the maximum value of the non-Gaussianties parameter $\left(f_{NL}\right)_{\rm max}\sim {10}^{-2}$. Further if we choose the stabilized Calabi-Yau volume ${\cal V}\sim{10}^{5}$ with $n^s\sim {\cal O}(1)$, we find $\epsilon\sim{0.0034}, |\eta|\sim {10}^{-4}$ with the number of e-foldings $N_e\sim{17}$ and the maximum possible $\left(f_{NL}\right)_{\rm max}\sim 3\times{10}^{-2}$. The above mentioned values of $\epsilon$ and $\eta$ parameters can be easily realized in our setup with the appropriate choice of
holomorphic isometric involution as part of the Swiss-Cheese orientifold. This way we have realized ${\cal O}({10}^{-2})$ non-gaussianities parameter $f_{NL}$ in slow-roll scenarios of our LVS Swiss-Cheese orientifold setup.

\subsection{Finite $f_{NL}$ in Beyond Slow-Roll Inflationary Scenarios}
We will now show that it is possible to obtain ${\cal O}(1)$ $f_{NL}$ while looking for non-Gaussianities in curvature perturbations beyond slow-roll case using the formalism developed in \cite{Yokoyama}. Before that let us summarize the results of \cite{Yokoyama} in which the authors analyze the non-Gaussianity of the primordial curvature perturbation generated on super-horizon scales in multi-scalar field inflation models {\it without imposing the slow-roll conditions} and using the $\delta N$ formalism of (\cite{Maldacenadeltan}) as the basic input.

Consider a model with $n$-component scalar field $\phi^a$.  Now consider the perturbations of the scalar fields in constant $N$ gauge as
\begin{eqnarray}
\delta \phi^{\cal A} (N)  \equiv \phi^{\cal A}( \lambda+\delta\lambda; N)-\phi^{\cal A}(\lambda; N),
\end{eqnarray}
where the short-hand notation of \cite{Yokoyama} is used - $X^{\cal A}\equiv X^a_{i}(i=1,2)=(X^a_1\equiv X^a,X^a_2\equiv\frac{dX^a}{dN})$, and where $\lambda^{\cal A}$'s are the $2n$ integral constants of the background field equations. After using the decomposition of the fields $\phi^{\cal A}$ up to second order in $\delta$ (defined through $\delta \tilde{\phi}^{\cal A}=\delta \tilde{\phi}_{(1)}^{\cal A}+{1 \over 2}\delta \tilde{\phi}_{(2)}^{\cal A}$; to preserve covariance under general coordinate transformation in the moduli space, the authors of \cite{Yokoyama} define: $(\delta \tilde{\phi}_{(1)})^a_1\equiv\frac{d\phi^a}{d\lambda}\delta\lambda$,\
$(\delta \tilde{\phi}_{(2)})^a_1\equiv\frac{D}{d\lambda}\frac{d\phi^a}{d\lambda}(\delta\lambda)^2$ and
$(\delta \tilde{\phi}_{(1)})^a_2\equiv\frac{D\phi^a_2}{d\lambda}\delta\lambda$, $(\delta \tilde{\phi}_{(2)})^a_2\equiv\frac{D^2\phi^a_2}{d\lambda^2}(\delta\lambda)^2$), one can solve the evolution equations for $\delta \tilde{\phi}_{(1)}^{\cal A}$ and $\delta \tilde{\phi}_{(2)}^{\cal A}$. The equation for $\delta \tilde{\phi}_{(2)}^{\cal A}$ is simplified with the choice of integral constants such that $\lambda^{\cal A}=\phi^{\cal A}(N_*)$ implying $\delta \tilde{\phi}^{\cal A}(N_*)=\delta \lambda^{\cal A}$ and hence $\delta\tilde{\phi}_{(2)}^{\cal A}(N)$ vanishing at $N_*$. Assuming $N_*$ to be a certain time soon after the relevant length scale crossed the horizon scale ($H^{-1}$), during the scalar dominant phase and $N_c$ to be a certain time after the complete convergence of the background trajectories has occurred and using the so called $\delta N$ formalism one gets
\begin{equation}
\zeta \simeq\delta N = \tilde{N}_{{\cal A}}\delta \tilde{\phi}^{\cal A} + {1 \over 2}\tilde{N}_{{\cal A}{\cal B}}\delta \tilde{\phi}^{\cal A} \delta \tilde{\phi}^{\cal B} +\cdots
\end{equation}
Now taking $N_f$ to be certain late time during the scalar dominant phase and using the solutions for $\delta \phi_{(1)}^{\cal A}$ and $\delta \phi_{(2)}^{\cal A}$ for the period $N_*<N<N_f$, one obtains the expressions for $\tilde{N}_{{\cal A}*}$ and $\tilde{N}_{{\cal A}{\cal B}*}$ (to be defined below) and finally writing the variance of $\delta \tilde{\phi}^{\cal A}_*$ (defined through
$\langle \delta \tilde{\phi}^{\cal A}_* \delta \tilde{\phi}^{\cal B}_*\rangle
\simeq A^{{\cal A}{\cal B}}{\left( \frac{H_\ast}{2\pi} \right)}^2$ including  corrections to the slow-roll terms in $A^{ab}$ based on \cite{Byrnes_Wands,Byrnestrispectrum}), and using the basic definition of the non- linear parameter $f_{NL}$ as the the magnitude of the bispectrum of the curvature perturbation $\zeta$, one arrives at a general expression for $f_{NL}$ (for beyond slow-roll cases)\cite{Yokoyama}. For our present interest, the expression for $f_{NL}$ for the beyond slow-roll case is given by
\begin{eqnarray}
\label{eq:fNL_defii}
& & -\frac{6}{5}f_{NL} = \nonumber\\
& & \frac{\tilde{N}^f_{{\cal A}{\cal B}}\Lambda^{\cal A}_{{\cal A}^\prime}(N_f,N_*)A^{{\cal A}^\prime {\cal A}^{\prime\prime}}\tilde{N}^*_{{\cal A}^{\prime\prime}} \Lambda^{\cal B}_{{\cal B}^\prime}A^{{\cal B}^\prime {\cal B}^{\prime\prime}}\tilde{N}^*_{{\cal B}^{\prime\prime}} + \int_{N_*}^{N_f}dN \tilde{N}_{\cal C}\tilde{Q}^{\cal C}_{{\cal D}{\cal F}}\Lambda^{\cal D}_{{\cal D}^\prime}A^{{\cal D}^\prime {\cal D}^{\prime\prime}}\tilde{N}^*_{{\cal D}^{\prime\prime}}
\Lambda^{\cal F}_{{\cal F}^\prime}A^{{\cal F}^\prime {\cal F}^{\prime\prime}}\tilde{N}^*_{{\cal F}^{\prime\prime}}}{(A^{{\cal K}{\cal L}}\tilde{N}^*_{\cal K}\tilde{N}^*_{\cal L})^2}\nonumber\\
\end{eqnarray}
where again the index ${\cal A}$ represents a pair of indices $^a_i$, $i=1$ corresponding to the field $b^a$ and $i=2$ corresponding to $\frac{db^a}{dN}$. Further,
\begin{eqnarray}
\label{eq:Adef}
& & A^{ab}_{11}={\cal G}^{ab}+\left(\sum_{m_1,m_2,m_3,m_4,m_5}^{<\infty}\left(||\frac{d\phi^a}{dN}||^2\right)^{m_1}
\left(\frac{1}{H}\frac{dH}{dN}\right)^{m_2}\epsilon^{m_3}\eta^{m_4}\right)^{ab},\nonumber\\
& & A^{ab}_{12}=A^{ab}_{21}=\frac{{\cal G}^{aa^\prime}\partial_{a^\prime}V{\cal G}^{bb^\prime}\partial_{b^\prime}}{V^2}-\frac{\partial_{a^\prime}({\cal G}^{aa^\prime}{\cal G}^{bb^\prime}\partial_{b^\prime}V)}{V},\nonumber\\
& & A^{ab}_{22}=\left(\frac{{\cal G}^{aa^\prime}\partial_{a^\prime}V\partial_cV}{V^2}-\frac{\partial_{a^\prime}({\cal G}^{aa^\prime}\partial_cV)}{V}\right)\left(\frac{{\cal G}^{cc^\prime}\partial_{c^\prime}V{\cal G}^{bb^\prime}\partial_{b^\prime}V}{V^2}-\frac{\partial_{c^\prime}({\cal G}^{cc^\prime}{\cal G}^{bb^\prime}\partial_{b^\prime}V)}{V}\right)\nonumber\\
\end{eqnarray}
where in $A^{ab}_{11}$, based on \cite{Byrnes_Wands,Byrnestrispectrum}, assuming the non-Gaussianity to be expressible as a finite-degree polynomial in higher order slow-roll parameter corrections.
In (\ref{eq:fNL_defii}), one defines:
\begin{equation}
\label{eq:Lambda}
\Lambda^{\cal A}_{\ {\cal B}}=\left(Te^{\int_{N_*}^{N_f}dN \tilde{P}(N)}\right)^{\cal A}_{\cal B};
\end{equation}
$\tilde{N}_{\cal A}, \tilde{N}_{{\cal A}{\cal B}},\tilde{P}^{\cal A}_{\cal B}$ and $\tilde{Q}^{\cal A}_{\ {\cal B}{\cal C}}$\footnote{We have modified the notations of \cite{Yokoyama} for purposes of simplification.} will be defined momentarily. The equations of motion
\begin{eqnarray}
\label{eq:eoms}
& & \frac{d^2b^a}{dN^2} + \Gamma^a_{bc}\frac{db^b}{dN}\frac{db^c}{dN} + \left(3+\frac{1}{H}\frac{dH}{dN}\right)\frac{db^a}{dN}+\frac{{\cal G}^{ab}\partial_bV}{H^2}=0,\nonumber\\
& & H^2=\frac{1}{3}\left(\frac{1}{2}H^2||\frac{db^a}{dN}||^2 + V\right)
\end{eqnarray}
yield
\begin{equation}
\label{eq:conseqEOM}
\frac{1}{H}\frac{dH}{dN}=\frac{6\left(\frac{2V}{H^2}-6\right) - 2\Gamma^a_{bc}\frac{db_a}{dN}\frac{db^b}{dN}\frac{db^c}{dN}}{12}.
\end{equation}
For slow-roll inflation, $H^2\sim\frac{V}{3}$; the Friedmann equation in (\ref{eq:eoms}) implies that
$H^2>\frac{V}{3}$ when slow-roll conditions are violated. The number of e-foldings away from slow-roll
is given by: $N\sim\int\frac{db^a}{||\frac{db^b}{dN}||}$, which using the Friedmann equation implies
\begin{equation}
\label{eq:NebeyondSlowRoll}
N_e^{\rm beyond \, \,slow-roll}\sim\int\frac{db^a}{\sqrt{1-\frac{V}{3H^2}}}.
\end{equation}
Assuming $\frac{V}{3H^2}\sim1-\frac{1}{6\times 3600}$, one still gets the number of e-foldings close to the required 60. We would require $\epsilon<<1$ and $|\eta|~{\cal O}(1)$ to correspond to beyond slow-roll case. Now, as the potential $V=V_0\frac{\sqrt{ln {\cal Y}}}{{\cal Y}^{2n^s+1}}\left(\sum_{m\in2{\bf Z}\pi}e^{-\frac{m^2}{2g_s}+\frac{n^s}{g_s}m.b+\frac{\kappa_{1ab}b^ab^b}{2g_s}}\right)^2$, One gets:
\begin{eqnarray}
\label{eq:slow_roll_viol_I}
& & \partial_{a}V\sim-\frac{g_s\sqrt{ln {\cal Y}}n^s}{{\cal Y}^{2n^s+1}}\sum_{\beta\in H_2^-(CY_3,{\bf Z})}\frac{n^0_\beta}{{\cal Y}}sin(nk.b+mk.c)k^ag_s^{\frac{3}{2}}\left(\sum_{m\in2{\bf Z}\pi}e^{-\frac{m^2}{2g_s}+\frac{n^s}{g_s}m.b+\frac{\kappa_{1ab}b^ab^b}{2g_s}}\right)^2 \nonumber\\
& & +
\frac{g_s\sqrt{ln {\cal Y}}}{{\cal Y}^{2n^s+1}}\left(\sum_{m\in2{\bf Z}\pi}e^{-\frac{m^2}{2g_s}+\frac{n^s}{g_s}m.b+\frac{\kappa_{1ab}b^ab^b}{2g_s}}\right)
\frac{n^s}{g_s}\sum_{m\in2{\bf Z}\pi}(m^a+\kappa_{1ab}b^b)\nonumber\\
& & \times \left(\sum_{m^a\in2{\bf Z}\pi}e^{-\frac{m^2}{2g_s}+\frac{n^s}{g_s}m.b+\frac{\kappa_{1ab}b^ab^b}{2g_s}}\right)\nonumber\\
\end{eqnarray}
which using
\begin{eqnarray}
& & \sum_{m\in2{\bf Z}\pi}e^{-\frac{m^2}{2g_s}+\frac{n^s}{g_s}m.b+\frac{\kappa_{1ab}b^ab^b}{2g_s}}\sim e^{\frac{\kappa_{1ab}b^ab^b}{2g_s}}\nonumber\\
& & \sum_{m\in2{\bf Z}\pi}(m^a+\kappa_{1ab}b^b)\left(\sum_{m^a\in2{\bf Z}\pi}e^{-\frac{m^2}{2g_s}+\frac{n^s}{g_s}m.b+\frac{\kappa_{1ab}b^ab^b}{2g_s}}\right)\sim\kappa_{1ab}b^b
e^{\frac{\kappa_{1ab}b^ab^b}{2g_s}}\nonumber\\
\end{eqnarray}
$\partial_{a}V=0$ implies:
\begin{equation}
\label{eq:slow_roll_viol_II}
\sum_{\beta\in H_2^-(CY_3,{\bf Z})}\frac{n^0_\beta}{{\cal Y}}sin(nk.b+mk.c)k^a\sim\frac{b^a}{g_s^{\frac{5}{2}}}.
\end{equation}
Slow-roll scenarios assumed that the LHS and RHS of (\ref{eq:slow_roll_viol_II}) vanished individually - the same will not be true for slow-roll violating scenarios. Near (\ref{eq:slow_roll_viol_II}), one can argue that:
\begin{eqnarray}
\label{eq:Ginv_Gamma_I}
& & {\cal G}^{ab}\sim\frac{b^ab^b - \sqrt{g_s^7\left(k^ak^b\right)^2\left(\frac{n^0_\beta}{\cal V}\right)^2 - b^2g_s^2k^2}}{b^2\sqrt{g_s^7\left(k^ak^b\right)^2\left(\frac{n^0_\beta}{\cal V}\right)^2 - b^2g_s^2k^2} + g_s^7\left(k^ak^b\right)^2\left(\frac{n^0_\beta}{\cal V}\right)^2 - b^2g_s^2k^2};\nonumber\\
& & \Gamma^{a}_{bc}\sim\frac{\left(\sqrt{g_s^7\left(k^ak^b\right)^2\left(\frac{n^0_\beta}{\cal V}\right)^2 - b^2g_s^2k^2} + g_s^2 + \frac{b^2g_s^2}{\sqrt{g_s^7\left(k^ak^b\right)^2\left(\frac{n^0_\beta}{\cal V}\right)^2 - b^2g_s^2k^2}}\right)}{b^2\sqrt{g_s^7\left(k^ak^b\right)^2\left(\frac{n^0_\beta}{\cal V}\right)^2 - b^2g_s^2k^2} + g_s^7\left(k^ak^b\right)^2\left(\frac{n^0_\beta}{\cal V}\right)^2 - b^2g_s^2k^2}.
\end{eqnarray}
Note, we no longer restrict ourselves to sub-Planckian axions - we only require $|b^a|<\pi$. For $g_s^7\left(k^ak^b\right)^2\left(\frac{n^0_\beta}{\cal V}\right)^2 - b^2g_s^2k^2\sim {\cal O}(1)$, and the holomorphic isometric involution, part of the Swiss-Cheese orientifolding, assumed to be such that the maximum degree of the holomorphic curve being summed over in the non-perturbative $\alpha^\prime$-corrections involving the genus-zero Gopakumar-Vafa invariants are such that $\sum_\beta\frac{n^0_\beta}{\cal V}\leq\frac{1}{60}, k\sim 3$, we see that (\ref{eq:slow_roll_viol_II}) is satisfied and
\begin{eqnarray}
\label{eq:Ginv_Gamma_II}
& & {\cal G}^{ab}\sim\frac{b^2+{\cal O}(1)}{b^2+{\cal O}(1)}\sim{\cal O}(1); \, \, \Gamma^{a}_{bc}\sim\frac{b^a\left(g_s^2+b^2g_s^2\right)}{b^2+{\cal O}(1)}\sim b^ag_s^2.
\end{eqnarray}
Hence, the affine connection components, for $b^a\sim {\cal O}(1)$, is of ${\cal O}(10)$; the curvature components $R^a_{\ bcd}$ will hence also be finite. Assuming $H^2\sim V$, the definitions of $\epsilon$ and $\eta$ continue to remain the same as those for slow-roll scenarios and one hence obtains:
\begin{eqnarray}
\label{eq:epsilon_eta_slow_roll_viol}
& & \epsilon\sim\frac{\left(n^s\right)^2e^{\frac{4\pi n^sb}{g_s}+\frac{b^2}{g_s}}\left(\pi+b\right)^2}{\cal V}\sim10^{-3},\nonumber\\
& & \eta\sim n^s(1+n^sb^2)-\frac{bg_s^2n^se^{\frac{4\pi n^sb}{g_s}+\frac{b^2}{g_s}}\left(\pi+b\right)}{\sqrt{\cal V}}\sim n^s(1+n^sb)\sim {\cal O}(1).
\end{eqnarray}
Finally, $\frac{db^a}{dN}\sim\sqrt{1-\frac{V}{3H^2}}\sim{\cal O}(1)$.

We now write out the various components of $\tilde{P}^{\cal A}_{\cal B}$, relevant to evaluation of
$\Lambda^{\cal A}_{\ {\cal B}}$ in (\ref{eq:Lambda}):
\begin{eqnarray}
\label{eq:P's}
& & \tilde{P}^{a1}_{1b}=0,\  \tilde{P}^{a2}_{1b}=\delta^a_b,\nonumber\\
& & \tilde{P}^{a1}_{2b}=-\frac{V}{H^2}\left(\frac{\partial_a({\cal G}^{ac}\partial_cV)}{V}
- \frac{{\cal G}^{ac}\partial_cV\partial_bV}{V^2}\right) - R^a_{\ bcd}\frac{db^c}{dN}\frac{db^d}{dN}\sim{\cal O}(1),\nonumber\\
& & \tilde{P}^{a2}_{b2}={\cal G}_{bc}\frac{db^a}{dN}\frac{db^c}{dN}+\frac{{\cal G}^{ac}\partial_cV}{V}{\cal G}_{bf}\frac{db^f}{dN}-\frac{V}{H^2}\delta^a_b\sim{\cal O}(1).
\end{eqnarray}
Similarly,
\begin{eqnarray}
\label{eq:N's}
(a) & & N^1_a\sim\frac{1}{||\frac{db^a}{dN}||}\sim{\cal O}(1), \  N^2_a=0;\nonumber\\
(b) & & N^{11}_{ab}=N^{12}_{ab}=N^{22}_{ab}=0,\  N^{21}_{ab}\sim\frac{{\cal G}_{bc}\frac{db^c}{dN}}{||\frac{db^d}{dN}||^3}\sim{\cal O}(1);\nonumber\\
(c) & & \tilde{N}^1_a\equiv N^1_a-N^2_b\Gamma^b_{ca}\frac{db^c}{dN}\sim\frac{1}{||\frac{db^a}{dN}||}\sim{\cal O}(1), \  \tilde{N}^2_a\equiv N^2_a=0;\nonumber\\
(d) & & \tilde{N}^{11}_{ab}\equiv N^{11}_{ab}+N^{22}_{ca}\Gamma^c_{ml}\Gamma^l_{nb}\frac{db^m}{dN}\frac{db^n}{dN}+(N^{12}_{ac} + N^{21}_{ac})\Gamma^c_{lb}\frac{db^l}{dN}\nonumber\\
& & -N^2_c(\bigtriangledown_a\Gamma^c_{lb})\frac{db^l}{dN}-N^2_c\Gamma^c_{al}\Gamma^l_{nb}\frac{db^n}{dN} \sim\frac{{\cal G}_{cd}\frac{db^d}{dN}}{||\frac{db^m}{dN}||^3}\Gamma^c_{lb}\frac{db^l}{dN}
\sim{\cal O}(1),\nonumber\\
& & \tilde{N}^{12}_{ab}\equiv N^{12}_{ab} - N^2_c\Gamma^c_{ab}-N^{22}_{cb}\Gamma^c_{al}\frac{db^l}{dN}=0, \ \tilde{N}^{22}_{ab}\equiv N^{22}_{ab}=0 \nonumber\\
& & \tilde{N}^{21}_{ab}\equiv N^{21}_{ab}-N^2_c\Gamma^c_{ab}-N^{22}_{ca}\Gamma^c_{bl}\frac{db^l}{dN}\sim\frac{{\cal G}_{bc}\frac{db^c}{dN}}{||\frac{db^m}{dN}||^3}\sim{\cal O}(1).
\end{eqnarray}
Finally,
\begin{eqnarray*}
& & \tilde{Q}^{a11}_{1bc}=-R^a_{\ bcd}\frac{db^d}{dN}\sim{\cal O}(1),\nonumber\\
& & \tilde{Q}^{a21}_{1bc}=\tilde{Q}^{a12}_{1bc}=\tilde{Q}^{a22}_{1bc}=0,\nonumber\\
& & \tilde{Q}^{a12}_{2bc}=\frac{\partial_a({\cal G}^{ad}\partial_dV)}{V}\frac{db^l}{dN}{\cal G}_{lc}-2R^a_{\ cbl}\frac{db^d}{dN}\sim{\cal O}(1),\nonumber\\
& & \tilde{Q}^{a22}_{2bc}=\delta^a_c{\cal G}_{bd}\frac{db^d}{dN}+\delta^a_b{\cal G}_{cd}\frac{db^d}{dN}
+{\cal G}_{bc}\left(\frac{db^a}{dN}+\frac{{\cal G}^{ad}\partial_aV}{V}\right)\sim{\cal O}(1),\nonumber\\
\end{eqnarray*}
\begin{eqnarray}
\label{eq:Q's}
& & \tilde{Q}^{a11}_{2bc}=-\frac{V}{H^2}\left(\frac{\partial_a\partial_b({\cal G}^{ad}\partial_dV)}{V}-\frac{\partial_b({\cal G}^{ad}\partial_dV)\partial_cV}{V^2}\right)
- (\bigtriangledown_cR^a_{\ mbl})\frac{db^m}{dN}\frac{db^l}{dN}\sim{\cal O}(1),\nonumber\\
& & \tilde{Q}^{a21}_{2bc}=\left(\frac{\partial_c({\cal G}^{df}\partial_fV)}{V}-\frac{{\cal G}^{ad}\partial_dV\partial_cV}{V^2}\right){\cal G}_{bd}\frac{db^d}{dN}-R^a_{\ lcb}\frac{db^l}{dN}\sim{\cal O}(1).
\end{eqnarray}
So, substituting (\ref{eq:Adef}), (\ref{eq:N's})-(\ref{eq:Q's}) into (\ref{eq:fNL_defii}), one sees that $f_{NL}\sim{\cal O}(1)$. After completion of this work, we were informed about \cite{largefNLloop} wherein observable values of $f_{NL}$ may be obtained by considering loop corrections.

\section{Finite Tensor-To-Scalar Ratio and Issue of Scale Invariance}

We now turn to looking for ``finite" values of ratio of ampltidues of tensor and scalar perturbations, ``$r$". Using the Hamilton-Jacobi formalism (See \cite{beyondslowroll} and references therein), which is suited to deal with beyond slow-roll approximations as well, the mode $u_k(y)$ - $y\equiv\frac{k}{aH}$ - corresponding to scalar perturbations, satisfies the following differential equation when one  does not assume slow roll conditions in the sense that even though $\epsilon$ and $\eta$ are still constants, but $\epsilon$ though less than unity need not be much smaller than unity and $|\eta|$ can even be of ${\cal O}(1)$ (See \cite{beyondslowroll}):
\begin{equation}
\label{eq:scalar1}
y^2(1-\epsilon)^2u^{\prime\prime}_k(y) + 2y\epsilon(\epsilon-\tilde{\eta})u_k^\prime(y)+\left(y^2 - 2\left(1+\epsilon-\frac{3}{2}\tilde{\eta}+\epsilon^2-2\epsilon\tilde{\eta}+\frac{\tilde{\eta}^2}{2}+\frac{\xi^2}{2}\right)\right)u_k(y)=0.
\end{equation}
In this section, to simplify calculations, we would be assuming that one continues to remain on the slow-roll locus $sin(nk.b+mk.c)=0$ implying that the axionic moduli space metric is approximately a constant and the axionic kinetic terms, and in particular the inflaton kinetic term, with a proper choice of basis - see \cite{axionicswisscheese} - can be cast into a diagonal form. Further following \cite{beyondslowroll}, we would be working with $\tilde{\eta}\equiv\eta-\epsilon$ instead of $\eta$ and we will be assuming that the slow parameter $\xi<<<1$. Further the calculations in this section are valid for slow-roll case, as in our setup, beyond slow-roll regime ``$\epsilon,\eta$" are non-constants making the above differential equation non-trivial to be solved. In order to get a the required Minkowskian free-field solution in the long wavelength limit - the following is the solution\footnote{We follow \cite{Linde_Book} and hence choose $H^{(2)}_{\tilde{\tilde{\nu}}}(\frac{y}{-1+\epsilon})$ as opposed to $H^{(1)}_{\tilde{\tilde{\nu}}}(-\frac{y}{-1+\epsilon})$}:
\begin{equation}
\label{eq:scalar2}
u_k(y)\sim c(k) y^{\frac{1-\epsilon^2+2\epsilon(-1+\tilde{\eta})}{2(-1+\epsilon)^2}}H^{(2)}_{\tilde{\tilde{\nu}}} \left(\frac{y}{(-1+\epsilon)}\right).
\end{equation}
where $\tilde{\tilde{\nu}}\equiv \frac{\sqrt{9+9\epsilon^4-12\tilde{\eta}+4\tilde{\eta}^2-4\epsilon^3(1+5\tilde{\eta})
-4\epsilon(3-3\tilde{\eta}+2\tilde{\eta}^2)+2\epsilon^2(1+6\tilde{\eta}+4\tilde{\eta}^2)}}{2(-1+\epsilon)^2}$ and the Henkel function is defined as: $H^{(2)}_\alpha\equiv J_\alpha-i\left(\frac{J_\alpha cos(\alpha\pi) - J_{-\alpha}}{sin(\alpha\pi)}\right)$.\footnote{One would be interested in taking the small-argument limit of the Bessel function. However, the condition for doing the same, namely $0<\left|\frac{y}{-1+\epsilon}\right|<<\sqrt{\tilde{\tilde{\nu}}+1}$ is never really satisfied. One can analytically continue the Bessel function by using the fact that   $J_{\tilde{\tilde{\nu}}}(\frac{y}{(-1+\epsilon)})$ can be related to the Hypergeometric function $\ _0F_1\left(\tilde{\tilde{\nu}}+1;-\frac{y^2}{4(1-\epsilon)^2}\right)$ as follows:
\begin{equation}
\label{eq:J0F1}
J_{\tilde{\tilde{\nu}}}(\frac{y}{(1-\epsilon)})=\frac{\left(\frac{y}{2(1-\epsilon)}\right)^{\tilde{\tilde{\nu}}}}
{\Gamma(\tilde{\tilde{\nu}}+1)}\ _0F_1\left(\tilde{\tilde{\nu}}+1;-\frac{y^2}{4(1-\epsilon)^2}\right).
\end{equation}
Now, the small-argument limit of (\ref{eq:scalar2}) can be taken only if
\begin{equation}
\label{eq:scalar3}
\left|\frac{y}{2(1-\epsilon)}\right|<1.
\end{equation}
This coupled with the fact that $\epsilon<1$ for inflation - see \cite{beyondslowroll} - and that (\ref{eq:scalar3}) will still be satisfied at $y=1$ - the horizon crossing - tells us that $\epsilon<0.5$. One can in fact, retain the $\left(\frac{y}{-1+\epsilon}\right)^{-a}$ prefactor for continuing beyond $\epsilon=0.5$ up to $\epsilon=1$, by using the following identity that helps in the analytic continuation of $\ _0F_1(a;z)$ to regions $|z|>1$ (i.e. beyond (\ref{eq:scalar3}))- see \cite{Wolfram}:
\begin{eqnarray*}
\label{eq:identity}
& & \frac{\ _0F_1(a;z)}{\Gamma(a)}=-\frac{e^{\frac{i\pi}{2}\left(\frac{3}{2} - a\right)}z^{\frac{1-2a}{4}}}{\sqrt{\pi}}\Biggl[sinh\left(\frac{\pi i}{2}\left(\frac{3}{2}-a\right)-2\sqrt{z}\right)
\sum_{k=0}^{\left[\frac{1}{4}(2|b-1|-1)\right]}\frac{(2k+|a-1|-\frac{1}{2})!}
{2^{4k}(2k)!(|a-1|-2k-\frac{1}{2})!z^k}\nonumber\\
& & +\frac{1}{\sqrt{z}}cosh\left(\frac{\pi i}{2}\left(\frac{3}{2}-a\right)-2\sqrt{z}\right)
\sum_{k=0}^{\left[\frac{1}{4}(2|b-1|-1)\right]}\frac{(2k+|a-1|-\frac{1}{2})!}
{2^{4k}(2k+1)!(|a-1|-2k-\frac{1}{2})!z^k}\Biggr],
\end{eqnarray*}
if $a-\frac{1}{2}\in{\bf Z}$.} The power spectrum of scalar perturbations is then given by:
\begin{equation}
\label{eq:scalar5}
P^{\frac{1}{2}}_R(k)\sim\left|\frac{u_k(y=1)}{z}\right|\sim\left|H^{(2)}_{\tilde{\tilde{\nu}}}
\left(\frac{1}{(\epsilon-1)}\right)\right|\frac{1}{\sqrt{\epsilon}},
\end{equation}
where $z\sim a\sqrt{\epsilon}$ (in $M_\pi=1$ units).

Next, the tensor perturbation modes $v_k(y)$ satisfy the following equation (See \cite{beyondslowroll}):
\begin{equation}
\label{eq:tensor1}
y^2(1-\epsilon)^2v^{\prime\prime}_k(y) + 2y\epsilon(\epsilon-\tilde{\eta})u_k^\prime(y)+ (y^2-(2-\epsilon))v_k(y)=0.
\end{equation}
Using arguments similar to ones given for scalar perturbation modes' solution, one can show that the solution to (\ref{eq:tensor1}) is again given by second order Henkle functions:
\begin{equation}
\label{eq:tensor2}
v_k(y)\sim y^{\frac{1-\epsilon^2+2\epsilon(-1+\tilde{\eta})}{2(-1+\epsilon)^2}}H^{(2)}_{\tilde{\nu}}\left(\frac{y}{(\epsilon-1}\right).
\end{equation}
where $\tilde{\nu}\equiv\frac{\sqrt{9+\epsilon^4+4\epsilon(-6+\tilde{\eta})-4\epsilon^3\tilde{\eta}
+2\epsilon^3(9-4\tilde{\eta}+2\tilde{\eta}^2)}}{2(-1+\epsilon)^2}$ and the power spectrum for tensor perturbations is given by:
\begin{equation}
\label{eq:tensor3}
P^{\frac{1}{2}}_g(k)\sim|v_k(y=1)|\sim\left|H^{(2)}_{\tilde{\nu}}\left(\frac{1}{-1+\epsilon}\right)\right|,
\end{equation}
Finally, we have the following ratio of the power spectra of tensor to scalar perturbations, given as:
\begin{equation}
\label{eq:r}
r\equiv\left(\frac{P^{\frac{1}{2}}_g(k)}{P^{\frac{1}{2}_R(k)}}\right)^2
\sim\epsilon\left|\frac{H^{(2)}_{\tilde{\nu}}\left(\frac{1}{(\epsilon-1)}\right)}{H^{(2)}_{\tilde{\tilde{\nu}}}
\left(\frac{1}{(\epsilon-1)}\right)}\right|^2,
\end{equation}which, for $\epsilon=0.0034, \tilde{\eta}\sim{10}^{-5}$ - a set of values which are realized with Calabi-Yau volume ${\cal V}\sim {10}^5$ and $D3$-instanton number $n^s\sim {\cal O}(1)$ for obtaining $f_{NL}\sim{10}^{-2}$ and are also consistent with ``freeze-out" of curvature perturbations at superhorizon scales (See (\ref{eq:freezeout})) - yields $r=0.003$.  One can therefore get a  ratio of tensor to scalar perturbations of ${\cal O}(10^{-2})$ in slow-roll inflationary scenarios in Swiss-Cheese compactifications. Further, one sees that the aforementioned choice of $\epsilon$ and $\tilde{\eta}$ implies choosing the holomorphic isometric involution as part of the Swiss-Cheese Calabi-Yau orientifolding, is such that the maximum degree of the genus-0 holomorphic curve to be such that $n^0_\beta\sim \frac{\cal V}{k^2 g_s^{\frac{5}{2}}}$, which can yield the number of e-foldings $N_e\sim {\cal O}(10)$ for $D3$-instanton number $n^s\sim{\cal O}(1)$ alongwith  the non-Gaussianties parameter $f_{NL}\sim{\cal O}(10^{-2})$ and tensor-to scalar ratio $r=0.003$.

Now we calculate the loss of scale invariance assuming the freeze-out of scalar power spectrum at super horizon scales and compare it with the known cosmological experimental bound. The expression for the scalar Power Spectrum at the super horizon scales i.e. near $y=0$ with $a(y) H(y)=$ constant, is given as:
\begin{equation}
\label{eq:freezeout4}
P^{(\frac{1}{2})}_R(y)\sim\frac{(1-\epsilon)^{\tilde{\tilde{\nu}}}y^{\frac{3}{2}-\tilde{\tilde{\nu}}}}
{H^{\nu-\frac{3}{2}}a^{\nu-\frac{1}{2}}\sqrt{\epsilon}}\sim A H(y) y^{\frac{3}{2}-\tilde{\tilde{\nu}}}
\end{equation}
where $\nu={\frac{1-\epsilon^2+2\epsilon(-1+\tilde{\eta})}{2(-1+\epsilon)^2}}$ and $A$ is some scale invariant quantity. Using $\frac{d ln H(y)}{d ln y}\equiv\frac{\epsilon}{1-\epsilon} $, we can see that scalar power spectrum will be frozen at superhorizon scales, i.e., $\frac{d ln P^{\frac{1}{2}}(y)}{d ln y}=0$ if the allowed values of $\epsilon$ and $\tilde\eta$ parameters satisfy the following constraint:
\begin{equation}
\label{eq:freezeout}
\frac{d ln H(y)}{d ln y}+ \frac{3}{2}-{\tilde{\tilde{\nu}}}\equiv \frac{\epsilon}{1-\epsilon}+ \frac{3}{2}-{\tilde{\tilde{\nu}}}\sim 0
\end{equation}
The loss of scale invariance is parameterized in terms of the spectral index which is:
\begin{equation}
\label{eq:freezeout6}
n_R - 1\equiv \frac{d ln P(k)}{d ln k} = 3 - 2 {\rm Re}(\tilde{\tilde{\nu}})
\end{equation}
which gives the value of spectral index $n_R-1=0.014$ for the allowed values e.g. say $(\epsilon=0.0034,\tilde{\eta}=0.000034)$ obtained with curvature fluctuations frozen of the order $10^{-2}$ at super horizon scales.

In a nutshell, for ${\cal V}\sim{10}^{5}$ and $n^s\sim {\cal O}(1)$ we have $\epsilon\sim{0.0034}, |\eta|\sim {0.000034}, N_e\sim{17}, |f_{NL}|_{max}\sim {10^{-2}}, r\sim 4\times{10}^{-3}$ and $|n_{R}-1|\sim{0.014}$ with super-horizon freezout condition's violation of ${\cal O}(10^{-3})$. Further if we try to satisfy the freeze-out condition more accurately, say we take the deviation from zero of the RHS of (\ref{eq:freezeout}) to be of ${\cal O}(10^{-4})$ then the respective set of values are: ${\cal V}\sim{10}^{6}$, $n^s\sim {\cal O}(1)$, $\epsilon\sim{0.00028}, |\eta|\sim {10}^{-6}, N_e\sim{60}, |f_{NL}|_{max}\sim {0.01}, r\sim{0.0003}$ and $|n_{R}-1|\sim{0.001}$. This way, we have realized $N_e\sim60$, $f_{NL}\sim {10}^{-2}$, $r\sim {10}^{-3}$ and an almost scale-invariant spectrum in the slow-roll case of our LVS Swiss-Cheese Calabi-Yau orientifold setup.

\section{Conclusion and Discussion}

In this chapter, we have generalized the idea  of obtaining a dS minimum (using perturbative and non-perturbative corrections to the K\"{a}hler potential and instanton corrections to the superpotential) without the addition of $\overline{D3}$-branes \cite{dSetal} by including the one-loop corrections to the K\"{a}hler potential and showed that the one-loop corrections are sub-dominant w.r.t. the perturbative and non-perturbative $\alpha^\prime$ corrections in the {\it LVS} limits.  Assuming the NS-NS and RR axions $b^a, c^a$'s to lie in the fundamental-domain and to satisfy: $\frac{|b^a|}{\pi}<1,\ \frac{|c^a|}{\pi}<1$, one gets a flat direction provided by the NS-NS axions for slow roll inflation to occur starting from a saddle point and proceeding to the nearest dS minimum. After a detailed calculation we find that for $\epsilon << 1$ in the {\it LVS} limit all along the slow roll. The ``$\eta$-problem" gets solved along the inflationary trajectory for some quantized values  of a linear combination of the NS-NS and RR axions; the slow-roll flat direction is provided by the NS-NS axions. A linear combination of the axions gets identified with the inflaton. Thus in a nutshell, we have shown the possibility of axionic slow roll inflation in the large volume limit of type IIB compactifications on orientifolds of Swiss Cheese Calabi-Yau's. As a linear combination of the NS-NS axions corresponds to the inflaton in our work, this corresponds to a discretized expansion rate and analogous to \cite{discreteinflation} may correspond to a CFT with discretized central charges.

Further, we argued that starting from large volume compactification of type IIB string theory involving orientifolds of a two-parameter Swiss-Cheese Calabi-Yau three-fold, for appropriate choice of the holomorphic isometric involution as part of the orientifolding and hence the associated Gopakumar-Vafa invariants  corresponding to the maximum degrees of the genus-zero rational curves , it is possible to obtain $f_{NL}$ - parameterizing non-Gaussianities in curvature perturbations - to be of ${\cal O}(10^{-2})$ in slow-roll and to be of ${\cal O}(1)$ in slow-roll violating scenarios alongwith the required 60 number of e-foldings. Using general considerations and some algebraic geometric assumptions as above, we show that requiring a ``freezeout" of curvature perturbations at super horizon scales, it is possible to get tensor-scalar ratio of ${\cal O}(10^{-3})$ in the same slow-roll Swiss-Cheese setup. We predict loss of scale invariance to be within the existing experimental bounds. In a nutshell, for Calabi-Yau volume ${\cal V}\sim{10}^{6}$ and $n^s\sim {\cal O}(1)$, we have realized $\epsilon\sim{0.00028}, |\eta|\sim {10}^{-6}, N_e\sim{60}, |f_{NL}|_{max}\sim {0.01}, r\sim{0.0003}$ and $|n_{R}-1|\sim{0.001}$ with a super-horizon-freezout condition's deviation (from zero) of ${\cal O}(10^{-4})$. Further we can see that with Calabi-Yau volume ${\cal V}\sim{10}^{5}$ and $n^s\sim {\cal O}(1)$ one can realize better values of non-Gaussienities parameter and ``r" ratio ($|f_{NL}|_{max}=0.03$ and $r=0.003$) but with number of e-foldings  less than $60$. Also in beyond slow-roll case, we have realized $f_{NL}\sim {\cal O}(1)$ with number of e-foldings $N_e\sim 60$ without worrying about the tensor-to-scalar ratio and $|n_R-1|$ parameter.

To conclude, we would like to make some curious observations pertaining to the intriguing possibility of dark matter  being modelled by the NS-NS axions presenting the interesting scenario of unification of inflation and dark matter and producing finite values of non-Gaussianities and tensor-scalar ratio. In (\ref{eq:nonpert21}), if one assumes:

\noindent (a) the degrees $k_a$'s of $\beta\in H_2^-(CY_3)$ are such that they are very close and large which can be quantified as $\frac{k_1^2-k_2^2}{k_1^2+k_2^2}\sim-{\cal O}\left(\frac{1}{2\sqrt{ln \cal V}{\cal V}^4}\right)$,

\noindent (b) one is close to the locus $sin(nk.b+mk.c)=0$, where the closeness is quantified as $sin(nk.b+mk.c)\sim{\cal O}(\frac{1}{\cal V})$, and

\noindent (c) the axions have sub-Planckian Vevs so that one can disregard quadratic terms in axions relative to terms linear in the same,

\noindent then the potential of (\ref{eq:nonpert21}) can then be written as
\begin{equation}
\label{eq:DM1}
V\sim V_0\left(\left(\sum_{m^a}e^{-\frac{m^2}{2g_s} + \frac{m_ab^a n^s}{g_s}}\right)^2 - 8\right).
\end{equation}
Now, the Jacobi theta function (``$\theta(\frac{i}{g_s},\frac{n^sb^a}{g_s})$") squared in (\ref{eq:DM1}) can be rewritten as:
\begin{equation}
\label{eq:DM2}
\sum_{{\cal M}_1^+,{\cal M}_1+^-;{\cal M}_2^+,{\cal M}_2^-}e^{-\frac{({\cal M}_1^+)^2+({\cal M}_1^-)^2}{2g_s}}e^{-\frac{({\cal M}_2^+)^2+({\cal M}_2^-)^2}{2g_s}}
e^{({\cal M}_1^+b^1+{\cal M}_2^+b^2)\frac{n^s}{g_s}}.
\end{equation}
Now, writing $m_1b^1+m_2b^2$ as $\frac{1}{2}({\cal M}_+(b^1+b^2)+{\cal M}_-(b^1-b^2))$ and noting that the inflaton ${\cal I}$, for $k_1\sim k_2$ can be identified with $b^1+b^2$ - see \cite{axionicswisscheese} - one sees that (\ref{eq:DM2}) can be written as
\begin{equation}
\label{eq:intermed}
\left(\sum_{{\cal M}_1^-}e^{-\frac{({\cal M}_1^-)^2}{4g_s}}\right)^2
\sum_{{\cal M}_+,{\cal M}_-}e^{-\frac{({\cal M}_+)^2+({\cal M}_-)^2}{2g_s}}e^{\frac{({\cal M}_+{\cal I} + {\cal M}_-{\cal I}^\perp) n^s}{2}},
\end{equation}
${\cal I}^\perp\sim b^1-b^2$ - for orthonormal axionic fields, ${\cal I}^\perp$ will be orthogonal to ${\cal I}$.  Now, assuming ${\cal I}^\perp$ has been stabilized to 0,
one sees that one could write
\begin{eqnarray}
\label{eq:DM3}
& & \left(\theta\left(\frac{i}{g_s},\frac{b^an^s}{g_s}\right)\right)^2\sim2\left(\sum_{{\cal M}_1^-}e^{-\frac{({\cal M}_1^-)^2}{2g_s}}\right)^3\sum_{{\cal M}_+\geq0}e^{-\frac{{\cal M}_+^2}{2g_s}}cosh\left(\frac{{\cal M}_+{\cal I}n^s}{2}\right),\nonumber\\
& & \sim 2\sum_{{\cal M}_1\geq0}e^{-\frac{{\cal M}_1^2}{4g_s}}cosh\left(\frac{{\cal M}_+{\cal I}n^s}{2}\right) \ \ {\rm for} \ g_s\ll1.
\end{eqnarray}
Thus the expression (\ref{eq:DM1}) for the potential in the weak coupling limit, yields:
\begin{equation}
\label{eq:DM4}
V\sim V_0\left(\sum_{{\cal M}\geq0}e^{-\frac{{\cal M}^2}{2g_s}}cosh\left(\frac{{\cal M}{\cal I}n^s}{2}\right) - \sum_{{\cal M}\geq0}e^{-\frac{{\cal M}^2}{2g_s}}\right).
\end{equation}
Once again, in the weak coupling limit, the sum in (\ref{eq:DM4}) can be assumed to be restricted to
${\cal M}$ proportional to 0 and 1. This hence gives:
\begin{equation}
\label{eq:DM5}
V\sim V_0\left(cosh\left(\frac{{\cal I}n^s}{2}\right) - 1\right).
\end{equation}
One sees that (\ref{eq:DM5}) is of the same form as the potential proposed in \cite{SahniWang}:
\[V=V_0\left(cosh(\lambda\phi)-1\right),\]
for cold dark matter! This, given the assumption of sub-Planckian axions, is by no means valid for all ${\cal I}$. However, in the given domain of validity, the fact that a string (SUGRA) potential can be recast into the form (\ref{eq:DM5}) is, we feel, quite interesting.

Alternatively, in the same spirit as \cite{gravwavesKallosh}, if one breaks the NS-NS axionic shift symmetry ``slightly" by restricting the symmetry group ${\bf Z}$ to ${\bf Z}_+\cup\{0\}$, then (\ref{eq:DM4}) can be rewritten as:
\begin{equation}
\label{eq:Q1}
\sum_{{\cal M}\geq0}e^{-\frac{{\cal M}^2}{2g_s}}e^{\frac{{\cal M}{\cal I}n^s}{2}} - \sum_{{\cal M}\geq0}e^{-\frac{{\cal M}^2}{2g_s}}\sim e^{-\frac{\pi^2}{2g_s}}e^{\frac{\pi{\cal I}n^s}{2}}
+ e^{-\frac{4\pi^2}{2g_s}}e^{\frac{4\pi{\cal I}n^s}{2}},
\end{equation}
which is similar to:
\[V=e^{\alpha_1+\alpha_2\phi}+e^{\beta_1+\beta_2\phi},\]
(where $\alpha_2,\beta_2$ are taken to be positive) that has been used to study quintessence models (in studies of dark energy) - see \cite{Q} -  in fact, as argued in \cite{Q}, one can even include ${\cal I}^\perp$.

%% file: chap4.tex
\chapter{Large Volume Swiss-Cheese D3/D7 Phenomenology}
\markboth{nothing}{\bf 4. Large Volume Swiss-Cheese D3/D7 Phenomenology}

\hskip1in{\it{``Some of nature's most exquisite handiwork is on a miniature scale, as anyone knows who has applied a magnifying glass to a snowflake."}} - Rachel Carson.

\section{Introduction}
From the point of view of ``testing" string theory in the laboratories, string phenomenology and string cosmology  have been the major areas of work and a lot of work has been done.  In  the context of Type IIB orientifold compactification in the L(arge) V(olume) S(cenarios) limit, a non-supersymmetric $AdS$ minimum was realized in \cite{Balaetal2} with the inclusion of perturbative ${\alpha^{\prime}}^3$ corrections  to the K\"{a}hler potential, which was then uplifted to $dS$ vacuum. Followed by this, it was shown in \cite{dSetal} that with the inclusion of  (non-)perturbative $\alpha^{\prime}$ corrections to the K\"{a}hler potential and instanton corrections to the superpotential, one can realize {\it non}-supersymmetric metastable $dS$ solution in a more natural way without having to add an uplifting term (e.g. with the inclusion of $\overline D3$-brane as in \cite{KKLT}).
However in order to put Cosmology as well as Particle Phenomenology together in the same string theoretic setup, there has been a tension between  LVS cosmology and LVS phenomenology studied so far. The scale required by cosmological/astrophysical experiments is nearly the same order as the GUT scale ($\sim 10^{16}$ GeV) while in LVS phenomenology, the supersymmetry-breaking at $TeV$ scale requires the string scale to be some intermediate scale of the order of $10^{11}$ GeV. In this way, there is a hierarchy in scales involved on both sides making it impossible to fulfill both requirements in the same string theory setup. Although LVS limits of Type IIB Swiss-Cheese orientifold compactifications have been exciting steps in the search for  realistic models on both cosmology as well as phenomenology sides, this hierarchy is reflected in LVS setups, as a hierarchy of compactification volume requirement from $ \cal{V}$ $\sim 10^6$ (for cosmology requirement, e.g. see \cite{kahlerinflation}) to $\cal{V}$ $\sim 10^{14}$ (for phenomenology requirement\footnote{In \cite{quevedoftheorysusy}, the authors have realized soft terms $\sim$ TeV with $ {\cal V}\sim O(10^6-10^7)$ in the context of String/F-theory models with SM supported on a del Pezzo surface, but with very heavy gravitino.}, e.g. see \cite{conloncal}) and  the tension has remained unresolved in a single string theoretic setup with the Calabi-Yau volume stabilized at a particular value\footnote {There has been a proposal \cite{tension1}, which involves a small CY volume for incorporating high-scale inflation and then evolves the volume modulus over a long range and finally stabilizes it in the large volume minimum with TeV gravitino mass after inflation.}. Now in the present LHC era equipped with PAMELA and PLANCK, string theoretic models with numbers, which could match with experimental-data are yet to come; and several phenomenologically motivated steps have also been initiated in this direction \cite{quevedojan09,abdussalam2,LHCpheno,kallosh1,KKLMMT,LargeVcons,largefNL_r}.

Also the study of LVS models in the context of ${\cal N}=1$ type IIB orientifold compactification in the presence of D7-branes, has been quite attractive and promising for the phenomenological purposes also because in such models, D7-brane wrapping the smaller cycle produces qualitatively similar gauge coupling as  that of the Standard Model and also with the magnetized D7-branes, the Standard Model chiral matter can be realized from strings stretching between stacks of D7-branes \cite{conloncal,Lustetal,conlonLVSsusy,ibanezfont,jockersetal}. In one of such models, RG evolutions  of soft-terms to the weak scale have been studied  to have a low energy spectra by using the RG equations of MSSM (assuming that only charged matter content below the string scale is the MSSM) and it was found that with D7 chiral matter fields, low energy supersymmetry breaking could be realized at a small hierarchy between the gravitino mass and soft supersymmetry breaking terms \cite{conloncal}. A much detailed study with fluxed $D3/D7$ branes has been done in the context of ${\cal N}=1$ type IIB orientifold compactification \cite{Lustetal,conlonLVSsusy,jockersetal} and it has been found that the ${\cal N}=1$ coordinates get modified with the inclusion of $D3$ and $D7$-branes. The gauge coupling of $D7$-brane wrapping a four-cycle  depends mainly on the size modulus of the wrapped four-cycle and also on the complex structure as well as axion-dilaton modulus  after including the loop-corrections, which in the diluted flux limit (without loop-corrections) was found to be dominated by the size modulus of the wrapping four-cycle \cite{conlonLVSsusy,BHP}.

We address phenomenological aspects of Swiss-Cheese Calabi-Yau orientifolds in Type IIB compactifications in this chapter, which is organized as follows. In section {\bf 2}, we start with extending our ``LVS Swiss-Cheese Cosmology" seup discussed in chapter {\bf 2} with the inclusion of a mobile spacetime filling $D3$-brane and stack(s) of $D7$-brane(s) wrapping the big divisor $\Sigma_B$ inside the Swiss-Cheese Calabi-Yau and discuss the consequent modifications in appropriate ${\cal N}=1$ coordinates on inclusion of D3/D7 in the setup. Section {\bf 3} has a detailed discussion on obtaining the geometric K\"{a}hler potential for the Calabi-Yau and in particular, for the above mentioned (``big" and "small") divisors using toric geometry, GLSM techniques and results by Umemura and Zhivkov. We also write out the complete moduli-space K\"{a}hler potential in terms of the closed-string moduli as well as the open-string moduli or matter fields, the latter being the position moduli of the spacetime filling mobile $D3$-brane and the Wilson-line moduli on the $D7$-brane(s). Section {\bf 4} is about resolution of a long-standing problem in large volume string phenomenology and cosmology - giving a geometric mechanism that would generate a $10^{12}GeV$ gravitino in the early inflationary epoch of the universe and then a $TeV$ gravitino at the present times to possibly be detected at the LHC, for the same value of the volume modulus of the Calabi-Yau at around $10^6$ (in $l_s = 1$ units). In section {\bf 5}, we discuss the construction of local involutively-odd harmonic one-forms on the aforementioned ``big" divisor to enable getting an ${\cal O}(1)$ $g_{YM}$ on the world-volume of a stack of $D7$-branes wrapping the divisor. Finally we summarize the chapter in section {\bf 6}.

\section{The Extended D3/D7 LVS Swiss-Cheese Setup}
The appropriate ${\cal N}=1$ coordinates in the presence of a single $D3$-brane and a single $D7$-brane wrapping the ``big" divisor $\Sigma^B$  along with $D7$-brane fluxes are given as under (See \cite{jockersetal,Jockers_thesis}):
\begin{eqnarray}
\label{eq:N=1_coords}
& & S = \tau + \kappa_4^2\mu_7{\cal L}_{A{\bar B}}\zeta^A{\bar\zeta}^{\bar B}\nonumber\\
& & {\cal G}^a = c^a - \tau {\cal B}^a\nonumber\\
& & T_\alpha=\frac{3i}{2}(\rho_\alpha - \frac{1}{2}\kappa_{\alpha bc}c^b{\cal B}^c) + \frac{3}{4}\kappa_\alpha + \frac{3i}{4(\tau - {\bar\tau})}\kappa_{\alpha bc}{\cal G}^b({\cal G}^c
- {\bar G}^c) \nonumber\\
& & + 3i\kappa_4^2\mu_7l^2C_\alpha^{I{\bar J}}a_I{\bar a_{\bar J}} + \frac{3i}{4}\delta^B_\alpha\tau Q_{\tilde{f}} + \frac{3i}{2}\mu_3l^2(\omega_\alpha)_{i{\bar j}} \Phi^i\left({\bar\Phi}^{\bar j}-\frac{i}{2}{\bar z}^{\tilde{a}}({\bar{\cal P}}_{\tilde{a}})^{\bar j}_l\Phi^l\right)\nonumber\\
& & \tau=l+ie^{-\phi},
\end{eqnarray}
where
\begin{itemize}
\item
for future reference in the remainder of the chapter, one defines: ${\cal T}_\alpha\equiv\frac{3i}{2}(\rho_\alpha - \frac{1}{2}\kappa_{\alpha bc}c^b{\cal B}^c) + \frac{3}{4}\kappa_\alpha + \frac{3i}{4(\tau - {\bar\tau})}\kappa_{\alpha bc}{\cal G}^b({\cal G}^c
- {\bar G}^c)$,
\item
\begin{equation}
  {\cal L}_{A{\bar B}}=\frac{\int_{\Sigma^B}\tilde{s}_A\wedge\tilde{s}_{\bar B}}{\int_{CY_3}\Omega\wedge{\bar\Omega}},
  \end{equation}
$\tilde{s}_A$ forming a basis for $H^{(2,0)}_{{\bar\partial},-}(\Sigma^B)$,
\item
the fluctuations of $D7$-brane in the $CY_3$ normal to $\Sigma^B$ are denoted by $\zeta\in H^0(\Sigma^B,N\Sigma^B)$, i.e., they are the space of global sections of the normal bundle $N\Sigma^B$,
\item
${\cal B}\equiv b^a - lf^a$, where $f^a$ are the components of elements of two-form fluxes valued in $i^*\left(H^2_-(CY_3)\right)$, the immersion map being defined as:
$i:\Sigma^B\hookrightarrow CY_3$,
\item
$C^{I{\bar J}}_\alpha=\int_{\Sigma^B}i^*\omega_\alpha\wedge A^I\wedge A^{\bar J}$, $\omega_\alpha\in H^{(1,1)}_{{\bar\partial},+}(CY_3)$ and $A^I$ forming a basis for $H^{(0,1)}_{{\bar\partial},-}(\Sigma^B)$,
\item
$a_I$ is defined via a Kaluza-Klein reduction of the $U(1)$ gauge field (one-form) $A(x,y)=A_\mu(x)dx^\mu P_-(y)+a_I(x)A^I(y)+{\bar a}_{\bar J}(x){\bar A}^{\bar J}(y)$, where $P_-(y)=1$ if $y\in\Sigma^B$ and -1 if $y\in\sigma(\Sigma^B)$,
\item
$z^{\tilde{a}}$ are $D=4$ complex scalar fields arising due to complex structure deformations of the Calabi-Yau orientifold defined via: $\delta g_{{\bar i}{\bar j}}(z^{\tilde{a}})=-\frac{i}{||\Omega||^2}z^{\tilde{a}}\left(\chi_{\tilde{a}}\right)_{{\bar i}jk}\left({\bar\Omega}\right)^{jkl}g_{l{\bar j}}$, where $\left(\chi_{\tilde{a}}\right)_{{\bar i}jk}$ are components of elements of $H^{(2,1)}_{{\bar\partial},+}(CY_3)$,
\item
$\left({\cal P}_{\tilde{a}}\right)^i_{\bar j}\equiv\frac{1}{||\Omega||^2}{\bar\Omega}^{ikl}\left(\chi_{\tilde{a}}\right)_{kl{\bar j}}$, i.e.,
${\cal P}:TCY_3^{(1,0)}\longrightarrow TCY_3^{(0,1)}$ via the transformation:
$\Phi\stackrel{\rm c.s.\ deform}{\longrightarrow}\Phi^i+\frac{i}{2}z^{\tilde{a}}\left({\cal P}_{\tilde{a}}\right)^i_{\bar j}{\bar\Phi}^{\bar j}$,
\item
$\Phi^i$ are scalar fields corresponding to geometric fluctuations of $D3$-brane inside the Calabi-Yau and defined via: $\Phi(x)=\Phi^i(x)\partial_i + {\bar\Phi}^{\bar i}({\bar x}){\bar\partial}_{\bar i}$,
and
\item
$Q_{\tilde{f}}\equiv l^2\int_{\Sigma^B}\tilde{f}\wedge\tilde{f}$, where $\tilde{f}\in\tilde{H}^2_-(\Sigma^B)\equiv{\rm coker}\left(H^2_-(CY_3)\stackrel{i^*}{\rightarrow}H^2_-(\Sigma^B)\right)$.
\end{itemize}
We will be working in the $x_2=1$-coordinate patch throughout this chapter with ``LVS Swiss-Cheese Cosmo-Pheno setup", for definiteness, we use the notation- $z_1= \frac{x_1}{x_2},\ z_2=\frac{x_3}{x_2},\ z_3= \frac{x_4}{x_2^6}$ and $z_4=\frac{x_5}{x_2^9}$.

\section{The Geometric K\"{a}hler Potential and Metric}
In this section we will derive the geometric K\"{a}ler potential for the Swiss-Cheese Calabi-Yau $\bf{ WCP}^4{[1,1,1,6,9]}$ because of a $D3$-brane present in our setup. This will enable us to determine the complete K\"{a}hler potential corresponding to the closed string moduli $\sigma^\alpha, {\cal G}^a$ as well as the open string moduli or matter fields: $z_i, a^I$.

The one-dimensional cones in the toric fan of the desingularized $\bf {WCP}^4{[1,1,1,6,9]}$ are given by the following vectors (See \cite{Candelasetal}):
\begin{eqnarray}
\label{eq:toric_fan}
& & v_1=(-1,-1,-6,-9)\nonumber\\
& & v_2=(1,0,0,0)\nonumber\\
& & v_3=(0,1,0,0)\nonumber\\
& & v_4=(0,0,1,0)\nonumber\\
& & v_5=(0,0,0,1)\nonumber\\
& & v_6({\rm Exceptional\ divisor})=(0,0,-2,-3).
\end{eqnarray}
The allowed charges under two ${\bf C}^*$ actions are given by solutions to:
\begin{equation}
\label{eq:chargesEqn}
\sum_{i=1}^6Q^a_iv_i=0, a=1,2.
\end{equation}
The solution to (\ref{eq:chargesEqn}) are of the type:
\begin{equation}
\label{eq:chargesSoln}
(q_1,q_1,q_1,2q_6+6q_1,3q_6+9q_1,q_6).
\end{equation}
The $({\bf C}^*)^2$ charges will be taken as under:
\begin{equation}
\label{eq:charges}
\begin{array}{c|cccccc}
&\Phi_1&\Phi_2&\Phi_3&\Phi_4&\Phi_5&\Phi_6 \\ \hline
Q^1&0&0&0&2&3&1\\
Q^2&1&1&1&0&0&-3
\end{array}
\end{equation}
Hence, one can construct the following inhomogeneous coordinates: $z_1=\frac{\Phi_1}{\Phi_2}, z_2=\frac{\Phi_3}{\Phi_2}, z_3=\frac{\Phi_4}{\Phi_\epsilon^2\Phi_2^6}, z_4=\frac{\Phi_5}{\Phi_\epsilon^3\Phi_2^9}$.

The NLSM for a two-dimensional ${\cal N}=2$ supersymmetric gauge theory whose target space
is the toric variety corresponding to (\ref{eq:toric_fan}) with (anti-)chiral superfields $({\bar\Phi_i})\Phi_i$,
Fayet-Iliopoulos parameters $r_a$  and vector superfields $V_a$,  is specified by the K\"{a}hler potential:
\begin{equation}
\label{eq:NLSM_K}
\int d^4\theta K=\int d^4\theta\left(\sum_{i=1}^6{\bar\Phi_i}e^{\sum_{a=1}^22Q^a_iV_a}\Phi_i-2r_aV_a\right).
\end{equation}
Substituting (\ref{eq:charges}) in (\ref{eq:NLSM_K}), one sees that:
\begin{equation}
\label{eq:K1}
K=\left(|\Phi_1|^2+|\Phi_2|^2+|\Phi_3|^2\right)e^{2V_2}+|\Phi_4|^2e^{4V_1}
+|\Phi_5|^2e^{6V_1}+|\Phi_\epsilon|^2e^{2V_1-6V_2}-2r_1V_1-2r_2V_2.
\end{equation}
Now, $\int d^4\theta K$ can be regarded as the IR limit of the GLSM Lagrangian - the gauge kinetic terms hence
decouple in this limit. One hence gets a supersymmetric NLSM Lagrangian ${\cal L}_{\rm NLSM}=\int d^4\theta K$ wherein the gauge superfields act as auxiliary fields and can be eliminated by their equations of motion - see \cite{Kimura}.
One can show that the variation of the NLSM Lagrangian w.r.t. the vector superfields $V_a$ yield:
\begin{eqnarray}
\label{eq:eoms1}
& & \frac{\partial{\cal L}_{\rm NLSM}}{\partial V_1}=0\Leftrightarrow 2|\Phi_1|^2e^{2V_1}+3|\Phi_5|^2e^{6V_1}+|\Phi_\epsilon|^2
e^{2V_1-6V_2}=r_1\nonumber\\
& & \frac{\partial{\cal L}_{\rm NLSM}}{\partial V_2}=0\Leftrightarrow |\Phi_1|^2e^{2V_2}+|\Phi_2|^2e^{2V_2}
+|\Phi_3|^2e^{2V_2}-3|\Phi_\epsilon|^2e^{2V_1-6V_2}=r_2.\nonumber\\
& &
\end{eqnarray}
Defining $x\equiv e^{2V_1}, y\equiv e^{2V_2}$, (\ref{eq:eoms1}) can be rewritten as:
\begin{eqnarray}
\label{eq:eoms2}
& & a_1x^2+b_1x^3+c_1xy^{-3}=r_1,\nonumber\\
& & a_2y+c_2xy^{-3}=r_2,
\end{eqnarray}
where
\begin{equation}
\label{eq:defs}
a_1\equiv2|\Phi_4|^2, c_1\equiv|\Phi_\epsilon|^2, a_2\equiv|\Phi_1|^2+|\Phi_2|^2+|\Phi_3|^2,
c_2\equiv-3|\Phi_\epsilon|^2.
\end{equation}
We would now be evaluating the K\"{a}hler potential for the divisor $D_5:\Phi_5=0$ or equivalently $z_4=0$, in
the large volume limit of the Swiss-Cheese Calabi-Yau. In the $\Phi_2=\Phi_\epsilon=1$-coordinate patch,
the defining hypersurface for $D_5$ is $1+z_1^{18}+z_2^{18}+z_3^3-3\phi z_1^6z_2^6=0$. In the LVS limit,
we would assume a scaling: $z_3\sim{\cal V}^{\frac{1}{6}}, z_{1,2}\sim{\cal V}^{\frac{1}{36}}$. Further,
the FI parameters, $r_{1,2}$ taken to scale like the big and small two-cycle areas $t_5$ and $t_4$ respectively,
i.e. like ${\cal V}^{\frac{1}{3}}$ and $\sqrt{ln {\cal V}}$.

The system of equations (\ref{eq:eoms2}) is equivalent to the following octic - we will not be careful with numerical factors in the following:
\begin{eqnarray}
\label{eq:octic}
& & P(z)\equiv|z_3|^2(1+|z_1|^2+|z_2|^2)^2y^8+|z_3|^2(1+|z_1|^2+|z_2|^2)r_2y^7 +|z_3|^2r_2^2y^6 \nonumber\\
& & +(1+|z_1|^2+|z_2|^2)y-4r_1=0.
\end{eqnarray}
Using Umemura's result \cite{Umemura} on expressing the roots of an algebraic polynomial in terms of Siegel theta functions
of genus $g>1$ -  $\theta\left[\begin{array}{c} \mu\\
\nu
\end{array}\right](z,\Omega)$ for $\mu,\nu\in{\bf R}^g, z\in {\bf C}^g$ and $\Omega$ being a complex symmetric
$g\times g$ period matrix with $Im(\Omega)>0$ defined as follows:
\begin{equation}
\label{eq:Siegel_theta}
\theta\left[\begin{array}{c} \mu\\
\nu
\end{array}\right](z,\Omega)=\sum_{n\in{\bf Z}^g}e^{i\pi(n+\mu)^T\Omega(n+\mu)+2i\pi(n+\mu)^T(z+\nu)}.
\end{equation}
The degree $n$ of the polynomial is related to the genus $g$ of the Riemann surface via
$g=\left[\frac{n+2}{2}\right]$. Hence for an octic, one needs to use Siegel theta functions of genus five.
The period matrix $\Omega$ will be defined as follows:
\begin{equation}
\label{eq:Omega}
\left(\begin{array}{ccccc}
\Omega_{11} & \Omega_{12} & \Omega_{13} & \Omega_{14} & \Omega_{15}\\
\Omega_{12} & \Omega_{22} & \Omega_{23} & \Omega_{24} & \Omega_{25}\\
\Omega_{13} & \Omega_{23} & \Omega_{33} & \Omega_{34} & \Omega_{35}\\
\Omega_{14} & \Omega_{24} & \Omega_{34} & \Omega_{44} & \Omega_{45}\\
\Omega_{15} & \Omega_{25} & \Omega_{35} & \Omega_{45} & \Omega_{55}\\
\end{array}
\right)=\left(\begin{array}{ccccc}
\sigma_{11} & \sigma_{12} & \sigma_{13} & \sigma_{14} & \sigma_{15} \\
\sigma_{21} & \sigma_{22} & \sigma_{23} & \sigma_{24} & \sigma_{25} \\
\sigma_{31} & \sigma_{32} & \sigma_{33} & \sigma_{34} & \sigma_{35} \\
\sigma_{41} & \sigma_{42} & \sigma_{43} & \sigma_{44} & \sigma_{45} \\
\sigma_{51} & \sigma_{52} & \sigma_{53} & \sigma_{54} & \sigma_{55} \\
\end{array}\right)^{-1}
\left(\begin{array}{ccccc}
\rho_{11} & \rho_{12} & \rho_{13} & \rho_{14} & \rho_{15} \\
\rho_{21} & \rho_{22} & \rho_{23} & \rho_{24} & \rho_{25} \\
\rho_{31} & \rho_{32} & \rho_{33} & \rho_{34} & \rho_{35} \\
\rho_{41} & \rho_{42} & \rho_{43} & \rho_{44} & \rho_{45} \\
\rho_{51} & \rho_{52} & \rho_{53} & \rho_{54} & \rho_{55} \\
\end{array}
\right),
\end{equation}
where $\sigma_{ij}\equiv\oint_{A_j}dz \frac{z^{i-1}}{\sqrt{z(z-1)(z-2)P(z)}}$ and
$\rho_{ij}\equiv\oint_{B_j}\frac{z^{i-1}}{\sqrt{z(z-1)(z-2)P(z)}}$,
$\{A_i\}$ and $\{B_i\}$ being a canonical basis of cycles satisfying: $A_i\cdot A_j=B_i\cdot B_j=0$ and
$A_i\cdot B_j=\delta_{ij}$. Umemura's result then is that a root of (\ref{eq:octic}) can be written as:
\begin{eqnarray}
\label{eq:soln_octic}
& & \frac{1}{2\left(\theta\left[\begin{array}{ccccc}
\frac{1}{2} & 0 & 0 & 0 & 0 \\
0 & 0 & 0 & 0 & 0  \end{array}\right](0,\Omega)\right)^4
\left(\theta\left[\begin{array}{ccccc}
\frac{1}{2} & \frac{1}{2} & 0 & 0 & 0 \\
0 & 0 & 0 & 0 & 0  \end{array}\right](0,\Omega)\right)^4}\nonumber\\
& & \times\Biggl[\left(\theta\left[\begin{array}{ccccc}
\frac{1}{2} & 0 & 0 & 0 & 0 \\
0 & 0 & 0 & 0 & 0  \end{array}\right](0,\Omega)\right)^4
\left(\theta\left[\begin{array}{ccccc}
\frac{1}{2} & \frac{1}{2} & 0 & 0 & 0 \\
0 & 0 & 0 & 0 & 0  \end{array}\right](0,\Omega)\right)^4\nonumber\\
& & + \left(\theta\left[\begin{array}{ccccc}
0 & 0 & 0 & 0 & 0 \\
0 & 0 & 0 & 0 & 0  \end{array}\right](0,\Omega)\right)^4
\left(\theta\left[\begin{array}{ccccc}
0 & \frac{1}{2} &  0 & 0 & 0 \\
0 & 0 & 0 & 0 & 0  \end{array}\right](0,\Omega)\right)^4\nonumber\\
& & - \left(\theta\left[\begin{array}{ccccc}
0 & 0 & 0 & 0 & 0 \\
\frac{1}{2} & 0 & 0 & 0 & 0  \end{array}\right](0,\Omega)\right)^4
\left(\theta\left[\begin{array}{ccccc}
0 & \frac{1}{2} & 0 & 0 & 0 \\
\frac{1}{2} & 0 & 0 & 0 & 0 \end{array} \right](0,\Omega)\right)^4\Biggr].\nonumber\\
& &
\end{eqnarray}
Now, if $|z_3|^2r_1^2y^6\sim r_1\sim\sqrt{ln {\cal V}}$, then this suggests that $y\sim \left(ln {\cal V}\right)^{\frac{1}{12}}{\cal V}^{-\frac{1}{6}}$.
Substituting this estimate for $y$ into the octic and septic terms, one sees that the same are of
${\cal O}\left(\left(ln {\cal V}\right)^{\frac{2}{3}}{\cal V}^{-\frac{8}{9}}\right)$ and
${\cal O}\left(\left(ln {\cal V}\right)^{\frac{7}{12}}{\cal V}^{-\frac{4}{9}}\right)$ respectively which
are both suppressed w.r.t. to the sextic term. Hence, in the LVS limit (\ref{eq:octic}) reduces to the
following sextic:
\begin{equation}
\label{eq:sextic}
y^6+\alpha y+\beta=0.
\end{equation}
Umemura's result would require the use of genus-four Siegel theta functions. However, using the results of
\cite{Zhivkov}, one can express the roots of a sextic in terms of genus-two Siegel theta functions as follows:
\begin{eqnarray*}
\label{eq:roots}
& & \left[\frac{\sigma_{22}\frac{d}{dz_1}\theta\left[\begin{array}{cc}
\frac{1}{2}&\frac{1}{2} \\
0&\frac{1}{2}
\end{array}\right]\left((z_1,z_2),\Omega\right)
- \sigma_{21}\frac{d}{dz_2}\theta\left[\begin{array}{cc}
\frac{1}{2}&\frac{1}{2} \\
0&\frac{1}{2}
\end{array}\right]\left((z_1,z_2),\Omega\right) }
{\sigma_{12}\frac{d}{dz_1}\theta\left[\begin{array}{cc}
\frac{1}{2}&\frac{1}{2} \\
0&\frac{1}{2}
\end{array}\right]\left((z_1,z_2),\Omega\right)
- \sigma_{12}\frac{d}{dz_2}\theta\left[\begin{array}{cc}
\frac{1}{2}&\frac{1}{2} \\
0&\frac{1}{2}
\end{array}\right]\left((z_1,z_2),\Omega\right)}\right]_{z_1=z_2=0}, \nonumber\\
\end{eqnarray*}
\begin{eqnarray}
\label{eq:roots}
& & \left[\frac{\sigma_{22}\frac{d}{dz_1}\theta\left[\begin{array}{cc}
0&\frac{1}{2} \\
0&\frac{1}{2}
\end{array}\right]\left((z_1,z_2),\Omega\right)
- \sigma_{21}\frac{d}{dz_2}\theta\left[\begin{array}{cc}
0&\frac{1}{2} \\
0&\frac{1}{2}
\end{array}\right]\left((z_1,z_2),\Omega\right) }
{\sigma_{12}\frac{d}{dz_1}\theta\left[\begin{array}{cc}
0&\frac{1}{2} \\
0&\frac{1}{2}
\end{array}\right]\left((z_1,z_2),\Omega\right)
- \sigma_{12}\frac{d}{dz_2}\theta\left[\begin{array}{cc}
0&\frac{1}{2} \\
0&\frac{1}{2}
\end{array}\right]\left((z_1,z_2),\Omega\right)}\right]_{z_1=z_2=0}, \nonumber\\
& & \left[\frac{\sigma_{22}\frac{d}{dz_1}\theta\left[\begin{array}{cc}
0&\frac{1}{2} \\
\frac{1}{2}&\frac{1}{2}
\end{array}\right]\left((z_1,z_2),\Omega\right)
- \sigma_{21}\frac{d}{dz_2}\theta\left[\begin{array}{cc}
0&\frac{1}{2} \\
\frac{1}{2}&\frac{1}{2}
\end{array}\right]\left((z_1,z_2),\Omega\right) }
{\sigma_{12}\frac{d}{dz_1}\theta\left[\begin{array}{cc}
0&\frac{1}{2} \\
\frac{1}{2}&\frac{1}{2}
\end{array}\right]\left((z_1,z_2),\Omega\right)
- \sigma_{12}\frac{d}{dz_2}\theta\left[\begin{array}{cc}
0&\frac{1}{2} \\
\frac{1}{2}&\frac{1}{2}
\end{array}\right]\left((z_1,z_2),\Omega\right)}\right]_{z_1=z_2=0}, \nonumber\\
& & \left[\frac{\sigma_{22}\frac{d}{dz_1}\theta\left[\begin{array}{cc}
\frac{1}{2}&0 \\
\frac{1}{2}&\frac{1}{2}
\end{array}\right]\left((z_1,z_2),\Omega\right)
- \sigma_{21}\frac{d}{dz_2}\theta\left[\begin{array}{cc}
\frac{1}{2}&0 \\
\frac{1}{2}&\frac{1}{2}
\end{array}\right]\left((z_1,z_2),\Omega\right) }
{\sigma_{12}\frac{d}{dz_1}\theta\left[\begin{array}{cc}
\frac{1}{2}&0 \\
\frac{1}{2}&\frac{1}{2}
\end{array}\right]\left((z_1,z_2),\Omega\right)
- \sigma_{12}\frac{d}{dz_2}\theta\left[\begin{array}{cc}
\frac{1}{2}&0 \\
\frac{1}{2}&\frac{1}{2}
\end{array}\right]\left((z_1,z_2),\Omega\right)}\right]_{z_1=z_2=0}, \nonumber\\
& & \left[\frac{\sigma_{22}\frac{d}{dz_1}\theta\left[\begin{array}{cc}
\frac{1}{2}&0 \\
\frac{1}{2}&0
\end{array}\right]\left((z_1,z_2),\Omega\right)
- \sigma_{21}\frac{d}{dz_2}\theta\left[\begin{array}{cc}
\frac{1}{2}&0 \\
\frac{1}{2}&0
\end{array}\right]\left((z_1,z_2),\Omega\right) }
{\sigma_{12}\frac{d}{dz_1}\theta\left[\begin{array}{cc}
\frac{1}{2}&0 \\
\frac{1}{2}&0
\end{array}\right]\left((z_1,z_2),\Omega\right)
- \sigma_{12}\frac{d}{dz_2}\theta\left[\begin{array}{cc}
\frac{1}{2}&0 \\
\frac{1}{2}&0
\end{array}\right]\left((z_1,z_2),\Omega\right)}\right]_{z_1=z_2=0}, \nonumber\\
& & \left[\frac{\sigma_{22}\frac{d}{dz_1}\theta\left[\begin{array}{cc}
\frac{1}{2}&\frac{1}{2} \\
\frac{1}{2}&0
\end{array}\right]\left((z_1,z_2),\Omega\right)
- \sigma_{21}\frac{d}{dz_2}\theta\left[\begin{array}{cc}
\frac{1}{2}&\frac{1}{2} \\
\frac{1}{2}&0
\end{array}\right]\left((z_1,z_2),\Omega\right) }
{\sigma_{12}\frac{d}{dz_1}\theta\left[\begin{array}{cc}
\frac{1}{2}&\frac{1}{2} \\
\frac{1}{2}&0
\end{array}\right]\left((z_1,z_2),\Omega\right)
- \sigma_{12}\frac{d}{dz_2}\theta\left[\begin{array}{cc}
\frac{1}{2}&\frac{1}{2} \\
\frac{1}{2}&0
\end{array}\right]\left((z_1,z_2),\Omega\right)}\right]_{z_1=z_2=0}.\nonumber\\
\end{eqnarray}
\begin{eqnarray}
\label{eq:dertheta}
& & {\rm Here,} \, \, \frac{d}{dz_i}\theta\left[\begin{array}{cc}
\mu_1 & \mu_2 \\
\nu_1 & \nu_2
\end{array}\right]\left((z_1,z_2),\Omega\right)_{z_1=z_2=0}=-2\pi\sum_{n_1,n_2\in{\bf Z}}
(-)^{2\nu_1n_1+2\nu_2n_2}(n_i+\mu_i)\nonumber\\
& & e^{i\pi\Omega_{11}(n_1+\mu_1)^2
+2i\pi\Omega_{12}(n_1+\mu_1)(n_2+\mu_2)+i\pi\Omega_{22}(n_2+\mu_2)^2},
\end{eqnarray}
where $\mu_i$ and $\nu_i$ are either 0 or $\frac{1}{2}$. The symmetric period matrix corresponding to the hyperelliptic
curve $w^2=P(z)$ is given by:
\begin{equation}
\label{eq:Omega_g=2}
\left(\begin{array}{cc}
\Omega_{11} & \Omega_{12} \\
\Omega_{12} & \Omega_{22}
\end{array}\right)=\frac{1}{\sigma_{11}\sigma_{22}-\sigma_{12}\sigma_{21}}\left(\begin{array}{cc}
\sigma_{22} & -\sigma_{12} \\
-\sigma_{21} & \sigma_{11}
\end{array}\right)\left(\begin{array}{cc}
\rho_{11} & \rho_{12} \\
\rho_{21} & \rho_{22}
\end{array}\right),
\end{equation}
where $\sigma_{ij}=\int_{z_*{A_j}}\frac{z^{i-1}dz}{\sqrt{P(z)}}$ and
$\rho_{ij}=\int_{z_*{B_j}}\frac{z^{i-1}dz}{\sqrt{P(z)}}$ where $z$ maps the $A_i$ and $B_j$ cycles to the
$z-$plane (See \cite{Zhivkov}). Now, for $y^6\sim\beta$ in (\ref{eq:sextic}), one can show that the term
$\alpha y\sim\frac{\beta}{{\cal V}^{\frac{1}{9}}}$ and hence can be dropped in the LVS limit. Further,
along $z_*(A_i)$ and $z_*(B_j)$, $y^6\sim\beta$ and thus:
\begin{eqnarray}
\label{eq:ints}
& & \int_{A_i\ {\rm or}\ B_j}\frac{dy}{\sqrt{y^6+\beta}}\sim\beta^{-\frac{1}{3}}
\ _2F_1\left(\frac{1}{3},\frac{1}{2};\frac{4}{3};1\right),\nonumber\\
& & \int_{A_i\ {\rm or}\ B_j}\frac{y dy}{\sqrt{y^6+\beta}}\sim\beta^{-\frac{1}{6}}
\ _2F_1\left(\frac{1}{6},\frac{1}{2};\frac{7}{6};1\right).
\end{eqnarray}
Hence,
\begin{equation}
\label{eq:period_appr}
\Omega\sim\beta^{\frac{1}{2}}\left(\begin{array}{cc}
\beta^{-\frac{1}{6}} & \beta^{-\frac{1}{3}} \\
\beta^{-\frac{1}{6}} & \beta^{-\frac{1}{3}}
\end{array}\right)\left(\begin{array}{cc}
\beta^{-\frac{1}{3}} & \beta^{-\frac{1}{3}} \\
\beta^{-\frac{1}{6}} & \beta^{-\frac{1}{6}}
\end{array}\right)
\sim\left(\begin{array}{cc}
{\cal O}(1) & {\cal O}(1) \\
{\cal O}(1) & {\cal O}(1)
\end{array}\right).
\end{equation}
Hence, one can {\it ignore the $D3-$brane moduli dependence of the period matrix $\Omega$ in the LVS
limit}. Substituting (\ref{eq:period_appr}) into (\ref{eq:dertheta}), one sees that
\begin{equation}
\label{eq:sol_y}
e^{2V_2}\sim\left(\zeta\frac{1}{r_1|z_3|^2}\right)^{\frac{1}{6}},
\end{equation}
in the LVS limit where $\zeta$ encodes the information about the {\it exact} evaluation of
the period matrix. Substituting (\ref{eq:sol_y}) into the second equation of (\ref{eq:eoms2}),
one obtains:
\begin{equation}
\label{eq:sol_x}
e^{2V_1}=\frac{\left(r_2 - \left(1+|z_1|^2+|z_2|^2\right)\left(\frac{\zeta}{
r_1|z_3|^2}\right)^{\frac{1}{6}}\right)\sqrt{\zeta}}{3\sqrt{r_1|z_3|^2}}.
\end{equation}
The geometric K\"{a}hler potential for the divisor $D_5$ in the LVS limit is hence given by:
\begin{eqnarray}
\label{eq:Kaehler_D_5}
& & K\biggl|_{\Sigma_B} =  r_2 - 4\frac{\left(r_2 - \left(1+|z_1|^2+|z_2|^2\right)\left(\frac{\zeta}{
r_1|z_3|^2}\right)^{\frac{1}{6}}\right)\sqrt{\zeta}}{3} -r_2 ln\left[\left(\zeta\frac{1}{r_1|z_3|^2}\right)^{\frac{1}{6}}\right]\nonumber\\
& & +|z_3|^2\left(\frac{\left(r_2 - \left(1+|z_1|^2+|z_2|^2\right)\left(\frac{\zeta}{
r_1|z_3|^2}\right)^{\frac{1}{6}}\right)\sqrt{\zeta}}{3\sqrt{r_1|z_3|^2}}\right)^2\nonumber\\
& & - r_1 ln\left[\frac{\left(r_2 - \left(1+|z_1|^2+|z_2|^2\right)\left(\frac{\zeta}{
r_1|z_3|^2}\right)^{\frac{1}{6}}\right)\sqrt{\zeta}}{3\sqrt{r_1|z_3|^2}}\right] \sim\frac{{\cal V}^{\frac{2}{3}}}{\sqrt{ln {\cal V}}}.\nonumber\\
\end{eqnarray}
Now the extremization of the NLSM Lagrangian w.r.t. the vector superfields corresponding to the divisor $D_4$ -  $z_3=0$ or equivalently $\Phi_4=0$ - yield the following pair of equations:
\begin{eqnarray}
\label{eq:D4_I}
& & b_1x^3 + c_1xy^{-3} = r_1\nonumber\\
& & a_2y + c_2xy^{-3} = r_2.
\end{eqnarray}
where $a_2\equiv|\Phi_1|^2+|\Phi_2|^2+|\Phi_3|^2$, $b_1\equiv3|\Phi_5|^2, c_1\equiv|\Phi_\epsilon|^2, c_2\equiv-3|\Phi_\epsilon|^2, x\equiv e^{2V_1}, y\equiv e^{2V_2}$. In the $\Phi_\epsilon=1$-patch, one gets the following degree-12 equation in $y$:
\begin{equation}
\label{eq:D4_II}
\frac{1}{9}|\Phi_5|^2\left(r_2^3y^9-3r_2^2a_2y^{10}+3r_2a_2^2y^{11}+a_2^3y^{12}\right)-\frac{1}{3}\left[r_2-\left(|\Phi_1|^2+|\Phi_2|^2+|\Phi_3|^2\right)y\right]
=r_1.
\end{equation}
Choosing a scaling (in $z_\epsilon=z_2=1$-patch): $z_4\sim z_{1,2}^9\sim\left(ln {\cal V}\right)^{\frac{1}{4}}$,
$r_1\sim\sqrt{ln {\cal V}}, r_2\sim{\cal V}^{\frac{1}{3}}$, (\ref{eq:D4_II}) would imply:
\begin{equation}
\label{eq:D4_III}
\frac{1}{9}\left({\cal V}y^9-3{\cal V}^{\frac{2}{3}}\left(ln {\cal V}\right)^{\frac{1}{18}}y^{10}+3{\cal V}^{\frac{1}{3}}\left(ln {\cal V}\right)^{\frac{1}{9}}y^{11}+\left(ln{\cal V}\right)^{\frac{1}{6}}y^{12}\right)
-\frac{1}{3}\left({\cal V}^{\frac{1}{3}}-(ln {\cal V})^{\frac{1}{18}}y\right)\sim\sqrt{ln {\cal V}}.
\end{equation}
Hence, if $y\sim\left[\left(ln {\cal V}\right)^{-\frac{1}{2}}{\cal V}^{-\frac{2}{3}}\right]^{\frac{1}{9}}$, i.e., if the $y^9$-term is the dominant term on the LHS of (\ref{eq:D4_III}), then the same is justified as the $y,y^{10}, y^{11}, y^{12}$-terms are respectively ${\cal V}^{-\frac{2}{27}},{\cal V}^{-\frac{2}{27}},{\cal V}^{-\frac{13}{27}}, {\cal V}^{-\frac{8}{9}}$, and hence are sub-dominant w.r.t. the $y^9$ terms and will hence be dropped. One thus obtains:
\begin{equation}
\label{eq:D4_V}
y=e^{2V_2}\sim\left[\frac{3}{r_2^2|z_1^{18}+z_2^{18}-3\phi z_1^6z_2^6|^2}\right]^{\frac{1}{9}},
\end{equation}
which hence yields:
\begin{eqnarray}
\label{eq:KD_4}
& & K\biggl|_{D_4}= \frac{3^{\frac{1}{9}}\,\left( 1 + |{z_1}|^2 +  |{z_2}|^2 \right) }{{\left( {{r_2}}^2\,
        |z_1^{18}+z_2^{18}-3\phi z_1^6z_2^6|^2 \right)  }^{\frac{1}{9}}}  - {\frac{{3^{\frac{1}{9}}\,\left( 1 + |{z_1}|^2 +
          |{z_2}|^2 \right) }{{\left( {{r_2}}^2\,
           |z_1^{18}+z_2^{18}-3\phi z_1^6z_2^6|^2 \right) }^{\frac{1}{9}}}}{3}} \nonumber\\
           & & +   \frac{r_2}{3} +  \frac{{\left( {r_2} - \frac{3^{\frac{1}{9}}\,  \left( 1 + |{z_1}|^2 +
             |{z_2}|^2 \right) }{{\left( {{r_2}}^2\,
              |z_1^{18}+z_2^{18}-3\phi z_1^6z_2^6|^2 \right) }^{\frac{1}{9}}} \right) }^3}{9\,{{r_2}}^2} -
   \frac{{r_2}\,\log \biggl(\frac{3}
       {{{r_2}}^2\,|z_1^{18}+z_2^{18}-3\phi z_1^6z_2^6|^2 }\biggr)}{9}\nonumber\\
       & & -   {r_1}\,\log \Biggl(\frac{3^{\frac{1}{3}}\,\left( {r_2} -
         \frac{3^{\frac{1}{9}}\,\left( 1 + |{z_1}|^2 +
              |{z_2}|^2 \right) }{{\left( {{r_2}}^2\,
               |z_1^{18}+z_2^{18}-3\phi z_1^6z_2^6|^2 \right) }^{\frac{1}{9}}} \right) }{{\left(
          {{r_2}}^2\,|z_1^{18}+z_2^{18}-3\phi z_1^6z_2^6|^2 \right) }^{\frac{1}{3}}}\Biggl)\sim{{\cal V}^{\frac{1}{3}}}{\sqrt{ln {\cal V}}}\nonumber\\
          \end{eqnarray}
The required derivatives of $K\biggl|_{\Sigma_B}$ and $K\biggl|_{\Sigma_S}$ are given in Appendix {\bf A.4}.
\section{The Complete Moduli Space K\"{a}hler Potential}
The complete K\"{a}hler potential is given as under:
\begin{eqnarray}
\label{eq:K}
& & K = - ln\left(-i(\tau-{\bar\tau})\right) - ln\left(i\int_{CY_3}\Omega\wedge{\bar\Omega}\right)\nonumber\\
 & & - 2 ln\Biggl[a\left(T_B + {\bar T}_B - \gamma K_{\rm geom}\right)^{\frac{3}{2}}-a\left(T_S + {\bar T}_S - \gamma K_{\rm geom}\right)^{\frac{3}{2}} + \frac{\chi}{2}\sum_{m,n\in{\bf Z}^2/(0,0)}
\frac{({\bar\tau}-\tau)^{\frac{3}{2}}}{(2i)^{\frac{3}{2}}|m+n\tau|^3}\nonumber\\
& &  - 4\sum_{\beta\in H_2^-(CY_3,{\bf Z})} n^0_\beta\sum_{m,n\in{\bf Z}^2/(0,0)}
\frac{({\bar\tau}-\tau)^{\frac{3}{2}}}{(2i)^{\frac{3}{2}}|m+n\tau|^3}cos\left(mk.{\cal B} + nk.c\right)\Biggr],
\end{eqnarray}
where $n^0_\beta$ are the genus-0 Gopakumar-Vafa invariants for the curve $\beta$ and
$k_a=\int_\beta\omega_a$ are the degrees of the rational curve. Further, to work out the moduli-space metric, one needs to complexify the Wilson line moduli via sections of $N\Sigma_B$ (See \cite{Kachruetal}). Allowing for the possibility of gaugino condensation requires $\Sigma_B$ to be rigid - we hence consider only zero sections of $N\Sigma_B$, i.e., we set $\zeta^A=0$. The complexified Wilson line moduli would then be ${\cal A}_I=ia_I$. For a stack of $N D7$-branes wrapping $D_5$, stricly speaking $\zeta^A$ and $a_I$ are $U(N)$ Lie algebra valued, which implies that they can be written as:
$\zeta^A=(\zeta^A_1)_a U^a + (\zeta_2^A)_{ab}e^{ab}$ and similarly for $a_I$ (See \cite{bifund_ferm}) where $U^a$ and $e^{ab}$ are the generators of the $U(N)$ algebra. Now, restricting the mobile $D3$ brane to $\Sigma_B$ guarantees nullification of the non-perturbative superpotential from gaugino condensation for all values of $N>1$. Hence, we are justified in setting ${\cal N}=1$ - ruling out gaugino condensation in our setup - $\zeta^A$ and $a_I$ are hence not matrix-valued. We then assume that $\zeta^A$  and all components save one of $a_I$ can be stabilized to a zero value; the non-zero component $a_1$ can be stabilized at around ${\cal V}^{-\frac{1}{4}}$. This is justified in a self-consistent manner, in Appendix {\bf A.3}.

There is the issue of using the modular completion of \cite{Grimm} for our setup which includes a $D7$ brane - or a stack of $D7$-branes. First, in our analysis, it is the large contribution from the world-sheet instantons - proportional to the genus-zero Gopakumar-Vafa invariants - that are relevant and not its appropriate form invariant (if  at all) under (a discrete subgroup of) $SL(2,{\bf Z})$ as in \cite{dSetal,Grimm}. Second, we could think of the $D7$ brane as a $(p,q,r)$ seven-brane satisfying the constraint: $pq=\left(\frac{r}{2}\right)^2$, which as per \cite{Bergshoeff_etal} would ensure $SL(2,{\bf Z})$ invariance.

Though the contribution from the matter fields ``$C_{37}$" coming from open strings stretched between the $D3$ and $D7$ branes wrapping $\Sigma_B(\equiv D_5$) for Calabi-Yau orientifolds is not known, but based on results for orientifolds of $\left(T^2\right)^3$ - see \cite{Lustetal} - we guess the following expression:
\begin{equation}
\label{eq:C_37}
\frac{|C_{37}|^2}{\sqrt{T_B}}\sim{\cal V}^{-\frac{1}{36}}|C_{37}|^2,
\end{equation}
which for sub-Planckian $C_{37}$ (implying that they get stabilized at ${\cal V}^{-c_{37}},\ c_{37}>0$) would be sub-dominant relative to contributions from world-sheet instantons, for instance, in (\ref{eq:K}). We will henceforth ignore (\ref{eq:C_37}).


\section{Resolution of the Tension between LVS Cosmology and LVS Phenomenology}

We need to figure a way of obtaining a TeV gravitino when dealing with LVS phenemenology and a $10^{12}$ GeV gravitino when dealing with LVS cosmology within the same setup for the same value of the volume modulus:
${\cal V}\sim10^6$ (in $l_s=1$ units). In this section we give a proposal to do the same.

The gravitino mass is given by:
$m_{\frac{3}{2}}=e^{\frac{K}{2}}W M_p\sim\frac{W}{{\cal V}}M_p$ in the LVS limit. We choose the complex-structure moduli-dependent superpotential to be such that $W\sim W_{n.p}$\footnote{As we have explained in chapter {\bf 3} on LVS cosmology, unlike usual LVS (for which $W_{c.s.}\sim{\cal O}(1)$) and similar to KKLT scenarios (for which $W_{c.s.}\ll 1$), we have $W_{c.s.}\ll 1$ in large volume limit; we would henceforth assume that the fluxes and complex structure moduli have been so fine tuned/fixed that $W_{c.s}\sim \pm W_{\rm ED3}(n^s=1)$ and hence implying $W\sim W_{n.p}$.}. Consider now a single $ED3-$instanton obtained by an $n^s$-fold wrapping of $\Sigma_S$ by a single $ED3-$brane. The holomorphic prefactor appearing in the non-perturbative superpotential that depends on the mobile $D3$ brane's position moduli, has to be a section class of the divisor bundle $[\Sigma_S]$ - and should have a zero of degree $n_s$ at the location of the $ED3$ instanton - see \cite{Ganor1_2}. This will contribute a superpotential of the type:
\begin{eqnarray*}
& & W\sim \left(1 + z_1^{18} + z_2^{18} + z_3^2 - 3\phi_0z_1^6z_2^6\right)^{n_s}e^{in^sT_s}\Theta_{n^s}({\cal G}^a,\tau)\nonumber\\
& & \hskip 0.2in \sim \frac{\left(1 + z_1^{18} + z_2^{18} + z_3^2 - 3\phi_0z_1^6z_2^6\right)^{n^s}}{{\cal V}^{n^s}}.
\end{eqnarray*}
The main idea will be that for a volume modulus fixed at ${\cal V}\sim10^6l_s^6$, during early stages of cosmological evolution, the geometric location of the mobile $D3$-brane on a non-singular elliptic curve embedded within the Swiss-Cheese $CY_3$ that we are considering was sufficient to guarantee that the gravitino was super-massive with a mass of $10^{12}$ GeV as required by cosmological data, e.g., density perturbations. Later, as the $D3$-brane moved  to another non-singluar elliptic curve within the $CY_3$ with the same value of the volume, in the present epoch, one obtains a TeV gravitino as required. Let $z_{i,(0)}$ denote the position moduli of the mobile $D3$-brane. Consider fluctuations about the same given by $\delta z_{i,(0)}$. Defining $P(\{z_{i,(0)}\})\equiv 1 + z_{1,(0)}^{18} + z_{2,(0)}^{18} + z_{3,(0)}^2 - 3\phi_0 z_{1,(0)}^6z_{2,(0)}^6$, one obtains:
\begin{eqnarray}
\label{eq:Wfluc}
& & W\sim\frac{\left(P(\{z_{i,(0)}\}) + \sum_{i=1,2}a_i\delta z_{i,(0)}\right)^{n^s}}{{\cal V}^{n^s}}e^{in^sT_s({\cal G}^a,{\bar{\cal G}}^a;\tau,{\bar\tau}) + i\mu_3l^2\left(z_{i,(0)}{\bar z}_{{\bar j},(0)}a_{i{\bar j}} + z_{i,(0)}z_{j,(0)}\tilde{a}_{ij}\right)}\nonumber\\
& & \times e^{i\mu_3l^2\left(\sum_i\alpha_i\delta z_{i,(0)} + \sum_{\bar i}\beta_{{\bar i}}\delta {\bar z}_{{\bar i}}\right)} \sim {\cal V}^{\alpha n^s - n^s}\left(1 + \frac{\sum_i a_i\delta z_{i,(0)}}{P(\{z_{i,(0)}\})}\right)^{n^s},
\end{eqnarray}
where one assumes $P(\{z_{i,(0)}\})\sim {\cal V}^\alpha$. This yields
$m_{\frac{3}{2}}\equiv e^{\frac{\hat{K}}{2}}|\hat{W}|\sim{\cal V}^{n^s(\alpha - 1) - 1}$.
\begin{enumerate}
\item
{\bf \underline{LVS Cosmology}:} Assume that one is a point in the Swiss-Cheese $CY_3:P(\{z_{i,(0)}\})\sim{\cal V}^{\alpha_{\rm cosmo}}$. Hence,  what we need is: $10^{18 +6(n^s\alpha_{\rm cosmo} - n^s - 1)}\sim10^{12}$, or $\alpha_{\rm cosmo} = 1$ ($n^s\geq2$ to ensure a metastable dS minimum in the LVS limit - see \cite{dSetal}).
Now, either $z_{1,2}^{18}\sim{\cal V}$, i.e., $z_{1,2}\sim{\cal V}^{\frac{1}{18}}<{\cal V}^{\frac{1}{6}}$(as $z_{1,2,3}\leq{\cal V}^{\frac{1}{6}}$) and is hence alright, or
$z_3^2\sim{\cal V}$, i.e., $z_3\sim\sqrt{\cal V}>{\cal V}^{\frac{1}{6}}$ and hence is impossible. Therefore, geometrically if one is at a point $(z_1,z_2,z_3)\sim({\cal V}^{\frac{1}{18}},{\cal V}^{\frac{1}{18}},z_3)$ where $z_3$ (in an appropriate coordinate patch) using (\ref{eq:hypersurface}) satisfies:
\begin{equation}
\label{eq:elliptic1}
\psi_0{\cal V}^{\frac{1}{9}}z_3z_4 - z_3^2 - z_4^3 \sim {\cal V},
\end{equation}
one can generate a $10^{12}$GeV gravitino at ${\cal V}\sim 10^6$!!! Note that (\ref{eq:elliptic1}) is a non-singular elliptic curve embedded in the Calabi-Yau. On redefining $iz_3\equiv y$ and $z_4\equiv x$,  one can compare (\ref{eq:elliptic1}) with the following elliptic
curve over C:
\begin{equation}
\label{eq:elliptic2}
y^2 + a_1xy+a_3y=x^3+a_2x^2+a_4x+a_6,
\end{equation}
for which the $j$-invariant is defined as: $j=\frac{(a_1^2+4a_2)^2 -
24(a_1a_3 + a_4)}{\Delta}$ where the discriminant $\Delta$ is defined as follows - see \cite{Husemoeller} -
\begin{eqnarray}
\label{eq:discdef}
& & \Delta\equiv
-(a_1^2+4a_2)^2(a_1^2a_6 - a_1a_3a_4 + a_2a_3^2+4a_2a_6 - a_4^2) +
9(a_1^2+4a_2)(a_1a_3+2a_4)\nonumber\\
& & (a_3^2+4a_6) - 8(a_1a_3+2a_4)^3-27(a_3^2+4a_6)^2.
\end{eqnarray}
 The discriminant works out to $-\psi_0^4{\cal V}^{\frac{37}{9}}  - 432{\cal V}^2\neq0$, implying that (\ref{eq:elliptic2}) is non-singular.
\item
{\bf \underline{LVS Phenomenology}:} A similar analysis would require: $6(n^s\alpha_{\rm pheno} - n^s - 1) + 18\sim 3$, or $\alpha_{\rm pheno}=1-\frac{3}{2n^s}$, which for $n^s=2$ ($n^s\geq2$ to ensure a metastable dS minimum in the LVS limit) yields $\alpha_{\rm pheno}=\frac{1}{4}$. So, either $z_{1,2}\sim{\cal V}^{\frac{1}{72}}<{\cal V}^{\frac{1}{6}}$ which is fine or $z_3\sim{\cal V}^{\frac{1}{8}}<{\cal V}^{\frac{1}{6}}$ and hence also alright. However, for ${\cal V}\sim10^7l_s^6$, one can show that one ends up with a non-singular elliptic curve embedded inside the Calabi-Yau given by: $\psi_0{\cal V}^{\frac{1}{21}}z_3z_4 - (z_3^3 + z_4^2)\sim{\cal V}^{\frac{3}{7}}$. It is hence more natural to thus choose $z_{1,2}\sim{\cal V}^{\frac{1}{72}}$ over $z_3\sim{\cal V}^{\frac{1}{8}}$. Hence, the mobile $D3$ brane moves to the elliptic curve embedded inside the Swiss-Cheese Calabi-Yau:
\begin{equation}
\label{eq:elliptic3}
\psi_0{\cal V}^{\frac{1}{36}}z_3z_4 - (z_3^2 + z_4^3)\sim{\cal V}^{\frac{1}{4}},
\end{equation}
one obtains a TeV gravitino. One can again see that the discriminant
corresponding to (\ref{eq:elliptic3}) is $-\psi_0^4{\cal V}^{\frac{13}{36}} - 432\sqrt{\cal V}\neq0$ implying that the elliptic curve (\ref{eq:elliptic3}) is non-singular.
\end{enumerate}
The volume of the Calabi-Yau can be extremized at one value - $10^6 \, l_s^6$ - for varying positions of the mobile $D3$-brane as discussed above for the following three reasons. Taking the small divisor's volume modulus and the Calabi-Yau volume modulus as independent variables, (a) the $D3$-brane position moduli enter the holomorphic prefactor - the section of the divisor bundle - and hence the overall potential will be proportional to the modulus square of the same and the latter does not influence the extremization condition of the volume modulus,  (b) in consistently taking the large volume limit as done in this chapter, the superpotential is independent of the Calabi-Yau volume modulus, and (c) $vol(\Sigma_S)\geq\mu_3{\cal V}^\beta$ for values of $\beta$ taken in this chapter corresponding to different positions of the $D3$-brane. Combining these three reasons, one can show that the extremization condition for the volume modulus is independent of the position of the $D3$-brane position moduli.

\section{Realizing Order one YM Couplings}
In order to realize order one YM couplings, we construct appropriate local  involutively-odd harmonic one-forms $\omega$ on the big divisor lying in the cokernel of the pullback of the immersion map applied to $H^{(1,0)}_-$ in the large volume limit, the Wilson line moduli provide a competing contribution to the gauge kinetic function as compared to the volume of the big divisor and the possible cancelation results in realizing order one gauge couplings $g_a\sim {\cal O}(1)$ .

Let $\omega=\omega_1(z_1,z_2)dz_1\in H^{(1,0)}_{\partial,-}(\Sigma_B:x_5=0)$ - this implies that $\omega_1(z_1\rightarrow-z_1,z_2\rightarrow z_2)=\omega_1(z_1,z_2)$. Then
$\partial(=dz_i\partial_i)\omega=0$ and $\omega$ must not be exact. Let
$\partial\omega=(1+z_1^{18}+z_2^{18}+z_3^3-\phi_0 z_1^6z_2^6)^2dz_1\wedge dz_2$ - it is exact on $\Sigma_B$ but not at any other point in the Calabi-Yau. This implies that restricted to $\Sigma_B$, $\frac{\partial\omega_1}{\partial z_2}|_{\Sigma_B}\sim(\phi_0 z_1^6z_2^6 - z_1^{18} - z_2^{18}-z_3^3)^2$. Taking $z_3$ to be around ${\cal V}^{\frac{1}{6}}$ - this would actually correspond to the location of the mobile $D3$-brane in the Calabi-Yau which for concrete calculations and its facilitation will eventually be taken to lie at $({\cal V}^{\frac{1}{36}}e^{i\theta_1},{\cal V}^{\frac{1}{36}}e^{i\theta_2},{\cal V}^{\frac{1}{6}}e^{i\theta_3})$ - one sees that in the LVS limit,
\begin{eqnarray}
\label{eq:harm1formD5}
& & \omega_1(z_1,z_2;z_3\sim{\cal V}^{\frac{1}{6}})\biggl|_{\Sigma_B}\nonumber\\
& & =-2\frac{\phi_0}{25}z_1^6z_2^{25}-(z_1^{18}+z_3^3)^2z_2 - \frac{z_2^{37}}{37} + \frac{\phi_0^2}{13}z_1^{12}z_2^{13} + 2(z_1^{18}+z_3^3)(\frac{z_2^{19}}{19}-\frac{\phi_0}{7}z_1^6z_2^7);\nonumber\\
\end{eqnarray}
this indeed does satisfy the required involutive property of being even. Now, the Wilson-line moduli term is: $i\kappa_4^2\mu_7\int_{\Sigma_B}i^*\omega\wedge A^I\wedge{\bar A}^{\bar J}a_I{\bar a}_{\bar J}$, where $\omega\in H^{(1,1)}_+(\Sigma_B)$ could be taken to be $i(dz_1\wedge d{\bar z_1} \pm dz_2\wedge d{\bar z_2})$. Hence,
\begin{eqnarray}
\label{eq:C11bar}
& &  C^{1{\bar 1}}\sim\int_{\{3\phi_0z_1^6z_2^6-z_1^{18}-z_2^{18}\sim\sqrt{\cal V}\}\subset\Sigma_B}|\omega_1|^2dz_1\wedge d{\bar z_1}\wedge dz_2\wedge d{\bar z_2}\bigg|_{|z_3|\sim{\cal V}^{\frac{1}{6}}}\nonumber\\
& & \sim \int_{\{3\phi_0z_1^6z_2^6-z_1^{18}-z_2^{18}\sim\sqrt{\cal V}\}\subset\Sigma_B} \ \ \ \ \bigg|-2\frac{\phi_0}{25}z_1^6z_2^{25}-(z_1^{18}+\sqrt{\cal V})^2z_2 - \frac{z_2^{37}}{37} + \frac{\phi_0^2}{13}z_1^{12}z_2^{13} \nonumber\\
& & + 2(z_1^{18}+\sqrt{\cal V})(\frac{z_2^{19}}{19}-\frac{\phi_0}{7}z_1^6z_2^7)\bigg|^2 dz_1\wedge d{\bar z_1}\wedge dz_2\wedge d{\bar z_2}\nonumber\\
& & \sim\bigg|\int_{z_1\sim{\cal V}^{\frac{1}{36}}e^{i\theta_1}}z_1^{18}dz_1\bigg|^2
\bigg|\int_{z_2\sim{\cal V}^{\frac{1}{36}}e^{i\theta_2}}z_2^{19}dz_2\bigg|^2 \sim{\cal V}^{\frac{3}{2}}{\rm vol}(\Sigma_B).\nonumber\\
\end{eqnarray}
 Hence, if the Wilson line modulus $a_1$ is stabilized at around ${\cal V}^{-\frac{1}{4}}$, then
$i\kappa_4^2\mu_7\int_{\Sigma_B}i^*\omega\wedge A^I\wedge{\bar A}^{\bar J}a_I{\bar a}_{\bar J}
\sim{\rm vol}(\Sigma_B)$. The fact that this is indeed possible, will be justified in Appendix {\bf A.3}. Even with a more refined evaluation of the integral in (\ref{eq:C11bar}) to obtain $C_{1{\bar 1}}$, the results on soft masses and soft SUSY parameters in the rest of the chapter, would qualitatively remain the same. This implies that the gauge couplings corresponding to the gauge theory living on a stack of $D7$-branes wrapping $\Sigma_B$ will be given by:
\begin{equation}
\label{eq:g_YM}
g_a^{-2}={\bf Re}(T_B)\sim \mu_3{\cal V}^{\frac{1}{18}}\sim ln{\cal V},
\end{equation}
implying a finite $g_a$. In the absence of $\alpha^\prime$-corrections, strictly speaking $g_a^{-2}={\bf Re}(T_B) - {\cal F}{\bf Re}(i\tau)$,
${\cal F}\equiv {\cal F}^\alpha {\cal F}^\beta \kappa_{\alpha\beta} + \tilde{\cal F}^a\tilde{\cal F}^b\kappa_{ab}$ (refer to \cite{V_D7_fl}) where ${\cal F}^\alpha$ and
${\cal F}^a$ are the components of the $U(1)$ two-form flux on the world-volume of $D7$-branes wrapping $\Sigma_B$ expanded in the basis
$i^*\omega_\alpha, \omega_\alpha\in H^{(1,1)}(CY_3)$ and $\tilde{\omega}_a\in coker\left(H^{(1,1)}_-(CY_3)\stackrel{i^*}{\rightarrow}H^{(1,1)}_-(\Sigma_B)\right)$, and $\kappa_{\alpha\beta}=\int_{\Sigma_B}i^*\omega_\alpha\wedge i^*\omega_\beta$ and $\kappa_{ab}=\int_{\Sigma_B}\tilde{\omega}_a\wedge\tilde{\omega}_b$. In the ``dilute flux approximation", we disregard the contribution coming from ${\cal F}$ as compared to the ${\cal V}^{\frac{1}{18}}$ coming from ${\bf Re}(T_B)$. Also, from the first reference in \cite{susyinitials}, the effective gauge couplings $g_a^{-2}$ for an observable gauge group $G_a$ including renormalization and string-loop corrections, to all orders, at an energy scale $\nu>>m_{\frac{3}{2}}$ satisfies the following equation:
\begin{eqnarray}
\label{eq:g_a-EXACT}
& & g_a^{-2}\left(\Phi^m,{\bar\Phi}^{\bar m};\nu\right) \nonumber\\
& & = {\bf Re} f_a + \frac{\sum_r n_rT_a(r) - 3 T_{\rm adj}(G_a)}{8\pi^2}ln\frac{M_p}{\nu} + \frac{\sum_r n_rT_a(r) -  T_{\rm adj}(G_a)}{16\pi^2} \hat{K}\left(\Phi^m,{\bar\Phi}^{\bar m}\right)\nonumber\\
 & & + \frac{T_{\rm adj}(G_a)}{8\pi^2} ln \left[g_a^{-2}\left(\Phi^m,{\bar\Phi}^{\bar m};\nu\right)\right]- \sum_r\frac{T_a(r)}{8\pi^2} ln\ {\rm det}\left[\hat{K}_{i{\bar j}}\left(\Phi^m,{\bar\Phi}^{\bar m}\right)\right].
\end{eqnarray}
$\Phi^m$ denoting the closed string moduli, $f_a$ being the gauge ($G_a$) kinetic function, $r$ denoting a representation for an observable gauge group $G_a$, $n_r$ denoting the number of matter fields transforming under the representation $r$ of $G_a$ and $T$ denoting the trace of the square of the generators in the appropriate representations. Given that we have been working in the approximation:
$\mu_3{\cal V}^{\frac{1}{18}}\sim ln{\cal V}$ (justified by ${\cal V}\sim10^6$), from (\ref{eq:K2}), (\ref{eq:Khat}) and (\ref{eq:detKhat}) one sees that the  third and fifth terms on the RHS of (\ref{eq:g_a-EXACT}) are proportional to $ln{\cal V}\sim\mu_3{\cal V}^{\frac{1}{18}}\sim f_a$, implying thereby that there are no major modifications in the tree-level results for the gauge couplings.

\section{Summary and Discussion}
In this chapter, we discussed several phenomenological issues in the context of LVS Swiss-Cheese orientifold compactifications of Type IIB with the inclusion of a single mobile space-time filling $D3$-brane and stack(s) of $D7$-brane(s) wrapping the ``big" divisor $\Sigma_B$ along with  supporting $D7$-brane fluxes (on two-cycles homologically non-trivial within the big divisor, and not the Calabi-Yau). Interestingly we realized many phenomenological implications different from the LVS studies done so far in the literature.

We started with the extension of our LVS Swiss-Cheese cosmology setup with the inclusion of a mobile spacetime filling $D3$-brane and stacks of $D7$-branes wrapping the ``big" divisor $\Sigma_B$ and on the geometric side to enable us to work out the complete K\"{a}hler potential, we calculated the geometric K\"{a}hler potential (of the two divisors $\Sigma_S$ and $\Sigma_B$) for Swiss-Cheese Calabi-Yau ${\bf{WCP}}^4[1,1,1,6,9]$ using its toric data and GLSM techniques in the large volume limit. The geometric K\"{a}hler potential is first expressed, using a general theorem due to Umemura, in terms of genus-five Siegel Theta functions or in the LVS limit genus-four Siegel Theta functions. Later using a result due to Zhivkov, for purposes of calculations for our chapter, we express the same in terms of derivatives of genus-two Siegel Theta functions.

Then we proposed a possible geometric resolution for the long-standing tension between LVS cosmology and LVS phenomenology : to figure out a way of obtaining a TeV gravitino when dealing with LVS phenemenology and a $10^{12}$ GeV gravitino when dealing with LVS cosmology in the early inflationary epoch of the universe, within the same setup.  The holomorphic pre-factor coming from the space-time filling mobile D3-brane position moduli  - section of (the appropriate) divisor bundle - plays a crucial role and we have shown that as the mobile space-time filling $D3$-brane moves from a particular non-singular elliptic curve embedded in the Swiss-Cheese Calabi-Yau to another non-singular elliptic curve, it is possible to obtain $10^{12}GeV$ gravitino during the primordial inflationary era supporting the cosmological/astrophysical data as well as a $TeV$ gravitino in the present era supporting the required SUSY breaking at $TeV$ scale within the same set up, for the same volume of the Calabi-Yau stabilized at around $10^6$ (in $l_s=1$ units). This way the string scale involved for our case is $\sim O(10^{15})$ GeV which is nearly of the same order as GUT scale. In the context of soft SUSY breaking, we have obtained the gravitino mass $m_{3/2}\sim  O(1-10^3)$ TeV  with ${\cal V}\sim 10^{6} {l_s}^6$ in our setup.

In the context of realizing the Standard Model (SM) gauge coupling $g_{YM}$ $ \sim O(1)$ in the LVS models with D7-branes, usually models with the D7-branes wrapping the smaller divisor have been proposed so far, as D7-branes wrapping the big divisor would produce very small gauge couplings. In our setup, we have realized $\sim O(1)$ $g_{YM}$ with D7-branes wrapping the big divisor in the rigid limit (i.e. considering zero sections of the normal bundle of the big divisor to prevent any obstruction to chiral matter resulting from adjoint matter - corresponding to fluctuations of the wrapped $D7$-branes within the Calabi-Yau - giving mass to open strings stretched between wrapped $D7$-branes)  implying the new possibility of supporting SM on D7-branes wrapping the big divisor. This has been possible because after constructing appropriate local  involutively-odd harmonic one-forms on the big divisor lying in the cokernel of the pullback of the immersion map applied to $H^{(1,0)}_-$ in the large volume limit, the Wilson line moduli provide a competing contribution to the gauge kinetic function as compared to the volume of the big divisor. This requires the complexified Wilson line moduli to be stabilized at around ${\cal V}^{-\frac{1}{4}}$ (which has been justified in Appendix {\bf A.3}). Note, similar to the case of local models corresponding to wrapping of $D7$-branes around the small divisor, our model is also local in the sense that the involutively-odd one-forms are constructed locally around the location of the mobile $D3$-brane restricted to (the rigid limit of) $\Sigma_B$. The detailed calculations after incorporating the effects of the motion of spacetime filling mobile $D3$-brane away from the big divisor $\Sigma_B$ and hence inclusion of subsequent induced gaugino-condensation effects might provide some more interesting string phenomenology in our LVS Swiss-Cheese setup.

%% file: chap5.tex
\chapter{Some Issues in D3/D7 Swiss-Cheese Phenomenology}
\markboth{nothing}{\bf 5. Some Issues in D3/D7 Swiss-Cheese Phenomenology}

\hskip1in{\it{`` While we would like to believe that the fundamental laws of Nature are symmetric,
a completely symmetric world would be rather dull, and as a matter of fact, the
real world is not perfectly symmetric. "}} - Anonymous\footnote{This has been taken from the book ``Quantum Field Theory in a Nutshell" - A. Zee.}\\

\section{Introduction}
In the context of embedding (MS)SM  and realizing its matter content from string phenomenology, the questions of supersymmetry breaking and its transmission to the visible sector are among the most outstanding challenges. The supersymmetry breaking is mainly controlled by the moduli potentials while the related information is transmitted  by the coupling of supersymmetry-breaking fields to the visible sector matter fields. The breaking of supersymmetry is supposed to occur in a hidden sector and is encoded in a set of soft terms. This is communicated to the visible sector (MS)SM via different mediation processes (e.g. gravity mediation, anomaly mediation, gauge mediation) among which although none is clearly preferred, gravity mediation is the most studied one due to its efficient computability. However, non-universality in gravity mediation of supersymmetry breaking to the visible sector has been a problematic issue that has been addressed in (see \cite{mirrormediation,conloncal}) with the arguments that the K\"{a}hler moduli sector (which controls the supersymmetry-breaking) and the complex structure moduli sector (which sources the flavor) are decoupled at least at the tree level resulting the flavor universal soft-terms, though it has been argued that the non-universality can appear at higher order. 

The study of supersymmetry-breaking in string theory context has been initiated long back \cite{susyinitials} and enormous amount of work has been done in this regard (see \cite{conloncal,Quevedosusy2,BBHL,towardsrealvacua,susybreakingKKLT,susykklt2} and references therein). A more controlled investigation of supersymmetry-breaking could be started only after all moduli got stabilized with inclusion of fluxes along with non-perturbative effects.  Since it is possible to embed the chiral gauge sectors (like that of the (MS)SM) in D-brane Models with fluxes, the study of $D$-brane Models have been fascinating  since the discovery of D-branes \cite{SMreview,SM2,QuevedoMSSM,SM3}. In a generic sense, the presence of fluxes generate the soft supersymmetry-breaking terms, the soft terms in various models in the context of gauge sectors realized on fluxed D-branes have been calculated \cite{susybreakingKKLT,susykklt2,granagrimm,ibanezuranga,Ibanez,Lustetal}. In the context of $dS$ realized in the KKLT setup, the uplifting term from the $\overline D3$- brane causes the soft supersymmetry-breaking; (also see \cite{susybreakingKKLT,susykklt2} for KKLT type models).

Similar to the context of string cosmology, the LVS models have been realized to be exciting steps towards realistic supersymmetry-breaking \cite{Balaetal2,conloncal,towardsrealvacua,Balaetal1,conlonLVSsusy,quevedojan09} with some natural advantages such as the large volume not only suppresses the string scale but also the gravitino mass and results in the hierarchically small scale of supersymmetry-breaking. Also unlike the KKLT models in which the anomaly mediated soft terms are equally important to that of the gravity mediated one \cite{susybreakingKKLT}, in some of the Large Volume models, it has been found that the gaugino mass contribution coming from gravity mediation dominates to the anomaly mediation one (the same being suppressed by the standard loop factor) \cite{conloncal,towardsrealvacua} and the same can be expected for the other soft masses as well. In the models having branes at  singularities, it has been argued that at the leading order, the soft terms vanish for the no-scale structure  which gets broken at higher orders with the inclusion of (non-)perturbative $\alpha^{\prime}$ and loop-corrections to the K\"{a}hler potential resulting in the non-zero soft-terms at higher orders. In the context of LVS phenomenology in such models with D-branes at singularities, it has been argued that all the leading order contributions to the soft supersymmetry-breaking (with gravity as well as anomaly mediation processes) still vanish and the non-zero soft terms have been calculated in the context of gravity mediation with inclusion of loop-corrections \cite{towardsrealvacua}. In the context of type IIB LVS Swiss-Cheese orientifold compactifications within $D3/D7$-branes setup, soft terms have been calculated in \cite{conlonLVSsusy}. The supersymmetry breaking with both D-term and F-term and some cosmological issues have been discussed in \cite{quevedojan09}.

As in the usual Higgs mechanism, fermion masses are generated by electroweak symmetry breaking through giving VEVs to Higgs(es) and there has been proposals for realizing fermion masses from a superstring inspired model using Higgs-like mechanism. In the context of realizing fermion masses in ${\cal N}=1$ type IIB orientifold compactifications, one has to introduce open string moduli and has to know the explicit K\"{a}hler potential and superpotential for matter fields, which makes the problem more complicated. Further the compelling evidences of non-zero neutrino masses and their mixing has attracted several minds for almost a decade, as it support the idea why one should think about physics beyond something which is experimentally well tested: the Standard Model. Also, the flavor conversion of solar, atmospheric, reactor, and accelerator neutrinos are convincing enough for nonzero masses of neutrinos and their mixing among themselves similar to that of quarks, provides the first evidence of new physics beyond the standard model.This has motivated enormous amount of activities not only towards particle physics side but also towards cosmology side (like dark energy and dark matter studies) which can be found in plenty of review articles \cite{Mohapatra_group}. Although there has been several aspects for theoretical realization of non-zero neutrino masses with its Dirac type (e.g. see \cite{dirac_neutrino2} and references therein) as well as Majorana type origin, however the models with sea-saw mechanism giving small Majorana neutrino masses has been among the most studied ones (see \cite{Mohapatra_group,conlon_neutrino} and reference therein). In the usual sea-saw mechanisms, a high intermediate scale of right handed neutrino (where some new physics starts) lying between TeV and GUT scale, is involved.
In fact, the mysterious high intermediate scale ($10^{11}-10^{15}$ GeV) required in generating small majorana neutrino masses via sea-saw mechanism has a natural geometric origin in the class of large volume models \cite{conlon_neutrino}.

The issue of proton stability which is a generic prediction of Grand unified theories, has been a dramatic outcome of Grand unified theories beyond SM. Although proton decay has not been experimentally observed, usually in Grand unified theories which provide an elegant explanation of various issues of real wold physics, the various decay channels are open due to higher dimensional operators violating baryon (B) numbers. However the life time of the proton (in decay channels) studied in various models has been estimated to be quite large (as $\tau_p\sim M_{X}$ with $M_X$ being some high scale) \cite{Prot_Decay_review}. Further, studies of dimension-five and dimension-six operators relevant to proton decay in SUSY GUT as well as String/M theoretic setups, have been of great importance in obtaining estimates on the lifetime of the proton (See \cite{Prot_Decay_review}).

So far, to our knowledge, a {\it single} framework which is able to reproduce the fermionic mass scales relevant to the quarks/leptons as well as the neutrinos and is able to demonstrate proton stability, has been missing and has remained a long-standing problem. It is the aim of this chapter to address these issues in a single string theoretic framework.

In this chapter, building on our  type IIB D3/D7 Swiss-Cheese orientifold large volume setup \cite{D3_D7_Misra_Shukla} discuss in previous chapter {\bf 4}, in section {\bf 2}, we start with the discussion of soft supersymmetry breaking and estimate gravitino/gaugino masses along with various supersymmetry  breaking parameters, like the masses of the matter fields, the $\mu$ and the physical $\hat{\mu}$ parameters, the Physical Yukawa couplings $\hat{Y}_{ijk}$ and the trilinear $A_{ijk}$-terms, and the $\hat{\mu}B$-parameters at an intermediate scale (which is about a tenth of the GUT scale). In Section {\bf 3} we discuss RG running of gaugino masses and one-loop RG running of  squark/slepton masses in mSUGRA-like models to the EW scale in the large volume limit. We identify open string or matter moduli with (MSSM) Higgses and squarks and sleptons. Based on identifications in section {\bf 3}, we explore on the possibility of realizing first two generation fermion masses and discuss the realization of non-zero ($\stackrel{<}{\sim} 1eV$) neutrino  masses by lepton number violating non-renormalizable dimension five operators along with an estimate of proton life time in the context of proton stability in section {\bf 4}. Finally we summarize the chapter with conclusions and related discussions in section {\bf 5}.

\section{Soft Supersymmetry Breaking Parameters}

The computation  of soft supersymmetry parameters are related to the expansion of the complete K\"{a}hler potential and the superpotential for open- and closed-string moduli as a power series in the open-string (the ``matter fields") moduli.  The power series is conventionally about zero values of the matter fields. In our setup, the matter fields - the mobile space-time filling $D3$-brane position moduli in the Calabi-Yau (restricted for convenience to the big divisor $\Sigma_B$) and the complexified Wilson line moduli arising due to the wrapping of $D7$-brane(s) around four-cycles - take values (at the extremeum of the potential) respectively of order
${\cal V}^{\frac{1}{36}}$ and ${\cal V}^{-\frac{1}{4}}$, which are finite. We will consider the soft supersymmetry parameters corresponding to expansions of the complete K\"{a}hler potential and the superpotential as a power series in fluctuations about the aforementioned extremum values of the open-string moduli.
The fluctuations around the extremum values of $z_{1,2}$ and ${\cal A}_1$ are: \begin{eqnarray}
\label{eq:fluctuations}
& & z_{1,2}={\cal V}^{\frac{1}{36}} + \delta z_{1,2},\ \ {\cal A}_1={\cal V}^{-\frac{1}{4}}+\delta {\cal A}_1.
\end{eqnarray}
Using (\ref{eq:K_fluc_1}) - similar to \cite{mirage} - and the appropriate cancelation between Wilson line moduli contribution and ``big" divisor volume;
\begin{eqnarray}
\label{eq:Xi_T}
& & {\cal T}_B(\sigma^B,{\bar\sigma ^B};{\cal G}^a,{\bar{\cal G}^a};\tau,{\bar\tau}) + \mu_3{\cal V}^{\frac{1}{18}} + i\kappa_4^2\mu_7C_{1{\bar 1}}{\cal V}^{-\frac{1}{2}} - \gamma\left(r_2 + \frac{r_2^2\zeta}{r_1}\right)\sim \mu_3{\cal V}^{\frac{1}{18}},\nonumber\\
& & {\cal T}_S(\sigma^S,{\bar\sigma ^S};{\cal G}^a,{\bar{\cal G}^a};\tau,{\bar\tau}) + \mu_3{\cal V}^{\frac{1}{18}}  - \gamma\left(r_2 + \frac{r_2^2\zeta}{r_1}\right)\sim\mu_3{\cal V}^{\frac{1}{18}}\sim ln{\cal V}.
\end{eqnarray}
one arrives at the following expression for the K\"{a}hler potential:
\begin{eqnarray}
\label{eq:K2}
& & K \biggl(\left\{\sigma^b,{\bar\sigma}^B;\sigma^S,{\bar\sigma}^S;{\cal G}^a,{\bar{\cal G}}^a;\tau,{\bar\tau}\right\};\left\{z_{1,2},{\bar z}_{1,2};{\cal A}_1,{\bar{\cal A}_1}\right\}\biggr)\nonumber\\
& & \sim - ln\left(-i(\tau-{\bar\tau})\right) - ln\left(i\int_{CY_3}\Omega\wedge{\bar\Omega}\right) - 2\ ln\ \Xi + \bigl(\left(|\delta z_1|^2 + |\delta z_2|^2 + \delta z_1{\bar\delta z_2} + \delta z_2{\bar\delta z_1}\right)\nonumber\\
& & K_{z_i{\bar z}_j} + ((\delta z_1)^2 + (\delta z_2)^2)Z_{z_iz_j} + c.c\bigr) + \bigl(|\delta{\cal A}_1|^2K_{{\cal A}_1\bar{\cal A}_1} + (\delta {\cal A}_1)^2 Z_{{\cal A}_1{\cal A}_1} + c.c \bigr)\nonumber\\
   & & + \bigl(\left(\delta z_1\delta{\bar{\cal A}_1} + \delta z_2\delta{\bar{\cal A}_1} \right)
K_{z_i\bar{\cal A}_1} + c.c \bigr) + \bigl((\delta z_1\delta{\cal A}_1 + \delta z_2\delta{\cal A}_1 ) Z_{z_i{\cal A}_1} +  c.c.\bigr) + ...,
\end{eqnarray}
where $\Xi\sim\sum_\beta n^0_\beta$ and $K_{z_i{\bar z}_j}, Z_{z_iz_j},K_{{\cal A}_1\bar{\cal A}_1},Z_{{\cal A}_1{\cal A}_1},K_{z_i\bar{\cal A}_1}$ and $Z_{z_i{\cal A}_1}$ are defined in Appendix {\bf A.5-A.7}. Using $\gamma\sim\kappa_4^2T_3$(See \cite{Maldaetal_Wnp_pref}) $\sim\frac{1}{\cal V}$, the K\"{a}hler matrix
\begin{equation}\hat{K}_{i{\bar j}}\equiv\frac{\partial^2 K \left(\left\{\sigma^b,{\bar\sigma}^B;\sigma^S,{\bar\sigma}^S;{\cal G}^a,{\bar{\cal G}}^a;\tau,{\bar\tau}\right\};\left\{\delta z_{1,2},{\bar\delta}{\bar z}_{1,2};\delta{\cal A}_1,{\bar\delta}{\bar{\cal A}_1}\right\}\right)}{\partial C^i{\bar\partial} {\bar C}^{\bar j}}|_{C^i=0}\end{equation} - the matter field fluctuations denoted by $C^i\equiv \delta z_{1,2},\delta{\cal A}_1$ -  is given by:
\begin{equation}
\label{eq:Khat}
\hat{K}_{i{\bar j}}\sim\left(\begin{array}{ccc}
A_{z_1z_1}\frac{{\cal V}^{\frac{1}{36}}}{\sum_\beta n^0_\beta} & A_{z_1z_2}\frac{{\cal V}^{\frac{1}{36}}}{\sum_\beta n^0_\beta} & A_{z_1a_1}\frac{{\cal V}^{\frac{11}{12}}}{\sum_\beta n^0_\beta}\\
{\bar A}_{z_1z_2}\frac{{\cal V}^{\frac{1}{36}}}{\sum_\beta n^0_\beta} & A_{z_2z_2}\frac{{\cal V}^{\frac{1}{36}}}{\sum_\beta n^0_\beta} & A_{z_2a_1}\frac{{\cal V}^{\frac{11}{12}}}{\sum_\beta n^0_\beta} \\
{\bar A}_{z_1a_1}\frac{{\cal V}^{\frac{11}{12}}}{\sum_\beta n^0_\beta} & {\bar A}_{z_2a_1}\frac{{\cal V}^{\frac{11}{12}}}{\sum_\beta n^0_\beta} & A_{a_1a_1}\frac{{\cal V}^{\frac{65}{36}}}{\sum_\beta n^0_\beta}
\end{array}\right).
\end{equation}
To work out the physical $\mu$ terms, Yukawa couplings, etc., one needs to diagonalize (\ref{eq:Khat}) and then work with corresponding diagonalized matter fields. To simplify, we have assumed $A_{z_iz_j}, A_{z_ia_1}$ to be real. One can show that (\ref{eq:Khat}) has the following two sets of eigenvalues (the second being two-fold degenerate) and respective eigenvectors in large volume limit:
\begin{equation}
\label{eq:ev1_2}
\underline{Eigenvalue\ 1}\sim\frac{1}{\sum_\beta n^0_\beta}\left(A_{a_1a_1}{\cal V}^{\frac{65}{36}} + \frac{\alpha_4}{3A_{a_1a_1}}\right).
\end{equation}$\underline{Eigenvalue\ 2 = Eigenvalue\ 3}$
\begin{eqnarray}
\label{eq:ev2_2}
\sim \frac{{\cal V}^{\frac{1}{36}}}{\sum_\beta n^0_\beta}\left(4(A_{z_1z_1} + A_{z_2z_2}) - \frac{2\left(-({A_{z_1z_1}}+{A_{z_2z_2}}) {A_{a_1a_1}}+3  {A_{z_1a_1}}^2+3 {A_{z_2a_1}}^2\right)}{A_{a_1a_1}}\right).\nonumber\\
\end{eqnarray}
$\underline{Eigenvectors}$: The eigenvector corresponding to the first eigenvalue (\ref{eq:ev1_2}) is:
   \begin{equation}
   \label{eq:eigenvector1}
   \left(\begin{array}{c} \beta_1{\cal V}^{-\frac{8}{9}} \\ \beta_2{\cal V}^{-\frac{8}{9}} \\ 1
   \end{array}\right) \, \, \, \, \, {\rm where \, \beta's \,  are \,  order \, one \, constants.}
   \end{equation}
which is already normalized to unity. Now for the two-fold degeneracy of the second eigenvalue (\ref{eq:ev2_2}), of the three equations implied by:
\begin{equation}
\label{eq:ev2_eqn1}
\hat{K}\left(\begin{array}{c}
\alpha_1 \\ \alpha_2 \\ \alpha_3 \end{array}\right)=\Lambda_2\frac{{\cal V}^{\frac{1}{36}}}{\sum_\beta n^0_\beta}\left(\begin{array}{c}
\alpha_1 \\ \alpha_2 \\ \alpha_3 \end{array}\right),
\end{equation}
only one equation is independent, say:
\begin{equation}
\label{eq:ev2_eqn2}
\alpha_1(A_{z_1z_1} - \Lambda_2) + \alpha_2 A_{z_1z_2} + \alpha_3 A_{z_1a_1} {\cal V}^{\frac{8}{9}} = 0.
\end{equation}
Two independent solutions to (\ref{eq:ev2_eqn2}) are:
\begin{eqnarray}
\label{eq:ev2_eqn3}
& & \alpha_1=0, \alpha_2=\frac{A_{z_1a_1}}{A_{z_1z_2}}{\cal V}^{\frac{8}{9}}\alpha_3;\, \, \alpha_2=0, \alpha_1=\frac{A_{z_1a_1}}{(A_{z_1z_1} - \Lambda_2)}{\cal V}^{\frac{8}{9}}\alpha_3.
\end{eqnarray}
Thus, the following are the remaining two linearly independent eigenvectors of (\ref{eq:Khat}):
\begin{equation}
\label{eq:eigenvectors2_3}
\left(\begin{array}{c}0 \\ 1\\ \lambda_1{\cal V}^{-\frac{8}{9}}\end{array}\right),\ \left(\begin{array}{c}
1\\ 0\\ \lambda_2{\cal V}^{-\frac{8}{9}}\end{array}\right) \, \, \, \, \, {\rm where \, \lambda's \,  are \,  order \, one \, constants}.
\end{equation}
In the LVS limit, (\ref{eq:eigenvector1}) and (\ref{eq:eigenvectors2_3}) form an orthonormal set of eigenvectors corresponding to $\hat{K}$. Hence, for evaluating the physical $\mu$ terms, Yukawa couplings, etc., we will work with the following set of (fluctuation) fields:
\begin{eqnarray}
\label{eq:diagonal}
& & \delta\tilde{{\cal A}_1}\equiv (\beta_1\delta z_1 + \beta_2\delta z_2){\cal V}^{-\frac{8}{9}} +
\delta{\cal A}_1,\nonumber\\
& & \delta{\cal Z}_1\equiv \delta z_1 + \lambda_2\delta{\cal A}_1{\cal V}^{-\frac{8}{9}},\nonumber\\
& & \delta{\cal Z}_2\equiv \delta z_2 + \lambda_1\delta{\cal A}_1{\cal V}^{-\frac{8}{9}}.
\end{eqnarray}
For purposes of evaluation of the physical $\mu$ terms, Yukawa couplings, etc., we will need the following expressions for the square-root of the elements of the diagonalized $\hat{K}$ in the basis of (\ref{eq:diagonal}):
\begin{eqnarray}
& & \sqrt{\hat{K}_{\tilde{{\cal A}_1}{\bar{\tilde{{\cal A}_1}}}}}\sim\sqrt{\frac{1}{\sum_\beta n^0_\beta}\left(A_{a_1a_1}{\cal V}^{\frac{65}{36}} + \frac{\alpha_4}{3A_{a_1a_1}}\right)}
\sim\frac{{\cal V}^{\frac{65}{72}}}{\sqrt{\sum_\beta n^0_\beta}}; \ \ \sqrt{\hat{K}_{Z_i{\bar Z}_i}}\nonumber\\
& & \sim\sqrt{\frac{{\cal V}^{\frac{1}{36}}}{\sum_\beta n^0_\beta}\left(4(A_{z_1z_1} + A_{z_2z_2}) - \frac{2\left(-({A_{z_1z_1}}+{A_{z_2z_2}}) {A_{a_1a_1}}+3  {A_{z_1a_1}}^2+3 {A_{z_2a_1}}^2\right)}{A_{a_1a_1}}\right)}\nonumber\\
   & & \sim\frac{{\cal V}^{\frac{1}{72}}}{\sqrt{\sum_\beta n^0_\beta}}; \ \  \ i\in\{1,2\}
\end{eqnarray}
From (\ref{eq:K2}), one sees that the coefficients of the ``pure" terms, $Z_{ij}$ are as given in (\ref{eq:Zcoeffs}) in Appendix {\bf A.5-A.7}. Quite interestingly, one can show that

\begin{equation}
\label{eq:Z2}
Z\sim\frac{{\cal V}^{\frac{1}{12}}}{\sum_\beta n^0_\beta}\left(\begin{array}{ccc}
{\cal O}(1)\frac{{\cal V}^{\frac{1}{36}}}{\sum_\beta n^0_\beta} & {\cal O}(1) \frac{{\cal V}^{\frac{1}{36}}}{\sum_\beta n^0_\beta} & {\cal O}(1) \frac{{\cal V}^{\frac{11}{12}}}{\sum_\beta n^0_\beta} \\
{\cal O}(1) \frac{{\cal V}^{\frac{1}{36}}}{\sum_\beta n^0_\beta} & {\cal O}(1) \frac{{\cal V}^{\frac{1}{36}}}{\sum_\beta n^0_\beta} & {\cal O}(1) \frac{{\cal V}^{\frac{11}{12}}}{\sum_\beta n^0_\beta} \\
{\cal O}(1) \frac{{\cal V}^{\frac{11}{12}}}{\sum_\beta n^0_\beta} & {\cal O}(1) \frac{{\cal V}^{\frac{11}{12}}}{\sum_\beta n^0_\beta} & {\cal O}(1) \frac{{\cal V}^{\frac{65}{36}}}{\sum_\beta n^0_\beta} \\
\end{array}\right).
\end{equation}
The eigenvectors corresponding to the diagonalized $Z_{ij}$ are the same as that for $\hat{K}_{i{\bar j}}$ - hence (\ref{eq:eigenvector1}) and (\ref{eq:eigenvectors2_3}) simultaneously diagonalize $\hat{K}_{i{\bar j}}$ and $Z_{ij}$! The eigenvalues of (\ref{eq:Z2}) corresponding to the diagonalized $Z$ are:
\begin{eqnarray}
\label{eq:diagonal_Z}
& & Z_{{\cal Z}_1{\cal Z}_1}=6\left(Z_{z_1z_1} + Z_{z_2z_2}\right) - \frac{6}{Z_{{\cal A}_1{\cal A}_1}}\left(Z^2_{z_1{\cal A}_1}+Z^2_{z_2{\cal A}_1}\right)\sim{\cal V}^{-\frac{17}{9}};\nonumber\\
& & Z_{\tilde{{\cal A}_1}\tilde{{\cal A}_1}}=Z_{{\cal A}_1{\cal A}_1}\sim{\cal V}^{-\frac{1}{9}}.
\end{eqnarray}
Before we proceed to read-off the soft SUSY breaking terms, we would like to recall the following: The non-perturbative superpotential corresponding to an $ED3-$instanton obtained as an
$n^s$-fold wrapping of $\Sigma_S$ by a single $ED3$-brane as well as a single $D7$-brane wrapping $\Sigma_B$ taking the rigid limit of the wrapping, along with a space-time filling $D3$-brane restricted for purposes of definiteness and calculational convenience to $\Sigma_B$, will be given by:
\begin{equation}
\label{eq:WD5}
W\sim\frac{\left[1 + z_1^{18} + z_2^{18} + \left(3\phi_0z_1^6z_2^6-z_1^{18}-z_2^{18}-1\right)^{\frac{2}{3}} - 3\phi_0z_1^6z_2^6\right]^{n^s}}{{\cal V}^{n^s}},
\end{equation}
which for $(z_1,z_2)\sim({\cal V}^{\frac{1}{36}},{\cal V}^{\frac{1}{36}})$, yields ${\cal V}^{-\frac{n^s}{2}}$. Hence, the gravitino mass
\begin{equation}
\label{eq:gravitino}
m_{\frac{3}{2}}=e^{\frac{K}{2}}W M_p\sim{\cal V}^{-\frac{n^s}{2} - 1}M_p,
\end{equation} which for $n^s=2, {\cal V}\sim10^7$ gives about $10TeV$. Substituting (\ref{eq:fluctuations}) into (\ref{eq:WD5}) (and again not being careful about ${\cal O}(1)$ numerical factors), one obtains:

\begin{eqnarray}
\label{eq:W_exp}
& & W\sim{\cal V}^{\frac{n^s}{2}}\Theta_{n^s}(\tau,{\cal G}^a)e^{in^sT(\sigma^S,{\bar\sigma^S};{\cal G}^a,{\bar{\cal G}^a};\tau,{\bar\tau})}\Biggl[1 + (\delta z_1 + \delta z_2)\Biggl\{n^s{\cal V}^{-\frac{1}{36}} + (in^s\mu_3)^3{\cal V}^{\frac{1}{36}}\Biggr\}\nonumber\\
& & +\delta\tilde{{\cal A}}_1\left\{-[\lambda_1+\lambda_2](in^s\mu_3){\cal V}^{-\frac{31}{36}} - n^s[\lambda_1+\lambda_2]{\cal V}^{-\frac{11}{12}}\right\}\Biggr]+ \left((\delta z_1)^2 + (\delta z_2)^2\right)\mu_{z_iz_i}  + \delta z_1\delta z_2\nonumber\\
& & \mu_{z_1z_2} + \left(\delta\tilde{{\cal A}}_1\right)^2\mu_{\tilde{\cal A}_1\tilde{\cal A}_1} + \delta z_1\delta\tilde{{\cal A}}_1\mu_{z_1\tilde{\cal A}_1} + \delta z_2\delta\tilde{{\cal A}}_1\mu_{z_2\tilde{\cal A}_1} + \left((\delta z_1)^3 + (\delta z_2)^3\right)Y_{z_iz_iz_i} + \nonumber\\
& & \left((\delta z_1)^2\delta z_2 + (\delta z_2)^2\delta z_1\right)Y_{z_iz_iz_j}+ (\delta z_1)^2\delta{\tilde{\cal A}}_1Y_{z_iz_i\tilde{\cal A}_1} + \delta z_1(\delta\tilde{\cal A}_1)^2Y_{z_i\tilde{\cal A}_1\tilde{\cal A}_1} \nonumber\\
& & + \delta z_1\delta z_2\delta\tilde{{\cal A}}_1Y_{z_1z_2\tilde{\cal A}_1} + \left(\delta\tilde{\cal A}_1\right)^3Y_{\tilde{\cal A}_1\tilde{\cal A}_1\tilde{\cal A}_1}+ .....\end{eqnarray}
The above expression which is a power series expansion of the superpotential in terms of open string moduli, is among the building blocks for computing soft parameters. The $\mu$ terms ($\mu_{ij}$) and the Yukawa couplings $Y_{ijk}$ appearing in the above expression are spelt out in subsections below.

\subsection{Gauginos' and Matter Field scalars' Masses}
In this subsection, we estimate the gaugino masses in our large volume $D3/D7$-setup. The gaugino masses are defined through the $F^m$ terms as below.
\begin{eqnarray}
M_{\tilde{g}}\equiv\frac{F^m\partial_m T_B}{2Re(T_B)},
\end{eqnarray}
where$\ F^m=e^{\frac{\hat{K}}{2}}\hat{K}^{m{\bar n}}{\bar D}_{\bar n}{\bar W}=e^{\frac{\hat{K}}{2}}\hat{K}^{m{\bar n}}\left({\bar\partial}_{\bar n}{\bar W} + {\bar W}{\bar\partial}_{\bar n}K\right)$ for which we first need to evaluate $\hat{K}^{m{\bar n}}$. Using (\ref{eq:K2}), in the LVS limit, one obtains:

\begin{equation}
\label{eq:Khat_metric}
\hat{K}_{m{\bar n}}\sim\left(\begin{array}{cccc}
\frac{1}{{\cal V}^{\frac{37}{36}}} & \frac{1}{{\cal V}^{\frac{35}{18}}} & \frac{\kappa_{B1a}}{({\cal G}^a,{\bar{\cal G}^a})}{{\cal V}^{\frac{37}{36}}} & \frac{\kappa_{B2a}}{({\cal G}^a,{\bar{\cal G}^a})}{{\cal V}^{\frac{37}{36}}}\\
 \frac{1}{{\cal V}^{\frac{35}{18}}} & \frac{1}{{\cal V}^{\frac{37}{36}}} & \frac{\kappa_{S1a}}{({\cal G}^a,{\bar{\cal G}^a})}{{\cal V}^{\frac{37}{36}}} & \frac{\kappa_{S2a}}{({\cal G}^a,{\bar{\cal G}^a})}{{\cal V}^{\frac{37}{36}}}\\
\frac{\kappa_{B1a}}{({\cal G}^a,{\bar{\cal G}^a})}{{\cal V}^{\frac{37}{36}}} & \frac{\kappa_{S1a}}{({\cal G}^a,{\bar{\cal G}^a})}{{\cal V}^{\frac{37}{36}}} & k_1^2 & k_1k_2 \\
\frac{\kappa_{B2a}}{({\cal G}^a,{\bar{\cal G}^a})}{{\cal V}^{\frac{37}{36}}} & \frac{\kappa_{S2a}}{({\cal G}^a,{\bar{\cal G}^a})}{{\cal V}^{\frac{37}{36}}} & k_1k_2 & k_2^2 \\
\end{array}\right),
\end{equation}
which - in the LVS limit - hence yields:
\begin{equation}
\label{eq:Khat_metric_inv}
\hat{K}^{m{\bar n}}\sim\left(\begin{array}{cccc}
{\cal V}^{\frac{37}{36}} & {\cal V}^{\frac{1}{9}} & ({\cal G}^a,{\bar{\cal G}^a}) & ({\cal G}^a,{\bar{\cal G}^a})\\
{\cal V}^{\frac{1}{9}} & {\cal V}^{\frac{37}{36}} & ({\cal G}^a,{\bar{\cal G}^a}) & ({\cal G}^a,{\bar{\cal G}^a})\\
({\cal G}^a,{\bar{\cal G}^a}) & ({\cal G}^a,{\bar{\cal G}^a}) & {\cal O}(1) & {\cal O}(1) \\
({\cal G}^a,{\bar{\cal G}^a}) & ({\cal G}^a,{\bar{\cal G}^a}) & {\cal O}(1) & {\cal O}(1) \\
\end{array}\right).
\end{equation}
Using above inputs the various $F$-terms are estimated to be:
\begin{eqnarray}
\label{eq:Fs}
& & F^{\sigma^B}\sim\frac{e^{-i\mu_3l^2{\cal V}^{\frac{1}{18}}} \, {\cal V}^{ - \frac{n^s}{2}}}{\cal V}\left({\cal V}^{\frac{1}{18}}\right)\sim{\cal V}^{-\frac{n^s}{2}-\frac{17}{18}};\nonumber\\
& & F^{\sigma^S}\sim\frac{e^{-i\mu_3l^2{\cal V}^{\frac{1}{18}}} \, {\cal V}^{ - \frac{n^s}{2}}}{\cal V}\left({\cal V}^{\frac{37}{36}}(n^s+{\cal V}^{-\frac{35}{36}}) + ({\cal G}^a,{\bar{\cal G}^a})\left[n^s(m^a + \frac{({\cal G}^a,{\bar{\cal G}^a})}{ln {\cal V}}) + {\cal V}^{-\frac{1}{6}}\right]\right)\nonumber\\
& & \sim n^s{\cal V}^{-\frac{n^s}{2}+\frac{1}{36}};\nonumber\\
& & F^{{\cal G}^a}\sim\frac{e^{-i\mu_3l^2{\cal V}^{\frac{1}{18}}} \, {\cal V}^{ - \frac{n^s}{2}}}{\cal V}\left(({\cal G}^a,{\bar{\cal G}^a})(n^s + {\cal V}^{-\frac{35}{36}}) + k^ak^b\left[n^s(m^b + \frac{({\cal G}^a,{\bar{\cal G}^a})}{ln {\cal V}}) + {\cal V}^{-\frac{1}{6}}\right]\right)\nonumber\\
& & \sim n^sk.mk^a{\cal V}^{-\frac{n^s}{2} - 1}
\end{eqnarray}
From (\ref{eq:Fs}), we conclude that the gaugino masses will of given by
\begin{eqnarray}
\label{eq:gaugino_mass}
& & M_{\tilde{g}}\equiv\frac{F^m\partial_m T_B}{2Re(T_B)}\sim\frac{{\cal V}^{-\frac{n^s}{2}}\left({\cal V}^{-\frac{17}{18}} + \frac{n^sk.mk.{\cal G}}{\cal V}\right)}{{\cal V}^{\frac{1}{18}}}\sim{\cal V}^{-\frac{n^s}{2}-1}\sim m_{\frac{3}{2}},
\end{eqnarray} where we have used the fact that the gravitino mass $m_{\frac{3}{2}}\sim{\cal V}^{-\frac{n^s}{2} - 1}$. Hence, what we see is that like the claims in the literature (See \cite{conloncal}, etc.), with the inclusion of $D3$- and $D7$-brane moduli, the gaugino masses are of the order of gravitino mass  - however, given that we are keeping the volume stabilized at around $10^7$ (in $l_s=1$-units) such that for $n^s=2$, $m_{\frac{3}{2}}\sim 10 TeV$. Finally, let's look at the anomaly-mediated gaugino masses which are given by (See \cite{Anomalymedisusy,Alwis}):

\begin{eqnarray}
\label{eq:gaugino_mass_anomaly_med_1}
& & \frac{M_{\tilde{g}}}{g_a^2}=-\frac{1}{8\pi^2}{\biggl[-\left(\sum_r n_rT_a(r) - 3 T_{\rm adj}(G_a)\right)m_{\frac{3}{2}} - \left(\sum_r n_rT_a(r) -  T_{\rm adj}(G_a)\right)F^m\hat{K}_m \biggr]}\nonumber\\
& & + \frac{1}{8\pi^2}{\sum_r 2 T_a(r)F^m\partial_m ln\ {\rm det}\left(\hat{K}_{i{\bar j}}\right)}.
\end{eqnarray}
Using $F^m\partial_m\hat{K}\sim m_{\frac{3}{2}}{\cal V}^{\frac{1}{18}}$, $g_a^2\sim{\cal V}^{-\frac{1}{18}}$ (from (\ref{eq:g_a-EXACT})) and (\ref{eq:F.dlndetKhat}), one sees that:
\begin{equation}
\label{eq:gaugino_mass_anomaly_med_2}
\frac{M_{\tilde{g}}}{g_a^2}\sim\frac{{\cal V}^{\frac{1}{18}}m_{\frac{3}{2}}}{8\pi^2},\ {\rm implying}\ M_{\tilde{g}}\sim {\frac{1}{8\pi^2}}m_{\frac{3}{2}}
\end{equation}
From (\ref{eq:gaugino_mass}) and (\ref{eq:gaugino_mass_anomaly_med_2}), one sees that similar to \cite{conloncal,towardsrealvacua}, the anomaly mediated gaugino masses are suppressed by the standard loop factor as compared to the gravity mediated gaugino masses.

Now, the open-string moduli or matter field scalars' masses, which are:
\begin{eqnarray}
\label{eq:matter_masses_1}
& & m_i^2 = m_{\frac{3}{2}}^2 + V_0 - F^m{\bar F}^{\bar n}\partial_m{\bar\partial}_{\bar n}ln \hat{K}_{i{\bar i}}\nonumber\\
& & = m_{\frac{3}{2}}^2 + V_0 + F^m{\bar F}^{\bar n}\left(\frac{1}{\hat{K}_{i{\bar i}}^2}\partial_m\hat{K}_{i{\bar i}}{\bar\partial}_{\bar n}\hat{K}_{i{\bar i}} - \frac{1}{\hat{K}_{i{\bar i}}}\partial_m{\bar\partial}_{\bar n}\hat{K}_{i{\bar i}}\right).
\end{eqnarray}
Using (\ref{eq:F_terms}), (\ref{eq:Fs}) and results of Appendix {\bf A.5 - A.7}, we arrive at:
\begin{eqnarray}
\label{eq:V_0}
& & V_0\sim e^KG^{\sigma^\alpha{\bar\sigma}^\alpha}D_{\sigma^\alpha}W{\bar D}_{{\bar\sigma}^\alpha}{\bar W}\nonumber\\
& & \sim\frac{(n^s)^2|W|^2{\cal V}^{\frac{19}{18}}}{{\cal V}^2}\sim{\cal V}^{-n^s - 2 + \frac{19}{18}}\sim {\cal V}^{\frac{19}{18}}m^2_{\frac{3}{2}}.
\end{eqnarray} and
\begin{eqnarray}
\label{eq:matter_masses_2}
& & F^m{\bar F}^{{\bar n}}\partial_m{\bar\partial}_{\bar n}\hat{K}_{{\cal Z}_i{\bar{\cal Z}}_i}\sim{\cal V}^{-n^s - \frac{1}{18}}\sim m^2_{\frac{3}{2}}{\cal V}^{\frac{35}{36}}; \nonumber\\
& & F^m{\bar F}^{{\bar n}}\partial_m{\bar\partial}_{\bar n}\hat{K}_{\tilde{\cal A}_1{\bar{\tilde{\cal A}}_1}}\sim{\cal V}^{-n^s + \frac{1}{36}}\sim{\cal V}^{\frac{73}{36}}m^2_{\frac{3}{2}}.\nonumber\\
\end{eqnarray}
Substituting (\ref{eq:V_0}) and (\ref{eq:matter_masses_2}) into (\ref{eq:matter_masses_1}), one obtains:
\begin{eqnarray}
\label{eq:matter_masses_3}
& & m_{{\cal Z}_i}^2\sim m^2_{\frac{3}{2}}\left(1 + {\cal V}^{\frac{19}{18}} + {\cal V}^{\frac{35}{36}}\right)\sim m^2_{\frac{3}{2}}{\cal V}^{\frac{19}{18}};\nonumber\\
& & m_{\tilde{\cal A}_1}^2\sim m^2_{\frac{3}{2}}\left(1 + {\cal V}^{\frac{19}{18}} +
{\cal V}^{\frac{73}{36}}\right)\sim {\cal V}^{\frac{73}{36}}m^2_{\frac{3}{2}},
\end{eqnarray} implying
\begin{equation}
\label{eq:matter_masses_4}
m_{{\cal Z}_i}\sim {\cal V}^{\frac{19}{36}}m_{\frac{3}{2}},\ m_{\tilde{\cal A}_1}\sim {\cal V}^{\frac{73}{72}}m_{\frac{3}{2}}.
\end{equation}

\subsection{Physical $\mu$ Terms}
To evaluate the canonical ``physical" $\mu$ terms - denoted by $\hat{\mu}$ - one needs to evaluate $F^m\partial_mZ_{{\cal Z}_i{\cal Z}_i}$ and
$F^m\partial_mZ_{{\cal A}_1{\cal A}_1}$. Therefore, using (\ref{eq:Fs}), (\ref{eq:dZ_z}) and (\ref{eq:dZ_a}) one obtains:
\begin{equation}
\label{eq:F.dZ_z}
F^m\partial_m Z_{{\cal Z}_i{\cal Z}_i}\sim{\cal V}^{-\frac{n^s}{2}-\frac{17}{9}}.
\end{equation}
\begin{equation}
\label{eq:F.dZ_a}
F^m\partial_mZ_{{\cal A}_1{\cal A}_1}\sim{\cal V}^{-\frac{n^s}{2} - \frac{19}{18}}.
\end{equation}
Now,
\begin{equation}
\label{eq:muhatdef}
\hat{\mu}_{ij}=\frac{\frac{{\bar{\hat{W}}}e^{\frac{\hat{K}}{2}}}{|\hat{W}|}\mu_{ij} + m_{\frac{3}{2}}Z_{ij}\delta_{ij} - {\bar F}^{\bar m}{\bar\partial}_{\bar m}Z_{ij}\delta_{ij}}{\sqrt{\hat{K}_{i{\bar i}}\hat{K}_{j{\bar j}}}}.
\end{equation}
From (\ref{eq:W_exp}), one obtains the following non-zero $\mu$-terms:
\begin{eqnarray}
\label{eq:mu}
& & \mu_{{\cal Z}_1{\cal Z}_2}\sim {\cal V}^{-\frac{n^s}{2}}\left[n^s\phi_0{\cal V}^{-\frac{2}{9}} + (in^s\mu_3)^3{\cal V}^{\frac{1}{18}}+ n^s(i\mu_3n^s)\right];\nonumber\\
& & \mu_{{\cal Z}_i{\cal Z}_i}\sim {\cal V}^{-\frac{n^s}{2}}\left[\left(\frac{n^s(n^s - 1)}{2} + n^s\right){\cal V}^{-\frac{1}{18}} + (in^s\mu_3)^3{\cal V}^{\frac{1}{18}} + n^s(i\mu_3n^s)\right];\nonumber\\
& & \mu_{\tilde{{\cal A}_1}\tilde{{\cal A}_1}}\sim {\cal V}^{-\frac{n^s}{2}}\biggl[{\cal V}^{-\frac{33}{18}}[\lambda_1^2+\lambda_2^2]\left[\frac{n^s(n^s-1)}{2} + n^s\right]
+ [\lambda_1^2+\lambda_2^2](in^s\mu_3)^2{\cal V}^{-\frac{31}{18}} \nonumber\\
& & \hskip 0.5in +[\lambda_1+\lambda_2]^2n^s(in^s\mu_3){\cal V}^{-\frac{16}{9}}\biggr];\nonumber\\
& & \mu_{{\cal Z}_i\tilde{{\cal A}_1}}\sim {\cal V}^{-\frac{n^s}{2}}\left[-\lambda_j{\cal V}^{-\frac{17}{18}}\Biggl\{n^s+\frac{n^s(n^s-1)}{2} + n^s(in^s\mu_3)[\lambda_1+\lambda_2]{\cal V}^{-\frac{8}{9}}\Biggr\} + \lambda_j(in^s\mu_3)^3{\cal V}^{-\frac{5}{6}}\right],\nonumber\\
\end{eqnarray}
where $j\neq i(=1,2)$ in the above equations. Finally one results in following physical $\hat{\mu}$-parameters:
\begin{eqnarray}
\label{eq:muhat}
& & \hat{\mu}_{{\cal Z}_i{\cal Z}_j}\sim{\cal V}^{-\frac{n^s}{2} - 1}{\cal V}^{\frac{35}{36}}\sim{\cal V}^{\frac{37}{36}}m_{\frac{3}{2}};
\nonumber\\
& & \hat{\mu}_{{\cal A}_1{\cal Z}_i}\sim{\cal V}^{-\frac{3}{4}}m_{\frac{3}{2}}; \ \ \hat{\mu}_{{\cal A}_1{\cal A}_1}\sim{\cal V}^{-\frac{33}{36}}m_{\frac{3}{2}}.
\end{eqnarray}

\subsection{Physical Yukawa Couplings ($\hat{Y}_{ijk}$) and $A$-Parameters ($A_{ijk}$) }

From (\ref{eq:W_exp}), one obtains the following unnormalized non-zero Yukawa couplings:
\begin{eqnarray*}
& & Y_{{\cal Z}_i{\cal Z}_i{\cal Z}_i}\sim {\cal V}^{-\frac{n^s}{2}}\left\{n^s{\cal V}^{\frac{1}{3}} + (in^s\mu_3)^3{\cal V}^{\frac{1}{12}} + n^s(in^s\mu_3)^2{\cal V}^{\frac{1}{36}} + \left[\frac{n^s(n^s-1)}{2}+n^s(in^s\mu_3){\cal V}^{-\frac{1}{36}}\right]\right\};\nonumber\\
& & Y_{{\cal Z}_i^2{\cal Z}_j}\sim {\cal V}^{-\frac{n^s}{2}}\Biggl\{{\cal V}^{-\frac{1}{12}}\left(\frac{n^s(n^s-1)(n^s-2)}{6} + \frac{n^s(n^s-1)}{2}\right) + (in^s\mu_3)^3{\cal V}^{\frac{1}{12}}\nonumber\\
& & + {\cal V}^{\frac{1}{36}}(in^s\mu_3)^2n^s + {\cal V}^{-\frac{1}{36}}(in^s\mu_3)\Biggl[n^s+\frac{n^s(n^s-1)}{2}\Biggr]\Biggr\};\nonumber\\
& & Y_{\tilde{\cal A}_1\tilde{\cal A}_1\tilde{\cal A}_1}\sim {\cal V}^{-\frac{n^s}{2}}\Biggl\{n^s{\cal V}^{-\frac{7}{3}}(\lambda_1^3+\lambda_2^3) + (in^s\mu_3)^3{\cal V}^{-\frac{31}{12}} + {\cal V}^{-\frac{95}{36}}n^s[\lambda-1+\lambda_2](in^s\mu_3)^2[\lambda_1^2+\lambda_2^2]\nonumber\\
 & & + {\cal V}^{-\frac{97}{36}}(in^s\mu_3)\Biggl[n^s+\frac{n^s(n^s-1)}{2}\Biggr][\lambda_1+\lambda_2][\lambda_1^2+\lambda_2^2]\Biggr\};\nonumber\\
& & Y_{{\cal Z}_i^2\tilde{{\cal A}}_1}\sim {\cal V}^{-\frac{n^s}{2}}\biggl\{-\lambda_2{\cal V}^{-\frac{5}{9}} - \lambda_2{\cal V}^{-\frac{29}{36}}(in^s\mu_3)^3+ \Biggl[n^s+\frac{n^s(n^s-1)}{2}\Biggr](in^s\mu_3)[\lambda_1+\lambda_2]{\cal V}^{-\frac{11}{12}}\nonumber\\
& & + {\cal V}^{-\frac{31}{36}}[\lambda_1+\lambda_2]n^s(in^s\mu_3)^2\biggr\};\nonumber\\
& & Y_{(\tilde{{\cal A}}_1)^2{\cal Z}_i}\sim {\cal V}^{-\frac{n^s}{2}}\Biggl\{-n^s\lambda_2^2{\cal V}^{-\frac{13}{9}} + (\lambda_{j(\neq i)}^2+2\lambda_1\lambda_2)(in^s\mu_3)^3{\cal V}^{-\frac{61}{36}} + {\cal V}^{-\frac{7}{4}}n^s(in^s\mu_3)^2[\lambda_1^2+\lambda_2^2]\nonumber\\
 & & + {\cal V}^{-\frac{65}{36}}[\lambda_1^2+\lambda_2^2]\Biggl[n^s+\frac{n^s(n^s-1)}{2}\Biggr](in^s\mu_3)\Biggr\};\nonumber\\
& & Y_{{\cal Z}_1{\cal Z}_2\tilde{{\cal A}}_1}\sim {\cal V}^{-\frac{n^s}{2}}\Biggl\{-2(\lambda_1+\lambda_2){\cal V}^{-\frac{35}{36}}\left[\frac{n^s(n^s-1)(n^s-2)}{6} + \frac{n^s(n^s-1)}{2}\right] - 2(\lambda_1+\lambda_2)\nonumber\\
 \end{eqnarray*}
\begin{eqnarray}
\label{eq:Ys}
& & \times {\cal V}^{-\frac{29}{36}}(in^s\mu_3)^3 +{\cal V}^{-\frac{39}{36}}(n^s\phi_0)(in^s\mu_3)[\lambda_1+\lambda_2] + {\cal V}^{-\frac{31}{36}}n^s[\lambda_1+\lambda_2](in^s\mu_3)^2\nonumber\\
& & + {\cal V}^{-\frac{11}{12}}(in^s\mu_3)\Biggl[n^s+\frac{n^s(n^s-1)}{2}\Biggr][\lambda_1+\lambda_2]\Biggr\}.
\end{eqnarray}
Given the following definition of the physical Yukawa couplings:
\begin{equation}
\label{eq:physicalYdef}
\hat{Y}_{ijk}=\frac{e^{\frac{\hat{K}}{2}}Y_{ijk}}{\sqrt{\hat{K}_{i{\bar i}}\hat{K}_{j{\bar j}}\hat{K}_{k{\bar k}}}},
\end{equation}
one obtains the following non-zero physical Yukawa couplings $\hat{Y}_{ijk}$s:
\begin{eqnarray}
\label{eq:physical_Ys}
& & \hat{Y}_{{\cal Z}_i{\cal Z}_i{\cal Z}_i}\sim {\cal V}^{\frac{19}{24}-\frac{n^s}{2}}, \, \, \hat{Y}_{{\cal Z}_i^2{\cal Z}_j}\sim {\cal V}^{\frac{13}{24}-\frac{n^s}{2}}, \, \, \hat{Y}_{{\cal Z}_i^2\tilde{\cal A}_1}\sim {\cal V}^{-\frac{71}{72}-\frac{n^s}{2}}\nonumber\\
& & \hat{Y}_{{\cal Z}_1{\cal Z}_2\tilde{\cal A}_1}\sim {\cal V}^{-\frac{89}{72}-\frac{n^s}{2}}, \, \, \hat{Y}_{\tilde{\cal A}_1^2{\cal Z}_i}\sim {\cal V}^{-\frac{199}{72}-\frac{n^s}{2}}, \, \, \hat{Y}_{\tilde{\cal A}_1\tilde{\cal A}_1\tilde{\cal A}_1}\sim {\cal V}^{-\frac{109}{24}-\frac{n^s}{2}}
\end{eqnarray}
The $A$-terms are defined as:
\begin{equation}
\label{eq:Adef}
A_{ijk}=F^m\left[\hat{K}_m + \partial_m ln Y_{ijk} - \partial_m ln\left(\hat{K}_{i{\bar i}}\hat{K}_{j{\bar j}}\hat{K}_{k{\bar k}}\right)\right].
\end{equation}
Using:
\begin{eqnarray}
\label{eq:dervs_Ys}
& & \partial_{\sigma^B}Y_{ijk}\sim0; \ \partial_{\sigma^S}Y_{ijk}\sim n^sY_{ijk};\ \partial_{{\cal G}^a}Y_{ijk}\sim({\cal G}^a,{\bar{\cal G}}^a)Y_{ijk},
\end{eqnarray}
and (\ref{eq:Fs}) one obtains:
\begin{equation}
\label{eq:F.dY}
F^m\partial_mY_{ijk}\sim n^s{\cal V}^{-\frac{n^s}{2}-\frac{35}{36}}Y_{ijk}\sim n^s{\cal V}^{\frac{1}{36}}m_{\frac{3}{2}}.
\end{equation}
Using:
\begin{eqnarray*}
& & \partial_{\sigma^\alpha}\hat{K}\sim \frac{\sqrt{{\cal T}_S(\sigma^\alpha,{\bar\sigma^\alpha};{\cal G}^a,{\bar{\cal G}^a};\tau,{\bar\tau}) + \mu_3{\cal V}^{\frac{1}{18}}  - \gamma\left(r_2 + \frac{r_2^2\zeta}{r_1}\right)}}{\Xi}\sim{\cal V}^{-\frac{35}{36}};\nonumber\\
\end{eqnarray*}
\begin{eqnarray}
\label{eq:ders_Khat}
& & \partial_{{\cal G}^a}\hat{K}\sim\frac{1}{\Xi}\times\Biggl[\sum_\beta k^an^0_\beta sin(...) + \left({\cal G}^a,{\bar{\cal G}}^a\right)\nonumber\\
& & \times\Biggl\{\sqrt{{\cal T}_B(\sigma^B,{\bar\sigma ^B};{\cal G}^c,{\bar{\cal G}^c};\tau,{\bar\tau}) + \mu_3{\cal V}^{\frac{1}{18}} + i\kappa_4^2\mu_7C_{1{\bar 1}}{\cal V}^{-\frac{1}{2}} - \gamma\left(r_2 + \frac{r_2^2\zeta}{r_1}\right)}\kappa_{Bac}\nonumber\\
& & -\sqrt{{\cal T}_S(\sigma^S,{\bar\sigma ^S};{\cal G}^a,{\bar{\cal G}^a};\tau,{\bar\tau}) + \mu_3{\cal V}^{\frac{1}{18}}  - \gamma\left(r_2 + \frac{r_2^2\zeta}{r_1}\right)}\kappa_{Sac}\Biggr\}\Biggr]\sim{\cal V}^{-\frac{1}{6}},
\end{eqnarray}
and (\ref{eq:Fs}), (\ref{eq:dKhat_z}) and (\ref{eq:dKhat_a}) one obtains:
\begin{eqnarray}
\label{eq:F.dKhat}
& & F^m\partial_m\hat{K}\sim n^s{\cal V}^{-\frac{n^s}{2}-\frac{17}{18}}\sim n^s{\cal V}^{\frac{1}{18}}m_{\frac{3}{2}}, \nonumber\\
& & F^m\partial_m ln\hat{K}_{{\cal Z}_i{\cal Z}_i}\sim n^s{\cal V}^{-\frac{n^s}{2}-\frac{1}{36}}\sim n^s  {\cal V}^{\frac{35}{36}}m_{\frac{3}{2}},\nonumber\\
& & F^m\partial_m ln\hat{K}_{\tilde{{\cal A}_1}{\bar{\tilde{\cal A}_1}}}\sim n^s{\cal V}^{-\frac{n^s}{2}-\frac{17}{18}}\sim n^s {\cal V}^{\frac{1}{18}}m_{\frac{3}{2}}.
\end{eqnarray}
Hence, substituting (\ref{eq:F.dY}), (\ref{eq:F.dKhat}) into (\ref{eq:Adef}), one obtains:
\begin{eqnarray}
\label{eq:A}
& & A_{{\cal Z}_i{\cal Z}_j{\cal Z}_k}\sim n^s{\cal V}^{\frac{37}{36}}m_{\frac{3}{2}}; \ \  A_{\tilde{{\cal A}_1}{\cal Z}_i{\cal Z}_j}\sim n^s{\cal V}^{\frac{37}{36}}m_{\frac{3}{2}}\nonumber\\
& & A_{\tilde{{\cal A}_1}\tilde{{\cal A}_1}\tilde{{\cal A}_1}}\sim n^s{\cal V}^{\frac{37}{36}}m_{\frac{3}{2}}; \ \  A_{\tilde{{\cal A}_1}^2{\cal Z}_i}\sim n^s{\cal V}^{\frac{37}{36}}m_{\frac{3}{2}}.
\end{eqnarray}

\subsection{The $\hat{\mu}B$ Parameters}
The $\hat{\mu}B$-parameters are defined as under:
\begin{eqnarray}
\label{eq:muhatB_def}
& & (\hat{\mu}B)_{ij}=\frac{1}{\sqrt{\hat{K}_{i{\bar i}}\hat{K}_{j{\bar j}}}}\times\Biggl\{\frac{{\bar{\hat{W}}}}{|\hat{W}|}e^{\frac{\hat{K}}{2}}
\left[F^m\left(\hat{K}_m\mu_{ij} + \partial_m\mu_{ij} - \mu_{ij}\partial_m ln\left(\hat{K}_{i{\bar i}}\hat{K}_{j{\bar j}}\right)\right) - m_{\frac{3}{2}}\mu_{ij}\right]\nonumber\\
& & + \left(2m^2_{\frac{3}{2}} + V_0\right)Z_{ij}\delta_{ij} - m_{\frac{3}{2}}{\bar F}^{\bar m}{\bar\partial}_{\bar n}Z_{ij}\delta_{ij} + m_{\frac{3}{2}}\delta_{ij}F^m\left[\partial_m Z_{ij} - Z_{ij}\partial_m ln\left(\hat{K}_{i{\bar i}}\hat{K}_{j{\bar j}}\right)\right]
\nonumber\\
& & - \delta_{ij}{\bar F}^{\bar m}F^n\left[{\bar\partial}_{\bar m}\partial_nZ_{ij} - {\bar\partial}_{\bar m} Z_{ij}\partial_n ln\left(\hat{K}_{i{\bar i}}\hat{K}_{j{\bar j}}\right)\right]\Biggr\},
\end{eqnarray}
where $\delta_{ij}$ has been put in before the $Z_{ij}$-dependent terms to indicate that we are working with the diagonalized matter fields (\ref{eq:diagonal}).

Now substituting (\ref{eq:F.dKhat}), (derivatives w.r.t the closed string moduli $\sigma^\alpha, {\cal G}^a$ of) (\ref{eq:mu}), (\ref{eq:Zcoeffs}), (\ref{eq:diagonal_Z}), (\ref{eq:Fs}), (\ref{eq:F.dZ_z}) and (\ref{eq:ddZ_z}):
\begin{eqnarray}
\label{eq:muhatB_zz_intermediate_1}
& & (a)\ \frac{{\bar{\hat{W}}}}{|\hat{W}|}e^{\frac{\hat{K}}{2}}F^m\partial_m\hat{K}\mu_{{\cal Z}_i{\cal Z}_i}\sim {\cal V}^{\frac{1}{9}}m_{\frac{3}{2}}^2, \ (b)\ \frac{{\bar{\hat{W}}}}{|\hat{W}|}e^{\frac{\hat{K}}{2}}F^m\partial_m\mu_{{\cal Z}_i{\cal Z}_i}\sim{\cal V}^{\frac{13}{12}}m_{\frac{3}{2}}^2,\nonumber\\
& & (c)\ \frac{{\bar{\hat{W}}}}{|\hat{W}|}e^{\frac{\hat{K}}{2}}\mu_{{\cal Z}_i{\cal Z}_i}F^m\partial_m ln\left(\hat{K}_{{\cal Z}_i{\bar{\cal Z}}_i}\right) \sim {\cal V}^{\frac{37}{36}}m_{\frac{3}{2}}^2,\ (d) \ \frac{{\bar{\hat{W}}}}{|\hat{W}|}e^{\frac{\hat{K}}{2}}\mu_{{\cal Z}_i{\cal Z}_i}\sim{\cal V}^{\frac{1}{18}}m_{\frac{3}{2}}^2,\nonumber\\
\end{eqnarray}
which gives:
\begin{eqnarray}
\label{eq:muhat_zz_intermediate_2}
& & \frac{(a)+(b)-2(c)-(d)}{\hat{K}_{{\cal Z}_i{\cal Z}_i}}\sim{\cal V}^{\frac{37}{18}}m_{\frac{3}{2}}^2.
\end{eqnarray}
Further, using (\ref{eq:F_terms}),
\begin{eqnarray}
\label{eq:muhatB_zz_intermediate_3}
& & \frac{\left(m_{\frac{3}{2}}^2+V_0\right)Z_{{\cal Z}_i{\cal Z}_i}}{\hat{K}_{{\cal Z}_i{\cal Z}_i}}\sim{\cal V}^{\frac{5}{36}}m_{\frac{3}{2}}^2, \ \frac{m_{\frac{3}{2}}{\bar F}^{\bar m}{\bar\partial}_{\bar m}Z_{{\cal Z}_i{\cal Z}_i}}{\hat{K}_{{\cal Z}_i{\cal Z}_i}}\sim{\cal V}^{\frac{1}{12}} m_{\frac{3}{2}}^2,\nonumber\\
& & \frac{m_{\frac{3}{2}}\left(F^m\partial_mZ_{{\cal Z}_i{\cal Z}_i} - 2Z_{{\cal Z}_i{\cal Z}_i}F^m\partial_m ln\left(\hat{K}_{{\cal Z}_i{\bar{\cal Z}}_i}\right)\right)}{\hat{K}_{{\cal Z}_i{\cal Z}_i}}\sim{\cal V}^{\frac{1}{12}}m_{\frac{3}{2}}^2,\nonumber\\
& & \frac{\left[{\bar F}^{\bar m}F^n{\bar\partial}_{\bar m}\partial_nZ_{{\cal Z}_i{\cal Z}_i} - 2 {\bar F}^{\bar m}F^n\left({\bar\partial}_{\bar m}Z_{{\cal Z}_i{\cal Z}_i}\right)\left(\partial_n ln\left(\hat{K}_{{\cal Z}_i{\cal Z}_i}\right)\right)\right]}{\hat{K}_{{\cal Z}_i{\cal Z}_i}}
\sim {\cal V}^{\frac{223}{108}}m_{\frac{3}{2}}^2.
\end{eqnarray}
Note, when substituting in the first equation of (\ref{eq:F_terms}) as the extremum value of the potential $V_0$ in (\ref{eq:muhatB_zz_intermediate_3}), we have assumed the following. For a non-supersymmetric configuration, from \cite{V_D7_fl} we see that the tadpole cancelation guarantees that the contributions to the potential from all the $D3$-branes and $O3$-planes as well as the $D7$-branes and $O7$-planes cancel out. However, there is still a $D$-term contribution from the
$U(1)$-fluxes on the world-volume of the $D7$-branes wrapped around $\Sigma_B$ of the form
$\frac{\left({\cal F}^\beta\kappa_{\alpha\beta}\partial_{T_\alpha}K\right)^2}{\left(Re(T_B) - {\cal F}Re(i\tau)\right)}$ - we drop the same in the dilute flux approximation.

From (\ref{eq:muhat_zz_intermediate_2}) and (\ref{eq:muhatB_zz_intermediate_3}), one obtains:
\begin{eqnarray}
\label{eq:muhatB_zz_final}
& & \left(\hat{\mu}B\right)_{{\cal Z}_i{\cal Z}_i}\sim{\cal V}^{\frac{223}{108}}m_{\frac{3}{2}}^2,\ \ \left(\hat{\mu}B\right)_{{\cal Z}_1{\cal Z}_2}\sim{\cal V}^{\frac{37}{18}}m_{\frac{3}{2}}^2.
\end{eqnarray}
Similarly,
\begin{eqnarray}
\label{eq:muhatB_aa_intermediate_1}
& & (a)\ \frac{{\bar{\hat{W}}}}{|\hat{W}|}e^{\frac{\hat{K}}{2}}F^m\partial_m\hat{K}\mu_{\tilde{\cal A}_1\tilde{\cal A}_1}\sim {\cal V}^{\frac{-5}{3}}m_{\frac{3}{2}}^2,
\nonumber\\
& & (b)\  \frac{{\bar{\hat{W}}}}{|\hat{W}|}e^{\frac{\hat{K}}{2}}F^m\partial_m\mu_{\tilde{\cal A}_1\tilde{\cal A}_1}\sim{\cal V}^{-\frac{25}{36}}m_{\frac{3}{2}}^2,\nonumber\\
& & (c)\ \frac{{\bar{\hat{W}}}}{|\hat{W}|}e^{\frac{\hat{K}}{2}}\mu_{\tilde{\cal A}_1\tilde{\cal A}_1}F^m\partial_m ln\left(\hat{K}_{\tilde{{\cal A}_1}{\bar{\tilde{\cal A}_1}}}\right)
\sim {\cal V}^{-\frac{5}{3}}m_{\frac{3}{2}}^2,\nonumber\\
& & (d)\ \frac{{\bar{\hat{W}}}}{|\hat{W}|}e^{\frac{\hat{K}}{2}}\mu_{\tilde{\cal A}_1\tilde{\cal A}_1}\sim{\cal V}^{-\frac{13}{18}}m_{\frac{3}{2}}^2,
\end{eqnarray}
which gives:
\begin{eqnarray}
\label{eq:muhat_aa_intermediate_2}
& & \frac{(a)+(b)-2(c)-(d)}{\hat{K}s_{\tilde{{\cal A}_1}{\bar{\tilde{\cal A}_1}}}}\sim{\cal V}^{-\frac{3}{2}}m_{\frac{3}{2}}^2.
\end{eqnarray}
Further,
\begin{eqnarray}
\label{eq:muhatB_aa_intermediate_3}
& & \frac{\left(m_{\frac{3}{2}}^2+V_0\right)Z_{\tilde{\cal A}_1\tilde{\cal A}_1}}{\hat{K}_{\tilde{{\cal A}_1}{\bar{\tilde{\cal A}_1}}}}\sim{\cal V}^{\frac{5}{36}}m_{\frac{3}{2}}^2,
\nonumber\\
& & \frac{m_{\frac{3}{2}}{\bar F}^{\bar m}{\bar\partial}_{\bar m}Z_{\tilde{\cal A}_1\tilde{\cal A}_1}}{\hat{K}_{\tilde{{\cal A}_1}{\bar{\tilde{\cal A}_1}}}}\sim{\cal V}^{\frac{31}{36}}
m_{\frac{3}{2}}^2,\nonumber\\
& & \frac{m_{\frac{3}{2}}\left(F^m\partial_mZ_{\tilde{\cal A}_1\tilde{\cal A}_1} - 2Z_{\tilde{\cal A}_1\tilde{\cal A}_1}F^m\partial_m ln\left(\hat{K}_{\tilde{{\cal A}_1}{\bar{\tilde{\cal A}_1}}}\right)\right)}{\hat{K}_{\tilde{{\cal A}_1}{\bar{\tilde{\cal A}_1}}}}\sim{\cal V}^{\frac{31}{36}}m_{\frac{3}{2}}^2,\nonumber\\
& & \frac{\left[{\bar F}^{\bar m}F^n{\bar\partial}_{\bar m}\partial_nZ_{\tilde{\cal A}_1\tilde{\cal A}_1} - 2 {\bar F}^{\bar m}F^n\left({\bar\partial}_{\bar m}Z_{\tilde{\cal A}_1\tilde{\cal A}_1}\right)\left(\partial_n ln\left(\hat{K}_{\tilde{\cal A}_1\tilde{\cal A}_1}\right)\right)\right]}{\hat{K}_{\tilde{{\cal A}_1}{\bar{\tilde{\cal A}_1}}}}
\sim {\cal V}^{\frac{1}{9}}m_{\frac{3}{2}}^2.
\end{eqnarray}
From (\ref{eq:muhat_aa_intermediate_2}) and (\ref{eq:muhatB_aa_intermediate_3}), one obtains:
\begin{equation}
\label{eq:muhatB_aa_final}
\left(\hat{\mu}B\right)_{\tilde{\cal A}_1\tilde{\cal A}_1}\sim{\cal V}^{\frac{5}{36}}m_{\frac{3}{2}}^2.
\end{equation}
Finally,
\begin{eqnarray}
\label{eq:muhatB_za_intermediate_1}
& & (a)\ \frac{{\bar{\hat{W}}}}{|\hat{W}|}e^{\frac{\hat{K}}{2}}F^m\partial_m\hat{K}\mu_{{\cal Z}_i\tilde{\cal A}_1}\sim {\cal V}^{-\frac{16}{9}}m_{\frac{3}{2}}^2,
\nonumber\\
& & (b)\  \frac{{\bar{\hat{W}}}}{|\hat{W}|}e^{\frac{\hat{K}}{2}}F^m\partial_m\mu_{{\cal Z}_i\tilde{\cal A}_1}\sim{\cal V}^{-\frac{29}{36}}m_{\frac{3}{2}}^2,\nonumber\\
& & (c)\ \frac{{\bar{\hat{W}}}}{|\hat{W}|}e^{\frac{\hat{K}}{2}}\mu_{{\cal Z}_i\tilde{\cal A}_1}F^m\partial_m ln\left(\hat{K}_{\tilde{{\cal A}_1}{\bar{\tilde{\cal A}_1}}}\hat{K}_{{\cal Z}_i{\bar{\cal Z}}_i}\right)
\sim {\cal V}^{-\frac{31}{36}}m_{
\frac{3}{2}}^2,\nonumber\\
& & (d)\ \frac{{\bar{\hat{W}}}}{|\hat{W}|}e^{\frac{\hat{K}}{2}}\mu_{{\cal Z}_i\tilde{\cal A}_1}\sim{\cal V}^{-\frac{11}{6}}m_{\frac{3}{2}}^2,
\end{eqnarray}
which gives:
\begin{eqnarray}
\label{eq:muhat_za_intermediate_2}
& & \frac{(a)+(b)-(c)-(d)}{\sqrt{\hat{K}_{\tilde{{\cal A}_1}{\bar{\tilde{\cal A}_1}}}\hat{K}_{{\cal Z}_i
{\bar{\cal Z}}_i}}}=\left(\hat{\mu}B\right)_{{\cal Z}_i\tilde{{\cal A}_1}}\sim{\cal V}^{-\frac{13}{18}}m_{\frac{3}{2}}^2.
\end{eqnarray}

It has been observed that due to competing contributions from the Wilson line moduli, there is a non-universality in the F-terms $F^{\sigma^B} \sim {\cal V}^{\frac{1}{18}}$ $ m_{\frac{3}{2}}$ which for ${\cal V}\sim10^6$ is approximately of the same order as $F^{{\cal G}^{a}} \sim m_{\frac{3}{2}}$; $F^{\sigma^S} \sim {\cal V}^{\frac{37}{36}}$ $ m_{\frac{3}{2}}$ - a reverse non-universality as compared to, e.g., \cite{Quevedo+Conlon_supp_gaugino_mass}. This is attributable to the cancelation between the divisor volume corresponding to $D_5$ and the Wilson line moduli contribution in $``T_B"$. Further, wherever there is a contribution from $F^{\sigma^S}$ to the soft parameters, there will be a hierarchy/non-universality.

The matter fields corresponding to the position moduli of the mobile $D3$-brane are heavier than the gravitino and show universality. However, Wilson line modulus mass is different. We obtain a hierarchy in the physical mu terms $\hat{\mu}$, the $\hat{\mu}B$-terms as well as the physical Yukawa couplings $\hat{Y}$; however we obtain a universality for the $A$-terms - larger than $m_{\frac{3}{2}}$ - for the $D3$-brane position moduli {\it and} the Wilson line moduli. However it can be easily seen from  {\bf Table A.4} that in the physical $\hat{\mu}, \hat{Y}$ and $\hat{\mu} B $ terms, that main part of the non-universality appears from the Wilson moduli contributions while there is an approximate universality in the $D3$-brane position moduli  components for which the physical $\hat{\mu}, \hat{Y}$ and $\hat{\mu} B $ are heavier than gravitino.  Also, as the string scale in our setup is nearly of the same order as the GUT scale and the open string moduli  are more massive as compared to the $\sim$ TeV gravitino (and gauginos), one can expect (e.g. see \cite{FCNC}) that the presence of non-universality will be consistent with the  low energy FCNC constraints.  Further we have found that ${\hat{\mu}}^2 \sim {\hat{\mu} B}$ for the $D3$-brane position moduli (which show universality of almost all the soft SUSY breaking parameters) consistent with the requirement of a stable vacuum spontaneously breaking supersymmetry - see \cite{mu2} - whereas ${\hat{\mu}}^2 \ll {\hat{\mu} B}$ for components with only Wilson line modulus as well as the same  mixed with the $D3$-brane position moduli. Also, the un-normalized physical mu-parameters for the $D3$-brane position moduli ($\hat{K}_{{\cal Z}_i{\bar{\cal  Z}}_i}\hat{\mu}_{{\cal Z}_i{\cal Z}_i}$) are $\sim$ TeV, as required for having correct electroweak symmetry breaking \cite{mu2,mu1}.   Our results are summarized in {\bf Table A.4}.

It will be interesting to see what happens to the couplings with the inclusion of higher derivative terms - one expects to include
$\frac{1}{48}\int_{\Sigma_B}\left(p_1\left(T\Sigma_B\right) - p_1\left(N\Sigma_B\right)\right)$ as an additive shift to ${\cal F}$ of section {\bf 2} (See \cite{V_D7_fl}).

\section{RG Flow of Squark and Slepton Masses to the EW Scale}
In the context of string phenomenology, the study of the origin and dynamics of SUSY breaking are among the most challenging issues and several proposals for the origin of SUSY breaking as well its transmission to the visible sector with a particular structure of soft parameters have been studied. For addressing realistic model-building issues, the gaugino masses as well as other soft SUSY-breaking parameters have to be estimated at low energy which requires the study of their running to electroweak scale using the respective RG-equations, imposing the low energy FCNC constraints. Further, ratio of gaugino masses to the square of gauge couplings ($\frac{M_a}{g_a^2}$), are well-known RG-invariants at one loop  as their RG-running up to two-loops are \cite{sparticlesreview}:

\begin{eqnarray}
& & \frac{dg_a}{dt} = \frac{{g_a}^3b^a}{16 {\pi}^2}  + \frac{{g_a}^3}{16 {\pi}^2}\biggl[\sum^3_{b=1}B^{(2)}_{ab}{g_b}^2- {\frac{1}{16 \pi^2}}\sum_{x=u,d,e,\nu}{C^a_x}/{16 \pi^2} Tr[{Y_x}^{\dagger}{Y_x}]\biggr] \nonumber\\
& & \frac{dM_a}{dt} = \frac{2{g_a}^2b^a M_a}{16 {\pi}^2}  + \frac{2{g_a}^2}{({16 {\pi}^2})^2}\biggl[\sum^3_{b=1}B^{(2)}_{ab}{g_b}^2 (M_a+M_b)\biggr]\nonumber\\
& & + \frac{2{g_a}^2}{({16 {\pi}^2})^2}\biggl[\sum_{x=u,d,e,\nu} {C^a_x}\left\{Tr[{Y_x}^{\dagger}{\tilde{A_x}}]-M_a Tr[{Y_x}^{\dagger}{Y_x}]\right\}\biggr]
\end{eqnarray}
\noindent where $t=ln({\frac{Q_{EW}}{{Q_0}}})$ defined in terms of ${Q_{EW}}$ which is the phenomenological low energy scale (of interest) and $Q_0$ some high energy scale alongwith the MSSM gauge coupling $\beta$-functions' given as $b^a=\{{{33}/{5},1,-3}\}$, $B^{(2)}_{ab}$ and $C^a_x$ being $3\times3$ and $4\times3$ matrices with ${\cal O}(1-10)$ components and ${\tilde{A_x}}$, $Y_x$ are trilinear $A$-term and Yukawa-coupling respectively. Further, the first term on the right hand sides of each of above equations represents one-loop effect while other terms in the square brackets are two-loop contributions to their RG running implying that $ d/dt\left[\frac{M_a}{{g_a}^2}\right]=0$ at one-loop.

RG equations of first family of squark and slepton masses result in the following set of equations  which represent the difference in their mass-squared values between  ${Q_{EW}}$ and  $Q_0$ at one-loop level {\cite{Martinreview,QuevedoLHC}}:
\begin{eqnarray}
\label{eq:RGsparticles}
& & {M^2_{\tilde{d}_L,\tilde{u}_L}}\bigg|_{Q_{EW}}-M^2_{\tilde{d}_L,\tilde{u}_L}\bigg|_{Q_0}={\cal K}_3+{\cal K}_2+{\frac{1}{36}}{\cal K}_1+ \tilde{\Delta}_{\tilde{d_L}}\nonumber\\
& & {M^2_{\tilde{d}_R}}\bigg|_{Q_{EW}}-M^2_{\tilde{d}_R}\bigg|_{Q_0}={\cal K}_3+{\frac{1}{9}}{\cal K}_1+ \tilde{\Delta}_{\tilde{d_R}}\nonumber\\
& & {M^2_{\tilde{u}_R}}\bigg|_{Q_{EW}}-M^2_{\tilde{u}_R}\bigg|_{Q_0}={\cal K}_3+{\frac{4}{9}}{\cal K}_1+ \tilde{\Delta}_{\tilde{u_R}}\nonumber\\
& & {M^2_{\tilde{e}_L}}\bigg|_{Q_{EW}}-M^2_{\tilde{e}_L}\bigg|_{Q_0}={\cal K}_2+{\frac{1}{4}}{\cal K}_1+ \tilde{\Delta}_{\tilde{e_L}}\nonumber\\
& & {M^2_{\tilde{e}_R}}\bigg|_{Q_{EW}}-M^2_{\tilde{e}_R}\bigg|_{Q_0}={\cal K}_1+ \tilde{\Delta}_{\tilde{e_R}}
\end{eqnarray}
where parameters ${\cal K}_a$ are the contributions to scalar masses RG running via gaugino massesare defined through integral (\ref{eq:Kintegrals}) below and the difference in the coefficients of ${\cal K}_1$ in the above set of solutions to respective RG equations is due to various weak hypercharge-squared values for each scalar:
\begin{equation}
\label{eq:Kintegrals}
{\cal K}_a\sim {{\cal O}({\frac{1}{10}})}\int_{ln Q_0}^{ln Q_{\rm EW}} dt g_a^2(t)M_a^2(t).
\end{equation}
Further, ${\tilde{\Delta}_{\tilde{x}}}$ (appearing in (\ref{eq:RGsparticles})), where ${\tilde{x}} \in \{{\tilde{d_L}},{\tilde{d_R}},{\tilde{u_L}},{\tilde{u_R}},{\tilde{e_L}},{\tilde{e_R}},{\tilde{\nu}}\}$ (i.e. the first family of squarks and sleptons)  are D-term contributions which are ``hyperfine" splitting in squark and slepton masses arising due to quartic interactions among squarks and sleptons with Higgs. These ${\tilde{\Delta}_{\tilde{x}}}$ contribution are generated via the neutral Higgs acquiring VEVs in electroweak symmetry breaking and are of the form \cite{Martinreview}:
$ {\tilde{\Delta}_{\tilde{x}}}\equiv[T_{3{\tilde{x}}}-Q_{{\tilde{x}}} {\rm Sin}^2(\theta_W)]{\rm Cos}(2\beta) m^2_Z$,
where $T_{3{\tilde{x}}}$ and $Q_{{\tilde{x}}}$ are third component of weak isospin and the electric charge of the respective left-handed chiral supermultiplet to which ${\tilde{x}}$ belong. The angle $\theta_W$ is electroweak mixing angle, $m_Z\sim100{\rm GeV}$ and  ${\rm tan}(\beta)$ is the ratio of vevs of the two Higgs  after electroweak symmetry breaking. Now, in our setup $Q_0\equiv{M_{\rm string}}={M_{\rm GUT}}/{10}\sim 10^{15}GeV$ and $Q_{\rm EW}\sim TeV$. As argued in \cite{sparticlesreview,Martinreview,QuevedoLHC}, up to one loop, ${d}/{dt}({M_a}/{g_a^2})=0$. Hence, the aforementioned ${\cal K}_a$-integrals (\ref{eq:Kintegrals}) can be written as:
\begin{equation}
{\cal K}_a\equiv {{\cal O}(\frac{1}{10})} \biggl(\frac{M_a}{g_a^2}\biggr)^2\bigg|_{Q_0}\biggl[g_a^4\bigg|_{Q_{EW}}-g_a^4\bigg|_{Q_0}\biggr]_{\rm 1-loop}.
\end{equation}
As argued in \cite{Kap_Louis}, the gauge couplings run as follows (up to one loop):
\begin{eqnarray}
\label{eq:RG_flow_1}
\frac{16\pi^2}{g_a^2(Q_{EW})}=\frac{16\pi^2}{g_a^2(Q_0)} + 2b_a ln\biggl[\frac{Q_0}{m_{{3}/{2}}}\biggr] + 2b_a^\prime ln\biggl[\frac{m_{{3}/{2}}}{Q_{EW}}\biggr] + \Delta_a^{\rm 1-loop},\nonumber\\
\end{eqnarray}
where $b_a, b_a^\prime$ are group-theoretic factors and $\Delta_a^{\rm 1-loop}\sim {\bf Tr} {ln ({{\cal M}}/{m_{{3}/{2}}})},$ and  ${\cal M}\equiv e^K\hat{K}^{-\frac{1}{2}}\mu^\dagger(\hat{K}^{-1})^T\mu\hat{K}^{-\frac{1}{2}}$ (See \cite{Kap_Louis}), in our setup, is to be evaluated for the $D7$ Wilson-line modulus ${\cal A}_1$'s. One can show that ${\cal M}_{{\cal A}_1}\sim{\cal V}^{-\frac{13}{6}}$ and  $m_{{3}/{2}}\sim{\cal V}^{-\frac{n^s}{2}-1}$. Hence, from (\ref{eq:RG_flow_1}) one obtains:
$$\frac{16\pi^2}{g_a^2(Q_{EW})}=\frac{16\pi^2}{g_a^2(Q_0)} + {\cal O}(10).$$
As argued in \cite{D3_D7_Misra_Shukla}, in anomaly-mediated scenarios, $\frac{M_a}{g_a^2}\bigg|_{Q_0}\sim\frac{{\cal V}^{\frac{1}{18}}m_{{3}/{2}}}{8\pi^2}$ which, bearing in mind that the same is generically suppressed relative to the gravity-mediation result by about ${1}/{8\pi^2}$ (See \cite{D3_D7_Misra_Shukla}), implies
$\frac{M_a}{g_a^2}\bigg|_{Q_0}\sim{\cal V}^{\frac{1}{18}}m_{{3}/{2}}$ as the gravity-mediation result. Further,
using ${1}/{g_a^2}\sim{\cal V}^{\frac{1}{18}}$, one obtains from (\ref{eq:Kintegrals}) :
\begin{eqnarray*}
& & K_a\sim {{\cal O}\bigg({\frac{1}{10}}\biggr)}{\cal V}^{\frac{1}{9}}m_{{3}/{2}}^2\Biggl(-{\cal V}^{-\frac{1}{9}}+\frac{(16\pi^2)^2}{\bigl[{\cal O}(10)+16\pi^2{\cal V}^{\frac{1}{18}}\bigr]^2}\Biggr)\sim m_{{3}/{2}}^2\biggl\{\frac{{\cal O}(1)}{340}\biggr\}
\end{eqnarray*}
for ${\cal V}\sim10^6$. Hence, for $m_{{3}/{2}}\sim10TeV$ (which can be realized in our setup - see \cite{D3_D7_Misra_Shukla}), one obtains $K_a\sim0.3(TeV)^2$ to be compared with $0.5(TeV)^2$ as obtained in \cite{QuevedoLHC}; an mSUGRA point on the ``SPS1a slope" has a value of around $(TeV)^2$. Further the ${\tilde{\Delta}_{\tilde{x}}}$ contributions, being proportional to $m^2_Z$, are suppressed as compared to ${\cal K}_a$-integrals at one-loop.

Further, as suggested by the phenomenological requirements, the first and second family of squarks and sleptons with given gauge quantum number are supposed to possess (approximate) universality in the soft parameters. However, the third family of squark and sleptons, feeling the effect of larger Yukawa's, can get normalized differently. For our setup, we have $D3$-brane position moduli and $D7$-brane Wilson line moduli, which could be suitable candidates to be identified with the Higgs, squarks and sleptons of the MSSM, given that they fulfil the phenomenological requirements. In our setup, one can see  that at a string scale of $10^{15}GeV$,  $D3$-brane position moduli masses are universal with a value of the order $10^4 \rm TeV$ (as ${m_{{\cal Z}_i}}\sim{\cal V}^{\frac{19}{36}} m_{\frac{3}{2}}\sim 10^4 TeV$ for ${\cal V}\sim 10^6$ corresponding to a $10 \rm TeV$ gravitino)  and may be identified with the two Higgses of MSSM spectrum. The Wilson line moduli have masses $\sim {\cal V}^{\frac{73}{72}} m_{\frac{3}{2}}\sim 10^7 TeV$ and are hence heavier than the $D3$-brane position moduli. Moreover, in our Swiss-Cheese orientifold setup the trilinear $A$-terms show universality and are calculated to be $\sim 10^7 {\rm TeV}$ along with the physical Yukawa-couplings which are found to be in the range from a negligible value $\sim{\cal V}^{-\frac{133}{24}}\sim 10^{-33}$ (for purely Wilson line moduli contributions) to a relatively high value $\sim{\cal V}^{-\frac{5}{24}}\sim 10^{-1}$ (for purely $D3$-brane position moduli contributions). As suggested by phenomenology, the first and second family of squarks and slepton masses involve negligible Yukawa-couplings. The Wilson line moduli in our setup could hence be identified with the first and second family of squarks and sleptons. Further, within the one-loop results and dilute flux approximation in our Swiss-Cheese LVS setup, gaugino masses are (nearly) universal with $m_{\rm gaugino}\sim m_{\frac{3}{2}}\sim 10{\rm TeV}$ at the string scale  $M_s\sim10^{15}GeV$ which being nearly the GUT scale would imply that the gauge couplings are almost unified.

\section{Realizing Fermion and Neutrino Masses}

\hskip1in {\it{``There is for me powerful evidence that there is something going on behind it all....It seems as though somebody has fine-tuned nature's numbers to make the Universe....The impression of design is overwhelming".}} - Paul Davies.\\

As we argued in in the previous section, the spacetime filling mobile D3-brane position moduli ${\cal Z}_i$'s and the Wilson line moduli ${\cal A}_{I}$'s could be respectively identified with the two-Higgses and sparticles (squarks and sleptons) of some (MS)SM like model. Now, we look at the fermion sector of the same Wilson line moduli (denoting the fermionic superpartners of ${\cal A}_I$'s as $\chi_I$'s). The fermion bilinear terms in the 4-dimensional effective action, which are generated from $\int d^4 x \, e^{\hat{K}/2} \partial_{\alpha}\partial_{\beta} {W} \psi^{\alpha} \psi^{\beta}$, can be given as:\be
\label{eq:fermion_bilinear}
\int d^4 x \,e^{\hat{K}/2} Y_{\alpha \beta \gamma} {\cal Z}^{\alpha} \chi^{\beta}
\chi^{\gamma} + e^{\hat{K}/2} \frac{{\cal O}_{\alpha \beta \gamma \delta}}{2 M_P}
{\cal Z}^{\alpha} {\cal Z}^{\beta} \chi^{\gamma} \chi^{\delta}
\ee
The first term in the above equation is responsible for generating fermion masses via giving some VEV to Higgs fields while the second term which is a lepton number violating term generates neutrino masses. We will elaborate on these issues in the respective subsections below.

\subsection{Fermion Masses} The relevant fermionic bilinear terms in the four dimensional effective action can be schematically written in terms of canonically normalized superfields ${\cal Z}^i$ and ${\cal A}^I$  as:
$\int d^4xd^2\theta \,\hat{Y}_{iIJ} {\cal Z}^i {\cal A}^I {\cal A}^J$. Now the fermionic masses are generated through Higgs mechanism by giving VEV to Higgs fields:
\begin{equation}
\label{eq: fermion_masses}
M_{IJ} = {\hat {Y}_{iIJ} <z_i }>
\end{equation}
where $\hat {Y}_{iIJ}$'s are  ``Physical Yukawas" defined as $\hat {Y}_{iIJ}= \frac{e^{\hat{K}/2} Y_{iIJ}}
 {\sqrt {K_{i{\bar i}}} \sqrt {K_{I{\bar I}}} \sqrt {K_{J{\bar J}}}}$ and the Higgs fields
 $z_i$'s are given a vev: $<z_i> \sim {{\cal V}^{\frac{1}{36}}} M_p$
\cite{D3_D7_Misra_Shukla}. 
Next, we discuss the possibility of realizing fermion masses in the range ${\cal O}({\rm MeV-GeV})$ in our setup, possibly corresponding to any of the masses $m_e=0.51$ MeV, $m_u=5$ MeV, $m_d=10$ MeV, $m_s=200$ MeV, $m_c=1.3$ GeV - the first two generation fermion masses \cite{fermionvaluesSM}.

Using (\ref{eq:W_exp}) the physical Standard Model-like ${\cal Z}_i{\cal A}_I^2$ Yukawa couplings  are \cite{D3_D7_Misra_Shukla}
\begin{equation}
\hat{Y}_{{\cal A}_1{\cal A}_1{\cal Z}_i}\sim{\cal V}^{-\frac{199}{72}-\frac{n^s}{2}}
\end{equation}
The leptonic/quark mass is given by:
${\cal V}^{-\frac{197}{72}-\frac{n^s}{2}}$ in units of $M_p$, which  implies a range of fermion mass $m_{\rm ferm}\sim{\cal O}({\rm MeV-GeV})$ for Calabi Yau volume ${\cal V}\sim {\cal O}(6\times10^5-10^6)$. For example, a mass of $0.5$ MeV could be realized with Calabi Yau volume ${\cal V}\sim 6.2\times 10^5, n^s=2$. In MSSM/2HDM models, up to one loop, the leptonic (quark) masses do not change (appreciably) under an RG flow from the intermediate string scale down to the EW scale (See \cite{Das_Parida}). This way, we show the possibility of realizing all fermion masses of first two generations in our setup. Although we do not have sufficient field content to identify all first two families' fermions, we believe that the same could be realized after inclusion of more Wilson line moduli in the setup. The above results also make the possible identification of Wilson line moduli with squarks and sleptons of first two families  \cite{Sparticles_MS}, more robust.

\subsection{Neutrino Mass} The non-zero neutrino masses are generated through the Wienberg (type-) dimension five operators arising from lepton number violating term of (\ref{eq:fermion_bilinear}) and are given as \cite{conlon_neutrino}
\be
m_{\nu}=\frac{v^2 sin^2\beta \hat{{\cal O}}_{{\cal Z}_i{\cal Z}_j{\cal Z}_k{\cal Z}_l}}{2M_p}
\ee where $\hat{{\cal O}}_{{\cal Z}_i{\cal Z}_i{\cal Z}_i{\cal Z}_i}$ is the coefficient of the physical/noramlized quartic in the $D3$-brane position moduli ${\cal Z}_i$ which are defined in terms of diagonal basis of K\"{a}hler potential in ${\cal Z}_i$'s and is given as
\be\hat{{\cal O}}_{{\cal Z}_i{\cal Z}_i{\cal Z}_i{\cal Z}_i}={\frac{{e^\frac{\hat{K}}{2}}{\cal O}_{{\cal Z}_i{\cal Z}_j{\cal Z}_k{\cal Z}_l}}{{\sqrt{\hat{K}_{{\cal Z}_i{\bar{\cal Z}}_{\bar i}}\hat{K}_{{\cal Z}_j{\bar{\cal Z}}_{\bar j}}\hat{K}_{{\cal Z}_k{\bar{\cal Z}}_{\bar k}}\hat{K}_{{\cal Z}_l{\bar{\cal Z}}_{\bar l}}}}}}\ee where
$v sin\beta$ being the vev of the $u$-type Higgs $H_u$ and $sin\beta$ defined via
$tan\beta={\langle H_u\rangle}/{\langle H_d\rangle}$.
Also in the simultaneous diagonal basis of ${K_{i\bar{j}}}$ and ${{Z}_{ij}}$, the fluctuations in D3-brane position moduli $\delta z_i$ (about $z_i\sim{\cal V}^{\frac{1}{36}}$) and Wilson line modolus $\delta {\cal A}_i$ (about ${\cal A}_I\sim{\cal V}^{-\frac{1}{4}}$) implies a small (negligible in LVS limit) mixing defined as ${\cal Z}_i = \delta z_i + \lambda_i \delta {\cal A}_i {\cal V}^{-\frac{8}{9}}$ and $\tilde{\cal A}_I=(\beta_1 \delta z_1 + \beta_2 \delta z_2){\cal V}^{-\frac{8}{9}} + \delta {\cal A}_I$, where $\lambda_is, \beta_is$ being some ${\cal O}(1)$ constants. 
Now expanding out superpotential (\ref{eq:W}) as a power series in ${\cal Z}_i$ one can show that the coefficient of unnormalized quartic comes out to be:
\begin{eqnarray}
\label{eq:O}
& & {\cal O}_{{\cal Z}_i{\cal Z}_j{\cal Z}_k{\cal Z}_l}\sim \frac{2^{n^s}}{24}10^2(\mu_3 n^s l^2)^4{\cal V}^{\frac{n^s}{2}+\frac{1}{9}} e^{-n^s {\rm vol}(\Sigma_S) + i n^s \mu_3l^2{\cal V}^{\frac{1}{18}}(\alpha+i\beta)}.\nonumber\\
\end{eqnarray}
where $\alpha,\beta\sim{\cal O}(1)$ constants and using $l=2\pi\alpha^\prime$, $\mu_3=\frac{1}{(2\pi)^3(\alpha^\prime)^2}$ and the results of \cite{D3_D7_Misra_Shukla}: 
$z_i=\gamma_i{\cal V}^{\frac{1}{36}}, i=1,2$ and ${\rm vol}(\Sigma_S)=\gamma_3 ln {\cal V}$ such that $\gamma_3 ln{\cal V} + \mu_3l^2\beta {\cal V}^{\frac{1}{18}} = ln {\cal V}$ (here $\gamma_i$'s and $\beta$ are order one constants) 
alongwith  $\hat{K}_{{\cal Z}_i{\bar{\cal Z}}_{\bar i}}\sim\frac{{\cal V}^{\frac{1}{72}}}{\sqrt{\sum_\beta n^0_\beta}}$, which assuming a holomorphic isometric involution $\sigma$ as part of the Swiss-Cheese orientifolding action $(-)^{F_L}\Omega\cdot \sigma$ such that $\sum_\beta n^0_\beta\sim\frac{{\cal V}}{10}$, the estimated ${\hat{\cal O}}_{{\cal Z}_i{\cal Z}_j{\cal Z}_k{\cal Z}_l}$ is given as above. 

Now, we will elaborate on running of the neutrino mass by first discussing the RG flow of $\langle H_u\rangle$ and then the coefficient ($\kappa_{ij}$) of dimension-five operator.
The RG flow of neutrino masses can be estimated through the running of coefficient $\kappa_{ij}$ of dimension-five operator $\kappa_{ij}L_iH.L_jH$. 
Unlike MSSM, usually there are two dimension-five operators in 2HDM corresponding to the Higgses, however as we have taken the two Higgses to be on the same footing in our setup and hence along this locus, there is only one type dimension-five operator in the 2HDM as well. In this limit, to have an estimate about the running of coefficient $\kappa_{ij}$ of dimension-five operator $\kappa_{ij}L_iH.L_jH$, the RG flow equation for $\kappa$ (the $\kappa^{(22)}$ of \cite{Babu_et_al}) is:
\begin{eqnarray}
\label{eq:RG_k_2HDM}
& & 8\pi^2\frac{d\kappa}{d(ln\mu)}=\left(tr\left(3Y^u Y^u\ ^\dagger\right)-4\pi\left(3\alpha_2+\frac{3}{5}\alpha_1\right)\right)\kappa
\nonumber\\
& & + \frac{1}{2}\left(\left(Y^e Y^e\ ^\dagger\right)\kappa + \kappa\left(Y^e Y^e\ ^\dagger\right)^T\right) + 2\lambda\kappa,
\end{eqnarray}
where $Y^u$, $Y^e$ are up quark and electron Yukawa coupling matrices, $\alpha_i$'s are the $U(1)$ and $SU(2)$ fine structure constants while $\lambda$ 
is the coefficient of $(\Phi^\dagger\Phi)^2$ in the Lagrangian. Assuming the $U(1)$ fine structure constant to be equal to $\lambda$ \footnote{Given that $\lambda~(n^s\mu_3l^2)^4\sim1/\pi^4$ in our setup this would imply, e.g., at the string scale $g_{U(1)}^2\sim0.02$, which is quite reasonable.} one then says that the 2HDM and MSSM RG flow equations $\kappa$ become identical. The analytic solution to RG running equation (\ref{eq:RG_k_2HDM}) is given by:
$$\kappa(M_s)
=\frac{\kappa_{\tau\tau}(M_s)}{\kappa_{\tau\tau}(M_{EW})}\left(\begin{array}{ccc}
\frac{I_e}{I_\tau}& 0 & 0\\
0 & \frac{I_\mu}{I_\tau} & 0 \\
0 & 0 & 1
\end{array}\right)\kappa(M_{EW})\left(\begin{array}{ccc}
\frac{I_e}{I_\tau}& 0 & 0\\
0 & \frac{I_\mu}{I_\tau} & 0 \\
0 & 0 & 1
\end{array}\right)$$
where
\begin{equation}
I_{e/\mu/\tau}=e^{\frac{1}{8\pi^2}\int_{ln M_{EW}}^{ln M_S}dt\hat{Y}_{e/\mu/\tau}^2}.
\end{equation}
and as in \cite{RG_neutrino_I}, for $$tan\beta=\frac{\langle z_1\rangle}{\langle z_2\rangle}<50,\, \, \frac{I_{e/\mu}}{I_\tau}\approx \left(1-\frac{\hat{Y}_\tau^2}{8\pi^2}ln\left(\frac{M_S}{M_{EW}}\right)\right).$$
For MSSM as well as 2HDM RG flows of fermionic masses, one sees that the same change very little with change of energy scales \cite{Das_Parida}. Hence, $\langle {H_u}\rangle_{M_S}\hat{Y}_\tau(M_S)\sim\langle {H_u}\rangle_{M_{EW}}$ $\hat{Y}_\tau(M_{EW})\sim m_\tau(M_{EW})\sim 10 GeV,$ which for $\langle {H_u}\rangle_{M_S}\sim M_p$ implies $Y_\tau(M_S)\sim 10^{-17}$. Hence, $I_{e/\mu/\tau}\approx 1$. As $\kappa_{\tau\tau}$ is an overall factor in $\tau$, one can argue that it can be taken to be scale-independent \cite{energy_MNS}. Hence, in MSSM and 2HDM, the coefficient of quartic term $\hat{{\cal O}}_{{\cal Z}_i{\cal Z}_i{\cal Z}_i{\cal Z}_i}$ does not run.

Next, using the one-loop solution to the $\langle H_u\rangle$ RG flow equation for the 2HDM \cite{Das_Parida} and subsequently using the one-loop RG flow solution for
$\alpha_i$ in the dilute flux approximation \cite{Sparticles_MS}, one obtains:
\begin{eqnarray}
\label{eq:sol_2HDM_Hu_I}
& & \langle H_u\rangle_{M_{EW}}=\langle H_u\rangle_{M_{s}}\Biggl[e^{\frac{3}{16\pi^2}\int_{ln(M_{EW})}^{\ln(M_S)}Y^2_t dt^\prime}
\bigl[{\frac{\alpha_1(M_s)}{\alpha_1(M_{EW})}}\bigr]^{\frac{-3}{56}}\nonumber\\
& & \bigl[{\frac{\alpha_2(M_s)}{\alpha_2(M_{EW})}}\bigr]^{\frac{3}{8}}\Biggr]\sim\langle H_u\rangle_{M_{s}}\left(1-\frac{{\cal O}(40)g^2(M_s)}{16\pi^2}\right)^{\frac{9}{28}}.
\end{eqnarray}
where exponential factor is order one as $Y_t(M_{EW})\sim {\cal O}(1)$ and there is a loop suppression factor. Hence, it can be seen that by requiring $g^2(M_s)$ to be sufficiently close to $[16\pi^2/{\cal O}(40)]\sim 4$, one can RG flow $\langle H_u\rangle_{M_s}$ to the required value $\langle H_u\rangle_{M_{EW}}\equiv (v sin\beta)_{M_{EW}}\sim246 GeV$ in the large $tan\beta$ regime. Finally, assuming that we are working in the large $\tan\beta$ regime and using above inputs pertaining to the RG flow of $\langle H_u\rangle$ and $\kappa$ along with ${\cal V}\sim10^6 l_s^6$ and $ n^s=2$, one obtains:
 \begin{equation}
 \label{eq:final_I}
 m_{\nu}\sim\frac{(v sin\beta)_{M_{EW}}^2\hat{O}_{{\cal Z}_i{\cal Z}_j{\cal Z}_k{\cal Z}_l}}{2M_p}\stackrel{<}{\sim} 1eV.
 \end{equation}

\section{Proton Decay}
\hskip3.8in {\it{``Protons Are Not Forever".}} \footnote{From the book ``Quantum Field Theory in a nutshell- A. Zee."}

The possibility of proton decay in Grand unified theories is caused by higher dimensional B-number-violating operators. In SUSY and SUGRA GUTs the most important contributions for proton decay come from dimension-four and dimension-five B-number-violating operators (which are model dependent), however for non-supersymmetric GUTs dimension-six operators are most important (see \cite{Prot_Decay_review} and references therein). Further in the supersymmetric GUTs-like theories, the contributions for proton decay coming from dimension-four operators are usually absent due to symmetries of model and the next crucial and potentially dangerous contributions are due to dimension-five and dimension-six operators. Also gauge dimension-six operators conserve $B-L$ and hence possible decay channels coming from these contributions are a meson and an antilepton, (e.g. $p\longrightarrow K^{+}{\bar\nu}$, $p\longrightarrow \pi^{+}{\bar\nu}$, $p\longrightarrow K^{0}{\bar e}$, $p\longrightarrow \pi^{0}{\bar e}$ etc. \cite{Prot_Decay_review}). Further the B-number-violating dimension-five operators in such (SUSY GUT-type) models relevant to proton decay are of the type: $({\rm squark})^2({\rm quark})({\rm lepton})$ or $({\rm squark})^2({\rm quark})^2$ (See \cite{Prot_Decay_review,EllisNanopRudaz}). This would correspond to ${\partial^2W}/{\partial{\cal A}_I^2}|_{\theta=0}(\chi^I)^2$, in our setup.  From (\ref{eq:W}), we see that as long as the mobile $D3-$brane is restricted to $\Sigma_B$, there is no ${\cal A}_I$-dependence of $W$ implying the stability of the proton up to dimension-five operators. Now, using the notations and technique of \cite{D3_D7_Misra_Shukla}, consider a holomorphic one-form $$A_2=\omega_2(z_1,z_2)dz_1+\tilde{\omega}_2(z_1,z_2)dz_2$$ where $\omega_2(-z_1,z_2)=\omega_2(z_1,z_2), \tilde{\omega}_2(-z_1,z_2)=-\tilde{\omega}_2(z_1,z_2)$ (under $z_1\rightarrow-z_1,z_{2,3}\rightarrow z_{2,3}$) and $\partial A_2=(1+z_1^{18}+z_2^{18}+z_3^3-\phi_0z_1^6z_2^6)dz_1\wedge dz_2$ (implying $dA_2|_{\Sigma_B}=0$).
Assuming $\partial_1\tilde{\omega}_2=-\partial_2\omega_2$, then
around $|z_3|\sim{\cal V}^{1/6}, |z_{1,2}|\sim{\cal V}^{1/36}$ - localized around the mobile $D3$-brane - one estimates $$\tilde{\omega}_2(z_1,z_2)\sim z_1^{19}/19+z_2^{18}z_1+\sqrt{\cal V}z_1-\phi_0/7z_1^7z_2^6$$ with $\omega_2(z_1,z_2)=-\tilde{\omega}_2(z_2,z_1)$ in the LVS limit, and utilizing the result of \cite{D3_D7_Misra_Shukla} pertaining to the $I=J=1$-term, one hence obtains: $$i\kappa_4^2\mu_7C_{I{\bar J}}a_I{\bar a}_{\bar J}\sim{\cal V}^{7/6}|a_1|^2+{\cal V}^{2/3}(a_1{\bar a}_{\bar 2}+c.c.)+{\cal V}^{1/6}|a_2|^2,$$ $a_2$ being another Wilson line modulus. Noting the large $n^0_\beta$'s and assuming $a_I$ to be stabilized at around ${\cal V}^{-1/4}$ (See \cite{D3_D7_Misra_Shukla}) and hence a partial cancelation between $vol(\Sigma_B)$ and
$i\kappa_4^2\mu_7C_{1{\bar 1}}|a_1|^2$ in $T_B$, consider fluctuation in $a_2$ about ${\cal V}^{-1/4}$: $a_2\rightarrow{\cal V}^{-1/4}+a_2$. The K\"{a}hler potential, in the LVS limit will then be of the form
\begin{eqnarray*}
& & K\sim-2 ln\Biggl[({\cal V}^{1/6}+{\cal V}^{5/12}({\cal A}_2+c.c.)+{\cal V}^{1/6}{\cal A}_2^\dagger{\cal A}_2)^{3/2} +\sum n^0_\beta(...)\Biggr]\end{eqnarray*}
 - $a_2$ promoted to the Wilson line modulus superfield ${\cal A}_2$. When expanded in powers of the canonically normalized $\hat{\cal A}_2$, the SUSY GUT-type four-fermion dimension-six proton decay operator obtained from $$\int d^2\theta d^2{\bar\theta}({\cal A}_2)^2
({\cal A}_2^\dagger)^2/M_p^2(\in K(\hat{\cal A}_I,\hat{\cal A}_I^\dagger,...))$$ will yield
$$\frac{\left({\cal V}^{5/4}/\sum n^0_\beta\right)\left(\chi_2^4/M_p^2\right)}{\left(\sqrt{\hat{K}_{{\cal A}_2{\bar{\cal A}}_2}}\right)^4}.$$ Like the single Wilson-line modulus case of \cite{D3_D7_Misra_Shukla}, 
 $$\sqrt{\hat{K}_{{\cal A}_2{\bar{\cal A}}_2}}\sim\frac{{\cal V}^{65/72}}{\sqrt{\sum_\beta n^0_\beta}}.$$ For ${\cal V}\sim 10^6, \sum n^0_\beta\sim\frac{{\cal V}}{10}$ (as in the previous section), the numerical factor approximates to $(10^{-9/2}/M_p)^2$. Using arguments of \cite{Prot_Decay_review} and \cite{Witten_Klebanov_proton_decay}, one expects the proton lifetime to be estimated at: $$\frac{{\cal O}(1)\times L_{\Sigma_B}^{-4/3}(10^{9/2}M_p)^4}{(\alpha^2(M_s)m_p^5)},$$ where $L_{\Sigma_B}$ is the Ray-Singer torsion of $\Sigma_B$. $L_{\Sigma_B}$ can in principle be calculated generalizing the large-volume limit of the metric of $\Sigma_B$ worked out in \cite{D3_D7_Misra_Shukla} using GLSM techniques, via the Donaldson algorithm (See \cite{Braun_et_al}). For the time being, we assume it to be ${\cal O}(1)$ and obtain an upper bound on the proton lifetime to be around $10^{61}$ years, in conformity with the very large sparticle masses in our setup.

\section{Summary and Discussion}
We estimated various soft supersymmetry breaking masses/parameters in the context of D3/D7 LVS Swiss-Cheese setup framed in the previous chapter {\bf 4} and realized order TeV gravitino and gaugino masses in the context of gravity mediated supersymmetry breaking. The anomaly mediated gaugino mass contribution was observed to be suppressed by a loop factor as compared to gravity mediated contribution. The the $D3$-brane position moduli and the $D7$-brane Wilson line moduli were found  to be heavier than gravitino. Further we observed a (near) universality in the masses, $\hat{\mu}$-parameters, Yukawa couplings  and the $\hat{\mu}B$-terms for the $D3$-brane position moduli - the two Higgses in our construction - and a hierarchy in the same set and a universality in the $A$ terms on inclusion of the $D7$-brane Wilson line moduli.  Based on phenomenological intuitions, we further argued that the Wilson line moduli could be be identified with the squarks/sleptons (at least the first and second families) of MSSM as the Yukawa couplings for the same were negligible; the non-universality in the Yukawa's for the Higgses and squarks, was hence desirable. Building up on some more phenomenological aspects of our setup, we discussed the RG flow of the slepton and squark masses to the EW scale and in the process showed that related integrals are close to the mSUGRA point on the ``SPS1a slope". Further, we showed the possibility of realizing fermions mass scales of first two generations along with order $eV$ neutrino masses for Calabi Yau volume ${\cal V}\sim(10^5-10^6)$ and D3-instanton number $n^s=2$. A detailed numerical analysis for solving the RG evolutions will definitely explore some more interesting phenomenology in the context of reproducing MSSM spectrum in this LVS Swiss-Cheese orientifold setup.

%% file: chap6.tex
\chapter{Some Other Implications in (Fluxed) Compactification Geometries}
\markboth{nothing}{\bf 6. Some Other Implications in (Fluxed) Compactification Geometries}
\hskip1in{\it{`` Everything we see hides another thing, we always want to see what is hidden by what we see."}}

\hskip4.5in - Rene Magritte.

\section{Introduction}
In the context of moduli stabilizations, inclusion of fluxes has been very crucial (See \cite{Grana}). All complex structure moduli along with axion-dilaton get stabilized by turning on fluxes, however for the K\"{a}hler moduli stabilization, non-perturbative effected have been required \cite{KKLT}. In the context of type II compactifications, it has been naturally interesting to look for examples wherein it may be possible to stabilize the complex structure moduli (and the axion-dilaton modulus) at different points of the moduli space that are finitely separated, for the {\it same} value of the fluxes. This phenomenon is referred to as ``area codes" that leads to formation of domain walls. Further,  there has been a close connection between flux vacua and black-hole attractors. As extremal black holes exhibit an interesting phenomenon - the attractor mechanism \cite{attractor} in which, the moduli get ``attracted" to some fixed values determined by the charges of the black hole, independent of the asymptotic values of the moduli and it has been extremely interesting to investigate the attractor behaviors via looking at the black hole solutions in effective low energy theories. Supersymmetric black holes at the attractor point, correspond to minimizing the central charge and the effective black hole potential, whereas non-supersymmetric attractors \cite{nonsusybh1}, at the attractor point, correspond to minimizing only the potential and not the central charge \cite{nonsusybh2}.

In this chapter, we address the issues in the context of moduli stabilization, like aspects of (non-)supersymmetric flux vacua and black holes in the context of type II compactifications on (orientifold) of the same Swiss-Cheese Calabi-Yau's which has multiple singular conifold loci. In section {\bf 2}, based on \cite{Candelasetal}, we perform a detailed analysis of the periods of the Swiss-Cheese Calabi-Yau three-fold we have been using, working out their forms in the symplectic basis for points away and close to the two singular conifold loci. We then discuss, in section {\bf 3}, stabilization of the complex structure moduli including the axion-dilaton modulus by extremizing the flux superpotential for points near and close to the two conifold loci, arguing the existence of ``area codes" and domain walls. In section {\bf 4},  we explicitly solve the ``inverse problem" using the techniques of \cite{VafaInverse}. In section {\bf 5}, using the techniques of \cite{Ceresole+Dall'agata} we show the existence of multiple superpotentials (including therefore ``fake superpotentials"). Finally, we summarize the results in section {\bf 6}.

\section{The Moduli Space Scan and the Periods}

In this section, based on results in \cite{Candelasetal}, we look at different regions in the moduli space of the Swiss-Cheese Calabi-Yau we have been using, and write out the explicit expressions for the periods. The explicit expressions, though cumbersome, will be extremely useful when studying complex structure moduli stabilization and existence of ``area codes" in section {\bf 3}, solving explicitly the ``inverse problem" in section {\bf 4} and showing explicitly the existence of ``fake superpotentials" in section {\bf 5} in the context of non-supersymmetric black hole attractors. More precisely,
based on \cite{Candelasetal}, the periods of the ``Swiss cheese" Calabi-Yau obtained as a resolution of the degree-18 hypersurface in ${\bf{WCP}}^4[1,1,1,6,9]$:
\begin{equation}
x_1^{18} + x_2^{18} + x_3^{18} + x_4^3 + x_5^2 - 18\psi \prod_{i=1}^5x_i - 3\phi x_1^6x_2^6x_3^6 = 0.\nonumber\\
\end{equation}
As discussed in chapter {\bf 2}, it is understood that only two complex structure
moduli $\psi$ and $\phi$ are retained in aforementioned hypersurface equation which are invariant under the group $G={\bf Z}_6\times{\bf Z}_{18}$ (${\bf Z}_6:(0,1,3,2,0,0); {\bf Z}_{18}:(1,-1,0,0,0)$, setting the other invariant complex structure moduli appearing at a higher order (due to invariance under $G$)
at their values at the origin. Further, defining $\rho\equiv (3^4.2)^{\frac{1}{3}}\psi$, the singular loci of the Swiss-Cheese Calabi Yau are in ${\bf{WCP}}^2[3,1,1]$ with homogenous coordinates $[1,\rho^6,\phi]$ and are given as under:
\begin{enumerate}
\item
${\it Conifold\ Locus 1}: \{(\rho,\phi)|(\rho^6+\phi)^3=1\}$
\item
${\it Conifold\ Locus 2}: \{(\rho,\phi)|\phi^3=1\}$
\item
${\it Boundary}: (\rho,\phi)\rightarrow\infty$
\item
${\it Fixed\ point\ of\ quotienting}$: The fixed point $\rho=0$ of ${\cal A}^3$ where
${\cal A}:(\rho,\phi)\rightarrow(\alpha\rho,\alpha^6\phi)$, where $\alpha\equiv e^{\frac{2\pi i}{18}}
$.
\end{enumerate}

We will be considering the following sectors in the $(\rho,\phi)$ moduli space:
\begin{itemize}
\item
$\underline{|\phi^3|>1, 0<arg\phi<\frac{2\pi}{3},{\rm large}\ \psi}$

The fundamental period $\varpi_0$, obtained by directly integrating the holomorphic three-form over the ``fundamental cycle" (See \cite{Candelasetal}), is given by:
\begin{eqnarray}
\label{eq:largephilargepsi1}
& & \varpi_0=\sum_{k=0}^\infty \frac{(6k)!}{k!(2k)!(3k)!}\left(\frac{-3}{18^6\psi^6}\right) U_k(\phi)\nonumber\\
& & =\sum_{k=0}^\infty\frac{(-)^k\Gamma(k+\frac{1}{6})\Gamma(k+\frac{5}{6})}{(k!)^2}\left(\frac{1}{\rho^{6k}}\right)
U_k(\phi),
\end{eqnarray}
where $U_\nu(\phi)\equiv \phi^\nu\ _3F_2(-\frac{\nu}{3},\frac{1-\nu}{3},\frac{2-\nu}{3};1,1;\frac{1}{\phi^3})$; the other components of the period vector are given by: $\varpi_i=\varpi_0(\alpha^i\psi,\alpha^{6i}\phi)$ where
$\alpha\equiv e^{\frac{2\pi i}{18}},\ i=1,2,3,4,5$.

\item
$\underline{|\phi^3|<1,\ {\rm large}\ \psi}$

The fundamental period is given by:
\begin{equation}
\label{eq:smallphilargepsi11}
\varpi_0=\sum_{m=0}^\infty\sum_{n=0}^\infty\frac{(18n + 6m)!(-3\phi)^m}{(9n + 3m)!(6n + 2m)!(n!)^3m!(18\psi)^{18n + 6m}},
\end{equation}
implying that around a suitable $\rho=\rho_0$ and $\phi=\phi_0$:
\begin{equation}
\label{eq:smallphilargepsi2}
\left(\begin{array}{c}
\varpi_0\\
\varpi_1\\
\varpi_2\\
\varpi_3\\
\varpi_4\\
\varpi_5\end{array}\right)
=\left(\begin{array}{ccc}P_1 & P_2 & P_3\end{array}\right)\left(\begin{array}{c} 1\\(\phi - \phi_0)\\(\rho - \rho_0)
\end{array}\right),
\end{equation}
where $P_{1,2,3}$ are given in Appendix {\bf A.8}.

\item
$\underline{|\frac{\rho^6}{\phi - \omega^{0,-1,-2}}|<1}$

\begin{equation}
\label{eq:awaycl11}
\varpi_{3a+\sigma}=\frac{1}{3\pi}\sum_{r=1,5}\alpha^{3ar}sin\left(\frac{\pi r}{3}\right)\xi^\sigma_r(\psi,\phi)[a=0,1;\sigma=0,1,2],
\end{equation}
where $\xi_r^\sigma(\psi,\phi)=\sum_{k=0}^\infty\frac{(\Gamma(k+\frac{r}{6})^2}{k!\Gamma(k+\frac{r}{3})}\rho^{6k+r}
U_{-(k+\frac{r}{6})}^\sigma(\phi), U^\sigma_\nu(\phi)=\omega^{-\nu\sigma}U_\nu(\omega^\sigma\phi)\ \footnote{The three values of $\sigma$ correspond to the three solutions to $(1-\phi^3)U_\nu^{\prime\prime\prime}(\phi)
+3(\nu-1)\phi^2U^{\prime\prime}_\nu(\phi) - (3\nu^2-3\nu+1)\phi U^\prime_\nu(\phi)+\nu^3 U_\nu(\phi)=0$; the
Wronskian of the three solutions is given by: $\frac{-27i}{2\pi^3}e^{-i\pi\nu}sin^2(\pi\nu)(1-\phi^3)^{\nu-1}$ -
the solutions are hence linearly independent except when $\nu\in{\bf Z}$}, \omega\equiv e^{\frac{2\pi i}{3}}$; for small $\phi$,
\begin{equation}
\label{eq:Usmallphi}
U_\nu(\phi)=\frac{3^{-1-\nu}}{\Gamma(-\nu)}\sum_{m=0}^\infty\frac{\Gamma(\frac{m-\nu}{3})(3\omega\phi)^m}{(\Gamma(1 - \frac{m-\nu}{3})^2m!}.
\end{equation}
Expanding about a suitable $\phi=\phi_0$ and $\rho=\rho_0$, one can show:
\begin{equation}
\label{eq:awaycl12}
\left(\begin{array}{c}
\varpi_0\\
\varpi_1\\
\varpi_2\\
\varpi_3\\
\varpi_4\\
\varpi_5\\
\end{array}\right)=\left(\begin{array}{ccc}
M_1 & M_2 & M_3
\end{array}\right)\left(\begin{array}{c}
1\\
(\rho - \rho_0)\\
(\phi - \phi_0)\\
\end{array}\right),
\end{equation}
where
$M_{1,2,3}$ are given in Appendix {\bf A.8}.

\item
$\underline{{\rm Near\ the\ conifold\ locus:}\ \rho^6 + \phi = 1}$

The periods are given by:
\begin{equation}
\label{eq:confl11}
\varpi_i = C_i g(\rho,\phi) ln(\rho^6 + \phi - 1) + f_i(\rho,\phi),
\end{equation}
where $C_i=(1,1,-2,1,0,0)$, $g(\rho,\phi)=\frac{i}{2\pi}(\varpi_1 - \varpi_0)\sim a(\rho^6 + \phi - 1)$ near $\rho^6 + \phi - 1\sim0$ where $a$ is a constant and $f_i$ are analytic in $\rho$ and $\psi$. The
analytic functions near the conifold locus are given by:
\begin{equation}
\label{eq:confl12}
f_{3a + \sigma}=\frac{1}{2\pi}\sum_{r=1,5}e^{\frac{i\pi ar}{3}}sin\left(\frac{\pi r}{3}\right) \xi_r^\sigma(\rho,\phi), a=0,1;\ \sigma=0,1,2.
\end{equation}
Defining $x\equiv(\rho^6 + \phi - 1)$, one can show that:
\begin{equation}
\label{eq:confl13}
\xi_r^\sigma=\sum_{k=0}^\infty\sum_{m=0}^\infty
\frac{3^{-1 + k + \frac{r}{6} + m}e^{\frac{2i\pi(\sigma + 1)}{3} + \frac{-i\pi(k + \frac{r}{6})}{3}}(-)^k(\Gamma(k + \frac{r}{6}))^2}{\Gamma(k + 1)\Gamma(k + \frac{r}{6})\Gamma(k + \frac{r}{6})(\Gamma(1 - \frac{m + k + \frac{r}{6}}{3}))^2m!}(x - \phi + 1)^{k + \frac{r}{6}}.
\end{equation}
One can hence see that:
\begin{equation}
\label{eq:confl14}
\left(\begin{array}{c}
f_0\\
f_1\\
f_2\\
f_3\\
f_4\\
f_5\end{array}\right)=\left(\begin{array}{ccc} N_1 & N_2 & N_3\end{array}\right)\left(\begin{array}{c}
1\\
x\\
\phi\end{array}\right),
\end{equation}
where $N_{1,2,3}$ are given in the Appendix {\bf A.8}.

\item
$\underline{{\rm Near}\ \phi^3=1,\ {\rm Large}\ \rho}$

From asymptotic analysis of the coefficients, one can argue:
\begin{eqnarray}
\label{eq:confl21}
& & U_\nu(\phi)\sim-\frac{\sqrt{3}}{2\pi(\nu+1)}\left[(\phi-1)^{\nu+1} - 2\omega(\phi-\omega^{-1}) + \omega^2(\phi-\omega^{-2})\right],\nonumber\\
& & \equiv y^0_\nu - 2y^1_\nu + y^2_\nu,
\end{eqnarray}
where $\omega\equiv e^{\frac{2i\pi}{3}}$. Defining $U^\sigma_\nu(\phi)=\sum_{\tau=0}^2\gamma_\nu^{\sigma,\tau}y_\nu^\tau(\phi)$, where
$\gamma^{\sigma,\tau}_\nu=\left(\begin{array}{ccc}
1 & -2 & 1 \\
e^{-2i\pi\nu} & 1 & -2 \\
-2e^{-2i\pi\nu} & e^{-2i\pi\nu} & 1\\
\end{array}\right)$, one can show that $U_\nu,\left(\frac{\sum_{\sigma=0}^2U^\sigma_\nu(\phi)}{1-e^{-2i\pi\nu}}=\right)y^0_\nu(\phi) - y^1_\nu(\phi)\equiv V_\nu(\phi),\left(\frac{3V_\nu(\phi) - 2U_\nu(\phi) - U^1_\nu(\phi)}{1-e^{-2i\pi\nu}}=\right)y^0_\nu(\phi)\equiv W_\nu(\phi)$ are linearly independent even for $\nu\in{\bf Z}$\footnote{The Wronskian of these three solutions
is given by $\frac{27i}{(2\pi)^3}e^{i\pi\nu}(1-\phi^3)^{\nu-1}\neq0,\nu\in{\bf Z}$.}.

For small $\rho$,
\begin{equation}
\label{eq:confl22}
\xi^\sigma_r=\int_\Gamma \frac{d\mu}{2i sin(\pi(\mu+\frac{r}{6}))}
\frac{(\Gamma(-\mu))^2}{\Gamma(-\mu+\frac{1}{6})\Gamma(-\mu+\frac{5}{6})}\rho^{-6\mu}U^\sigma_\mu(\phi),
\end{equation}
where the contour $\Gamma$ goes around the Im$(\mu)<0$ axis. To deform the contour to a contour $\Gamma^\prime$ going around the Im$(\mu)>0$ axis, one sees that one can do so for $\sigma=0$ but not for $\sigma=1,2$. For the latter, one modifies $U^\sigma_\mu(\phi)$ by adding a function which does not contribute to the poles and has simple zeros at integers as follows:
\begin{equation}
\label{eq:confl23}
U^\sigma_\mu(\phi)\rightarrow\tilde{U^\sigma_{\mu,r}(\phi)}\equiv U^\sigma_\mu(\phi) - e^{\frac{i\pi r}{6}}\frac{sin(\pi(\mu+\frac{r}{6}))}{sin(\pi\mu)}f^\sigma_\mu(\phi),
\end{equation}
where
\begin{eqnarray}
\label{eq:confl231}
& & f^0_\mu(\phi)=0,\nonumber\\
& & f^1_\mu(\phi)=-(1-e^{-2i\pi\nu})y^0_\mu(\phi),\nonumber\\
& & f^2_\mu(\phi)=(1-e^{-2i\pi\nu})V_\nu(\phi) + (1-e^{-2i\pi\nu})W_\mu(\phi).
\end{eqnarray}
One can then deform the contour $\Gamma$ to the contour $\Gamma^\prime$ to evaluate the periods.

Expanding about $\phi=\omega^{-1}$, and a large $\rho=\rho_0$, one gets the following periods:
\begin{equation}
\label{eq:confl210}
\left(\begin{array}{c}
\varpi_0\\
\varpi_1\\
\varpi_2\\
\varpi_3\\
\varpi_4\\
\varpi_5\\
\end{array}\right)=\left(\begin{array}{c}
A^\prime_0 + B^\prime_{01}x + C^\prime_0(\rho - \rho_0)\\
A^\prime_1 + B^\prime_{11}x + C^\prime_1(\rho - \rho_0)\\
A^\prime_2 + B^\prime_{21}x + C^\prime_2(\rho - \rho_0)\\
A^\prime_3 + B^\prime_{31}x + B^\prime_{32}\ x\ ln x + C^\prime_3(\rho - \rho_0)\\
A^\prime_4 + B^\prime_{41}x + B^\prime_{42}\ x\ ln x + C^\prime_4(\rho - \rho_0)\\
A^\prime_5 + B^\prime_{51}x + B^\prime_{52}\ x\ ln x + C^\prime_5(\rho - \rho_0)\\
\end{array}\right),
\end{equation}.
\end{itemize}
where $x\equiv(\phi - \omega^{-1})$. The equations (\ref{eq:confl24}) and (\ref{eq:confl210}) will get used to arrive at (\ref{eq:Wconfl11}) and finally (\ref{eq:Wconfl161}) and (\ref{eq:Wconfl21}). The Picard-Fuchs basis of periods evaluated above can be transformed to a symplectic basis as under
(See \cite{Candelasetal}):
\begin{equation}
\label{eq:PFbasis1}
\Pi=\left(\begin{array}{c}
F_0\\
F_1\\
F_2\\
X^0\\
X^1\\
X^2
\end{array}\right) = M \left(\begin{array}{c}
\varpi_0\\
\varpi_1\\
\varpi_2\\
\varpi_3\\
\varpi_4\\
\varpi_4
\end{array}\right),
\end{equation}
where
\begin{equation}
\label{eq:PFbasis2}
M=\left(\begin{array}{cccccc}
-1&1&0&0&0&0\\
1&3&3&2&1&0\\
0&1&1&1&0&0\\
1&0&0&0&0&0\\
-1&0&0&1&0&0\\
2&0&0&-2&1&1
\end{array}\right).
\end{equation}
In the next section, we use information about the periods evaluated in this section, in looking for ``area codes".

\section{Extremization of Superpotential and Existence of ``Area Codes"}

In this section, we argue the existence of area codes, i.e., points in the moduli space close to and away from the two singular conifold loci that are finitely
separated where for the same large values (and hence not necessarily integral) of RR and NS-NS fluxes, one can extremize the (complex structure and axion-dilaton) superpotential (for different values of the complex structure and axion-dilaton moduli)\footnote{For techniques in special geometry relevant to this work, see \cite{Mohaupt,Giryavets} for a recent review; see \cite{Curio+Spillner} for moduli-stablization calculations as well.}.

The axion-dilaton modulus $\tau$ gets stabilized (from $D_\tau W_{c.s.}=0$, $W_{c.s.}$ being the Gukov-Vafa-Witten complex structure superpotential $\int (F_3 - \tau H_3)\wedge\Omega=(2\pi)^2\alpha^\prime(f-\tau h)\cdot\Pi$, $F_3$ and $H_3$ being respectively the NS-NS and RR three-form field strengths, and are given by:
$F_3=(2\pi)^2\alpha^\prime\sum_{a=1}^3(f_a\beta_a + f_{a+3}\alpha_a)$ and
$H_3=(2\pi)^2\alpha^\prime\sum_{a=1}^3(h_a\beta_a+h_{a+3}\alpha_a)$;
$\alpha_a,\beta^a$,  $a=1,2,3$,
form an integral cohomology basis) at a value given by:
\begin{equation}
\label{eq:taufix1}
\tau=\frac{f^T.{\bar\Pi_0}}{h^T.{\bar\Pi_0}},
\end{equation}
where $f$ and $h$ are the fluxes corresponding to the NS-NS and RR fluxes; it is understood that the complex structure moduli appearing in (\ref{eq:taufix1}) are already fixed from $D_iW=0,\ i=1,2$.

\begin{itemize}
\item
$\underline{{\rm Near\ the\ conifold\ locus:}\ \phi^3=1,\ {\rm Large}\ \psi}$

The period vector in the symplectic basis can be simplified to:
\begin{eqnarray}
\label{eq:Wconfl11}
& & \Pi=\nonumber\\
& & \equiv \left(\begin{array}{c} A_0 + B_{01} x + C_0 (\rho - \rho_0) \\ A_1 + B_{11} x + B_{12} x ln x + C_1 (\rho - \rho_0) \\ A_2 + B_{21} x + B_{22} x ln x + C_2 (\rho - \rho_0) \\ A_3 + B_{31} x + C_3 (\rho - \rho_0) \\ A_4 + B_{41} x + B_{42} x ln x + C_4 (\rho - \rho_0)\\ A_5 + B_{51} x + B_{52} x ln x + C_5 (\rho - \rho_0) \end{array} \right).
\end{eqnarray}
The tree-level K\"{a}hler potential is given by:
\begin{equation}
\label{eq:Wconfl12}
K = -ln\left(-i(\tau - {\bar\tau})\right) - ln\left(-i\Pi^\dagger\Sigma\Pi\right),
\end{equation}
where the symplectic metric $\Sigma=\left(\begin{array}{cc} 0 & {\bf 1}_3\\ -{\bf 1}_3 & 0 \end{array}\right)$. Near $x=0$, one can evaluate $\partial_xK, \tau$ and $\partial_xW_{c.s.}$ - this is done in Appendix {\bf A.9}. Using (\ref{eq:Wconfl11}) - (\ref{eq:Wconfl12}) and (\ref{eq:Wconfl13})-(\ref{eq:Xisdefs}), one gets the following (near $x=0,\rho - \rho_0=0$):
\begin{eqnarray}
\label{eq:Wconfl161}
& & D_xW_{c.s.}\approx ln x\Biggl({\cal A}_1 + {\cal B}_1x + {\cal C}_1 x ln x + {\cal D}_1 (\rho - \rho_0)
+ {\cal B}_1^\prime{\bar x} + {\cal C}_1^\prime{\bar x} ln {\bar x} + {\cal D}_1^\prime ({\bar\rho} - {\bar\rho_0})\Biggr)=0,\nonumber\\
& & D_{\rho - \rho_0}W_{c.s.}\approx {\cal A}_2 + {\cal B}_2 x + {\cal C}_2 x ln x + {\cal D}_2 (\rho - \rho_0)
+ {\cal B}_2^\prime{\bar x} + {\cal C}_2^\prime{\bar x} ln {\bar x} + {\cal D}_2^\prime ({\bar\rho} - {\bar\rho_0})=0.\nonumber\\
\end{eqnarray}

\item
$\underline{{\rm Near}\ \rho^6 + \phi= 1}$

Near $y\equiv \rho^6 + \rho - 1 =0$ and a small $\rho=\rho_0^\prime$, one can follow a similar analysis
as (\ref{eq:Wconfl11}) - (\ref{eq:Wconfl161}) and arrive at similar equations:
\begin{eqnarray}
\label{eq:Wconfl21}
& & D_yW_{c.s.}\approx ln y({\cal A}_3 + {\cal B}_3y + {\cal C}_3 y ln y + {\cal D}_3 (\rho - \rho_0^\prime)
+ {\cal B}_3^\prime{\bar y} + {\cal C}_3^\prime{\bar y} ln {\bar y} + {\cal D}_3^\prime ({\bar\rho} - {\bar\rho_0^\prime})=0,\nonumber\\
& & D_{\rho - \rho_0^\prime}W_{c.s.}\approx {\cal A}_4 + {\cal B}_4 y + {\cal C}_4 y ln y + {\cal D}_4 (\rho - \rho_0^\prime)
+ {\cal B}_4^\prime{\bar y} + {\cal C}_4^\prime{\bar y} ln {\bar y} + {\cal D}_4^\prime ({\bar\rho} - {\bar\rho_0^\prime})=0.\nonumber\\
\end{eqnarray}

\item
$\underline{\rm Points\ away\ from\ both\ conifold\ loci}$

It can be shown, again following an analysis similar to the one carried out in (\ref{eq:Wconfl11}) -
(\ref{eq:Wconfl21}), that one gets the following set of equations from extremization of the complex-structure moduli superpotential:
\begin{equation}
\label{eq:Wawayconfl}
{\cal A}_i + {\cal B}_i\psi + {\cal C}_i\phi + {\cal B^\prime}_i{\bar\psi} + {\cal C^\prime}_i{\bar\phi} = 0,
\end{equation}
where $i$ indexes the different regions in the moduli space away from the two conifold loci.
\end{itemize}

Therefore, to summarize,
\begin{eqnarray}
\label{eq:DW=0}
& & \underline{{\rm Near}\ \phi=\omega^{-1}}:\nonumber\\
& & {\cal A}_1 + {\cal B}_1 (\phi_1 - \omega^{-1}) + {\cal C}_1 (\phi_1 - \omega^{-1}) ln (\phi_1 - \omega^{-1}) + {\cal D}_1 (\rho_1 - \rho_0)\nonumber\\
& & + {\cal B}_1^\prime({\bar\phi_1} - \omega) + {\cal C}_1^\prime({\bar\phi} - \omega) ln ({\bar\phi} - \omega) + {\cal D}_1^\prime ({\bar\rho_1} - {\bar\rho_0})=0,\nonumber\\
& & {\cal A}_2 + {\cal B}_2 (\phi_1 - \omega^{-1}) + {\cal C}_2 (\phi_1 - \omega^{-1})
ln (\phi_1 - \omega^{-1}) + {\cal D}_2 (\rho_1 - \rho_0)\nonumber\\
& & + {\cal B}_2^\prime({\bar\phi_1} - \omega) + {\cal C}_2^\prime({\bar\phi_1} - \omega) ln ({\bar\phi_1} - \omega) + {\cal D}_2^\prime ({\bar\rho_1} - {\bar\rho_0})=0,\nonumber\\
& & \tau_1=\frac{\Xi[f_i;{\bar\phi_1} - \omega,\rho_1 - \rho_0]}{\sum_{i=0}^5h_i{\bar A_i}}\left[1 - \frac{\Xi[h_i;{\bar\phi_1} - \omega,\rho_1 - \rho_0]}{\sum_{j=0}^5h_i{\bar A_i}}\right];\nonumber\\
\end{eqnarray}
\begin{eqnarray}
& & \underline{{\rm Near}\ \rho^6 + \phi - 1 = 0}:\nonumber\\
& & {\cal A}_3 + {\cal B}_3 (\rho_2^6+\phi-1) + {\cal C}_3 (\rho_2^6+\phi_2-1) ln (\rho_2^6+\phi_2-1) + {\cal D}_3 \phi_2
+ {\cal B}_3^\prime({\bar\rho_2^6}+{\bar\phi}-1)\nonumber\\
 & & + {\cal C}_3^\prime({\bar\rho_2^6}+{\bar\phi}-1) ln ({\bar\rho_2^6}+{\bar\phi}-1) + {\cal D}_3^\prime{\bar\phi}=0,\nonumber\\
& &  {\cal A}_4 + {\cal B}_4 (\rho_2^6+\phi_2-1) + {\cal C}_4 (\rho_2^6+\phi_2-1)
ln (\rho_2^6+\phi_2-1) + {\cal D}_4 \phi_2\nonumber\\
& & + {\cal B}_4^\prime({\bar\rho}^6-{\bar\phi}-1) + {\cal C}_4^\prime({\bar\rho}^6+{\bar\phi}-1) ln ({\bar\rho}^6+{\bar\phi}-1)+ {\cal D}_4^\prime {\bar\phi_2}=0,\nonumber\\
& & \tau_2=\frac{\Xi[f_i;{\bar\rho}^6+{\bar\phi}-1,\phi_2 ]}{\sum_{i=0}^5h_i{\bar A_i^\prime}}\left[1 - \frac{\Xi[h_i;{\bar\rho}^6+{\bar\phi}-1,\phi_2 ]}{\sum_{j=0}^5h_i{\bar A_i^\prime}}\right]\nonumber\\
\end{eqnarray}
\begin{eqnarray}
& & \underline{|\phi^3|<1,\ {\rm Large}\ \psi}:\nonumber\\
& & {\cal A}_5 + {\cal B}_5(\phi_3 - \phi_0^{\prime\prime}) + {\cal C}_5(\rho_3 - \rho_0^{\prime\prime})
{\cal B}^\prime_5({\bar\phi}_3 - {\bar\phi}_0^{\prime\prime})
+ {\cal C}^\prime_5({\bar\rho}_3 - {\bar\rho}_0^{\prime\prime})= 0,\nonumber\\
& & {\cal A}_6 + {\cal B}_6(\phi_3 - \phi_0^{\prime\prime}) + {\cal C}_6(\rho_3 - \rho_0^{\prime\prime})
+ {\cal B}^\prime_6({\bar\phi}_3 - {\bar\phi}_0^{\prime\prime})
+ {\cal C}^\prime_6({\bar\rho}_3 - {\bar\rho}_0^{\prime\prime}) = 0,\nonumber\\
& & \tau_3=\frac{\tilde{\Xi}[f_i;{\bar\phi_3},\rho_3 ]}{\sum_{i=0}^5h_i{\bar A_i^{\prime\prime}}}\left[1 - \frac{\tilde{\Xi}[h_i;{\bar\phi_3},\rho_3]}{\sum_{j=0}^5h_i{\bar A_i^{\prime\prime}}}\right];\nonumber\\
\end{eqnarray}
\begin{eqnarray}
& & \underline{\left|\frac{\rho^6}{\phi - \omega^{0,-1,-2}}\right|<1}:\nonumber\\
& & {\cal A}_7 + {\cal B}_7(\phi_4 - \phi_0^{\prime\prime\prime}) + {\cal C}_7(\rho_4 - \rho_0^{\prime\prime\prime})
+ {\cal B}^\prime_7({\bar\phi}_4 - {\bar\phi}_0^{\prime\prime\prime}) + {\cal C}^\prime_7({\bar\rho}_4 - {\bar\rho}_0^{\prime\prime\prime}) = 0,\nonumber\\ & & {\cal A}_7 + {\cal B}_7(\phi_4 - \phi_0^{\prime\prime\prime}) + {\cal C}_7(\rho_4 - \rho_0^{\prime\prime\prime})
+ + {\cal B}^\prime_7({\bar\phi}_4 - {\bar\phi}_0^{\prime\prime\prime}) + {\cal C}^\prime_7({\bar\rho}_4 - {\bar\rho}_0^{\prime\prime\prime}) = 0,\nonumber\\ & & \tau_4=\frac{\tilde{\Xi}[f_i;{\bar\phi_4},\rho_4 ]}{\sum_{i=0}^5h_i{\bar A_i^{\prime\prime\prime}}}\left[1 - \frac{\tilde{\Xi}[h_i;{\bar\phi_4},\rho_4]}{\sum_{j=0}^5h_i{\bar A_i^{\prime\prime\prime}}}\right],\nonumber\\
\end{eqnarray}
where on deleting the $ln$ terms in $\Xi$ one gets the form of $\tilde{\Xi}$ in (\ref{eq:DW=0}). Given that the Euler characteristic of
the elliptically-fibered Calabi-Yau four-fold to which, according to the Sen's construction \cite{Sen}, the orientifold of the Calabi-Yau three-fold of (\ref{eq:hypersurface}) corresponds to, will be very large\footnote{See \cite{DDF} - $\chi(CY_4)=6552$ where the $CY_4$ for the ${\bf{WCP}^4}[1,1,1,6,9]$-model, is the resolution of a Weierstrass over a three-fold B with $D_4$ and $E_6$ singularities along two sections, with the three-fold a ${\bf{CP}^1}$-fibration over ${\bf{CP}^2}$ with the two divisors contributing to the instanton superpotential \`{a} la Witten being sections thereof.}, and further assuming the absence of $D3$-branes, this would imply that one is allowed to take a large value of $f^T.\Sigma.h$, and hence
the fluxes - therefore, similar to the philosophy of \cite{VafaInverse}, we would disregard the integrality of fluxes. Without doing the numerics, we will now give a plausibility argument about the existence of solution to any one of the four sets of equations in (\ref{eq:DW=0}). As one can drop $x$ as compared to $x ln x$ for $x\sim0$, the equations in (\ref{eq:DW=0}) pair off either as:
\begin{itemize}
\item
Near either of the two conifold loci:
\begin{eqnarray}
\label{eq:DWsimp1}
& & A_i + (B_i cos \alpha_i + B_i^\prime sin \alpha_i) \epsilon_i ln \epsilon_i + C_i\beta_i + C_i^\prime {\bar\beta_i} = 0,\nonumber\\
& & \tilde{a}_i + (\tilde{b}_i cos \alpha_i + \tilde{b}_i^\prime sin \alpha_i) \epsilon_i ln \epsilon_i + \tilde{c}_i\beta_i + \tilde{c}_i^\prime {\bar\beta_i} = 0,
\end{eqnarray}
or
\item
Away from both the conifold loci:
\begin{eqnarray}
\label{eq:DWsimp2}
& & A_i + B_i \gamma_i + C_i\delta_i + B_i^\prime {\bar\gamma}_i + C_i^\prime {\bar\delta}_i = 0,\nonumber\\
& & \tilde{A}_i + \tilde{B}_i \gamma_i + \tilde{C}_i\delta_i + \tilde{B}_i^\prime {\bar\gamma}_i
+ \tilde{C}_i^\prime {\bar\delta}_i = 0,
\end{eqnarray}
where $\epsilon_i,\alpha_i$ correspond to the magnitude and phase of the extremum values of either $\phi - \omega^{-1}$ or $\rho^6 + \phi - 1$, and $\gamma_i,\delta_i$ are different (functions of) extremum values of $\phi,\psi$ near and away, respectively, from the two conifold loci, and both sets are understood to be ``close to zero" each.
 \end{itemize}
 From the point of view of practical calculations, let us rewrite, e.g., (\ref{eq:DWsimp1}) as the equivalent four real equations:
 \begin{eqnarray}
 \label{eq:DWsimp12}
 & & {\cal A}_i + {\cal B}_i cos\alpha_i \epsilon_i ln\epsilon_i + {\cal B}_i^\prime sin\alpha_i \epsilon_i ln\epsilon_i + {\cal C}_i Re(\beta_i) + {\cal C}_i^\prime Im(\beta_i) = 0,\nonumber\\
 & & \widetilde{{\cal A}_i} + \widetilde{{\cal B}_i} cos\alpha_i \epsilon_i ln\epsilon_i +
 \widetilde{{\cal B}_i}^\prime sin\alpha_i \epsilon_i ln\epsilon_i + \widetilde{{\cal C}_i} Re(\beta_i) + \widetilde{\cal C}_i^\prime Im(\beta_i) = 0,\nonumber\\
 & & {\nu}_i + {\chi}_i cos\alpha_i \epsilon_i ln\epsilon_i + {\chi}_i^\prime sin\alpha_i \epsilon_i ln\epsilon_i + {\vartheta}_i Re(\beta_i) + {\vartheta}^\prime Im(\beta_i) = 0,\nonumber\\
 & & \widetilde{{\nu}_i} + \widetilde{{\chi}_i} cos\alpha_i \epsilon_i ln\epsilon_i +
 \widetilde{{\chi}_i}^\prime sin\alpha_i \epsilon_i ln\epsilon_i + \widetilde{{\vartheta}_i} Re(\beta_i) + \widetilde{\vartheta}_i^\prime Im(\beta_i) = 0.
 \end{eqnarray}
  In (\ref{eq:DW=0}), by ``close to zero", what we would be admitting are, e.g., $\epsilon_i,|\beta_i|\sim e^{-5}\approx 7\times10^{-3}$ implying that
 $\epsilon_i ln\epsilon_i\approx10^{-2}$. Let us choose the moduli-independent constants in (\ref{eq:DWsimp12}),
  after suitable rationalization, to be $7\times {\cal O}(1)$, the coefficients of the
 $\epsilon_i ln\epsilon_i$-terms to be $7\times10^2$ and the coefficients of $Re(\beta_i)$ and $Im(\beta_i)$
 to be $\sim10^3$. On similar lines, for (\ref{eq:DWsimp2}), we could
  take the moduli to be $\sim e^{-5}$ and the moduli-independent and moduli-dependent constants to be
  $7\times {\cal O}(1)$ and $\sim10^3$ respectively. Now, the constants appearing in (\ref{eq:DWsimp12})
  (and therefore (\ref{eq:DW=0})) are cubic in the fluxes (more precisely, they are of the type
  $h^2f$ in obvious notations), which for  (\ref{eq:hypersurface}) would be $\sim10^3$ (See \cite{DDF}).
  In other words, {\it for the same choice of the NS-NS and RR fluxes} - 12 in number - one gets 6 or 9 or 12 complex
(inhomogenous [in $\psi,\phi$] algebraic/transcendetal) constraints (coming from (\ref{eq:DW=0}))
on the 6 or 9 or 12 extremum values of the complex structure moduli ($\phi_i,\psi_i,\tau_i;\ i=1,2,3,4$) finitely separated from each other in the moduli space. In principle, as long as one keeps $f^T.\Sigma.h$ fixed, one should be able to tune the fluxes $f_i,h_i;\ i=0,...,5$ to be able to solve these equations.
  Therefore, the expected estimates of the values of the constants and the moduli tuned by the
  algebraic-geometric inputs of the periods in the different regions of the moduli space as
  discussed in section {\bf 2}, are reasonable implying the possibility of existence of ``area codes",
  and the interpolating domain walls \cite{CDGKL}. Of course, complete numerical calculations,
  which will be quite involved, will be needed to see explicitly everything working out.

\section{The Inverse Problem for Extremal Black Holes}

We now switch gears and address two issues in this and the subsequent sections, related to supersymmetric and non-supersymmetric black hole attractors\footnote{See \cite{Mohaupt} for a nice review of special geometry relevant to sections {\bf 4} and {\bf 5}.}. In this section, using the techniques discussed in  \cite{VafaInverse}, we explicitly solve the ``inverse problem" for extremal black holes in type II compactifications on (the mirror of) (\ref{eq:hypersurface}) - given a point in the moduli space, to find the charges $(p^I,q_I)$ that would satisfy $\partial_iV_{BH}=0$,  $V_{BH}$ being the black-hole potential. In the next section, we address the issue of existence of ``fake superpotentials" in the same context.

We will now summarize the ``inverse problem" as discussed in \cite{VafaInverse}). Consider $D=4, {\cal N}=2$ supergravity coupled to $n_V$ vector multiplets in the absence of higher derivative terms. The black-hole potential can be written as \cite{nonsusybh1}:
 \begin{equation}
 \label{eq:BHinv1}
 V_{BH} = -\frac{1}{2}(q_I - {\cal N}_{IK} p^K)\left((Im {\cal N})^{-1}\right)^{IJ}(q_J - {\bar{\cal N}}p^L),
 \end{equation}
 where the $(n_V + 1)\times(n_V + 1)$ symmetric complex matrix, ${\cal N}_{IJ}$, the vector multiplet moduli space metric, is defined as:
 \begin{equation}
 \label{eq:BHinv2}
 {\cal N}_{IJ} \equiv {\bar F}_{IJ} + \frac{2i Im(F_{IK}) X^K Im (F_{IL}) X^L}{Im(F_{MN}) X^M X^N},
 \end{equation}
 $X^I,F_J$ being the symplectic sections and $F_{IJ}\equiv\partial_IF_J=\partial_JF_I$. The black-hole potential (\ref{eq:BHinv1}) can be rewritten (See \cite{VafaInverse}) as:
 \begin{equation}
 \label{eq:BHinv3}
 \tilde{V}_{BH} = \frac{1}{2}{\cal P}^I Im({\cal N}_{IJ}){\bar{\cal P}}^J - \frac{i}{2}{\cal P}^I(q_I - {\cal N}_{IJ}p^J)
 + \frac{i}{2}{\bar{\cal P}}^I(q_I - {\bar{\cal N}}_{IJ}p^J).
 \end{equation}
 The variation of (\ref{eq:BHinv3}) w.r.t. ${\cal P}^I$ gives:
 \begin{equation}
 \label{eq:BHinv4}
 {\cal P}^I=-i\left((Im {\cal N})^{-1})^{IJ}\right)(q_J - {\cal N}_{IJ}p^J),
 \end{equation}
 which when substituted back into (\ref{eq:BHinv3}), gives (\ref{eq:BHinv1}). From (\ref{eq:BHinv4}), one
 gets:
\begin{eqnarray}
\label{eq:BHinv5}
& & p^I = Re({\cal P}^I)\nonumber\\
& &  q_I = Re({\cal N}_{IJ}{\cal P}^J).
\end{eqnarray}
Extremizing $\tilde{V}_{BH}$ gives:
\begin{equation}
\label{eq:BHinv6}
{\cal P}^I{\bar{\cal P}}^J\partial_i Im({\cal N}_{IJ}) + i({\cal P}^I\partial_i {\cal N}_{IJ} - {\bar{\cal P}}^J\partial_i{\bar{\cal N}}_{IJ})p^J = 0,
\end{equation}
which using (\ref{eq:BHinv5}) yields:
\begin{equation}
\label{eq:BHinv7}
\partial_i Im({\cal P}^I{\cal N}_{IJ}{\cal P}^J) = 0.
\end{equation}
As in section {\bf 3}, one uses the semi-classical approximation and disregards the integrality of the electric and magnetic charges taking them to be large. The inverse problem is not straight forward to define as all sets of charges $(p^I,q_I)$ which are related to each other by an $Sp(2n_V + 2,{\bf Z})$-transformation, correspond to the same point in the moduli space. This is because the $V_{BH}$ (and $\partial_iV_{BH}$) is (are) symplectic invariants. Further, $\partial_iV_{BH}=0$ give $2n_V$ real equations in $2n_V+2$ real variables $(p^I,q_I)$. To fix these two problems, one looks at critical values of $V_{BH}$ in a fixed gauge $W=w\in{\bf C}$. In other words,
\begin{equation}
\label{eq:BHinv8}
W=\int_M\Omega\wedge H = q_I X^I - p^I F_I = X^I(q_I - {\cal N}_{IJ}p^J) = w,
\end{equation}
which using (\ref{eq:BHinv5}), gives:
\begin{equation}
\label{eq:BHinv9}
X^I Im({\cal N}_{IJ}){\bar{\cal P}}^J = w.
\end{equation}
Thus, the inverse problem boils down to solving:
\begin{eqnarray}
\label{eq:BHinv10}
& & p^I = Re({\cal P}^I),\ q_I=Re({\cal N}_{IJ}{\cal P}^J);\nonumber\\
 & & \partial_i({\cal P}^I{\cal N}_{IJ}{\cal P}^J)=0,\ X^I{\cal N}_{IJ}{\bar{\cal P}}^J = iw.
\end{eqnarray}
One solves for ${\cal P}^I$s from the last two equations in (\ref{eq:BHinv10}) and substitutes the result
into the first two equations of (\ref{eq:BHinv10}).

We will now solve the last two equations of (\ref{eq:BHinv10}) for (\ref{eq:hypersurface}). As an example,
we work with points in the moduli space close to one of the two conifold loci: $\phi^3=1$. We need to work out the matrix $F_{IJ}$ so that one can work out the matrix ${\cal N}_{IJ}$. From the symmetry of $F_{IJ}$
w.r.t. $I$ and $J$, one sees that the constants appearing in (\ref{eq:confl210}) must satisfy some
constraints (which must be borne out by actual numerical computations). To summarize, near $x=0$ and using
(\ref{eq:confl24})-(\ref{eq:confl210}):
\begin{eqnarray}
\label{eq:BHinv11}
& & F_{01}=F_{10}\Leftrightarrow ln x \frac{B_{01}}{B_{31}} + \frac{C_1}{C_3}
= \frac{B_{01}}{B_{41} + B_{42} (ln x + 1)} + \frac{C_0}{C_4}\Rightarrow B_{12}=0,\ \frac{C_1}{C_3}=\frac{C_0}{C_4};\nonumber\\
& & F_{02}=F_{20}\Leftrightarrow ln x \frac{B_{22}}{B_{31}} + \frac{C_2}{C_3} = \frac{B_{01}}{B_{51} + B_{52} (ln x + 1)} + \frac{C_0}{C_5}\Rightarrow B_{22}=0,\ \frac{C_2}{C_3}=\frac{C_0}{C_5};\nonumber\\
& & F_{12}=F_{21}\Leftrightarrow \frac{B_{22}}{B_{42}} + \frac{C_2}{C_4} = \frac{B_{11}}{B_{51} + B_{52} (ln x + 1)} + \frac{C_1}{C_5}\Rightarrow\frac{C_2}{C_4}=\frac{C_1}{C_5}.
\end{eqnarray}
In (\ref{eq:BHinv11}), the constants $A_i, B_{ij},C_k$ are related to the constants $A_i, B_{ij}, C_k$
via matrix elements of $M$ of (\ref{eq:PFbasis2}). Therefore, one gets the following form of $F_{IJ}$:
\begin{equation}
\label{eq:BHinv12}
F_{IJ}=\left(\begin{array}{ccc} \frac{B_{01}}{C_3} + \frac{C_0}{C_3} & \frac{C_1}{C_3} & \frac{C_2}{C_3}\\
\frac{C_1}{C_3} & \frac{C_1}{C_4} & \frac{C_2}{C_4}\\
\frac{C_2}{C_3} & \frac{C_2}{C_4} & \frac{C_2}{C_5}
\end{array}\right)
\end{equation}
Using (\ref{eq:BHinv12}), one can evaluate $X^I Im(F_{IJ}) X^J$ - this is done in the Appendix {\bf A.10}. Further using (\ref{eq:BHinv12}), (\ref{eq:BHinv13}) - (\ref{eq:BHinv14}), one gets:
\begin{eqnarray}
\label{eq:BHinv15}
\left({\cal N}\right)_{ij} = {a}_{ij}+{b}^{(1)}_{ij}x +{b}^{(2)}_{ij} x ln x + {c}_{ij}(\rho-\rho_0); \ i,j\in\left\{0,1,2\right\}
\end{eqnarray}
The constants $a_{ij}, b^{(1),(2)}_{jk},c_{lm}$ are constrained by relations, e.g.,
\begin{equation}
F_I={\cal N}_{IJ}X^J,
\end{equation}
which, e.g., for $I=0$ would imply:
\begin{eqnarray}
\label{eq:BHinv16}
& & a_{00} A_3 + a_{01} A_4 + a_{02} A_5 = A_0\nonumber\\
& & a_{00} B_{31} + b^{(1)}_{00} A_3 + a_{01} B_{41} + A_4 b^{(1)}_{01} + a_{02} B_{51} + b^{(1)}_{02} A_5
= B_{01}\nonumber\\
& & b^{(2)}_{00} A_3 + a_{01} B_{42} + b^{(2)}_{01} A_4 + a_{02} B_{52} + A_5 b^{(2)}_{02} = 0\nonumber\\
& & a_{00} C_3 + c_{00} A_3 + a_{01} C_4 + c_{01} A_4 + a_{02} C_5 + c_{02} A_5 = C_0.
\end{eqnarray}

So, substituting (\ref{eq:BHinv15}) into the last two equations of (\ref{eq:BHinv10}), one gets:
\begin{eqnarray}
\label{eq:BHinv17}
& & \partial_x({\cal P}^I{\cal N}_{IJ}{\cal P}^J)=0\Rightarrow \nonumber\\
& & ln x\left[({\cal P}^0)^2b^{(2)}_{00} + ({\cal P}^1)^2b^{(2)}_{11} + ({\cal P}^2)^2b^{(2)}_{22} + 2{\cal P}^0{\cal P}^1b^{(2)}_{01} + 2{\cal P}^0{\cal P}^2b^{(2)}_{02} + 2{\cal P}^1{\cal P}^2b^{(2)}_{12}\right]=0;\nonumber\\
& & \partial_{\rho-\rho_0}({\cal P}^I{\cal N}_{IJ}{\cal P}^J)=0\Rightarrow \nonumber\\
& & ({\cal P}^0)^2c^{(2)}_{00} + ({\cal P}^1)^2c_{11} + ({\cal P}^2)^2c_{22} + 2{\cal P}^0{\cal P}^1c_{01} + 2{\cal P}^0{\cal P}^2c_{02} + 2{\cal P}^1{\cal P}^2c_{12}=0,
\end{eqnarray}
and ${\bar X}^I Im({\cal N}_{IJ}){\cal P}^J=-iw$ implies:
\begin{eqnarray}
\label{eq:BHinv18}
& & {\bar A}_I(a_{IJ}-{\bar a}_{IJ}){\cal P}^J+{\bar x}[{\bar B}_{I1}(a_{IJ} - {\bar a}_{IJ}){\cal P}^J
- {\bar b^{(1)}}_{IJ}{\bar A}_I{\cal P}^J] + x[b^{(1)}_{IJ}{\bar A}_I{\cal P}^J]
+ x ln x[{\bar A}_I b^{(2)}_{IJ}{\cal P}^J] \nonumber\\
& & + (\rho-\rho_0)[{\bar A}_I c_{IJ}{\cal P}^J]  + ({\bar\rho} - {\bar\rho_0})[{\bar C}_I(a_{IJ} - {\bar a}_{IJ}){\cal P}^J - {\bar c}_{IJ}A_I{\cal P}^J]
+ {\bar x} ln{\bar x}[B_{I2}a_{IJ}{\cal P}^J]=-2{\bar w}\nonumber\\
& & {\rm or}\nonumber\\
& & \sum_{I=0}^2\Upsilon^I(x,{\bar x}, x ln x, {\bar x} ln {\bar x};\rho-\rho_0,{\bar\rho}-{\bar\rho_0}){\cal P}^I={\bar w}.
\end{eqnarray}
Using (\ref{eq:BHinv18}), we eliminate ${\cal P}^2$ from (\ref{eq:BHinv17}) to get:
\begin{eqnarray}
\label{eq:BHinv19}
& & \alpha_1({\cal P}^0)^2 + \beta_1({\cal P}^1)^2 + \gamma_1{\cal P}^0{\cal P}^1 = \lambda_1,\nonumber\\
& & \alpha_2({\cal P}^0)^2 + \beta_2({\cal P}^1)^2 + \gamma_2{\cal P}^0{\cal P}^1 = \lambda_2.
\end{eqnarray}
The equations (\ref{eq:BHinv19}) can be solved and yield four solutions which are:
\begin{eqnarray}
\label{eq:BHinv20}
& &
{\cal P}^0=
\frac{1}{2\,{\sqrt{2}}\,\Biggl( {\alpha_2}\,{\lambda_1} - {\alpha_1}\,{\lambda_2} \Biggr) }\Biggl( {\gamma_2}\,{\lambda_1} - {\gamma_1}\,{\lambda_2} +
           \sqrt{Y} \Biggr)\sqrt{X} \,\nonumber\\
& &    {{\cal P}^1}=-\frac{\sqrt{X}}{\sqrt{2}};\nonumber\\
& & {\cal P}^0=
-\frac{1}{2\,{\sqrt{2}}\,\Biggl( {\alpha_2}\,{\lambda_1} - {\alpha_1}\,{\lambda_2} \Biggr) }\Biggl( {\gamma_2}\,{\lambda_1} - {\gamma_1}\,{\lambda_2} +
           \sqrt{Y} \Biggr)\sqrt{X} \,\nonumber\\
& &    {{\cal P}^1}=\frac{\sqrt{X}}{\sqrt{2}};\nonumber\\
& & {\cal P}^0=
\frac{1}{2\,{\sqrt{2}}\,\Biggl( {\alpha_2}\,{\lambda_1} - {\alpha_1}\,{\lambda_2} \Biggr) }\Biggl( {\gamma_2}\,{\lambda_1} - {\gamma_1}\,{\lambda_2} -
           \sqrt{Y} \Biggr)\sqrt{X} \,\nonumber\\
& &    {{\cal P}^1}=-\frac{\sqrt{X}}{\sqrt{2}};\nonumber\\
& & {\cal P}^0=
-\frac{1}{2\,{\sqrt{2}}\,\Biggl( {\alpha_2}\,{\lambda_1} - {\alpha_1}\,{\lambda_2} \Biggr) }\Biggl( {\gamma_2}\,{\lambda_1} - {\gamma_1}\,{\lambda_2} -
           \sqrt{Y} \Biggr)\sqrt{X} \,\nonumber\\
& &    {{\cal P}^1}=\frac{\sqrt{X}}{\sqrt{2}}
                         \end{eqnarray}
                         where
                         \begin{eqnarray}
                         \label{eq:BHinv21}
                         & & X\equiv\frac{1}{{{\alpha_2}}^2\,{{\beta_1}}^2 +
               {\alpha_2}\,\Bigl[ -2\,{\alpha_1}\,{\beta_1}\,{\beta_2} +
                  {\gamma_1}\,\Bigl( {\beta_2}\,{\gamma_1} - {\beta_1}\,{\gamma_2} \Bigr)  \Bigr]  +
               {\alpha_1}\,\Bigl[ {\alpha_1}\,{{\beta_2}}^2 +
                  {\gamma_2}\,\Bigl( -{\beta_2}\,{\gamma_1}   + {\beta_1}\,{\gamma_2} \Bigr)
                  \Bigr] }\nonumber\\
                  & & \times\Bigl[2\,{{\alpha_2}}^2\,{\beta_1}\,{\lambda_1} +
               {\alpha_1}\,\Bigl( {{\gamma_2}}^2\,{\lambda_1} + 2\,{\alpha_1}\,{\beta_2}\,{\lambda_2} -
                  {\gamma_2}\,\Bigl( {\gamma_1}\,{\lambda_2} +
                     {\sqrt{X_1}}
                     \Bigr)  \Bigr)\Bigr]; \ Y\equiv{{\gamma_2}}^2\,{{\lambda_1}}^2 - 2\,{\gamma_1}\,{\gamma_2}\,{\lambda_1}\,{\lambda_2} \nonumber\\
                     & & +                4\,{\alpha_2}\,{\lambda_1}\,\Bigl( -{\beta_2}\,{\lambda_1}  +
                  {\beta_1}\,{\lambda_2} \Bigr)  +
               {\lambda_2}\,\Bigl( 4\,{\alpha_1}\,{\beta_2}\,{\lambda_1} -
                  4\,{\alpha_1}\,{\beta_1}\,{\lambda_2} + {{\gamma_1}}^2\,{\lambda_2} \Bigr).\end{eqnarray}
                                                \begin{eqnarray}
                         \label{eq:BHinv22}
                         & & X_1\equiv Y + {\alpha_2}\,
                \Bigl[ -2\,{\alpha_1}\,\Bigl( {\beta_2}\,{\lambda_1} + {\beta_1}\,{\lambda_2} \Bigr)  +
                  {\gamma_1}\,\Bigl( -{\gamma_2}\,{\lambda_1}  + {\gamma_1}\,{\lambda_2}+\nonumber\\
                  & & \sqrt{{{\gamma_2}}^2\,{{\lambda_1}}^2 -
                         2\,{\gamma_1}\,{\gamma_2}\,{\lambda_1}\,{\lambda_2}  +       4\,{\alpha_2}\,{\lambda_1}\,
                          ( -{\beta_2}\,{\lambda_1}   + {\beta_1}\,{\lambda_2})  +
                         {\lambda_2}\,( 4\,{\alpha_1}\,{\beta_2}\,{\lambda_1} -
                            4\,{\alpha_1}\,{\beta_1}\,{\lambda_2} + {{\gamma_1}}^2\,{\lambda_2} )}\Bigr)\Bigr].\nonumber\\
                                 \end{eqnarray}
One can show that one does get ${\cal P}^I\sim X^I$ as one of the solutions - this corresponds to a supersymmetric black hole, and the other solutions correspond to non-supersymmetric black holes.

\section{Fake Superpotentials}

In this section, we show the existence of ``fake superpotentials" corresponding to black-hole solutions for type II compactification on (\ref{eq:hypersurface}).

As argued in \cite{Ceresole+Dall'agata}, dS-curved domain wall solutions in gauged supergravity and non-extremal black hole solutions in Maxwell-Einstein theory  have the same effective action. In the context of domain wall solutions, if there exists a ${\cal W}(z^i,{\bar z}^i)\in{\bf R}: V_{DW}(\equiv{\rm Domain\ Wall\ Potential})=-{\cal W}^2 + \frac{4}{3\gamma^2}g^{i{\bar j}}\partial_i{\cal W}\partial_{\bar j}{\cal W}$, $z^i$ being complex scalar fields, then the solution to the second-order equations for domain walls, can also be derived from the following first-order flow equations: $U^\prime=\pm e^U\gamma(r){\cal W};\ (z^i)^\prime = \mp e^U\frac{2}{\gamma^2}g^{i{\bar j}}\partial_{\bar j}{\cal W}$, where
$\gamma\equiv\sqrt{1 + \frac{e^{-2U}\Lambda}{{\cal W}^2}}$.

Now, spherically symmetric, charged, static and asymptotically flat black hole solutions of Einstein-Maxwell theory coupled to complex scalar fields have the form: $dz^2 = - e^{2U(r)} dt^2 + e^{-2U(r)}\biggl[\frac{c^4}{sinh^4(cr)} dr^2 $ $+ \frac{c^2}{sinh^2(cr)}(d\theta^2 + sin^2\theta d\phi^2)\biggr]$, where the non-extremality parameter
$c$ gets related to the positive cosmological constant $\Lambda>0$ for domain walls. For non-constant scalar fields, only for $c=0$ that corresponds to extremal black holes, one can write down first-order flow equations in terms of a ${\cal W}(z^i,{\bar z}^i)\in{\bf R}$: $U^\prime=\pm e^U{\cal W};\ (z^i)^\prime=\pm2e^U g^{i{\bar j}}\partial_{\bar j}{\cal W},$ and the potential $\tilde{V}_{BH}\equiv {\cal W}^2 + 4g^{i{\bar j}}\partial_i{\cal W}\partial_{\bar j}{\cal W}$ can be compared with the ${\cal N}=2$ supergravity black-hole potential $V_{BH}=|Z|^2+g^{i{\bar j}}D_iZD_{\bar j}{\bar Z}$ by identifying
${\cal W}\equiv|Z|$. For non-supersymmetric theories or supersymmetric theories where the black-hole constraint equation admits multiple solutions which may happen because several ${\cal W}$s may correspond to the same $\tilde{V}_{BH}$ of which only one choice of ${\cal W}$ would correspond to the true central charge, one hence talks about ``fake superpotential" or ``fake supersymmetry" - a ${\cal W}:\partial_i{\cal W}=0$ would correspond to a stable non-BPS black hole. Defining ${\cal V}\equiv e^{2U}V(z^i,{\bar z}^i),
{\cal\bf W}\equiv e^U{\cal W}(z^i,{\bar z}^i)$, one sees that ${\cal V}(x^A\equiv U,z^i,{\bar z}^i)=g^{AB}\partial_A{\bf W}(x)\partial_B{\bf W}(x)$, where $g_{UU}=1$ and $g_{Ui}=0$. This illustrates the fact that one gets the same potential ${\cal V}(x)$ for all vectors $\partial_A{\cal\bf W}$ with the same norm. In other words, ${\bf W}$ and $\tilde{\bf W}$ defined via: $\partial_A{\bf W}=R_A^{\ B}(z,{\bar z})\partial_B\tilde{\bf W}$ correspond to the same ${\cal V}$ provided: $R^TgR=g$.
For ${\cal N}=2$ supergravity, the black hole potential $V_{BH}=Q^T{\cal M}Q$ where $Q=(p^\Lambda,q_\Lambda)$ is an $Sp(2n_v+2,{\bf Z})$-valued vector ($n_V$ being the number of vector multiplets) and ${\cal M}\in Sp(2n_V+2,{\bf Z})$ is given by:
\begin{equation}
\label{eq:FakeW1}
{\cal M}=\left(\begin{array}{cc}
A&B\\
C&D
\end{array}\right),
\end{equation} where \begin{eqnarray}
\label{eq:FakeW2}
& & A\equiv Re {\cal N} (Im {\cal N})^{-1}, \ B\equiv-Im {\cal N} - Re {\cal N} (Im {\cal N})^{-1} Re {\cal N} \nonumber\\
& & C\equiv (Im {\cal N})^{-1}, \  D = -A^T = - (Im {\cal N}^{-1})^T (Re {\cal N})^T.
\end{eqnarray}
Defining $M: {\cal M}={\cal I}M$ where
\begin{eqnarray}
\label{eq:FakeW3}
& & M\equiv\left(\begin{array}{cc}
D&C\\
B&A
\end{array}\right); \ \ {\cal I}\equiv\left(\begin{array}{cc}
0&-{\bf 1}_{n_V+1}\\
{\bf 1}_{n_V+1}&0
\end{array}\right).
\end{eqnarray}
The central charge $Z=e^{\frac{K}{2}}(q_\Lambda X^\Lambda - p^\Lambda F_\lambda)$, a symplectic invariant
is expressed as a symplectic dot product of $Q$ and covariantly holomorphic sections: ${\cal V}\equiv e^{\frac{K}{2}}(X^\Lambda,F_\Lambda)=(L^\Lambda,M_\Lambda) (M_\Lambda={\cal N}_{\Lambda\Sigma}L^\Sigma)$,
and hence can be written as
\begin{equation}
\label{eq:FakeW4}
Z = Q^T{\cal I}{\cal V} = L^\Lambda q_\Lambda - M_\lambda p^\Lambda.
\end{equation}
Now, the black-hole potential $V_{BH}=Q^T{\cal M}Q$ (being a symplectic invariant) is invariant under:
\begin{eqnarray}
\label{eq:FakeW5}
& & Q\rightarrow SQ; \ \  S^T{\cal M}S={\cal M}.
\end{eqnarray}
As $S$ is a symplectic matrix, $S^T{\cal I}={\cal I}S^{-1}$, which when substituted in (\ref{eq:FakeW5})
yields:
\begin{equation}
\label{eq:fakeW6}
[S,M]=0.
\end{equation}
In other words, if there exists a constant symplectic matrix $S:[S,M]=0$, then there exists a fake superpotential $Q^TS^T{\cal I}{\cal V}$ whose critical points, if they exist, describe non-supersymmetric black holes. We now construct an explicit form of $S$. For concreteness, we work at the point in the moduli space for (\ref{eq:hypersurface}): $\phi^3=1$ and large $\psi$ near $x=0$ and $\rho=\rho_0$. Given the form of ${\cal N}_{IJ}$ in (\ref{eq:BHinv17}), one sees that:
\begin{eqnarray}
\label{eq:FakeW7}
\left({\cal N}^{-1}\right)_{ij} = \tilde{a}_{ij}+\tilde{b}^{(1)}_{ij}x + \tilde{b}^{(2)}_{ij} x ln x + \tilde{c}_{ij}(\rho-\rho_0); \ i,j\in\left\{0,1,2\right\}
\end{eqnarray}
which as expected is symmetric (and hence so will $Re {\cal N}$ and $(Im {\cal N})^{-1}$). One can therefore write
\begin{equation}
\label{eq:fakeW8}
M\equiv\left(\begin{array}{cc}
U & V \\
X & -U^T
\end{array}\right),
\end{equation}
where $V^T=V,\ X^T=X$ and $U, V, X$ are $3\times 3$ matrices constructed from $Re {\cal N}$ and
$(Im {\cal N})^{-1}$. Writing
\begin{equation}
\label{eq:FakeW9}
S=\left(\begin{array}{cc}
{\cal A} & {\cal B} \\
{\cal C} & {\cal D}
\end{array}\right),
\end{equation}
(${\cal A}, {\cal B}, {\cal C}, {\cal D}$ are $3\times3$ matrices) and given that $S\in Sp(6)$, implying:
\begin{equation}
\label{eq:FakeW10}
\left(\begin{array}{cc}
{\cal A}^T & {\cal C}^T\\
{\cal B}^T & {\cal D}^T
\end{array}\right)\left(\begin{array}{cc}
0 & -{\bf 1}_3 \\
{\bf 1}_3 & 0
\end{array}\right)\left(\begin{array}{cc}
{\cal A} & {\cal B} \\
{\cal C} & {\cal D}
\end{array}\right)=\left(\begin{array}{cc}
0 & -{\bf 1}_3 \\
{\bf 1}_3 & 0
\end{array}\right),
\end{equation}
which in turn implies the following matrix equations:
\begin{eqnarray}
\label{eq:FakeW11}
& & -{\cal A}^T{\cal C} + {\cal C}^T{\cal A} = 0, \ \  -{\cal B}^T{\cal D} + {\cal D}^T{\cal B} = 0,\nonumber\\
& & -{\cal A}^T{\cal D} + {\cal C}^T{\cal B} = - {\bf 1}_3, \  \  -{\cal B}^T{\cal C} + {\cal D}^T{\cal A} = {\bf 1}_3.
\end{eqnarray}
Now, $[S,M]=0$ implies:
\begin{equation}
\label{eq:FakeW12}
\left(\begin{array}{cc}
{\cal A} U + {\cal B} X & {\cal A} V - {\cal B} U^T \\
{\cal C} U + {\cal D} X & {\cal C} V - {\cal D} U^T
\end{array}\right) = \left(\begin{array}{cc}
U {\cal A} + V {\cal C} & U {\cal B} + V {\cal D} \\
X {\cal A} - U^T {\cal C} & X {\cal B} - U^T {\cal D}
\end{array}\right).
\end{equation}
The system of equations (\ref{eq:FakeW11}) can be satisfied, e.g., by the following choice of ${\cal A}, {\cal B}, {\cal C}, {\cal D}$:
\begin{equation}
\label{eq:FakeW13}
{\cal B} = {\cal C}=0;\ {\cal D} = ({\cal A}^{-1})^T.
\end{equation}
To simplify matters further, let us assume that ${\cal A}\in O(3)$ implying that $({\cal A}^{-1})^T = {\cal A}$. Then (\ref{eq:FakeW12}) would imply: \begin{eqnarray}
\label{eq:FakeW14}
& & [{\cal A},V] = 0,\ \ [{\cal A},X] = 0,\nonumber\\
& & [{\cal A}^{-1},U] = 0, \ \  [{\cal A},U] = 0.
\end{eqnarray}
For points near the conifold locus $\phi=\omega^{-1},\rho=\rho_0$, using (\ref{eq:confl24})-(\ref{eq:confl210}) and (\ref{eq:BHinv12}) and dropping the moduli-dependent terms in (\ref{eq:Wconfl11}), one can show:
\begin{eqnarray}
\label{eq:FakeW141}
& & \left(Im {\cal N}^{-1}\right)_{0I}\left(Re {\cal N}\right)_{IK}=0, \left(Im {\cal N}^{-1}\right)_{0K}=0, \nonumber\\
& & \left(Im {\cal N}\right)_{0K} + \left(Re {\cal N}\right)_{0I}\left(Im {\cal N}^{-1}\right)_{IJ}
\left(Re{\cal N}\right)_{JK} = 0,\ K=1,2.
\end{eqnarray}
This is equivalent to saying that the first two and the last equations in (\ref{eq:FakeW14}) can be satisfied by:
\begin{equation}
\label{eq:FakeW15}
{\cal A} = \left(\begin{array}{ccc}
1&0&0\\
0&-1&0\\
0&0&-1
\end{array}\right).
\end{equation}
The form of $A$ chosen in (\ref{eq:FakeW15}) also satisfies the third equation in (\ref{eq:FakeW14}) - similar solutions were also
considered in \cite{Ceresole+Dall'agata}. Hence,
\begin{equation}
\label{eq:FakeW16}
S = \left(\begin{array}{cccccc}
1&0&0&0&0&0\\
0&-1&0&0&0&0\\
0&0&-1&0&0&0\\
0&0&0&1&0&0\\
0&0&0&0&-1&0\\
0&0&0&0&0&-1
\end{array}\right).
\end{equation}
We therefore see that the non-supersymmetric black-hole corresponding to the fake superpotential
$Q^TS^T{\cal I}{\cal V}$, $S$ being given by (\ref{eq:FakeW16}), corresponds to the change of sign of
two of the three electric and magetic charges as compared to a supersymmetric black hole. The symmetry
properties of the elements of ${\cal M}$ and hence $M$ may make it generically possible to find a constant
$S$ like the one in (\ref{eq:FakeW16}) for two-paramater Calabi-Yau compactifications.

\section{Conclusions and Discussions}

We looked at several aspects of complex structure moduli stabilization for a two-parameter ``Swiss-Cheese" Calabi-Yau three-fold of a projective variety expressed as a (resolution of a) hypersurface in a complex weighted projective space, with multiple conifold loci in its moduli space. As regards ${\cal N}=1$ type IIB compactifications on orientifold of the aforementioned Calabi-Yau, we argued the existence of (extended) ``area codes" wherein for the same values of the RR and NS-NS fluxes, one is able to stabilize the complex structure and axion-dilaton moduli at points away from and close to the two singular conifold loci. It would be nice to explicitly work out the numerics and find the set of fluxes corresponding to the aforementioned area codes (whose existence we argued), as well as the flow of the moduli corresponding to the domain walls arising as a consequence of such area codes. As regards supersymmetric and non-supersymmetric black-hole attractors in ${\cal N}=2$ type II compactifications on the same Calabi-Yau three-fold, we explicitly solve the ``inverse problem" of determining the electric and magnetic charges of an extremal black hole given the extremum values of the moduli. In the same context, we also show explicitly the existence of ``fake superpotentials" as a consequence of non-unique superpotentials for the same black-hole potential corresponding to reversal of signs of some of  the electric and magnetic charges.

%% file: chap7.tex
\chapter{Summary and Future Directions}
\markboth{nothing}{\bf 7. Summary and Future Directions}

\hskip1in{\it{``Nature does not hurry, yet everything is accomplished"}} - Lao Tzu

\section{Summary: Overall Conclusions}
After providing some general motivations and relevant literature in the very first chapter ({\bf 1}), we built up our large volume Swiss-Cheese setup in chapter {\bf 2}, where we considered Type IIB compactified on an orientifold of a Swiss-Cheese Calabi Yau and included (non-)perturbative corrections (along with perturbative string one-loop correction, which we showed to be subdominant in the L(arge) V(olume) S(cenarios) limit) to the K\"{a}hler potential and non-perturbative contribution coming from $ED3$-instantons in the superpotential along with flux superpotential. We also considered modular completions of the K\"{a}hler potential and the superpotential and utilized the LVS limit.

In chapter {\bf 3}, we generalized the idea  of obtaining  de-Sitter solutions (\`{a} la KKLT or LVS-type models in which some uplifting mechanism is needed) without the need of addition of any $\overline{D3}$-branes \cite{dSetal}. The same has been  done naturally with the inclusion of non-perturbative $\alpha^\prime$-corrections to the K\"{a}hler potential coming from world sheet instanton, in addition to the earlier LVS-setup of \cite{Balaetal2}. Assuming the NS-NS and RR axions $b^a, c^a$'s to lie in the fundamental-domain and satisfying: $\frac{|b^a|}{\pi}<1,\ \frac{|c^a|}{\pi}<1$, we realized a flat direction provided by the NS-NS axions for slow roll inflation to occur starting from a saddle point and proceeding to the nearest dS minimum. After a detailed calculation we found that for $\epsilon << 1$, in the {\it LVS} limit all along a slow roll trajectory, $sin(n k_a b^a + m k_a c^a)=0$. The ``$\eta$-problem" gets solved at the saddle point along slow-roll trajectory for some quantized values  of a linear combination of the NS-NS and RR axions. As the slow-roll flat direction is provided by the NS-NS axions, a linear combination of the axions gets identified with the inflaton. Thus in a nutshell, we showed the possibility of axionic slow roll inflation in the large volume limit of type IIB compactifications on orientifolds of Swiss Cheese Calabi-Yau's. As a linear combination of the NS-NS axions - the inflaton in our work, corresponds to a discretized expansion rate, analogous to \cite{discreteinflation}, the same may correspond to a CFT with discretized central charges.

Further, we argued that starting from large volume compactification of type IIB string theory involving orientifolds of a two-parameter Swiss-Cheese Calabi-Yau three-fold, for appropriate choice of the holomorphic isometric involution as part of the orientifolding and hence the associated Gopakumar-Vafa invariants  corresponding to the maximum degrees of the genus-zero rational curves , it is possible to obtain $f_{NL}$ - parameterizing non-Gaussianities in curvature perturbations - to be of ${\cal O}(10^{-2})$ in slow-roll and to be of ${\cal O}(1)$ in beyond slow-roll case along with the required 60 number of e-foldings. Moreover, using general considerations and some algebraic geometric assumptions as above, we showed that requiring a ``freezeout" of curvature perturbations at super horizon scales, it was possible to get tensor-scalar ratio of ${\cal O}(10^{-3})$ in the same slow-roll Swiss-Cheese setup. We predicted a loss of scale invariance to be within the existing experimental bounds. To be specific about values, for Calabi-Yau volume ${\cal V}\sim{10}^{6}$ and $n^s\sim {\cal O}(1)$, we realized $\epsilon\sim{0.00028}, |\eta|\sim {10}^{-6}, N_e\sim{60}, |f_{NL}|_{max}\sim {0.01}, r\sim{0.0003}$ and $|n_{R}-1|\sim{0.001}$ with a super-horizon-freezout condition's deviation (from zero) of ${\cal O}(10^{-4})$. Further we observed that with Calabi-Yau volume ${\cal V}\sim{10}^{5}$ and $n^s\sim {\cal O}(1)$ one could realize better values of non-Gaussienities parameter and ``r" ratio ($|f_{NL}|_{max}=0.03$ and $r=0.003$) but with number of e-foldings  less than $60$. Also in the beyond slow-roll case, for $n^s\sim {\cal O}(1)$, we realized $f_{NL}\sim {\cal O}(1)$ with number of e-foldings $N_e\sim 60$. We do not evaluate the tensor-to-scalar ratio and $|n_R-1|$, as the differential equations for scalar and tensor perturbations are highly non-trivial due to non-linearity appearing after $\epsilon$ and $\eta$ becoming non-constant while deviating from the slow-roll trajectory. We closed the chapter on LVS string cosmology by giving some arguments to show the possibility of identifying the inflaton, responsible for slow-roll inflation, to also be a dark matter candidate as well as a quintessence field.

In chapter {\bf 4}, we discussed several phenomenological issues in the context of LVS Swiss-Cheese orientifold compactifications of type IIB with the inclusion of a single mobile space-time filling $D3$-brane  and stack(s) of $D7$-brane(s) wrapping the ``big" divisor along with  supporting $D7$-brane fluxes (on two-cycles homologically non-trivial within the big divisor, and not the Calabi-Yau). Interestingly we found several phenomenological implications which are different from the LVS studies done so far in the literature.

We started with the extension of our LVS Swiss-Cheese cosmology setup with the inclusion of a mobile spacetime filling $D3$-brane and stacks of $D7$-branes wrapping the ``big" divisor $\Sigma_B$ and on the geometric side to enable us to work out the complete K\"{a}hler potential, we calculated the geometric K\"{a}hler potentials of the two divisors $\Sigma_S$ and $\Sigma_B$ of the Swiss-Cheese Calabi-Yau in ${\bf{WCP}}^4[1,1,1,6,9]$ using its toric data and GLSM techniques in the large volume limit. The geometric K\"{a}hler potential was first expressed, using a general theorem due to Umemura, in terms of genus-five Siegel Theta functions or in the LVS limit genus-four Siegel Theta functions. Later using a result due to Zhivkov, for purposes of calculations for our chapter, we expressed the same in terms of derivatives of genus-two Siegel Theta functions.

Then we proposed a possible geometric resolution for a long-standing tension between LVS cosmology and LVS phenomenology : to figure out a way of obtaining a TeV gravitino when dealing with LVS phenemenology and a $10^{12}$ GeV gravitino when dealing with LVS cosmology in the early inflationary epoch of the universe, within the same setup.  The holomorphic pre-factor coming from the space-time filling mobile D3-brane position moduli  - section of (the appropriate) divisor bundle - was found to play a crucial role and we showed that as the mobile space-time filling $D3$-brane moves from a particular non-singular elliptic curve embedded in the Swiss-Cheese Calabi-Yau to another non-singular elliptic curve, it was possible to obtain $10^{12}GeV$ gravitino during the primordial inflationary era supporting the cosmological/astrophysical data as well as a $TeV$ gravitino in the present era supporting the required SUSY breaking at $TeV$ scale within the same set up, for the same volume of the Calabi-Yau stabilized at around $10^6$ (in $l_s=1$ units). This way the string scale involved for our case is $\sim O(10^{15})$ GeV which is nearly of the same order as GUT scale. In the context of soft SUSY breaking, we obtained the gravitino mass $m_{3/2}\sim  O(1-10^3)$ TeV  with ${\cal V}\sim 10^{6} {l_s}^6$ in our setup.

While realizing the Standard Model (SM) gauge coupling $g_{YM}$ $ \sim O(1)$ in the LVS models with D7-branes, usually models with the D7-branes wrapping the smaller divisor have been proposed so far, as D7-branes wrapping the big divisor would produce very small gauge couplings. In our setup, we realized $\sim O(1)$ $g_{YM}$ with D7-branes wrapping the big divisor in the ``rigid limit" i.e. considering zero sections of the normal bundle of the big divisor implying the new possibility of supporting SM on D7-branes wrapping the big divisor. The rigid limit of wrapping is to prevent any obstruction to chiral matter resulting from adjoint matter - corresponding to fluctuations of the wrapped $D7$-branes within the Calabi-Yau - giving mass to open strings stretched between wrapped $D7$-branes. Realizing $g_{YM}$ $ \sim O(1)$ became possible because after constructing appropriate local  involutively-odd harmonic one-forms on the big divisor lying in the cokernel of the pullback of the immersion map applied to $H^{(1,0)}_-$ in the large volume limit, the Wilson line moduli provided a competing contribution to the gauge kinetic function as compared to the volume of the big divisor. This required the complexified Wilson line moduli to be stabilized at around ${\cal V}^{-\frac{1}{4}}$ (which was justifiedby extremization of the potential). Note, similar to the case of local models corresponding to wrapping of $D7$-branes around the small divisor, our model is also local in the sense that the involutively-odd one-forms are constructed locally around the location of the mobile $D3$-brane restricted to (the rigid limit of) $\Sigma_B$.

In chapter {\bf 5}, we estimated various soft supersymmetry breaking parameters, couplings and open string moduli masses in the context of our D3/D7 LVS Swiss-Cheese setup framed in the previous chapter {\bf 4} and realized order TeV gravitino and gaugino masses in the context of gravity mediated supersymmetry breaking. It was observed that anomaly mediated gaugino mass contribution was suppressed by the standard loop factor as compared to gravity mediated contribution. The $D3$-brane position moduli and the $D7$-brane Wilson line moduli were found  to be heavier than gravitino. Further, we observed a (near) universality in masses, $\hat{\mu}$-parameters, Yukawa couplings  and the $\hat{\mu}B$-terms for the $D3$-brane position moduli - the two Higgses in our construction - and a hierarchy in the same set and a universality in the $A$ terms on inclusion of the $D7$-brane Wilson line moduli.  Based on phenomenological intuitions, we further argued that the Wilson line moduli could be identified with squarks/sleptons (of at least the first and second generations) of MSSM as the Yukawa couplings for the same were negligible; the non-universality in the Yukawa's for the Higgses and squarks, was hence desirable. Building up on some more phenomenological aspects of our setup, we discussed the RG flow of the slepton and squark masses to the EW scale and in the process show that related integrals were found to be close to the mSUGRA point on the ``SPS1a slope". 

Further, we showed the possibility of realizing fermions mass scales of first two generations along with order $eV$ neutrino mass scales for Calabi Yau volume ${\cal V}\sim(10^5-10^6)$ and D3-instanton number $n^s=2$. We also argued the absence of SUSY GUT-type dimension-five operators and estimate an upper bound on the proton lifetime to be around $10^{61}$ years from a SUSY GUT-type dimension-six operator. A detailed numerical analysis for solving the RG evolutions will definitely explore some more interesting phenomenology in the context of reproducing MSSM spectrum in this LVS Swiss-Cheese orientifold setup. Large scalar masses and their respective small fermionic superpartners' masses realized in our setup provides clue for the possibility of realizing ``spit-supersymmetry" scenarios in our setup and a detailed exploration on the same is in progress \cite{DM_SplitSusy}. Some interesting work related to aforementioned phenomenological issues mentioned in this paragraph can be found in \cite{Sarkar1,DebajyotiAndMukhopadhyaya,Sarkar2,SenguptaAndMukhopadhyaya,Majumdar1,Mukhopadhyaya,Dienes3,Antoniadis2}.

In chapter {\bf 6}, we looked at several aspects of (complex structure) moduli stabilization  with the same two-parameter ``Swiss cheese" Calabi-Yau which has multiple conifold loci in its moduli space. As regards ${\cal N}=1$ type IIB orientifold compactifications in our Swiss-Cheese setup, we argued the existence of ``area codes" wherein for the same values of the RR and NS-NS fluxes, one could be able to stabilize the complex structure and axion-dilaton moduli at points away from and close to the two singular conifold loci. It would be nice to explicitly work out the numerics and find the explicit set of fluxes corresponding to the aforementioned area codes (whose existence we argued), as well as the flow of the moduli corresponding to the domain walls arising as a consequence of such area codes. Regarding the supersymmetric and non-supersymmetric black-hole attractors in ${\cal N}=2$ type II compactifications on the same Calabi-Yau three-fold, we explicitly solved the ``inverse problem" of determining the electric and magnetic charges of an extremal black hole given the extremum values of the moduli. In the same context, we also showed explicitly the existence of ``fake superpotentials" as a consequence of non-unique superpotentials for the same black-hole potential corresponding to reversal of signs of some of  the electric and magnetic charges after constructing a constant symplectic matrix for our two-paramater Swiss-Cheese Calabi-Yau. There may be interesting connection between the existence of such fake superpotentials and works like \cite{EH}.

Now we provide some of future directions in the context of our LVS Swiss-Cheese setup as below.

\section{Future Directions}
Recently in \cite{PS_FTLVS}\footnote{This work is not included in my Ph.D. thesis.}, again in the context of Type IIB compactified on an orientifold of a large volume Swiss-Cheese Calabi Yau in ${\bf{WCP}}^4[1,1,1,6,9]$, in the presence of a mobile space-time filling $D3$-brane and stack(s) of fluxed $D7$-brane(s) wrapping the ``big" divisor $\Sigma_B$, we explored various implications of moduli dynamics and discussed their couplings and decay into MSSM(-like) matter fields early in the history of universe to reach thermal equilibrium. Like finite temperature effects in O'KKLT, we  observed that the local minimum of zero-temperature effective scalar potential is stable against any finite temperature corrections (up to two-loops) in large volume scenarios as well. Moreover, we found the moduli to be heavy enough to avoid any cosmological moduli problem. Also, interestingly it has been shown to realize split susy scenarios in our LVS Swiss-Cheese setup \cite{DM_SplitSusy}. Thus, based on the cosmological/phenomenological implications of our LVS Swiss-Cheese setup, it is not surprising to believe that our setup is rich enough in realistic implications and a lot of interesting physics can be extracted. It will be interesting to work on various aspects of Type IIB compactifications for cosmology as well as phenomenology model building in our large volume Swiss-Cheese setup(s) in a single compactification scheme.
%
\subsection{Embedding Dark Energy/Matter in our LVS Setup}

The universe in which we live is not only expanding but also the expansion is accelerating and one of the reasons for the same has been supposed to be dark energy. As not much work has been done in the area of realizing dark energy and dark matter in the context of string theory, the same is an extremely interesting as well as challenging topic to work on. Although LVS class of models have been quite exciting for realistic model building for cosmology as well as particle physics, the explicit construction of a model embedding dark energy and dark matter in the context of large volume scenarios has been missing. Further, as we have observed in one of our previous work \cite{axionicswisscheese,largefNL_r} that in some corner(s) of moduli space, the ${\cal N}=1$ scalar potential in our setup, takes a similar form which has been used for building (cold) dark matter \cite{SahniWang} as well as dark energy models \cite{Q}. As part of my future plan, I plan to work on exploring the possibility of explicit embedding of dark matter and dark energy scenarios in our type IIB large volume Swiss-Cheese orientifold setup.

In an investigation of the possibility of dark energy solutions in string theory framework \cite{Sami} with the inclusion of perturbative $\alpha^{\prime}$-corrections in the four-dimensional effective action of Type II, heterotic, and bosonic strings, it has been observed that all these respond differently to dark energy. Further, it has been concluded that a dark energy solution exists in the case of the bosonic string, while
the other two theories do not lead to realistic dark energy universes. Hence, it will be interesting to explore the possibility of realizing dark energy solutions with the inclusion of non-perturbative corrections to the K\"{a}hler potential and the superpotential (and hence ${\cal N}=1$ scalar potential) in the context of our type IIB LVS Swiss-Cheese orientifold compactifications. In this context, we will be proceeding with realizing a very small energy scale $10^{-3}$ eV to be comparable with the cosmological constant and study the ``Equations of State" (EOS) $w=p/\rho$ (ratio of pressure and energy density) investigating which kind of scenarios will be applicable (vacuum energy $w=-1$ or Quintessence $w > -1$) for our LVS setup. Starting with the ${\cal N}=1$ type IIB supergravity action, we will calculate the energy-momentum tensor to estimate and study the EOS and various implications thereof. Finally, it will also be interesting to look at the possible modifications in the realized dark energy solution(s) with the inclusion of thermal corrections incorporating the same in  the ${\cal N}=1$ scalar potential. The general form of scalar potential to start with, is given as:
\begin{equation}
V_{\rm Tot} = V_{T=0} + V_T^{1-loop} + V_T^{2-loop} +.....,
\end{equation} where $V_{T=0}$ is the zero temperature contribution coming from $\alpha^{\prime}$- corrections, string loop corrections and non-perturbative instanton contributions, and other terms are finite temperature corrections which recently, have been shown to be subdominant in our LVS setup \cite{PS_FTLVS}.\\


\subsection{Issues in Beyond Standard Model Physics with our $D3/D7$ LVS Setup}

In our setup, we have shown a novel implication of LVS models wherein it is possibile to realize the first two families' fermion mass scales identifying the fermions with the fermionic superpartners of the $D7$-brane Wilson line moduli; for computational simplicity we included a single Wilson line modulus \cite{Sparticles_MS}. Further, in the context of realizing (MS)SM spectrum, the fermionic superpartner of the fluctuations of the wrapped $D7$-branes normal to $\Sigma^B$ (spanning the space of global sections of the normal bundle $N\Sigma^B$), are also possible candidates for (MS)SM fermions.

For a check of the robustness as well as realizing more realistic features of our LVS $D3/D7$ setup in the context of string phenomenology, introducing more Wilson line moduli (after constructing local odd harmonic one-forms in Cohomology of ``big" divisor) and with the inclusion of fermionic superpartnters of null sections (to implement chirality of the spectrum)  of the normal bundle of the ``big" divisor, it will be interesting to explore on realizing the complete (MS)SM spectrum in our large volume Swiss-Cheese setup. The same could be possible with the explicit construction of appropriate local involutively-odd harmonic one-forms on the big divisor lying in the cokernel of the pullback of the immersion map applied to $H^{(1,0)}_-$ in the large volume limit. At least one of such Wilson line moduli (after constructing such one-form in \cite{D3_D7_Misra_Shukla}) provide a competing contribution to the gauge kinetic function as compared to the volume of the big divisor and the possible cancelation results in realizing order one gauge couplings $g_a\sim {\cal O}(1)$.
The inclusion of more Wilson line moduli  would imply a modification in the ${\cal N}=1$ coordinate ``$T_\alpha$" via $i\kappa_4^2\mu_7\int_{\Sigma_B}i^*\omega\wedge A^I\wedge{\bar A}^{\bar J}a_I{\bar a}_{\bar J}
$ (where $\kappa_4$ is related to four-dimensional Newton's constant, $\mu_7$ is the $D7$-brane tension and $a_I$'s are defined through KK reduction of $U(1)$ gauge field). The same along with the inclusion of fermionic superpartners of the moduli corresponding to fluctuations of wrapped D7-branes normal to $\Sigma_B$  will result in various interesting implications, e.g. identification of possibly all SM fermions with the aforementioned open string moduli superpartners, new couplings etc. which could improve our understanding of supersymmetry breaking in our setup. The same could facilitate identification with all three squark/slepton generations, induce more higher dimensional operators which could be exciting to address other issues of exploring some new physics beyond MSSM, like studying neutrino oscillations, proton stability, etc. (See \cite{MS_LVSfermions,Prot_Decay_review,Witten_Klebanov_proton_decay}) identifying possibly Dark matter and Dark energy candidate in the same setup.


\subsection{Implications of Moduli Redefinitions in LVS}

In the context of string compactifications, moduli are redefined at one-loop level due to possible redefinition in the chiral superfields through compactification geometries. Recently, the effect of these moduli redefinitions has been studied in LVS models in the context of moduli stabilization and supersymmetry breaking scenarios in \cite{Conlonmoduliredef} and it has been observed that redefinitions of the small moduli do not alter the basic structure of the large volume minimum leaving it at the same location and at an exponentially large volume while for redefinitions of the overall volume, the modified K\"{a}hler potential
gives a scalar potential that actually leads to runaway behavior and delocalizes the large volume
minimum. The results on SUSY breaking side can also be modified significantly after the redefinitions. It will be interesting to answer a curious question as to what the effects of these moduli redefinitions are on moduli stabilization  as well as on supersymmetry breaking parameters in our $D3/D7$ Swiss-Cheese setup and investigate the effects of moduli redefinitions on various cosmological and phenomenological aspects.

\vskip 1in

\newpage

I would like to close this review article with:\\

\hskip1in{\it{`` The time will come when diligent research over long periods will bring to light things that now lie hidden. A single life time, even though entirely devoted to research, would not be enough for the investigation of so vast a subject. . . . And so this knowledge will be unfolded through long successive ages. There will come a time when our descendants will be amazed that we we did not know things that are so plain to them. . . . Many discoveries are reserved for ages still to come, when memory of us will have been effaced. Our universe is a sorry little affair unless it has in it something for every age to investigate . . . . Nature does not reveal her mysteries once and for all."}}

\hskip3.7in -Seneca

\hskip2.5in (Natural Questions Book 7, c. first century.)

%% file: Appendix.tex
\chapter{Appendix}
\markboth{nothing}{\it Appendix}
\section{Constructing a Basis: $Dim_{\bf{Re}} \, H^{1,1}_-(CY_3,{\bf Z})=2$}
\setcounter{equation}{0}
\seceqaa
Here we construct a basis of $H^{1,1}_-(CY_3, {\bf Z})$ with real dimensionality 2, in justification of what we have been using in our setup. Consider $B_{i{\bar j}}dz^i\wedge d{\bar z}^{{\bar j}}$ with only $B_{1{\bar 2}}$ and $B_{1{\bar 3}}$ non-zero. Reality of $B_2$-filed implies:
$$B_{2{\bar 1}}=-\overline{B_{1{\bar 2}}},\ B_{1{\bar 2}}=-\overline{B_{2{\bar 1}}}; B_{3{\bar 1}}=-\overline{B_{1{\bar 3}}},\ B_{1{\bar 3}}=-\overline{B_{3{\bar 1}}},$$
which assuming $B_{i{\bar j}}\in{\bf R}$ implies $B_{1{\bar 2}}=-B_{2{\bar 1}}, B_{1{\bar 3}}=-B_{3{\bar 1}}$. Consider:
$\{\omega_1^{-},\omega_2^{-}\}=\{(dz^1\wedge d{\bar z}^{\bar 2} - dz^2\wedge d{\bar z}^{\bar 1}),(dz^1\wedge d{\bar z}^{\bar 3} - dz^3\wedge d{\bar z}^{\bar 1})\}$. If the $\omega^a_-$s form a basis for
$H^{(1,1)}_-(CY_3,{\bf Z})$ - a real subspace of  $H^{1,1}(CY_3,{\bf Z})$  then:
$$\int_{CY_3}\omega_a^-\wedge *_6\overline{\omega_b^-}=\int_{CY_3}\omega_a^-\wedge \tilde{\omega}^b_-=\delta^b_a,$$
where $\tilde{\omega}^a_-$'s form a basis for $H^{2,2}_-(CY_3,{\bf Z})$. As $\omega_a^-\in{\bf R}$, there is no need to complex conjugate the Hodge dual of the same when taking the inner product of two such $(1,1)$-forms in implementing the completeness requirement for $\omega_a^-$.

Now, as $*_n:H^{p,q}\rightarrow H^{n-q,n-p}$, we have,
\begin{eqnarray}
\label{eq:6D_complex *}
& & *_n\omega_{i_1...i_p,{\bar j}_1,...,{\bar j}_q}dz^{i_1}\wedge ...\wedge dz^{i_p}\wedge d{\bar z}^{\bar j_1}\wedge ...\wedge d{\bar z}^{\bar j_q}\nonumber\\
& & \sim\sqrt{g}\epsilon^{i_1...i_p}\ _{{\bar i}_{p+1}...{\bar i}_n}
\epsilon^{{\bar j}_1...{\bar j}_q}\ _{j_{q+1}...j_n}\omega_{i_1...i_p,{\bar j}_1,...,{\bar j}_q}
d{\bar z}^{{\bar i}_{p+1}}\wedge...d{\bar z}^{{\bar i}_n}\wedge dz^{j_{q+1}}\wedge...dz^{j_n}.
\end{eqnarray}
Hence, locally assuming a diagonal Calabi-Yau metric,
\begin{eqnarray*}
\label{eq:Hodge_duals}
& & *_6\omega_1^-\sim\sqrt{g}\epsilon^1\ _{{\bar 2}{\bar 3}}\epsilon^{\bar 2}\ _{31}d{\bar z}^{\bar 2}\wedge d{\bar z}^{\bar 3}\wedge dz^3\wedge dz^1 - \epsilon^2\ _{{\bar 3}{\bar 1}}\epsilon^{\bar 1}\ _{23}d{\bar z}^{\bar 3}\wedge d{\bar z}^{\bar 1}\wedge dz^2\wedge dz^3,
\end{eqnarray*}
implying
\begin{eqnarray*}
\label{eq:completeness_1}
& & \omega_1^-\wedge*_6\overline{\omega_1^-}= \omega_1^-\wedge*_6\omega_1^-\sim2\sqrt{g}dz^1\wedge d{\bar z}^{\bar 1}\wedge dz^2\wedge d{\bar z}^{\bar 2}\wedge dz^3\wedge d{\bar z}^{\bar 3}\sim{\rm volume-form},
\end{eqnarray*}
as well as:
\begin{eqnarray*}
\label{eq:completeness_2}
& & \omega_2^-\wedge*_6\overline{\omega_1^-}=\omega_2^-\wedge*_6\omega_1^-=0.
\end{eqnarray*}
Similarly, one can argue $\omega_2^-\wedge*_6\omega_2^-\sim$ volume-form. For a more exact calculation, one can show that:
\begin{eqnarray}
& & \omega_1^-\wedge*_6\omega_1^- \sim2\sqrt{g}\Biggl[\left(g_{2{\bar 3}}g_{3{\bar 1}}-g_{2{\bar 1}}g_{3{\bar 3}}\right)\left(g_{2{\bar 3}}g_{3{\bar 1}}-g_{2{\bar 1}}g_{3{\bar 3}}\right)-\left(g_{2{\bar 2}}g_{3{\bar 3}}-g_{2{\bar 3}}g_{3{\bar 2}}\right)\nonumber\\
& & \left(g_{1{\bar 3}}g_{3{\bar 1}}-g_{1{\bar 1}}g_{3{\bar 3}}\right)+c.c.\Biggr] \times\sqrt{g}dz^1\wedge d{\bar z}^{\bar 1}\wedge dz^2\wedge d{\bar z}^{\bar 2}\wedge dz^3\wedge d{\bar z}^{\bar 3}\nonumber\\
\end{eqnarray}
whereas:
\begin{eqnarray}
\label{eq:eq:completeness_exact 1}
& & \omega_2^-\wedge*_6\omega_1^- \sim2\Biggl[\left(g_{2{\bar 1}}g_{3{\bar 2}}-g_{2{\bar 2}}g_{3{\bar 1}}\right)\left(g_{2{\bar 3}}g_{3{\bar 1}}-g_{2{\bar 1}}g_{3{\bar 3}}\right)-\left(g_{2{\bar 2}}g_{3{\bar 3}}-g_{2{\bar 3}}g_{3{\bar 2}}\right)\nonumber\\
& & \left(g_{1{\bar 3}}g_{2{\bar 1}}-g_{1{\bar 1}}g_{2{\bar 3}}\right)+c.c.\Biggr]\times\sqrt{g}dz^1\wedge d{\bar z}^{\bar 1}\wedge dz^2\wedge d{\bar z}^{\bar 2}\wedge dz^3\wedge d{\bar z}^{\bar 3}.
\end{eqnarray}
To get some idea about $g_{i{\bar j}}$, we will look at the LVS limit of the geometric K\"{a}hler potential of $\Sigma_B$ obtained using GLSM techniques. One sees that:
\begin{equation}
\label{eq:metric-Big}
g_{i{\bar j}}|_{\Sigma_B}\sim \left(\begin{array}{cc}
-\frac{\left(\frac{\zeta }{\log ^{\frac{1}{2}}(z)}\right)^{7/6}}{6 \sqrt[9]{2}
   }{\cal V}^{\frac{2}{9}}-\frac{\left(\frac{\zeta }{\log ^{\frac{1}{2}}({\cal V})}\right)^{7/6}}{3 \sqrt[9]{2}
   }{\cal V}^{\frac{5}{18}} & -\frac{\left(\frac{\zeta }{\log ^{\frac{1}{2}}(z)}\right)^{7/6}}{6 \sqrt[9]{2}
   }{\cal V}^{\frac{2}{9}}+\frac{\left(\frac{\zeta }{\log ^{\frac{1}{2}}({\cal V})}\right)^{7/6}}{3 \sqrt[9]{2}
   }{\cal V}^{\frac{5}{18}}\\
-\frac{\left(\frac{\zeta }{\log ^{\frac{1}{2}}(z)}\right)^{7/6}}{6 \sqrt[9]{2}
   }{\cal V}^{\frac{2}{9}}+\frac{\left(\frac{\zeta }{\log ^{\frac{1}{2}}({\cal V})}\right)^{7/6}}{3 \sqrt[9]{2}
   }{\cal V}^{\frac{5}{18}}& -\frac{\left(\frac{\zeta }{\log ^{\frac{1}{2}}(z)}\right)^{7/6}}{6 \sqrt[9]{2}
   }{\cal V}^{\frac{2}{9}}-\frac{\left(\frac{\zeta }{\log ^{\frac{1}{2}}({\cal V})}\right)^{7/6}}{3 \sqrt[9]{2}
   }{\cal V}^{\frac{5}{18}}
   \end{array}\right).
\end{equation}
For ${\cal V}\sim10^{5-6}$, $g_{i\neq j}/g_{i{\bar i}}<1$. We expect this to hold for the full Calabi-Yau.
Hence, \begin{equation}
\frac{\int_{CY_3}\omega_a^-\wedge *_6\omega_b^-}{\int_{CY_3}\omega_a^-\wedge *_6\omega_a^-}<1, a\neq b\end{equation} implying that the completeness relation is approximately satisfied (in the LVS limit).

\section{Inverse Metric Components}
\setcounter{equation}{0}
\seceqbb
\subsection{With the Inclusion of (Non-)Perturbative $\alpha^\prime$-Corrections to the K\"{a}hler Potential}
From the K\"{a}hler potential (without loop-correction), one can show that the corresponding K\"{a}hler metric of (\ref{eq:nonpert13}) is given by:
\begin{eqnarray}
\label{eq:nonpert131}
& & {\cal G}_{A{\bar B}} = \left(\begin{array}{cccc}
{\cal G}_{\rho_s{\bar\rho_s}}
&{\cal G}_{\rho_s{\bar\rho_b}}
& {\cal G}_{\rho_s{\bar G_1}}
& {\cal G}_{\rho_s{\bar G_2}}  \\
\overline{{\cal G}_{\rho_s{\bar\rho_b}}}
& {\cal G}_{\rho_b{\bar\rho_b}}
& {\cal G}_{\rho_b{\bar G_1}}
& {\cal G}_{\rho_b{\bar G_2}}  \\
\overline{{\cal G}_{\rho_1{\bar G_1}}}
& \overline{{\cal G}_{\rho_2{\bar G_1}}}
& k_1^2{\cal X}_1 & k_1k_2{\cal X}_1 \\
\overline{{\cal G}_{\rho_1{\bar G_2}}}
& \overline{{\cal G}_{\rho_2{\bar G_2}}}
& k_1k_2 {\cal X}_1 & k_2^2{\cal X}_1
\end{array}\right),
\end{eqnarray}
where

\begin{eqnarray*}
& & {\cal G}_{\rho_s{\bar\rho_s}}=\frac{1}{4}\left(\frac{1}{6\sqrt{2}}\frac{1}{\sqrt{{\bar\rho}_s - \rho_s}{\cal Y}} + \frac{1}{18}\frac{({\bar\rho}_s - \rho_s)}{{\cal Y}^2}\right),\nonumber\\
& & {\cal G}_{\rho_s{\bar\rho_b}}=\frac{1}{144}\left(\frac{\sqrt{({\bar\rho}_s - \rho_s)({\bar\rho}_b - \rho_b)}}{{\cal Y}^2}\right),\nonumber\\
& & {\cal G}_{\rho_s{\bar G^1}}=\frac{-ie^{-\frac{3\phi_0}{2}}\sqrt{{\bar\rho}_s - \rho_s}{\cal Z}(\tau)}{6\sqrt{2}{\cal Y}^2},\nonumber\\
& & {\cal G}_{\rho_s{\bar G^2}}=\frac{-ie^{-\frac{3\phi_0}{2}}\sqrt{{\bar\rho}_s - \rho_s}{\cal Z}(\tau)}{6\sqrt{2}{\cal Y}^2},\nonumber\\
& & {\cal G}_{\rho_b{\bar\rho_b}}= \frac{1}{4}\left(\frac{1}{6\sqrt{2}}\frac{\sqrt{{\bar\rho}_b - \rho_b}}{{\cal Y}}+ \frac{1}{18}\frac{\sqrt{{\bar\rho}_b - \rho_b}}{{\cal Y}^2}\right),\nonumber\\
& & {\cal G}_{\rho_b{\bar G^1}}=frac{-ie^{-\frac{3\phi_0}{2}}\sqrt{{\bar\rho}_b - \rho_b}{\cal Z}(\tau)}{6\sqrt{2}{\cal Y}^2},\nonumber\\
& & {\cal G}_{\rho_b{\bar G^2}}=\frac{-ie^{-\frac{3\phi_0}{2}}\sqrt{{\bar\rho}_b - \rho_b}{\cal Z}(\tau)}{6\sqrt{2}{\cal Y}^2}
  \end{eqnarray*}
\begin{eqnarray*}
& & {\cal Z}(\tau)\equiv \sum_c\sum_{m,n}A_{n,m,n_{k^c}}(\tau) sin(nk.b + mk.c),\ A_{n,m,n_{k^c}}(\tau)\equiv \frac{(n+m\tau)n_{k^c}}{|n+m\tau|^3}\nonumber\\
&& {\cal Y}\equiv {\cal V}_E + \frac{\chi}{2}\sum_{m,n\in{\bf Z}^2/(0,0)}
\frac{(\tau - {\bar\tau})^{\frac{3}{2}}}{(2i)^{\frac{3}{2}}|m+n\tau|^3}
- 4\sum_{\beta\in H_2^-(CY_3,{\bf Z})}n^0_\beta\sum_{m,n\in{\bf Z}^2/(0,0)}
\frac{(\tau - {\bar\tau})^{\frac{3}{2}}}{(2i)^{\frac{3}{2}}|m+n\tau|^3}\nonumber\\
& & \hskip 1cm \times cos\left((n+m\tau)k_a\frac{(G^a-{\bar G}^a)}{\tau - {\bar\tau}}
 - mk_aG^a\right)
 \end{eqnarray*}
The inverse metric is given as:
\begin{equation}
\label{eq:nonpert15}
{\cal G}^{-1}=\left(\begin{array}{cccc}
({\cal G})^{\rho_s{\bar\rho_s}} & ({\cal G})^{\rho_s{\bar\rho_b}} &
({\cal G})^{\rho_s{\bar G^1}} & 0 \\
\overline{({\cal G})^{\rho_s{\bar\rho_b}}} & ({\cal G})^{\rho_b{\bar\rho_b}} & (
{\cal G})^{\rho_b{\bar G^1}} & 0 \\
\overline{({\cal G})^{\rho_s{\bar G^1}}} & \overline{({\cal G})^{\rho_b{\bar G^1}}} & \frac{1}{(k_1^2-k_2^2){\cal X}_1}
& \frac{k_2}{(k_1k_2^2-k_1^3){\cal X}_1} \\
0 & 0 & \frac{k_2}{(k_1k_2^2-k_1^3){\cal X}_1} & \frac{1}{(k_1^2-k_2^2){\cal X}_1}
\end{array}\right),
\end{equation}
where the components of the inverse of the metric (\ref{eq:nonpert13}) are given as under:

\begin{eqnarray*}
& & ({\cal G}^{-1})^{\rho_1{\bar\rho_1}}=\frac{1}{\Delta}\Biggl[
144\,{\cal Y}^2\,{\sqrt{-{\rho_1} + {\bar\rho_1}}}\,
  \biggl( 2\,{\rho_2}\,{\cal X}^2\,{\sqrt{-{\rho_2} + {\bar\rho_2}}}  -
    \left( 2\,{\cal X}^2 + e^{3\,\phi}\,{\cal X}_1\,{\cal Y}^2 \right) \nonumber\\
  & & \hskip 0.9in \times{\bar\rho_2}\,
     {\sqrt{-{\rho_2} + {\bar\rho_2}}}
    e^{3\,\phi}\, + {\cal X}_1\,{\cal Y}^2\,\left( 3\,{\sqrt{2}}\,{\cal Y} +
       {\rho_2}\,{\sqrt{-{\rho_2} + {\bar\rho_2}}} \right)  \biggr)\Biggr],\nonumber\\
& & ({\cal G}^{-1})^{\rho_1{\bar\rho_2}}=\frac{1}{\Delta}\Biggl[
144\,{\cal Y}^2\,\left( -2\,{\cal X}^2 + e^{3\,\phi}\,{\cal X}_1\,{\cal Y}^2 \right) \,
  \left( {\rho_1} - {\bar\rho_1} \right) \,
  \left( {\rho_2} - {\bar\rho_2} \right)\Biggr],\nonumber\\
  & & ({\cal G}^{-1})^{\rho_1{\bar G^1}}=\frac{1}{\Delta}
24\,i  \,e^{\frac{3\,\phi}{2}}\,{\cal X}\,{\cal Y}^2\,\left( {\rho_1} - {\bar\rho_1} \right) \,
  \left( 3\,{\cal Y} + {\sqrt{2}}\,{\rho_2}\,{\sqrt{-{\rho_2} + {\bar\rho_2}}} -
    {\sqrt{2}}\,{\bar\rho_2}\,{\sqrt{-{\rho_2} + {\bar\rho_2}}}
    \right)\nonumber\\
& & ({\cal G}^{-1})^{\rho_2{\bar\rho_2}}= \frac{1}{\Delta}
144\,{\cal Y}^2\,\biggl[ -2\,{\rho_1}\,{\cal X}^2\,{\sqrt{-{\rho_1} + {\bar\rho_1}}} +
    \left( 2\,{\cal X}^2 + e^{3\,\phi}\,{\cal X}_1\,{\cal Y}^2 \right) \,{\bar\rho_1}\,
     {\sqrt{-{\rho_1} + {\bar\rho_1}}} \nonumber\\
     & & \hskip 0.9in + e^{3\,\phi}\,{\cal X}_1\,{\cal Y}^2\,\left( 3\,{\sqrt{2}}\,{\cal Y} -
       {\rho_1}\,{\sqrt{-{\rho_1} + {\bar\rho_1}}} \right)  \biggr] \,
  {\sqrt{-{\rho_2} + {\bar\rho_2}}},\nonumber\\
& & ({\cal G}^{-1})^{\rho_2{\bar G^1}}=\frac{1}{\Delta}\Bigl[
-24\,i  \,e^{\frac{3\,\phi}{2}}\,{\cal X}\,{\cal Y}^2\,\left( 3\,{\cal Y} -
    {\sqrt{2}}\,{\rho_1}\,{\sqrt{-{\rho_1} + {\bar\rho_1}}} +
    {\sqrt{2}}\,{\bar\rho_1}\,{\sqrt{-{\rho_1} + {\bar\rho_1}}}
    \right) \,\left( {\rho_2} - {\bar\rho_2} \right)\Bigr],\nonumber\\
& & ({\cal G}^{-1})^{G^1{\bar G^1}}=\frac{1}{\Delta}\Biggl[
18\,e^{3\,\phi}\,{{k1}}^2\,{\cal X}_1\,{\cal Y}^4 -
  6\,{\sqrt{2}}\,{{k2}}^2\,{\rho_1}\,{\cal X}^2\,{\cal Y}\,
   {\sqrt{-{\rho_1} + {\bar\rho_1}}} -
  3\,{\sqrt{2}}\,e^{3\,\phi}\,{{k1}}^2\,{\rho_1}\,{\cal X}_1\,{\cal Y}^3\nonumber\\
   & & \hskip 0.6in \,
   {\sqrt{-{\rho_1} + {\bar\rho_1}}}  +
  6\,{\sqrt{2}}\,{{k_2}}^2\,{\rho_2}\,{\cal X}^2\,{\cal Y}\,
   {\sqrt{-{\rho_2} + {\bar\rho_2}}} +
  3\,{\sqrt{2}}\,e^{3\,\phi}\,{{k_1}}^2\,{\rho_2}\,{\cal X}_1\,{\cal Y}^3\,
   {\sqrt{-{\rho_2} + {\bar\rho_2}}} \nonumber\\
   & & \hskip 0.6in -   8\,{{k_2}}^2\,{\rho_1}\,{\rho_2}\,{\cal X}^2 \,
   {\sqrt{-{\rho_1} + {\bar\rho_1}}}\,
   {\sqrt{-{\rho_2} + {\bar\rho_2}}}  -
  \bigl( 3\,{\sqrt{2}}\,e^{3\,\phi}\,{{k_1}}^2\,{\cal X}_1\,{\cal Y}^3 +
     2\,{{k_2}}^2\,{\cal X}^2\nonumber\\
   & & \hskip 0.6in \,(3\,{\sqrt{2}}\,{\cal Y} -
        4\,{\rho_1}\,{\sqrt{-{\rho_1} + {\bar\rho_1}}})  \bigr)\,
   {\bar\rho_2}\,{\sqrt{-{\rho_2} + {\bar\rho_2}}} +
  {\bar\rho_1}\,{\sqrt{-{\rho_1} + {\bar\rho_1}}} \,
   \bigl( 3\,{\sqrt{2}}\,e^{3\,\phi}\,{{k_1}}^2\nonumber\\
   & & \hskip 0.6in \,{\cal X}_1\,{\cal Y}^3 -
     8\,{{k_2}}^2\,{\cal X}^2\,{\bar\rho_2}\,
      {\sqrt{-{\rho_2} + {\bar\rho_2}}}+
     2\,{{k_2}}^2\,{\cal X}^2\,\left( 3\,{\sqrt{2}}\,{\cal Y} +
        4\,{\rho_2}\,{\sqrt{-{\rho_2} + {\bar\rho_2}}} \right)  \bigr)\Biggr],
        \end{eqnarray*}
with:
\begin{eqnarray*}
& & \Delta=
-18\,e^{3\,\phi}\,{\cal X}_1\,Y^4 + 6\,{\sqrt{2}}\,{\rho_1}\,{\cal X}^2\,{\cal Y}\,
   {\sqrt{-{\rho_1} + {\bar\rho_1}}} +
  3\,{\sqrt{2}}\,e^{3\,\phi}\,{\rho_1}\,{\cal X}_1\,{\cal Y}^3\,
   {\sqrt{-{\rho_1} + {\bar\rho_1}}}\nonumber\\
   & &  -
  6\,{\sqrt{2}}\,{\rho_2}\,{\cal X}^2\,{\cal Y}\,{\sqrt{-{\rho_2} + {\bar\rho_2}}} -
  3\,{\sqrt{2}}\,e^{3\,\phi}\,{\rho_2}\,{\cal X}_1\,{\cal Y}^3\,
   {\sqrt{-{\rho_2} + {\bar\rho_2}}} +
  8\,{\rho_1}\,{\rho_2}\,X^2\,{\sqrt{-{\rho_1} + {\bar\rho_1}}}\nonumber\\
   & & \,
   {\sqrt{-{\rho_2} + {\bar\rho_2}}}  +
  \left( 3\,{\sqrt{2}}\,e^{3\,\phi}\,{\cal X}_1\,{\cal Y}^3 +
     {\cal X}^2\,\left( 6\,{\sqrt{2}}\,{\cal Y} - 8\,{\rho_1}\,{\sqrt{-{\rho_1} + {\bar\rho_1}}}
        \right)  \right) \,{\bar\rho_2}\,
   {\sqrt{-{\rho_2} + {\bar\rho_2}}} -
  {\bar\rho_1}\nonumber\\
   & & \,{\sqrt{-{\rho_1} + {\bar\rho_1}}}\,
   \left( 3\,{\sqrt{2}}\,e^{3\,\phi}\,{\cal X}_1\,{\cal Y}^3 -
     8\,{\cal X}^2\,{\bar\rho_2}\,{\sqrt{-{\rho_2} + {\bar\rho_2}}} +
     {{\cal X}}^2\,( 6\,{\sqrt{2}}\,{\cal Y} + 8\,{\rho_2}\,
         {\sqrt{-{\rho_2} + {\bar\rho_2}}})  \right);\nonumber\\
         & & \nonumber\\
         & & {\cal X}\equiv \sum_c\sum_{(n,m)\in{\bf Z}^2/(0,0)}A_{n,m,n_{k^c}}(\tau)sin(nk.b + mk.c).
         \end{eqnarray*}
         \begin{eqnarray*}
& & {\cal X}_1\equiv\frac{\sum_{\beta\in H_2^-(CY_3,{\bf Z})}n^0_\beta\sum_{m,n\in{\bf Z}^2/(0,0)}e^{-\frac{3\phi_0}{2}}|n+m\tau|^3|
A_{n,m,n_{k^c}}(\tau)|^2cos(nk.b + mk.c)}{{\cal Y}}
 \nonumber\\
& & + \frac{|\sum_{\beta\in H_2^-(CY_3,{\bf Z})}n^0_\beta\sum_{m,n\in{\bf Z}^2/(0,0)}e^{-\frac{3\phi_0}{2}}|n+m\tau|^3A_{n,m,n_{k^c}}(\tau)sin(nk.b + mk.c)|^2}{{\cal Y}^2},
\nonumber\\
& & {\cal X}_2\equiv \sum_{\beta\in H_2^-(CY_3,{\bf Z})}n^0_\beta\sum_{m,n\in{\bf Z}^2/(0,0)}|n+m\tau|^3|
A_{n,m,n_{k^c}}(\tau)|^2cos(nk.b + mk.c).\nonumber\\
\end{eqnarray*}

\subsection{With the Inclusion of String Loop-Corrections along with (Non-)Perturbative $\alpha^\prime$-Corrections to the K\"{a}hler Potential}

Based on (\ref{eq:nonpert81}), the inverse metric (not been careful as regards numerical factors in the numerators and denominators) is given by:
\begin{eqnarray}
\label{eq:nonpert151}
& & {\cal G}^{-1}=\left(\begin{array}{cccc}
{\cal G}^{\rho_s{\bar\rho_s}} & {\cal G}^{\rho_s{\bar\rho_b}} & {\cal G}^{\rho_s{\bar G^1}} & 0 \\
\overline{{\cal G}^{\rho_s{\bar\rho_b}}} & {\cal G}^{\rho_b{\bar\rho_b}} & {\cal G}^{\rho_b{\bar G^1}} & 0 \\
\overline{{\cal G}^{\rho_s{\bar G^1}}} & \overline{{\cal G}^{\rho_b{\bar G^1}}} & {\cal G}^{G^1{\bar G^1}}
&  {\cal G}^{G^1{\bar G}^2}\\
0 & 0 & \overline{{\cal G}^{G^1{\bar G}^2}} & {\cal G}^{G^2{\bar G}^2}
\end{array}\right),
\end{eqnarray}
where
\begin{eqnarray*}
& & {\cal G}^{\rho_s{\bar\rho_s}}=\frac{\frac{\tau-{\bar\tau}}{2i}{\cal Y}(ln {\cal Y})^{\frac{3}{2}}}{\frac{ln {\cal Y}}{ \left(\frac{\tau-{\bar\tau}}{2i}\right)}+\frac{C^{KK\ (1)}_s}{{\cal T}}}\nonumber\\
& & {\cal G}^{\rho_s{\bar\rho_b}} = \frac{\frac{{\cal Y}^{\frac{2}{3}}(ln {\cal Y})^2}{\left(\frac{\tau-{\bar\tau}}{2i}\right)}+\frac{C^{KK\ (1)}_s {\cal Y}^{\frac{2}{3}} ln {\cal Y}}{{\cal T}}}{ln {\cal Y} \left(\frac{\tau-{\bar\tau}}{2i}\right)^{-1}+\frac{C^{KK\ (1)}_s}{{\cal T}}}\nonumber\\
& & {\cal G}^{\rho_s{\bar G^1}} = \frac{i{\cal Z}{\cal Y}^{-1}}{\frac{\tau-{\bar\tau}}{2i}}\frac{\left(-\left(\frac{\tau-{\bar\tau}}{2i}\right)(ln {\cal Y})^2+\frac{C^{KK\ (1)}_s  ln {\cal Y}}{{\cal T}}\right)}{\frac{\chi_1ln {\cal Y}} {\left(\frac{\tau-{\bar\tau}}{2i}\right)}+\frac{C^{KK\ (1)}_s}{{\cal T}}}\nonumber\\
\end{eqnarray*}

\begin{eqnarray*}
& & {\cal G}^{\rho_b{\bar\rho_b}} = \frac{{\cal Y}^{\frac{4}{3}}\left(\frac{ln {\cal Y}}{\left(\frac{\tau-{\bar\tau}}{2i}\right)}
- \frac{C^{KK\ (1)}_s}{{\cal T}}\right)}{\frac{ln {\cal Y}}{\left(\frac{\tau-{\bar\tau}}{2i}\right)}-\frac{C^{KK\ (1)}_s}{{\cal T}}}\nonumber\\
& & {\cal G}^{\rho_b{\bar G^1}} = \frac{i{\cal Z}{\cal Y}^{-\frac{1}{3}}\left(\frac{\tau-{\bar\tau}}{2i}\right)}{k_1{\cal X}_1}\frac{\left(ln {\cal Y}-\left(\frac{\tau-{\bar\tau}}{2i}\right)\frac{C^{KK\ (1)}_s
 }{{\cal T}}\right)}{\frac{ln {\cal Y}}{\left(\frac{\tau-{\bar\tau}}{2i}\right)}-\frac{C^{KK\ (1)}_s}{{\cal T}}}\nonumber\\
 & & {\cal G}^{G^1{\bar G^1}} = \frac{1}{(k_1^2-k_2^2)\chi_1}\frac{\left(-\frac{ln {\cal Y}}{\left(\frac{\tau-{\bar\tau}}{2i}\right)}
 + \frac{C^{KK\ (1)}_s}{{\cal T}}\right)}{\frac{ln {\cal Y}}{\left(\frac{\tau-{\bar\tau}}{2i}\right)}-\frac{C^{KK\ (1)}_s}{{\cal T}}} \nonumber\\
 & & {\cal G}^{G^1{\bar G}^2} = \frac{k_2}{(k_1k_2^2-k_1^3)\chi_1}\nonumber\\
 & & {\cal G}^{G^2{\bar G}^2} = \frac{1}{\chi_1(k_1^2-k_2^2)}.
\end{eqnarray*}

and
\begin{equation}
\label{eq:defs}
{\cal T}\equiv \sum_{(m,n)\in{\bf Z}^2/(0,0)}\frac{\frac{(\tau-{\bar\tau})}{2i}}{|m+n\tau|^2}.
\end{equation}

\section{Justification behind ${\cal A}_I\sim{\cal V}^{-\frac{1}{4}}$}
\setcounter{equation}{0}
\seceqcc
In this section we justify that the Wilson line moduli can be stabilized, in a self-consistent manner, at values of the order of ${\cal V}^{-\frac{1}{4}}$. We evaluate the complete moduli space metric for arbitrary Wilson line moduli but close to ${\cal V}^{-\frac{1}{4}}$ - for simplicity we assume only one such modulus. This implies that we replace ${\cal T}_B(\sigma^B,{\bar\sigma ^B};{\cal G}^a,{\bar{\cal G}^a};\tau,{\bar\tau}) + \mu_3{\cal V}^{\frac{1}{18}} + i\kappa_4^2\mu_7C_{1{\bar 1}}{\cal V}^{-\frac{1}{2}} - \gamma\left(r_2 + \frac{r_2^2\zeta}{r_1}\right)$ with ${\cal V}^{\frac{1}{18}}$ (and the same for
${\cal T}_S(\sigma^S,{\bar\sigma ^B};{\cal G}^a,{\bar{\cal G}^a};\tau,{\bar\tau}) + \mu_3{\cal V}^{\frac{1}{18}} + i\kappa_4^2\mu_7C_{1{\bar 1}}{\cal V}^{-\frac{1}{2}} - \gamma\left(r_2 + \frac{r_2^2\zeta}{r_1}\right)$) with the understanding that there is a cancelation between the big divisor's volume and the quadratic term in the Wilson line moduli. This is only to simplify the calculation of the metric for arbitrary values of the Wilson line modulus - we would arrive at the same conclusion by starting out with a completely arbitrary value of the Wilson line modulus and stabilizing it by extremizing the potential. We assume that all the remaining moduli have been stabilized (the complex structure and axion-dilaton moduli via the covariant constancy of the superpotential, the closed string K\"{a}hler and the open string mobile $D3$ brane position moduli via extremization of the potential). We then show that the potential is identically an extremum for all values of the Wilson line modulus close to ${\cal V}^{-\frac{1}{4}}$.

As we are considering the rigid limit of wrapping of the $D7$-brane around $\Sigma_B$ (to ensure that there is no obstruction to a chiral matter spectrum), there will be no superpotential generated due to the fluxes on the world volume of the $D7$-brane \cite{jockersetal} - the same is given by $\kappa_4^2\mu_7 l\zeta^A\int_{\Sigma_B}\tilde{s}_A\wedge \tilde{\cal F}$, $\tilde{s}_A\in H^2_{{\bar\partial},-}(\Sigma_B)$ and vanishes when $\zeta^A=0$. Further,  by restricting the mobile $D3$-brane to $\Sigma_B$, possible contribution to the non-perturbative superpotential due to gaugino condensation in the presence of a stack of $D7$-branes wrapping (a rigid) $\Sigma_B$, will be nullified. The reason is that the contribution to the non-perturbative superpotential due to gaugino condensation on a stack of $N$ $D7$-branes wrapping $D_5$ will be proportional to $\left(1+z_1^{18}+z_2^{18}+z_3^3-3\phi_0z_1^6z_2^6\right)^{\frac{1}{N}}$, which according to \cite{Ganor1_2}, vanishes whenever the mobile $D3$-brane touches the wrapped $D7$-brane. Hence, when the mobile $D3$-brane is restricted to $D_5$, the aforementioned contribution to the non-perturbative superpotential goes to zero. It is for this reason that we are justified in considering a single  wrapped $D7$-brane, which anyway can not effect gaugino condensation. As discussed in chapter {\bf 3}, unlike usual LVS (for which $W_{c.s.}\sim{\cal O}(1)$) and similar to KKLT scenarios (for which $W_{c.s.}\ll 1$), in either of the cases for us, we have $W_{c.s.}\ll 1$ in large volume limit; we would henceforth assume that the fluxes and complex structure moduli have been so fine tuned/fixed that $W_{c.s}\sim \pm W_{\rm ED3}(n^s=1)$, hence the superpotential will be given by $W\sim W_{n.p.}$:

\begin{equation}
\label{eq:W}
W\sim\left(1+z_1^{18}+z_2^{18}+z_3^2-3\phi_0z_1^6z_2^6\right)^{n^s}\sum_{m^a}\frac{e^{i\frac{\tau m^2}{2}+in^sm_aG^a+in^sT_s}}{f(\tau)},
\end{equation}
where $f(\tau)$ is some appropriate modular function, which we do not know. In the following, we assume that the complexified Wilson line moduli are given entirely in terms of the Wilson line moduli and verify this in a self-consistent manner by extremization of the potential.

To evaluate the potential, we would need to evaluate the inverse of the moduli space metric. As also stated in {\bf 3.2}, we then show then in a self-consistent manner that one can set all components of sections of $N\Sigma_B$ and all components save one of the Wilson line moduli ${\cal A}_1$ to zero - the non-zero Wilson line modulus can be consistently stabilized to ${\cal V}^{-\frac{1}{4}}$.
Now, the derivatives of $K$ relevant to the calculation of the moduli space metric $G_{A{\bar B}}$, assuming ${\cal A}_1$ to be in the neighborhood of ${\cal V}^{-\frac{1}{4}}$,  are given below:

{\it Single Derivatives}
\begin{eqnarray}
\label{eq:singleder_z}
& & \frac{\partial K}{\partial z_i}=
-\frac{2}{\cal Y}\Biggl[\frac{3a}{2}\left(2\tau_b + \mu_3l^2{\cal V}^{\frac{1}{18}} + ... -\gamma K_{\rm geom}\right)^{\frac{1}{2}}\Bigl\{3i\mu_3\l^2(\omega_B)_{i{\bar j}}{\bar z}^{\bar j}+\frac{3}{4}\mu_3l^2\bigl((\omega_B)_{i{\bar j}}{\bar z}^{\tilde{a}}\nonumber\\
& & ({\cal P}_{\tilde{a}})^{\bar j}_lz^l + (\omega_B)_{l{\bar j}}z^l{\bar z}^{\tilde{a}}({\cal P})^{\bar j}_i\bigr)\-\gamma(ln {\cal V})^{-\frac{7}{12}}{\cal V}^{\frac{29}{36}}\Bigr\}
- \frac{3a}{2}\left(2\tau_s + \mu_3l^2{\cal V}^{\frac{1}{18}} + ... -\gamma K_{\rm geom}\right)^{\frac{1}{2}}\nonumber\\
& & \Bigl\{3i\mu_3\l^2(\omega_S)_{i{\bar j}}{\bar z}^{\bar j}+\frac{3}{4}\mu_3l^2\left((\omega_S)_{i{\bar j}}{\bar z}^{\tilde{a}}({\cal P}_{\tilde{a}})^{\bar j}_lz^l + (\omega_S)_{l{\bar j}}z^l{\bar z}^{\tilde{a}}({\cal P})^{\bar j}_i\right) -\gamma(ln {\cal V})^{-\frac{7}{12}}{\cal V}^{\frac{29}{36}}\Bigr\} \Biggr]\nonumber\\
\end{eqnarray}
\begin{equation}
\label{eq:singleder_sigma}
\hskip -1.3in \frac{\partial K}{\partial\sigma^\alpha}=-\frac{2}{\cal Y}\left[\frac{3a}{2}(2\tau_\alpha + \mu_3l^2{\cal V}^{\frac{1}{18}} + ... -\gamma K_{\rm geom})\right]^{\frac{1}{2}}, \ \ \ \alpha \in\{ B,S\}.
\end{equation}

\begin{eqnarray}
\label{eq:singleder_G}
& & \frac{\partial K}{\partial {\cal G}^a}=-\frac{2}{{\cal Y}}\Biggl[-\frac{3a}{2}\frac{\left(2\tau_b + \mu_3l^2{\cal V}^{\frac{1}{18}} + ... -\gamma K_{\rm geom}\right)^{\frac{1}{2}}}{(\tau-{\bar\tau})}\kappa_{Bac}({\cal G}^c-{\bar{\cal G}}^c)\nonumber\\
& & +\frac{3a}{2}\frac{\left(2\tau_s + \mu_3l^2{\cal V}^{\frac{1}{18}} + ... -\gamma K_{\rm geom}\right)^{\frac{1}{2}}}{(\tau-{\bar\tau})}\kappa_{Sac}({\cal G}^c-{\bar{\cal G}}^c)\nonumber\\
& & + 4\sum_{\beta\in H_2^-(CY_3,{\bf Z})} n^0_\beta\sum_{m,n\in{\bf Z}^2/(0,0)}
\frac{({\bar\tau}-\tau)^{\frac{3}{2}}}{(2i)^{\frac{3}{2}}|m+n\tau|^3}sin\left(mk.{\cal B} + nk.c\right)\frac{\tau nk^a + mk^a}{(\tau-{\bar\tau}^a)}\Biggr]\nonumber\\
\end{eqnarray}
\begin{equation}
\label{eq:singleder_a}
\frac{\partial K}{\partial {\cal A}^I}=-\frac{2}{{\cal Y}}\left[\left(2\tau_b + \mu_3l^2{\cal V}^{\frac{1}{18}} + ... -\gamma K_{\rm geom}\right)^{\frac{1}{2}}.6i\kappa^4\mu_7(C_B)^{I{\bar K}}{\bar {\cal A}}_{\bar K} \right].
\end{equation}

{\it Double Derivatives}
\begin{eqnarray}
\label{eq:double_sigmasigma}
& & \frac{\partial^2K}{{\bar\partial}{\bar\sigma}^\alpha\partial\sigma^\alpha} =\frac{2}{{\cal Y}^2} \left[\frac{3a}{2}\sqrt{2\tau_\alpha + \mu_3l^2{\cal V}^{\frac{1}{18}} +..-\gamma K_{\rm geom}}\right]^2 -\frac{3a}{2{\cal Y}}\frac{1}{\sqrt{2\tau_\alpha + \mu_3l^2{\cal V}^{\frac{1}{18}} +..-\gamma K_{\rm geom}}}\nonumber\\
& & \hskip 1.5in \sim\frac{\mu_3l^2}{{\cal V}^{\frac{35}{36}}}.
\end{eqnarray}
\begin{eqnarray}
\label{eq:double_sigmaBsigmaS}
& & \frac{\partial^2K}{{\bar\partial}{\bar\sigma}^S\partial\sigma^B} =\frac{2}{{\cal Y}^2}
\left[\frac{3a}{2}\sqrt{2\tau_s + \mu_3l^2{\cal V}^{\frac{1}{18}}+..-\gamma K_{\rm geom}}\right]+ \left[\frac{3a}{2}\sqrt{2\tau_b + \mu_3l^2{\cal V}^{\frac{1}{18}}+..-\gamma K_{\rm geom}}\right]\nonumber\\
& & \hskip 1.5in \sim\frac{\mu_3l^2}{{\cal V}^{\frac{35}{36}}}.
\end{eqnarray}
\begin{eqnarray*}
& & \frac{\partial^2K}{{\bar\partial}{{\bar\sigma}^B}{\partial}{{\cal A}^I}}=\frac{2}{{\cal Y}^2}\left(2\tau_b + \mu_3l^2{\cal V}^{\frac{1}{18}} +..-\gamma K_{\rm geom}\right)^{\frac{1}{2}}
\Biggl[\frac{3a}{2}\left(2\tau_b + \mu_3l^2{\cal V}^{\frac{1}{18}}+..-\gamma K_{\rm geom}\right)^{\frac{1}{2}}\nonumber\\
& & \times 6i\kappa^2\mu_7(c_B)^{I{\bar J}}{\cal A}_{\bar J}\Biggr] -\frac{2}{{\cal Y}}\Biggl[\frac{3a}{4}\frac{6i\kappa^2\mu_7(c_B)^{I{\bar K}}{\bar {\cal A}}_{\bar K}}{\left(2\tau_b + \mu_3l^2{\cal V}^{\frac{1}{18}} + ... -\gamma K_{\rm geom}\right)^{\frac{1}{2}}}\Biggr] \sim\frac{{\cal V}^{\frac{5}{36}}}{\sqrt{\mu_3l^2}}{\cal A}_1;\nonumber\\
\end{eqnarray*}
\begin{eqnarray}
\label{eq:Double_BSa}
& & \ {\rm Similarly},\ \frac{\partial^2K}{{\bar\partial}{{\bar\sigma}^S}{\partial}{{\cal A}^I}}
\sim\frac{{\cal V}^{\frac{5}{36}}}{\sqrt{\mu_3l^2}}{\cal A}_1.
\end{eqnarray}
\begin{eqnarray}
\label{eq:Double_aa}
& & \frac{\partial^2K}{\partial{a^I}{\bar\partial}{\bar{\cal A}_I}}=-\frac{2}{{\cal Y}}\Biggl[\frac{3a}{4}\frac{(6i\kappa_4^2\mu_7)^2
(c_B)^{I{\bar K}}{\cal A}_{\bar K}(c_B)^{L{\bar J}}{\cal A}_L}{(2\tau_b + \mu_3l^2{\cal V}^{\frac{1}{18}} +..-\gamma K_{\rm geom})^{\frac{1}{2}}}
+ \frac{3a}{2}(2\tau_b + \mu_3l^2{\cal V}^{\frac{1}{18}}+..-\gamma K_{\rm geom})^{\frac{1}{2}}\nonumber\\
& & \times 6i\kappa^2\mu_7(c_B)^{I{\bar J}}\Biggr]  +\frac{2}{{\cal Y}^2}\Biggl[\frac{3a}{2}\left(2\tau_b + \mu_3l^2{\cal V}^{\frac{1}{18}} + ... -\gamma K_{\rm geom}\right)^{\frac{1}{2}}.6i\kappa^2\mu_7(c_B)^{I{\bar K}}{\bar a}_{\bar K}
\Biggr] \nonumber\\
& & \times\Biggl[\frac{3a}{2}\left(2\tau_b + \mu_3l^2{\cal V}^{\frac{1}{18}} + ... -\gamma K_{\rm geom}\right)^{\frac{1}{2}}.6i\kappa^2\mu_7(c_B)^{L{\bar J}}
{\cal A}_L \Biggr] \sim\frac{{\cal V}^{\frac{47}{36}}}{\sqrt{\mu_3 l^2}}|{\cal A}_1|^2
\end{eqnarray}
\begin{eqnarray}
\label{eq:double_GG}
& & \frac{\partial^2K}{\partial{{\cal G}^a}{\bar\partial}{{\cal G}^b}}
=\frac{2}{{\cal Y}^2}
\Biggl[-\frac{3a}{2}\frac{\sqrt{2\tau_b + \mu_3l^2{\cal V}^{\frac{1}{18}} + ... -\gamma K_{\rm geom}}}{(\tau-{\bar\tau})}
\kappa_{Bac}({\cal G}^c-{\bar{\cal G}}^c)\nonumber\\
& & +\frac{3a}{2}\frac{\sqrt{2\tau_s + \mu_3l^2{\cal V}^{\frac{1}{18}} + ... -\gamma K_{\rm geom}}}{(\tau-{\bar\tau})}
\kappa_{Sac}({\cal G}^c-{\bar{\cal G}}^c)\nonumber\\
& & - 4\sum_{\beta\in H_2^-(CY_3,{\bf Z})} n^0_\beta\sum_{m,n\in{\bf Z}^2/(0,0)}
\frac{({\bar\tau}-\tau)^{\frac{3}{2}}}{(2i)^{\frac{3}{2}}|m+n\tau|^3}sin\left(mk.{\cal B} + nk.c\right)
\frac{{\bar\tau} nk^a + mk^a}{(\tau-{\bar\tau})}\Biggr]\nonumber\\
& & \times\Biggl[-\frac{3a}{2}\frac{\sqrt{2\tau_b + \mu_3l^2{\cal V}^{\frac{1}{18}} + ... -\gamma K_{\rm geom}}}{(\tau-{\bar\tau})}
\kappa_{Bad}({\cal G}^d-{\bar{\cal G}}^d)\nonumber\\
& & +\frac{3a}{2}\frac{\sqrt{2\tau_s + \mu_3l^2{\cal V}^{\frac{1}{18}} + ... -\gamma K_{\rm geom}}}{(\tau-{\bar\tau})}
\kappa_{Sad}({\cal G}^d-{\bar{\cal G}}^d)\nonumber\\
& & - 4\sum_{\beta\in H_2^-(CY_3,{\bf Z})} n^0_\beta\sum_{m,n\in{\bf Z}^2/(0,0)}
\frac{({\bar\tau}-\tau)^{\frac{3}{2}}}{(2i)^{\frac{3}{2}}|m+n\tau|^3}sin\left(mk.{\cal B} + nk.c\right)
\frac{\tau nk^a + mk^a}{(\tau-{\bar\tau})}\Biggr]\nonumber\\
& & -\frac{2}{{\cal Y}}\Biggl[\frac{3a}{2}\frac{\sqrt{2\tau_b + \mu_3l^2{\cal V}^{\frac{1}{18}} + ... -\gamma K_{\rm geom}}}{(\tau-{\bar\tau})}\kappa_{Bab}
+ \frac{3a}{2}\frac{\sqrt{2\tau_s + \mu_3l^2{\cal V}^{\frac{1}{18}} + ... -\gamma K_{\rm geom}}}{(\tau-{\bar\tau})}
\kappa_{Sac}\nonumber\\
& & - 4\sum_{\beta\in H_2^-(CY_3,{\bf Z})} n^0_\beta\sum_{m,n\in{\bf Z}^2/(0,0)}
\frac{({\bar\tau}-\tau)^{\frac{3}{2}}}{(2i)^{\frac{3}{2}}|m+n\tau|^3}cos\left(mk.{\cal B} + nk.c\right)
\frac{{\bar\tau} nk^a + mk^a}{(\tau-{\bar\tau})}
\frac{{\bar\tau} nk^b + mk^b}{(\tau-{\bar\tau})}\Biggr]\nonumber\\
& & \sim
\frac{\sum n^0_\beta cos\left(mk.{\cal B} + nk.c\right)}{{\cal V}}
\end{eqnarray}
\begin{eqnarray}
\label{eq:Double_GBS}
& & \frac{\partial^2K}{\partial{\sigma^B}{\bar\partial}{{\cal G}^a}}
=\frac{3a}{{\cal Y}^2}\sqrt{2\tau_b + \mu_3l^2{\cal V}^{\frac{1}{18}} + ... -\gamma K_{\rm geom}}
\Biggl[-\frac{3a}{2}\frac{\sqrt{2\tau_b + \mu_3l^2{\cal V}^{\frac{1}{18}} + ... -\gamma K_{\rm geom}}}{(\tau-{\bar\tau})}\nonumber\\
& & \hskip 0.5in \kappa_{Bac}({\cal G}^c-{\bar{\cal G}}^c)
+\frac{3a}{2}\frac{\sqrt{2\tau_s + \mu_3l^2{\cal V}^{\frac{1}{18}} + ... -\gamma K_{\rm geom}}}{(\tau-{\bar\tau})}
\kappa_{Sac}({\cal G}^c-{\bar{\cal G}}^c)\nonumber\\
& & \hskip 0.5in - 4\sum_{\beta\in H_2^-(CY_3,{\bf Z})} n^0_\beta\sum_{m,n\in{\bf Z}^2/(0,0)}
\frac{({\bar\tau}-\tau)^{\frac{3}{2}}}{(2i)^{\frac{3}{2}}|m+n\tau|^3}sin\left(mk.{\cal B} + nk.c\right) \frac{{\bar\tau} nk^a + mk^a}{(\tau-{\bar\tau})}\Biggr]\nonumber\\
& & \hskip 0.5in + \frac{9ai}{2\sqrt{2\tau_b + \mu_3l^2{\cal V}^{\frac{1}{18}} + ... -\gamma K_{\rm geom}}}\kappa_{Bac}\left({\cal G}^a-{\bar{\cal G}}^a\right)
\sim\frac{{\cal V}^{-\frac{37}{36}}\kappa_{Bac}{\cal G}^c}{\sqrt{\mu_3l^2}}.
\end{eqnarray}


\begin{eqnarray}
\label{eq:Double_zz}
& & \frac{\partial^2K}{\partial z_i{\bar\partial} {\bar z}_{\bar j}}=
\frac{2}{{\cal Y}^2}\Biggl[\frac{3a}{2}\left(2\tau_b + \mu_3l^2{\cal V}^{\frac{1}{18}} + ... -\gamma K_{\rm geom}\right)^{\frac{1}{2}}\Biggl\{3i\mu_3\l^2
(\omega_B)_{i{\bar k}}{\bar z}^{\bar k}+\frac{3}{4}\mu_3l^2\bigl((\omega_B)_{i{\bar k}}{\bar z}^{\tilde{a}}\nonumber\\
& & ({\cal P}_{\tilde{a}})^{\bar k}_lz^l + (\omega_B)_{l{\bar k}}z^l{\bar z}^{\tilde{a}}({\cal P})^{\bar k}_i\bigr) -\gamma(ln {\cal V})^{-\frac{7}{12}}{\cal V}^{\frac{29}{36}}\Biggr\}
- \frac{3a}{2}\left(2\tau_s + \mu_3l^2{\cal V}^{\frac{1}{18}} + ... -\gamma K_{\rm geom}\right)^{\frac{1}{2}}\nonumber\\
& & \times \Biggl\{3i\mu_3\l^2(\omega_S)_{i{\bar k}}
{\bar z}^{\bar k}+\frac{3}{4}\mu_3l^2\left((\omega_S)_{i{\bar k}}{\bar z}^{\tilde{a}}({\cal P}_{\tilde{a}})^{\bar j}_lz^l
+ (\omega_S)_{l{\bar k}}z^l{\bar z}^{\tilde{a}}({\cal P})^{\bar k}_i\right) -\gamma(ln {\cal V})^{-\frac{7}{12}}{\cal V}^{\frac{29}{36}}\Biggr\} \Biggr]\nonumber\\
& & \times\Biggl[\frac{3a}{2}\left(2\tau_b + \mu_3l^2{\cal V}^{\frac{1}{18}} + ... -\gamma K_{\rm geom}\right)^{\frac{1}{2}}\Biggl\{-3i\mu_3\l^2
(\omega_B)_{k{\bar j}}z^k-\frac{3}{4}\mu_3l^2\bigl((\omega_B)_{k{\bar j}}z^{\tilde{a}}\nonumber\\
& & ({\cal P}_{\tilde{a}})^k_{\bar l}{\bar z}^l + (\omega_B)_{{\bar l}k}{\bar z}^{\bar l}z^{\tilde{a}}
({\cal P})^k_{\bar i}\bigr) -\gamma(ln {\cal V})^{-\frac{7}{12}}{\cal V}^{\frac{29}{36}}\Biggr\}
- \frac{3a}{2}\left(2\tau_s + \mu_3l^2{\cal V}^{\frac{1}{18}} + ... -\gamma K_{\rm geom}\right)^{\frac{1}{2}}\nonumber\\
& & \times \Biggl\{-3i\mu_3\l^2(\omega_S)_{k{\bar j}}
z^k-\frac{3}{4}\mu_3l^2\left((\omega_S)_{k{\bar j}}z^{\tilde{a}}({\cal P}_{\tilde{a}})^j_{\bar l}{\bar z}^{\bar l}
+ (\omega_S)_{k{\bar l}}{\bar z}^{\bar l}z^{\tilde{a}}({\cal P})^k_{\bar i}\right)-\gamma(ln {\cal V})^{-\frac{7}{12}}{\cal V}^{\frac{29}{36}}\Biggr\} \Biggr]\nonumber\\
& & -\frac{2}{{\cal Y}}\Biggl[\frac{3a}{2}\left(2\tau_b + \mu_3l^2{\cal V}^{\frac{1}{18}} + ... -\gamma K_{\rm geom}\right)^{\frac{1}{2}}\left\{
3i\mu_3l^2(\omega_B)_{i{\bar j}}-\gamma\left(ln {\cal V}\right)^{-\frac{7}{12}}{\cal V}^{\frac{5}{18}}\right\}\Biggr]\nonumber\\
& & \sim\frac{(\mu_3l^2)^3\left(\left\{\omega_B-\omega_S\right\}_{i{\bar j}}\xi^{\bar j}\right)^2}{{\cal V}^{\frac{17}{18}}}
\end{eqnarray}


\begin{eqnarray}
\label{eq:Double_az}
& & \frac{\partial^2K}{\partial{{\cal A}^I}{\bar\partial}{\bar z_i}}=\frac{2}{{\cal Y}^2}
\Biggl[\frac{3a}{2}\left(2\tau_b + \mu_3l^2{\cal V}^{\frac{1}{18}} + ... -\gamma K_{\rm geom}\right)^{\frac{1}{2}}\Biggl\{3i\mu_3\l^2
(\omega_B)_{i{\bar k}}{\bar z}^{\bar k}+\frac{3}{4}\mu_3l^2\bigl((\omega_B)_{i{\bar k}}{\bar z}^{\tilde{a}}\nonumber\\
& & ({\cal P}_{\tilde{a}})^{\bar k}_lz^l + (\omega_B)_{l{\bar k}}z^l{\bar z}^{\tilde{a}}({\cal P})^{\bar k}_i\bigr) -\gamma(ln {\cal V})^{-\frac{7}{12}}{\cal V}^{\frac{29}{36}}\Biggr\}
- \frac{3a}{2}\left(2\tau_s + \mu_3l^2{\cal V}^{\frac{1}{18}} + ... -\gamma K_{\rm geom}\right)^{\frac{1}{2}}\nonumber\\
& & \Biggl\{3i\mu_3\l^2(\omega_S)_{i{\bar k}}
{\bar z}^{\bar k}+\frac{3}{4}\mu_3l^2\left((\omega_S)_{i{\bar k}}{\bar z}^{\tilde{a}}({\cal P}_{\tilde{a}})^{\bar j}_lz^l
+ (\omega_S)_{l{\bar k}}z^l{\bar z}^{\tilde{a}}({\cal P})^{\bar k}_i\right) -\gamma(ln {\cal V})^{-\frac{7}{12}}{\cal V}^{\frac{29}{36}}\Biggr\} \Biggr]\nonumber\\
& & \hskip-0.3cm -\Biggl[\frac{3a}{2}
\frac{\Biggl\{3i\mu_3\l^2
(\omega_B)_{i{\bar k}}{\bar z}^{\bar k}+\frac{3}{4}\mu_3l^2\left((\omega_B)_{i{\bar k}}{\bar z}^{\tilde{a}}
({\cal P}_{\tilde{a}})^{\bar k}_lz^l + (\omega_B)_{l{\bar k}}z^l{\bar z}^{\tilde{a}}({\cal P})^{\bar k}_i\right)
-\gamma(ln {\cal V})^{-\frac{7}{12}}{\cal V}^{\frac{29}{36}}\Biggr\}}{\left(2\tau_b + \mu_3l^2{\cal V}^{\frac{1}{18}} + ... -\gamma K_{\rm geom}\right)^{\frac{1}{2}}}\Biggr]\nonumber\\
& & \times \Biggl[\frac{2}{{\cal Y}}6i\kappa_4^2\mu_7(c_B)^{I{\bar J}}{\bar {\cal A}}_{\bar J}\Biggr]\sim{\cal V}^{\frac{1}{6}}\sqrt{\mu_3 l^2}\left\{\omega_B-\omega_S\right\}_{i{\bar j}}\xi^{\bar j}{\cal A}_I
\end{eqnarray}


\begin{eqnarray}
\label{eq:Double_Gz}
& & \frac{\partial^2K}{\partial{\bar z}^{\bar i}{\bar\partial}{{\cal G}^a}}=
 \frac{2}{{\cal Y}^2}\left[\sqrt{2\tau_b + \mu_3l^2{\cal V}^{\frac{1}{18}} + ... -\gamma K_{\rm geom}}
\left\{\mu_3l^2{\cal V}^{\frac{1}{18}}\left\{\omega_B-\omega_S\right\}_{i{\bar j}}\xi^{\bar j}-\gamma\left(ln {\cal V}\right)^{-\frac{7}{12}}{\cal V}^{\frac{29}{36}}\right\}\right]\nonumber\\
& & \Biggl[-\frac{3a}{2}\frac{\sqrt{2\tau_b + \mu_3l^2{\cal V}^{\frac{1}{18}} + ... -\gamma K_{\rm geom}}}{(\tau-{\bar\tau})}
\kappa_{Bac}({\cal G}^c-{\bar{\cal G}}^c)
+\frac{3a}{2}\frac{\sqrt{2\tau_s + \mu_3l^2{\cal V}^{\frac{1}{18}} + ... -\gamma K_{\rm geom}}}{(\tau-{\bar\tau})}\nonumber\\
& & \kappa_{Sac}({\cal G}^c-{\bar{\cal G}}^c) - 4\sum_{\beta\in H_2^-(CY_3,{\bf Z})} n^0_\beta\sum_{m,n\in{\bf Z}^2/(0,0)} \frac{({\bar\tau}-\tau)^{\frac{3}{2}}}{(2i)^{\frac{3}{2}}|m+n\tau|^3}sin\left(mk.{\cal B} + nk.c\right)
\frac{{\bar\tau} nk^a + mk^a}{(\tau-{\bar\tau})}\Biggr]\nonumber\\
& & -\frac{2}{{\cal Y}}\Biggl[-\frac{3a}
{2(\tau-{\bar\tau})}\left(\frac{\kappa_{Bac}}{\sqrt{2\tau_b + \mu_3l^2{\cal V}^{\frac{1}{18}} + ... -\gamma K_{\rm geom}}}-\frac{\kappa_{Sac}}
{\sqrt{2\tau_s + \mu_3l^2{\cal V}^{\frac{1}{18}} + ... -\gamma K_{\rm geom}}}
\right)\Biggr]\nonumber\\
& & \sim\frac{\left\{\omega_B-\omega_S\right\}_{i{\bar j}}\xi^{\bar j}\sum_\beta k^a n^0_\beta sin(...)(\mu_3l^2)^{\frac{3}{2}}}{{\cal V}^{\frac{35}{18}}}
\end{eqnarray}
\begin{eqnarray}
\label{eq:Double_aG}
& & \frac{\partial^2K}{\partial {\cal A}^I{\bar\partial}{{\cal G}^a}}=-\frac{2}{\cal Y}
\Biggl[\frac{6i\kappa^2\mu_7(c_B)^{I{\bar K}}}{\sqrt{2\tau_b + \mu_3l^2{\cal V}^{\frac{1}{18}} + ... -\gamma K_{\rm geom}}}
\left(-\frac{3i}{(\tau-{\bar\tau})}
\kappa_{Bac}({\cal G}^c-{\bar G}^{\bar c})\right)\Biggr]\nonumber\\
& & +\frac{2}{{\cal Y}}\Biggl[-\frac{3a}{2}\frac{\sqrt{2\tau_b + \mu_3l^2{\cal V}^{\frac{1}{18}} + ... -\gamma K_{\rm geom}}}{(\tau-{\bar\tau})}
\kappa_{Bac}({\cal G}^c-{\bar{\cal G}}^c)\nonumber\\
& & +\frac{3a}{2}\frac{\sqrt{2\tau_s + \mu_3l^2{\cal V}^{\frac{1}{18}} + ... -\gamma K_{\rm geom}}}{(\tau-{\bar\tau})}
\kappa_{Sac}({\cal G}^c-{\bar{\cal G}}^c)\nonumber\\
& & - 4\sum_{\beta\in H_2^-(CY_3,{\bf Z})} n^0_\beta\sum_{m,n\in{\bf Z}^2/(0,0)}
\frac{({\bar\tau}-\tau)^{\frac{3}{2}}}{(2i)^{\frac{3}{2}}|m+n\tau|^3}sin\left(mk.{\cal B} + nk.c\right)
\frac{\tau nk^a + mk^a}{(\tau-{\bar\tau})}\Biggr]\nonumber\\
& & \times\Biggl[\frac{3a}{2}\left(2\tau_b + \mu_3l^2{\cal V}^{\frac{1}{18}} + ... -\gamma K_{\rm geom}\right)^{\frac{1}{2}}
.6i\kappa^2\mu_7(c_B)^{I{\bar K}}
{\bar {\cal A}}_{\bar K}\Biggr]\sim\frac{{\cal V}^{\frac{5}{36}}\kappa_{Bab}{\cal G}^b}{\sqrt{\mu_3l^2}}{\cal A}_1.\nonumber\\
\end{eqnarray}
\nonumber\\
Similar to (\ref{eq:Double_GBS}) $$,\ \frac{\partial^2K}{\partial{\sigma^S}{\bar\partial}{{\cal G}^a}}\sim\frac{{\cal V}^{-\frac{37}{36}}\kappa_{Sac}{\cal G}^c}{\sqrt{\mu_3l^2}}.$$

\begin{eqnarray}
\label{eq:Double_BSz}
& &  \frac{\partial^2K}{{\bar\partial}{\bar z_i}\partial\sigma^\alpha}=
\frac{2}{{\cal Y}^2}\frac{3a}{2}\left(2\tau_b + \mu_3l^2{\cal V}^{\frac{1}{18}} + ... -\gamma K_{\rm geom}\right)^{\frac{1}{2}} \Biggl[
\frac{3a}{2}\left(2\tau_b + \mu_3l^2{\cal V}^{\frac{1}{18}} + ... -\gamma K_{\rm geom}\right)^{\frac{1}{2}}\nonumber\\
& & \times \Biggl\{3i\mu_3\l^2
(\omega_B)_{i{\bar k}}{\bar z}^{\bar k}+\frac{3}{4}\mu_3l^2\left((\omega_B)_{i{\bar k}}{\bar z}^{\tilde{a}}
({\cal P}_{\tilde{a}})^{\bar k}_lz^l + (\omega_B)_{l{\bar k}}z^l{\bar z}^{\tilde{a}}({\cal P})^{\bar k}_i\right)-\gamma\left(ln {\cal V}
\right)^{-\frac{7}{12}}{\cal V}^{\frac{29}{36}}\Biggr\}\nonumber\\
 & &  - \frac{3a}{2}\left(2\tau_s + \mu_3l^2{\cal V}^{\frac{1}{18}} + ... -\gamma K_{\rm geom}\right)^{\frac{1}{2}}
\Biggl\{3i\mu_3\l^2(\omega_S)_{i{\bar k}}
{\bar z}^{\bar k}+\frac{3}{4}\mu_3l^2\bigl((\omega_S)_{i{\bar k}}{\bar z}^{\tilde{a}}({\cal P}_{\tilde{a}})^{\bar j}_lz^l\nonumber\\
& & + (\omega_S)_{l{\bar k}}z^l{\bar z}^{\tilde{a}}({\cal P})^{\bar k}_i\bigr)-\gamma\left(ln {\cal V}
\right)^{-\frac{7}{12}}{\cal V}^{\frac{29}{36}}\Biggr\}\Biggr]\nonumber\\
& & -\frac{3a}{2{\cal Y}}\frac{\Biggl\{3i\mu_3\l^2
(\omega_\alpha)_{i{\bar k}}{\bar z}^{\bar k}+\frac{3}{4}\mu_3l^2\left((\omega_\alpha)_{i{\bar k}}{\bar z}^{\tilde{a}}
({\cal P}_{\tilde{a}})^{\bar k}_lz^l + (\omega_\alpha)_{l{\bar k}}z^l{\bar z}^{\tilde{a}}({\cal P})^{\bar k}_i\right)-\gamma\left(ln {\cal V}
\right)^{-\frac{7}{12}}{\cal V}^{\frac{29}{36}}\Biggr\}}{\left(2\tau_\alpha + \mu_3l^2{\cal V}^{\frac{1}{18}} + ... -\gamma K_{\rm geom}\right)^{\frac{1}{2}}}\nonumber\\
& & \sim\left\{\omega_B-\omega_S\right\}_{i{\bar j}}\xi^{\bar j}\frac{\mu_3l^2}{{\cal V}}.
\end{eqnarray}


Hence, the combined closed- and open-string (matter field) moduli-space metric (in large volume limit) is given as under:

\begin{equation}
\label{eq:G_wilson}
G_{A{\bar B}}\sim\left(
\begin{array}{lllllll}
 \frac{1}{{\cal V}^{35/36}} & \frac{1}{{\cal V}^{35/36}} & \frac{{A_{\sigma^B{\cal G}^1}}}{{\cal V}^{37/36}} & \frac{{A_{\sigma^B{\cal G}^2}}}{{\cal V}^{37/36}} &
   \frac{{A_{\sigma^Bz_1}}}{{\cal V}} & \frac{{A_{\sigma^Bz_2}}}{{\cal V}} & {A_{\sigma^\alpha{\cal A}_1}} {{\cal A}_1} {\cal V}^{5/36} \\
 \frac{1}{{\cal V}^{35/36}} & \frac{1}{{\cal V}^{35/36}} & \frac{{A_{\sigma^S{\cal G}^1}}}{{\cal V}^{37/36}} & \frac{{A_{\sigma^S{\cal G}^2}}}{{\cal V}^{37/36}} &
   \frac{{A_{\sigma^Sz_1}}}{{\cal V}} & \frac{{A_{\sigma^Sz_2}}}{{\cal V}} & {A_{\sigma^\alpha{\cal A}_1}} {{\cal A}_1} {\cal V}^{5/36} \\
 \frac{{A_{\sigma^B{\cal G}^1}}}{{\cal V}^{37/36}} & \frac{{A_{\sigma^S{\cal G}^1}}}{{\cal V}^{37/36}} & {A_{{\cal G}^1{\cal G}^1}} & {A_{{\cal G}^1{\cal G}^2}} & \frac{{A_{{\cal G}^1z_1}}}{{\cal V}^{10/9}} &
   \frac{{A_{{\cal G}^1z_2}}}{{\cal V}^{10/9}} & {A_{{\cal G}^1{\cal A}_1}} {{\cal A}_1} {\cal V}^{5/36} \\
 \frac{{A_{\sigma^B{\cal G}^2}}}{{\cal V}^{37/36}} & \frac{{A_{\sigma^S{\cal G}^2}}}{{\cal V}^{37/36}} & {A_{{\cal G}^1{\cal G}^2}} & {A_{{\cal G}^2{\cal G}^2}} & \frac{{A_{{\cal G}^2z_1}}}{{\cal V}^{10/9}} &
   \frac{{A_{{\cal G}^2z_2}}}{{\cal V}^{10/9}} & {A_{{\cal G}^2{\cal A}_1}} {{\cal A}_1} {\cal V}^{5/36} \\
 \frac{{A_{\sigma^Bz_1}}}{{\cal V}} & \frac{{A_{\sigma^Sz_1}}}{{\cal V}} & \frac{{A_{{\cal G}^1z_1}}}{{\cal V}^{10/9}} & \frac{{A_{{\cal G}^2z_1}}}{{\cal V}^{10/9}} &
   \frac{{A_{z_1z_1}}}{{\cal V}^{17/18}} & \frac{{A_{z_1z_2}}}{{\cal V}^{17/18}} & {A_{z_1{\cal A}_1}} {{\cal A}_1} \sqrt[6]{{\cal V}} \\
 \frac{{A_{\sigma^Bz_2}}}{{\cal V}} & \frac{{A_{\sigma^Sz_2}}}{{\cal V}} & \frac{{A_{{\cal G}^1z_2}}}{{\cal V}^{10/9}} & \frac{{A_{{\cal G}^2z_2}}}{{\cal V}^{10/9}} &
   \frac{{A_{z_1z_2}}}{{\cal V}^{17/18}} & \frac{{A_{z_2z_2}}}{{\cal V}^{17/18}} & {A_{z_2{\cal A}_1}} {{\cal A}_1} \sqrt[6]{{\cal V}} \\
 \frac{{A_{\sigma^\alpha{\cal A}_1}} {{\cal A}_1}}{ {\cal V}^{-5/36}} &\frac{{A_{\sigma^\alpha{\cal A}_1}} {{\cal A}_1}}{ {\cal V}^{-5/36}} & \frac{{A_{{\cal G}^1{\cal A}_1}} {{\cal A}_1}} {{\cal V}^{-5/36}} & \frac{{A_{{\cal G}^2{\cal A}_1}} {{\cal A}_1}}{ {\cal V}^{-5/36}} & \frac{{A_{z_1{\cal A}_1}} {{\cal A}_1}} {{\cal V}^{-1/6}} & \frac{{A_{z_2{\cal A}_1}} {{\cal A}_1}} {{{\cal V}}^{-1/6}} & \frac{{{\cal A}_1}^2}{{\cal V}^{-47/36}}
\end{array}
\right)
\end{equation}

We have assumed that the holomorphic, isometric involution $\sigma$ is such that
\begin{equation}
\frac{\sum_{\beta\in H_2^-(CY_3,{\bf Z})}n^0_\beta sin(...)}{{\cal V}^{\frac{1}{3}}}\sim{\cal V}^k,\nonumber\\
\end{equation} where $k\left(\in\left(0,\frac{2}{3}\right)\right)=\frac{1}{2}$ and $\sum_\beta n^0_\beta cos(...)\sim{\cal V}$. The components of $(G^{-1})^{A{\bar B}}$ are given as follows:
\begin{eqnarray}
\label{eq:inv_comps}
& & (G^{-1})^{\sigma^\alpha{\bar\sigma^\beta}}\sim{\cal V}^{\frac{19}{18}};\nonumber\\
& & (G^{-1})^{\sigma^\alpha{\bar z}^{\bar i}}\sim{\cal V};\nonumber\\
  & & (G^{-1})^{\sigma^\alpha{\bar G}^a}\sim{\cal V}^{\frac{1}{36}};\nonumber\\
  & & (G^{-1})^{\sigma^\alpha{\bar {\cal A}}_1}\sim\frac{{\cal V}^{\frac{5}{36}}}{{\cal A}_1};\nonumber\\
   & & (G^{-1})^{{\cal G}^a{\bar{\cal G}^b}}\sim{\cal V}^0;\nonumber\\
   & & (G^{-1})^{{\cal G}^a{\bar z}^{\bar i}}\sim{\cal V}^{-\frac{1}{36}};\nonumber\\
   & & (G^{-1})^{{\cal G}^a{\bar {\cal A}}_1}\sim \frac{{\cal V}^{-\frac{7}{6}}}{{\cal A}_1};\nonumber\\
   & & (G^{-1})^{z^i{\bar z^{\bar j}}}\sim{\cal V}^{\frac{17}{18}};\nonumber\\& & (G^{-1})^{z^i{\bar {\cal A}^1}}\sim \frac{{\cal V}^{-\frac{7}{36}}}{{\cal A}_1};\nonumber\\
   & & (G^{-1})^{{\cal A}_1{\bar {\cal A}_1}}\sim\frac{{\cal V}^{-\frac{47}{36}}}{{\cal A}_1^2}.
\end{eqnarray}

Now, restricted to $\Sigma_B$, using (\ref{eq:inv_comps}), (\ref{eq:W}), assuming that the complexified Wilson line moduli can be stabilized around ${\cal A}_1\sim{\cal V}^{-\frac{1}{4}}$ and:
\begin{eqnarray}
\label{eq:dersKW}
& & \partial_{\sigma^\alpha}K\sim\sqrt{\mu_3l^2}{\cal V}^{-\frac{35}{36}},\ \partial_{\sigma^B}W\sim0,\ \partial_{\sigma^S}W\sim n^sW;\nonumber\\
& & \partial_{{\cal G}^a}K\sim\frac{\sum_\beta n^0_\beta sin(...)}{\cal V}\sim{\cal V}^{-\frac{1}{6}},\
\partial_{{\cal G}^a}W\sim n^s(m_a+\frac{{\cal G}^a}{ln{\cal V}})W;\nonumber\\
& & \partial_{z^i}K|_{D_5}\sim \frac{(\mu_3l^2)^{\frac{3}{2}}\left\{\omega_B-\omega_S\right\}_{i{\bar j}}\xi^{\bar j}}{{\cal V}^{\frac{17}{18}}},\ \partial_{z^i}W|_{D_5}\sim \mu_3l^2\left\{\omega_S\right\}_{i{\bar j}}{\cal V}^{\frac{1}{36}}W;\nonumber\\
& & \partial_{{\cal A}_1}K\sim\frac{\sqrt{\mu_3l^2}(i\kappa_4^2\mu_7C_{1{\bar 1}})}{\cal V}{\cal A}_1\sim{\cal V}^{\frac{7}{36}}{\cal A}_1,\ \partial_{{\cal A}_1}W\sim0,
\end{eqnarray}
one obtains the following F-terms:
\begin{eqnarray}
\label{eq:F_terms}
& & e^KG^{\sigma^\alpha{\bar\sigma^{\bar\alpha}}}D_{\sigma^\alpha}W{\bar D}_{\bar\sigma^{\bar\alpha}}{\bar W}\sim\frac{(n^s)^2|W|^2{\cal V}^{\frac{19}{18}}}{{\cal V}^2}\equiv{\rm most\ dominant}\sim V_0(\equiv{\rm extremum\ value});\nonumber\\
& & e^KG^{{\cal G}^a{\bar{\cal G}^b}}D_{{\cal G}^a}WD_{\bar{\cal G}^b}{\bar W}\sim\frac{(n^s)^2m_am_b|W|^2}{{\cal V}^2};\nonumber\\
& & e^KG^{\sigma^\alpha{\bar{\cal G}^a}}D_{\sigma^\alpha}WD_{\bar{\cal G}^a}{\bar W}
\sim\frac{(n^s)^2m_a|W|^2{\cal V}^{\frac{1}{36}}}{{\cal V}^2};\nonumber\\
& & e^KG^{\sigma^\alpha{\bar z^{\bar i}}}D_{\sigma^\alpha}WD_{\bar z^{\bar i}}{\bar W}\sim
\frac{|W|^2n^s\mu_3l^2\left(\omega_\alpha\right)_{i{\bar j}}\xi^{\bar j}{\cal V}^{\frac{37}{36}}}{{\cal V}^2};\nonumber\\
& & e^KG^{\sigma^\alpha{\bar {\cal A}_1}}D_{\sigma^\alpha}WD_{\bar {\cal A}_1}{\bar W}\sim\frac{n^s|W|^2{\cal V}^{\frac{1}{18}}{\cal A}_1}{{\cal V}^2{\cal A}_1},\nonumber\\
& & e^KG^{{\cal G}^a{\bar z^{\bar i}}}D_{{\cal G}^a}WD_{\bar z^i}{\bar W}\sim
\frac{|W|^2n^s\left(m^a+\frac{{\cal G}^a}{ln {\cal V}}\right)\mu_3l^2\left(\omega_S\right)_{i{\bar j}}\xi^{\bar j}}{{\cal V}^2};\nonumber\\
& & e^KG^{{\cal G}^a{\bar {\cal A}_1}}D_{{\cal G}^a}WD_{\bar {\cal A}_1}{\bar W}\sim
\frac{n^sm^a|W|^2{\cal V}^{-\frac{35}{36}}{\cal A}_1}{{\cal V}^2{\cal A}_1},\nonumber\\
& & e^KG^{z^i{\bar z^j}}D_{z^i}WD_{\bar z^j}{\bar W}\sim\frac{|W|^2{\cal V}(\mu_3l^2)^2\left(\omega_S\right)_{i{\bar k}}\xi^{\bar k}\left(\omega_S\right)_{{\bar j}l}\xi^l}{{\cal V}^2},\nonumber\\
& & e^KG^{z^i{\bar {\cal A}_1}}D_{z^i}WD_{\bar {\cal A}_1}{\bar W}\sim
\frac{|W|^2\mu_3l^2{\cal V}^{\frac{1}{36}}\left(\omega_S\right)_{i{\bar j}}\xi^{\bar j}{\cal A}_1}{{\cal V}^2{\cal A}_1},\nonumber\\
& & e^KG^{{\cal A}_1{\bar {\cal A}_1}}D_{{\cal A}_1}WD_{\bar {\cal A}_1}{\bar W}\sim
\frac{{\cal V}^{-\frac{11}{12}}|{\cal A}_1|^2}{{\cal V}^2|{\cal A}_1|^2}.
\end{eqnarray}
We thus see the independence of the ${\cal N}=1$ potential in the LVS limit in a self-consistent way on ${\cal A}_1$ assuming it to be around ${\cal V}^{-\frac{1}{4}}$. This justifies our assumption that one can take the Wilson line moduli to be stabilized around ${\cal V}^{-\frac{1}{4}}$; we hence do get a competing contribution of the order of the volume of $\Sigma_B$ in $T_B$, which would hence guarantee  ${\cal O}(1)$
Yang-Mills coupling constant corresponding to the non-abelian gauge theory living on a stack of $D7$-branes wrapping $\Sigma_B$. Note that ${\cal A}_I=0$ is also an allowed extremum, which is in conformity with switching off of all but one Wilson line moduli for our analysis.

\section{Derivatives of $K|_{D_5}$ and $K|_{D_4}$}
\setcounter{equation}{0}
\seceqdd

One needs the first and second derivatives of the geometric K\"{a}hler potential with respect to the position moduli of the mobile $D3$ brane, restricted for convenience, to $D_5$. We also give, for completeness and for future work, the same for the geometric K\"{a}hler potential restricted to $D_4$.

The first and (mixed) second order derivatives of $K|_{D_5}$ are as follows:\begin{itemize}
\item
$\partial_{z_1}K|_{D_5}=
\frac{-3\,{r2}\,}{3\,\left( z_3^3 \right) }\left( {{z_1}}^{17} - \phi\,{{z_1}}^5\,{{z_2}}^6 \right)  +
    4\,{\left( \frac{\zeta}
         {r_1\,|z_3|^2} \right) }^{\frac{1}{6}}\,
     ( \left( -2\,\phi\,{{z_1}}^6\,{{z_2}}^6 + {{z_2}}^{18} \right) \,
        {\bar z_1} $

        $- {{z_1}}^5\,
        \left( {{z_1}}^{12} - \phi\,{{z_2}}^6 \right) \,
        \left( 1 + |{z_2}|^2 \right) ) + 3\,r_1\,{\left( \frac{\zeta}
           {r_1\,|z_3|^2} \right) }^{\frac{1}{6}}$

        $ \times     \frac{\left( 4\,{{z_1}}^{17} - 4\,\phi\,{{z_1}}^5\,{{z_2}}^6 +
         3\,{{z_1}}^{18}\,{\bar z_1} -
         \phi\,{{z_1}}^6\,{{z_2}}^6\,{\bar z_1} -
         {{z_2}}^{18}\,{\bar z_1} +
         4\,{{z_1}}^{17}\,{z_2}\,{\bar z_2} -
         4\,\phi\,{{z_1}}^5\,{{z_2}}^7\,{\bar z_2} -
         \frac{3\,r_2\,{{z_1}}^{17}}
          {{\left( \frac{\zeta}
               {r_1\,|z_3|^2} \right) }^{\frac{1}{6}}} +
         \frac{3\,\phi\,r_2\,{{z_1}}^5\,{{z_2}}^6}
          {{\left( \frac{\zeta}
               {r_1\,|z_3|^2} \right) }^{\frac{1}{6}}} \right) }{-
         r_2 + {\left( \frac{\zeta}
           {r_1\,|z_3|^2} \right) }^{\frac{1}{6}} +
       {z_1}\,{\bar z_1}\,
        {\left( \frac{\zeta}
            {r_1\,|z_3|^2} \right) }^{\frac{1}{6}} +
       {z_2}\,{\bar z_2}\,
        {\left( \frac{\zeta}
            {r_1|z_3|^2} \right) }^{\frac{1}{6}}}$

            $ +
    \frac{2\,\zeta\,{\left( \frac{\zeta}
           {r_1\,|z_3|^2} \right) }^{\frac{1}{6}}\,
       \left( \left( 2\,\phi\,{{z_1}}^6\,{{z_2}}^6 - {{z_2}}^{18} \right) \,
          {\bar z_1} + {{z_1}}^5\,
          \left( {{z_1}}^{12} - \phi\,{{z_2}}^6 \right) \,
          \left( 1 + |{z_2}|^2 \right)  \right) \,
       \left( r_2 - \left( 1 + |{z_1}|^2 +
            |{z_2}|^2 \right) \,
          {\left( \frac{\zeta}
              {r_1\,|z_3|^2} \right) }^{\frac{1}{6}} \right) }{
       r_1}$

$\sim {\cal V}^{\frac{11}{36}}
+ \left(ln {\cal V}\right)^{-\frac{1}{12}}{\cal V}^{\frac{17}{36}}
+ \sqrt{ln {\cal V}}{\cal V}^{\frac{17}{36}}
+ \left(ln {\cal V}\right)^{-\frac{7}{12}}{\cal V}^{\frac{29}{36}}
 \sim \left(ln {\cal V}\right)^{-\frac{7}{12}}{\cal V}^{\frac{29}{36}}$

\item
$\partial_{z_2}K_{\rm geom}=\partial_{z_1}K_{\rm geom}(z_1\leftrightarrow z_2)\sim \left(ln {\cal V}\right)^{-\frac{7}{12}}{\cal V}^{\frac{29}{36}}$

\item
$\partial_{z_1}{\bar\partial_{\bar z_1}}K_{\rm geom}=\frac{-1}{3\,
    \left( z_3^3 \right) }{\left( \frac{\zeta}{r_1\,|z_3|^2} \right) }^{\frac{1}{6}}\,
    \Biggl[ 4\,\left( -2\,\phi\,{{z_1}}^6\,{{z_2}}^6 + {{z_2}}^{18} \right) + 4\,\left( {{\bar z_1}}^{17} -
           \phi\,{{\bar z_1}}^5\,{{\bar z_2}}^6 \right) \,$

    $ \times  \frac{
         \left( \left( 2\,\phi\,{{z_1}}^6\,{{z_2}}^6 - {{z_2}}^{18} \right) \,
            {\bar z_1} + {{z_1}}^5\,
            \left( {{z_1}}^{12} - \phi\,{{z_2}}^6 \right) \,
            \left( 1 + |{z_2}|^2 \right)  \right) }{-3z_3^3} + 2\,\zeta\,{\left( \frac{\zeta}
             {r_1\,|z_3|^2} \right) }^{\frac{1}{6}}\,$

            $ \times
      \frac{  \left( ( 2\phi{{z_1}}^6{{z_2}}^6 - {{z_2}}^{18} ){\bar z_1} + {{z_1}}^5 ( {{z_1}}^{12} - \phi{{z_2}}^6 )
            ( 1 + |{z_2}|^2 )\right)
         \left( 2\,\phi\,{z_1}\,{{\bar z_1}}^6\,{{\bar z_2}}^6 -
           {z_1}\,{{\bar z_2}}^{18} +
           {{\bar z_1}}^{17}\,
            \left( 1 + |{z_2}|^2 \right)  -
           \phi\,{{\bar z_1}}^5\,{{\bar z_2}}^6\,
            \left( 1 + |{z_2}|^2 \right)  \right) }{r_1\,
         \left( {\bar z_3}^3 \right) }$

         $ -
      \frac{ \left( 4\,{{z_1}}^{17} - 4\,\phi\,{{z_1}}^5\,{{z_2}}^6 +
           3\,{{z_1}}^{18}\,{\bar z_1} -
           \phi\,{{z_1}}^6\,{{z_2}}^6\,{\bar z_1} -
           {{z_2}}^{18}\,{\bar z_1} +
           4\,{{z_1}}^{17}\,{z_2}\,{\bar z_2} -
           4\,\phi\,{{z_1}}^5\,{{z_2}}^7\,{\bar z_2} -
           \frac{3\,r_2\,{{z_1}}^{17}}
            {{\left( \frac{\zeta}
                 {r_1\,|z_3|^2} \right) }^{\frac{1}{6}}} +
           \frac{3\,\phi\,r_2\,{{z_1}}^5\,{{z_2}}^6}
            {{\left( \frac{\zeta}
                 {r_1\,|z_3|^2} \right) }^{\frac{1}{6}}} \right) }
         {\left( -{\bar z_3}^3 \right) \,
         \left( -r_2 + {\left( \frac{\zeta}
               {r_1\,|z_3|^2} \right) }^{\frac{1}{6}} +
           |{z_1}|^2\,
            {\left( \frac{\zeta}
                {r_1\,|z_3|^2} \right) }^{\frac{1}{6}} +
           |{z_2}|^2\,
            {\left( \frac{\zeta}
                {r_1\,|z_3|^2} \right) }^{\frac{1}{6}} \right) } $

                $\times\left( 3\,r_1\,{{\bar z_1}}^5\,
         \left( {{\bar z_1}}^{12} - \phi\,{{\bar z_2}}^6 \right)\right) \, -
      \frac{3\,r_1\,\left( 2\,\phi\,{z_1}\,{{\bar z_1}}^6\,
            {{\bar z_2}}^6 - {z_1}\,{{\bar z_2}}^{18} +
           {{\bar z_1}}^{17}\,
            \left( 1 + |{z_2}|^2 \right)  -
           \phi\,{{\bar z_1}}^5\,{{\bar z_2}}^6\,
            \left( 1 + |{z_2}|^2 \right)  \right) \,
         \left( \Sigma_1 \right) }{
         \left(-{\bar z_3}^3 \right) \,
         {\left( -r_2 + {\left( \frac{\zeta}
                 {r_1\,|z_3|^2} \right) }^{\frac{1}{6}} +
             |{z_1}|^2\,
              {\left( \frac{\zeta}
                  {r_1\,|z_3|^2} \right) }^{\frac{1}{6}} +
             |{z_2}|^2\,
              {\left( \frac{\zeta}
                  {r_1\,|z_3|^2} \right) }^{\frac{1}{6}} \right) }^2} $

                  $+
       \frac{2\,\left( 2\,\phi\,{{z_1}}^6\,{{z_2}}^6 - {{z_2}}^{18} \right) \,\zeta\,
         \left( r_2 - \left( 1 + |{z_1}|^2 +
              |{z_2}|^2 \right) \,
            {\left( \frac{\zeta}
                {r_1\,|z_3|^2} \right) }^{\frac{1}{6}} \right) }{
         r_1}$

         $ - \frac{2\,\zeta\,{{\bar z_1}}^5\,
         \left( {{\bar z_1}}^{12} - \phi\,{{\bar z_2}}^6 \right) \,
         \left( \left( 2\,\phi\,{{z_1}}^6\,{{z_2}}^6 - {{z_2}}^{18} \right) \,
            {\bar z_1} + {{z_1}}^5\,
            \left( {{z_1}}^{12} - \phi\,{{z_2}}^6 \right) \,
            \left( 1 + |{z_2}|^2 \right)  \right) \,
         \left( r_2 - \left( 1 + |{z_1}|^2 +
              |{z_2}|^2 \right) \,
            {\left( \frac{\zeta}
                {r_1\,|z_3|^2} \right) }^{\frac{1}{6}} \right) }{
         r_1\,\left( -{\bar z_3}^3\right) }$

         $ +
      \frac{3\,r_1\,\left( 3\,{{z_1}}^{18} - \phi\,{{z_1}}^6\,{{z_2}}^6 -
           {{z_2}}^{18} - \frac{3\,r_2\,{{z_1}}^{17}\,
              \left( {{\bar z_1}}^{17} -
                \phi\,{{\bar z_1}}^5\,{{\bar z_2}}^6 \right) }{
              {\left( \frac{\zeta}
                  {r_1\,|z_3|^2} \right) }^{\frac{1}{6}}\,
              \left( -{\bar z_3}^3\right) } -
           \frac{3\,\phi\,r_2\,{{z_1}}^5\,{{z_2}}^6\,
              \left( -{{\bar z_1}}^{17} +
                \phi\,{{\bar z_1}}^5\,{{\bar z_2}}^6 \right) }{
              {\left( \frac{\zeta}
                  {r_1\,|z_3|^2} \right) }^{\frac{1}{6}}\,
              \left( {{\bar z_1}}^{18} -
                3\,\phi\,{{\bar z_1}}^6\,{{\bar z_2}}^6 +
                {{\bar z_2}}^{18} \right) } \right) }{-r_2 +
         {\left( \frac{\zeta}
             {r_1\,|z_3|^2} \right) }^{\frac{1}{6}} +
         |{z_1}|^2\,
          {\left( \frac{\zeta}
              {r_1\,|z_3|^2} \right) }^{\frac{1}{6}} +
         |{z_2}|^2\,
          {\left( \frac{\zeta}
              {r_1\,|z_3|^2} \right) }^{\frac{1}{6}}} \Biggr] $

$\sim \left(ln {\cal V}\right)^{-\frac{1}{12}}{\cal V}^{-\frac{5}{9}}\Biggl(
\sqrt{\cal V}
+ \sqrt{\cal V}
+ \left(ln {\cal V}\right)^{-\frac{7}{12}}{\cal V}^{-\frac{1}{36}}
+ \left(ln {\cal V}\right)^{\frac{7}{12}}\sqrt{\cal V}
+ \sqrt{{\cal V} ln {\cal V}}
+ \frac{1}{\sqrt{ln {\cal V}}}{\cal V}^{\frac{5}{6}}
+ \frac{1}{\sqrt{ln {\cal V}}}{\cal V}^{\frac{5}{6}}
+ \left(ln {\cal V}\right)^{\frac{7}{12}}\sqrt{\cal V}
\Biggr)\sim \left(ln {\cal V}\right)^{-\frac{7}{12}}{\cal V}^{\frac{5}{18}}$

\item
$
\partial_{z_2}{\bar\partial_{\bar z_2}}K_{\rm geom}=\partial_{z_1}{\bar\partial_{\bar z_1}}K_{\rm geom}(z_1\leftrightarrow z_2)\sim \left(ln {\cal V}\right)^{-\frac{7}{12}}{\cal V}^{\frac{5}{18}}$

\item
$\partial_{z_1}{\bar\partial_{\bar z_2}}K_{\rm geom}=-\frac{1}{3\,
    \left( -z_3^3 \right) }\Biggl[ 4\,{{z1}}^5\,{z_2}\,\left( {{z_1}}^{12} - \phi\,{{z_2}}^6 \right) \,
       {\left( \frac{\zeta}
           {r_1\,|z_3|^2} \right) }^{\frac{1}{6}} +4\,\left( \phi\,{{\bar z_1}}^6\,{{\bar z_2}}^5 -
           {{\bar z_2}}^{17} \right) \,
         $

             $ \times  \frac{{\left( \frac{\zeta}
             {r_1\,|z_3|^2} \right) }^{\frac{1}{6}}\,
         \left( \left( 2\,\phi\,{{z1}}^6\,{{z_2}}^6 - {{z_2}}^{18} \right) \,
            {\bar z_1} + {{z_1}}^5\,
            \left( {{z_1}}^{12} - \phi\,{{z_2}}^6 \right) \,
            \left( 1 + |{z_2}|^2 \right)  \right) }{-{\bar z_3}^3}+ 2\,\zeta\,{\left( \frac{\zeta}
             {r_1\,|z_3|^2} \right) }^{\frac{1}{3}}\,$

            $ \times  \frac{
         \left( {z_2}\,{{\bar z_1}}^{18} +
           \phi\,{z_1}\,{{\bar z_1}}^7\,{{\bar z_2}}^5 -
           {{\bar z_2}}^{17} -
           {z1}\,{\bar z_1}\,{{\bar z_2}}^{17} +
           \phi\,{{\bar z_1}}^6\,{{\bar z_2}}^5\,
            \left( 1 - 2\,{z2}\,{\bar z_2} \right)  \right) \,
         \left( \left( 2\,\phi\,{{z_1}}^6\,{{z_2}}^6 - {{z_2}}^{18} \right) \,
            {\bar z_1} + {{z_1}}^5\,
            \left( {{z_1}}^{12} - \phi\,{{z_2}}^6 \right) \,
            \left( 1 + |{z_2}|^2 \right)  \right) }{r_1\,
         \left( -{\bar z_3}^3\right) } $

         $+
      \frac{\left( 4\,{{z_1}}^{17} - 4\,\phi\,{{z_1}}^5\,{{z_2}}^6 +
           3\,{{z_1}}^{18}\,{\bar z_1} -
           \phi\,{{z_1}}^6\,{{z_2}}^6\,{\bar z_1} -
           {{z_2}}^{18}\,{\bar z_1} +
           4\,{{z_1}}^{17}\,{z2}\,{\bar z_2} -
           4\,\phi\,{{z_1}}^5\,{{z_2}}^7\,{\bar z_2} -
           \frac{3\,r_2\,{{z_1}}^{17}}
            {{\left( \frac{\zeta}
                 {r_1\,|z_3|^2} \right) }^{\frac{1}{6}}} +
           \frac{3\,\phi\,r_2\,{{z1}}^5\,{{z_2}}^6}
            {{\left( \frac{\zeta}
                 {r_1\,|z_3|^2} \right) }^{\frac{1}{6}}} \right) }
         {\left( -{\bar z_3}^3 \right) \,
         \left( -r_2 + {\left( \frac{\zeta}
               {r_1\,|z_3|^2} \right) }^{\frac{1}{6}} +
           |{z_1}|^2\,
            {\left( \frac{\zeta}
                {r_1\,|z_3|^2} \right) }^{\frac{1}{6}} +
           |{z_2}|^2\,
            {\left( \frac{\zeta}
                {r_1\,|z_3|^2} \right) }^{\frac{1}{6}} \right) } $

                $\times\left(3\,r_1\,{{\bar z_2}}^5\,
         \left( -\left( \phi\,{{\bar z_1}}^6 \right)  + {{\bar z_2}}^{12}
           \right) \,{\left( \frac{\zeta}
             {r_1\,|z_3|^2} \right) }^{\frac{1}{6}}\,\right)$

             $
      +\frac{3\,r_1\,{\left( \frac{\zeta}
             {r_1\,|z_3|^2} \right) }^{\frac{1}{6}}\,
         \left( -\left( {z_2}\,{{\bar z_1}}^{18} \right)  -
           \phi\,{z_1}\,{{\bar z_1}}^7\,{{\bar z_2}}^5 +
           {{\bar z_2}}^{17} +
           {z_1}\,{\bar z_1}\,{{\bar z_2}}^{17} +
           \phi\,{{\bar z_1}}^6\,{{\bar z_2}}^5\,
            \left( -1 + 2\,|{z_2}|^2 \right)  \right) \,
         \left( \Sigma_1\right) }{
         \left( -{\bar z_3}^3\right) \,
         {\left( -r_2 + {\left( \frac{\zeta}
                 {r_1\,|z_3|^2} \right) }^{\frac{1}{6}} +
             {z_1}\,{\bar z_1}\,
              {\left( \frac{\zeta}
                  {r_1\,|z_3|^2} \right) }^{\frac{1}{6}} +
             {z_2}\,{\bar z_2}\,
              {\left( \frac{\zeta}
                  {r_1\,|z_3|^2} \right) }^{\frac{1}{6}} \right) }^2} $

                  $\hskip -1cm +
       \frac{3\,r_1\,{{z_1}}^5\,\left( {{z_1}}^{12} - \phi\,{{z_2}}^6 \right) \,
         \left( -3\,\phi\,r_2\,{{\bar z_1}}^6\,{{\bar z_2}}^5 +
           3\,r_2\,{{\bar z_2}}^{17} -
           4\,{z_2}\,{{\bar z_1}}^{18}\,
            {\left( \frac{\zeta}
                {r_1\,|z_3|^2} \right) }^{\frac{1}{6}} +
           12\,\phi\,{z_2}\,{{\bar z_1}}^6\,{{\bar z_2}}^6\,
            {\left( \frac{\zeta}
                {r_1\,|z_3|^2} \right) }^{\frac{1}{6}} -
           4\,{z_2}\,{{\bar z_2}}^{18}\,
            {\left( \frac{\zeta}
                {r_1\,|z_3|^2} \right) }^{\frac{1}{6}} \right) }{
         \left( -{\bar z_3}^3\right) \,
         \left( -r_2 + {\left( \frac{\zeta}
               {r_1\,|z_3|^2} \right) }^{\frac{1}{6}} +
           |{z_1}|^2\,
            {\left( \frac{\zeta}
                {r_1\,|z_3|^2} \right) }^{\frac{1}{6}} +
           |{z_2}|^2\,
            {\left( \frac{\zeta}
                {r_1\,|z_3|^2} \right) }^{\frac{1}{6}} \right) }$

                $ \hskip -1cm -
      \frac{2\,{{z_1}}^5\,z_2\,\left( {{z_1}}^{12} - \phi\,{z_2}^6 \right) \,\zeta\,
         {\left( \frac{\zeta}
             {r_1\,|z_3|^2} \right) }^{\frac{1}{6}}\,
         \left( r_2 - \left( 1 + |{z_1}|^2 +
              |{z_2}|^2 \right) \,
            {\left( \frac{\zeta}
                {r_1\,|z_3|^2} \right) }^{\frac{1}{6}} \right) }{
         r_1} + 2\,\zeta\,{{\bar z_2}}^5\,
         \left( -\left( \phi\,{{\bar z_1}}^6 \right)  + {{\bar z_2}}^{12}
           \right) \,$

         $\hskip -1cm \times \frac{{\left( \frac{\zeta}
             {r_1\,|z_3|^2} \right) }^{\frac{1}{6}}\,
         \left( \left( 2\,\phi\,{{z_1}}^6\,{z_2}^6 - {z_2}^{18} \right) \,
            {\bar z_1} + {{z_1}}^5\,
            \left( {{z_1}}^{12} - \phi\,{z_2}^6 \right) \,
            \left( 1 + |{z_2}|^2 \right)  \right) \,
         \left( r_2 - \left( 1 + |{z_1}|^2 +
              |{z_2}|^2 \right) \,
            {\left( \frac{\zeta}
                {r_1\,|z_3|^2} \right) }^{\frac{1}{6}} \right) }{
         r_1\,\left( -{\bar z_3}^3 \right) } \Biggr], $

$\hskip -1cm \sim\frac{1}{\sqrt{{\cal V}}}\Biggl(
\left(ln {\cal V}\right)^{-\frac{1}{12}}{\cal V}^{\frac{11}{18}}
+ \left(ln {\cal V}\right)^{-\frac{1}{12}}{\cal V}^{\frac{4}{9}}
+ \left(ln {\cal V}\right)^{-\frac{2}{3}}{\cal V}^{\frac{4}{9}}
+ \left(ln {\cal V}\right)^{\frac{1}{12}}{\cal V}^{\frac{4}{9}}
+ \left(ln {\cal V}\right)^{\frac{5}{12}}{\cal V}^{\frac{4}{9}}
+ \sqrt{ln {\cal V}}{\cal V}^{\frac{4}{9}}
+ \left(ln {\cal V}\right)^{-\frac{7}{12}}{\cal V}^{\frac{7}{9}}
+ \left(ln {\cal V}\right)^{-\frac{7}{12}}{\cal V}^{\frac{7}{9}}
\Biggr) \sim \left(ln {\cal V}\right)^{-\frac{7}{12}}{\cal V}^{\frac{5}{18}}$

where

$\hskip -1cm \Sigma_1\equiv 3\,r_2\,{{z_1}}^{17} - 3\,\phi\,r_2\,{{z_1}}^5\,{{z_2}}^6 -
           4\,{{z_1}}^{17}\,{\left( \frac{\zeta}  {r_1\,|z_3|^2} \right) }^{\frac{1}{6}} +
           4\,\phi\,{{z_1}}^5\,{{z_2}}^6\, {\left( \frac{\zeta}
                {r_1\,|z_3|^2} \right) }^{\frac{1}{6}} -
           3\,{{z_1}}^{18}\,{\bar z_1}\,
            {\left( \frac{\zeta}
                {r_1\,|z_3|^2} \right) }^{\frac{1}{6}}$

                $\hskip -1cm +
           \phi\,{{z_1}}^6\,{{z_2}}^6\,{\bar z_1}\,
            {\left( \frac{\zeta}
                {r_1\,|z_3|^2} \right) }^{\frac{1}{6}} +
           {{z_2}}^{18}\,{\bar z_1}\,
            {\left( \frac{\zeta}
                {r_1\,|z_3|^2} \right) }^{\frac{1}{6}} -
           4\,{{z_1}}^{17}\,|{z_2}|^2\,
            {\left( \frac{\zeta}
                {r_1\,|z_3|^2} \right) }^{\frac{1}{6}} +
           4\,\phi\,{{z_1}}^5\,{{z_2}}^7\,{\bar z_2}\,
            {\left( \frac{\zeta}
                {r_1\,|z_3|^2} \right) }^{\frac{1}{6}}$

                $\hskip -1cm \sim{\cal V}^{\frac{29}{36}},$
and $\Sigma_2=\Sigma_1(z_1\leftrightarrow z_2)\sim{\cal V}^{\frac{29}{36}}$

\hskip -1cm Hence, in the LVS limit, the $D_5$-metric components will scale with ${\cal V}$ as
follows:
\begin{equation}
\label{eq:D_5_metric}
\hskip -1cm G_{i{\bar j}}|_{D_5}(z_1,z_2)=\left(\begin{array}{cc}
\partial_{z_1}{\bar\partial_{\bar z_1}}K_{\rm geom} &
\partial_{z_1}{\bar\partial_{\bar z_2}}K_{\rm geom}\\
\partial_{z_2}{\bar\partial_{\bar z_1}}K_{\rm geom} &
\partial_{z_2}{\bar\partial_{\bar z_2}}K_{\rm geom}
\end{array}\right)\sim
 \left(ln {\cal V}\right)^{-\frac{7}{12}}{\cal V}^{\frac{5}{18}}\left(\begin{array}{cc}
{\cal O}(1) & {\cal O}(1) \\
{\cal O}(1) & {\cal O}(1)
\end{array}\right).
\end{equation}
\end{itemize}

The mixed double derivatives of the K\"{a}hler potential restricted to $D_4$ are given as under:

$\partial_2{\bar\partial_2}K|_{D_4}=
\frac{1}{3}\Biggl\{\frac{-12\,3^{\frac{1}{9}}\,{{r_2}}^2\,
       \left( \phi\,{{z_1}}^6\,{{z_2}}^5 - {{z_2}}^{17} \right) \,
       {{\bar z_2}}^5\,\left( 1 + |{z_1}|^2 +
         |{z_2}|^2 \right) \,
       \left( -\left( \phi\,{{\bar z_1}}^6 \right)  + {{\bar z_2}}^{12}
         \right) }{{\left( {{r_2}}^2\,|z_1^{18}+z_2^{18}-3\phi z_1^6z_2^6|^2 \right) }^{\frac{10}{9}}}$

            $ -
    \frac{2\,3^{\frac{1}{9}}\,{{r_2}}^2\,{{\bar z_2}}^5\,
       \left( -2\,\phi\,{{z_1}}^6\,{{z_2}}^5 + 2\,{{z_2}}^{17} +
         \left( -2\,\phi\,{{z_1}}^7\,{{z_2}}^5 + 2\,{z_1}\,{{z_2}}^{17} \right) \,
          {\bar z_1} + \left( -{{z_1}}^{18} + \phi\,{{z_1}}^6\,{{z_2}}^6 +
            {{z_2}}^{18} \right) \,{\bar z_2} \right) \,
       \left( -\left( \phi\,{{\bar z_1}}^6 \right)  + {{\bar z_2}}^{12}
         \right) }{{\left( {{r_2}}^2\,|z_1^{18}+z_2^{18}-3\phi z_1^6z_2^6|^2\right) }^{\frac{10}{9}}}$

            $- \frac{6\,3^{\frac{1}{9}}\,{{r_2}}^2\,\left( {{z_1}}^{18} - 3\,\phi\,{{z_1}}^6\,{{z_2}}^6 +
         {{z_2}}^{18} \right) \,{\bar z_2}\,
       \left( -\left( \phi\,{{\bar z_1}}^6\,{{\bar z_2}}^5 \right)  +
         {{\bar z_2}}^{17} \right) }{{\left( {{r_2}}^2\,
          |z_1^{18}+z_2^{18}-3\phi z_1^6z_2^6|^2 \right) }^{\frac{10}{9}}} +
    \frac{3\,3^{\frac{1}{9}}}
     {{\left( {{r_2}}^2\,|z_1^{18}+z_2^{18}-3\phi z_1^6z_2^6|^2  \right) }^{\frac{1}{9}}}$

            $-
    \frac{6\,3^{\frac{1}{9}}\,{z_2}\,\left( -\left( \phi\,{{z_1}}^6\,{{z_2}}^5 \right)  +
         {{z_2}}^{17} \right) }{\left( {{z_1}}^{18} - 3\,\phi\,{{z_1}}^6\,{{z_2}}^6 +
         {{z_2}}^{18} \right) \,{\left( {{r_2}}^2\,
           |z_1^{18}+z_2^{18}-3\phi z_1^6z_2^6|^2\right) }^{\frac{1}{9}}} + \frac{3^{\frac{1}{9}}\,\left( -{{z_1}}^{18} + \phi\,{{z_1}}^6\,{{z_2}}^6 + {{z_2}}^{18} \right) }{\left( {{z_1}}^{18} - 3\,\phi\,{{z_1}}^6\,{{z_2}}^6 + {{z_2}}^{18} \right) \,
       {\left( {{r_2}}^2\,|z_1^{18}+z_2^{18}-3\phi z_1^6z_2^6|^2\right) }^{\frac{1}{9}}}$

             $-
    \frac{3\,{r_1}\,\left( 3^{\frac{1}{9}}\,
          \left( {{z_1}}^{18} + 5\,\phi\,{{z_1}}^6\,{{z_2}}^6 - 7\,{{z_2}}^{18} \right)  -
         \frac{12\,{{r_2}}^3\,{{z_2}}^5\,
            \left( -\left( \phi\,{{z_1}}^6 \right)  + {{z_2}}^{12} \right) \,
            \left( {{z_1}}^{18} - 3\,\phi\,{{z_1}}^6\,{{z_2}}^6 + {{z_2}}^{18} \right) \,
            \left( \phi\,{{\bar z_1}}^6\,{{\bar z_2}}^5 -
              {{\bar z_2}}^{17} \right) }{{\left( {{r_2}}^2\,
               |z_1^{18}+z_2^{18}-3\phi z_1^6z_2^6|^2\right) }^{\frac{8}{9}}} \right) }{
       \left( {{z_1}}^{18} - 3\,\phi\,{{z_1}}^6\,{{z_2}}^6 + {{z_2}}^{18} \right) \,
       \left( 3^{\frac{1}{9}} + 3^{\frac{1}{9}}\,|{z_1}|^2 +
         3^{\frac{1}{9}}\,|{z_2}|^2 -
         {r_2}\,{\left( {{r_2}}^2\,
              |z_1^{18}+z_2^{18}-3\phi z_1^6z_2^6|^2 \right) }^{\frac{1}{9}} \right) }$

                $-
    \frac{2\,3^{\frac{1}{9}}\,\left( 2\,\phi\,{{z_1}}^6\,{{z_2}}^5 - 2\,{{z_2}}^{17} +
         2\,\left( \phi\,{{z_1}}^7\,{{z_2}}^5 - {z_1}\,{{z_2}}^{17} \right) \,
          {\bar z_1} + \left( {{z_1}}^{18} - \phi\,{{z_1}}^6\,{{z_2}}^6 -
            {{z_2}}^{18} \right) \,{\bar z_2} \right) \,
       \left( {{\bar z_1}}^{18} -
         3\,\phi\,{{\bar z_1}}^6\,{{\bar z_2}}^6 +
         {{\bar z_2}}^{18} \right) \,
       \left(\Sigma_2\right) \,
       \left( \Sigma_1 \right) }{{\left( {{r_2}}^2\,
          |z_1^{18}+z_2^{18}-3\phi z_1^6z_2^6|^2\right) }^{\frac{4}{3}}}$

            $+
    \frac{6\,3^{\frac{1}{9}}\,{{\bar z_2}}^5\,
       \left( 2\,\phi\,{{z_1}}^6\,{{z_2}}^5 - 2\,{{z_2}}^{17} +
         2\,\left( \phi\,{{z_1}}^7\,{{z_2}}^5 - {z_1}\,{{z_2}}^{17} \right) \,
          {\bar z_1} + \left( {{z_1}}^{18} - \phi\,{{z_1}}^6\,{{z_2}}^6 -
            {{z_2}}^{18} \right) \,{\bar z_2} \right) \,
       \left( -\left( \phi\,{{\bar z_1}}^6 \right)  + {{\bar z_2}}^{12}
         \right) \,{\left( \Sigma_1 \right) }^2}{{\left(
          {{r_2}}^2\,|z_1^{18}+z_2^{18}-3\phi z_1^6z_2^6|^2\right) }^{\frac{4}{3}}}$

            $-
    \frac{3^{\frac{1}{9}}\,\left( {{z_1}}^{18} - \phi\,{{z_1}}^6\,{{z_2}}^6 - {{z_2}}^{18}
         \right) \,\left( {{\bar z_1}}^{18} -
         3\,\phi\,{{\bar z_1}}^6\,{{\bar z_2}}^6 +
         {{\bar z_2}}^{18} \right) \,
       {\left( \Sigma_1\right) }^2}{{\left(
          {{r_2}}^2\,|z_1^{18}+z_2^{18}-3\phi z_1^6z_2^6|^2\right) }^{\frac{4}{3}}} +
    \frac{1}{
       \left( {{z_1}}^{18} - 3\,\phi\,{{z_1}}^6\,{{z_2}}^6 + {{z_2}}^{18} \right) \,
       {\left( \Sigma \right) }^2}$

       $\times\Biggl[3\,{r_1}\,\Biggl( \Sigma_2\Biggr) \,
       \Biggl( 8\,3^{\frac{1}{9}}\,{z_1}\,{{z_2}}^5\,
          \left( \phi\,{{z_1}}^6 - {{z_2}}^{12} \right) \,{\bar z_1} +
         3^{\frac{1}{9}}\,\left( {{z_1}}^{18} + 5\,\phi\,{{z_1}}^6\,{{z_2}}^6 -
            7\,{{z_2}}^{18} \right) \,{\bar z_2}$

            $-
         2\,{{z_2}}^5\,\left( -\left( \phi\,{{z_1}}^6 \right)  + {{z_2}}^{12} \right) \,
          \left( 4\,3^{\frac{1}{9}} - 3\,{r_2}\,
             {\left( {{r_2}}^2\,|z_1^{18}+z_2^{18}-3\phi z_1^6z_2^6|^2\right) }^{\frac{1}{9}} \right)  \Biggr) \Biggr]\Biggr\}
$

$\stackrel{LVS}{\sim}\frac{3\,{r_1}\,\left( 3^{\frac{1}{9}}\,
          \left( {{z_1}}^{18} + 5\,\phi\,{{z_1}}^6\,{{z_2}}^6 - 7\,{{z_2}}^{18} \right)  -
         \frac{12\,{{r_2}}^3\,{{z_2}}^5\,
            \left( -\left( \phi\,{{z_1}}^6 \right)  + {{z_2}}^{12} \right) \,
            \left( {{z_1}}^{18} - 3\,\phi\,{{z_1}}^6\,{{z_2}}^6 + {{z_2}}^{18} \right) \,
            \left( \phi\,{{\bar z_1}}^6\,{{\bar z_2}}^5 -
              {{\bar z_2}}^{17} \right) }{{\left( {{r_2}}^2\,
               |z_1^{18}+z_2^{18}-3\phi z_1^6z_2^6|^2\right) }^{\frac{8}{9}}} \right) }{
       \left( {{z_1}}^{18} - 3\,\phi\,{{z_1}}^6\,{{z_2}}^6 + {{z_2}}^{18} \right) \,
       \left( 3^{\frac{1}{9}} + 3^{\frac{1}{9}}\,|{z_1}|^2 +
         3^{\frac{1}{9}}\,|{z_2}|^2 -
         {r_2}\,{\left( {{r_2}}^2\,
              |z_1^{18}+z_2^{18}-3\phi z_1^6z_2^6|^2\right) }^{\frac{1}{9}} \right) }$

                $+\frac{1}{
       \left( {{z_1}}^{18} - 3\,\phi\,{{z_1}}^6\,{{z_2}}^6 + {{z_2}}^{18} \right) \,
       {\left( \Sigma_1 \right) }^2}$

       $\times\Biggl[3\,{r_1}\,\Biggl( \Sigma_2\Biggr) \,
       \Biggl( 8\,3^{\frac{1}{9}}\,{z_1}\,{{z_2}}^5\,
          \left( \phi\,{{z_1}}^6 - {{z_2}}^{12} \right) \,{\bar z_1} +
         3^{\frac{1}{9}}\,\left( {{z_1}}^{18} + 5\,\phi\,{{z_1}}^6\,{{z_2}}^6 -
            7\,{{z_2}}^{18} \right) \,{\bar z_2} -$

            $
         2\,{{z_2}}^5\,\left( -\left( \phi\,{{z_1}}^6 \right)  + {{z_2}}^{12} \right) \,
          \left( 4\,3^{\frac{1}{9}} - 3\,{r_2}\,
             {\left( {{r_2}}^2\,|z_1^{18}+z_2^{18}-3\phi z_1^6z_2^6|^2\right) }^{\frac{1}{9}} \right)  \Biggr) \Biggr]$

         $\sim\sqrt{ln{\cal V}} {\cal V}^{-\frac{1}{18}}.$
where

$\Sigma_1\equiv 3^{\frac{1}{9}} + 3^{\frac{1}{9}}\,|{z_1}|^2 +
         3^{\frac{1}{9}}\,|{z_2}|^2 -
         {r_2}\,{\left( {{r_2}}^2\,
              |z_1^{18}+z_2^{18}-3\phi z_1^6z_2^6|^2\right) }^{\frac{1}{9}},$

               $\Sigma_2\equiv  3^{\frac{1}{9}}\,{z_2} + \frac{2\,{{r_2}}^3\,
            \left( {{z_1}}^{18} - 3\,\phi\,{{z_1}}^6\,{{z_2}}^6 + {{z_2}}^{18} \right) \,
            \left( \phi\,{{\bar z_1}}^6\,{{\bar z_2}}^5 -
              {{\bar z_2}}^{17} \right) }{{\left( {{r_2}}^2\,
               |z_1^{18}+z_2^{18}-3\phi z_1^6z_2^6|^2\right) }^{\frac{8}{9}}} $

$\partial_1{\bar\partial_2}K_s=$

$
\frac{1}{3}\Biggl\{\frac{12\,3^{\frac{1}{9}}\,{{r_2}}^2\,
       \left( {{z_1}}^{17} - \phi\,{{z_1}}^5\,{{z_2}}^6 \right) \,
       {{\bar z_2}}^5\,\left( 1 + |{z_1}|^2 +
         |{z_2}|^2 \right) \,
       \left( -\left( \phi\,{{\bar z_1}}^6 \right)  + {{\bar z_2}}^{12}
         \right) }{{\left( {{r_2}}^2\,|z_1^{18}+z_2^{18}-3\phi z_1^6z_2^6|^2 \right) }^{\frac{10}{9}}} +
    \frac{6\,3^{\frac{1}{9}}\,{{r_2}}^2\,\left( {{z_1}}^{18} - 3\,\phi\,{{z_1}}^6\,{{z_2}}^6 +
         {{z_2}}^{18} \right) \,{\bar z_1}\,
       \left( \phi\,{{\bar z_1}}^6\,{{\bar z_2}}^5 -
         {{\bar z_2}}^{17} \right) }{{\left( {{r_2}}^2\,
          |z_1^{18}+z_2^{18}-3\phi z_1^6z_2^6|^2 \right) }^{\frac{10}{9}}} $

            $+
    \frac{2\,3^{\frac{1}{9}}\,{{z_1}}^5\,{z_2}\,\left( {{z_1}}^{12} - \phi\,{{z_2}}^6 \right) }
     {\left( {{z_1}}^{18} - 3\,\phi\,{{z_1}}^6\,{{z_2}}^6 + {{z_2}}^{18} \right) \,
       {\left( {{r_2}}^2\,|z_1^{18}+z_2^{18}-3\phi z_1^6z_2^6|^2\right) }^{\frac{1}{9}}} +
    \frac{6\,3^{\frac{1}{9}}\,{z_2}\,\left( -{{z_1}}^{17} + \phi\,{{z_1}}^5\,{{z_2}}^6 \right) }
     {\left( {{z_1}}^{18} - 3\,\phi\,{{z_1}}^6\,{{z_2}}^6 + {{z_2}}^{18} \right) \,
       {\left( {{r_2}}^2\,|z_1^{18}+z_2^{18}-3\phi z_1^6z_2^6|^2\right) }^{\frac{1}{9}}} $

       $-\frac{2\,3^{\frac{1}{9}}\,{{r_2}}^2\,{{\bar z_2}}^5\,
       \left( -\left( \phi\,{{\bar z_1}}^6 \right)  + {{\bar z_2}}^{12}
         \right) \,\left( \left( {{z_1}}^{18} + \phi\,{{z_1}}^6\,{{z_2}}^6 - {{z_2}}^{18}
            \right) \,{\bar z_1} +
         2\,{{z_1}}^5\,\left( {{z_1}}^{12} - \phi\,{{z_2}}^6 \right) \,
          \left( 1 + |{z_2}|^2 \right)  \right) }{{\left( {{r_2}}^2\,
          |z_1^{18}+z_2^{18}-3\phi z_1^6z_2^6|^2 \right) }^{\frac{10}{9}}} $

            $+
    \frac{6\,{r_1}\,{{z_1}}^5\,\left( {{z_1}}^{12} - \phi\,{{z_2}}^6 \right) \,
       \left( 4\,3^{\frac{1}{9}}\,{z_2} + \frac{6\,{{r_2}}^3\,
            \left( {{z_1}}^{18} - 3\,\phi\,{{z_1}}^6\,{{z_2}}^6 + {{z_2}}^{18} \right) \,
            \left( \phi\,{{\bar z_1}}^6\,{{\bar z_2}}^5 -
              {{\bar z_2}}^{17} \right) }{{\left( {{r_2}}^2\,
               |z_1^{18}+z_2^{18}-3\phi z_1^6z_2^6|^2\right) }^{\frac{8}{9}}} \right) }{
       \left( {{z_1}}^{18} - 3\,\phi\,{{z_1}}^6\,{{z_2}}^6 + {{z_2}}^{18} \right) \,
       \left( \Sigma_1\right) }$

        $+\frac{2\,3^{\frac{1}{9}}\,{{z_1}}^5\,{z_2}\,\left( {{z_1}}^{12} - \phi\,{{z_2}}^6 \right) \,
       \left( {{\bar z_1}}^{18} -
         3\,\phi\,{{\bar z_1}}^6\,{{\bar z_2}}^6 +
         {{\bar z_2}}^{18} \right) \,
       {\left( \Sigma_1 \right) }^2}{{\left(
          {{r_2}}^2\,|z_1^{18}+z_2^{18}-3\phi z_1^6z_2^6|^2\right) }^{\frac{4}{3}}}$

                $+
    \frac{2\,3^{\frac{1}{9}}\,\left( {{\bar z_1}}^{18} -
         3\,\phi\,{{\bar z_1}}^6\,{{\bar z_2}}^6 +
         {{\bar z_2}}^{18} \right) \,
       \left( \left( {{z_1}}^{18} + \phi\,{{z_1}}^6\,{{z_2}}^6 - {{z_2}}^{18} \right) \,
          {\bar z_1} + 2\,{{z_1}}^5\,
          \left( {{z_1}}^{12} - \phi\,{{z_2}}^6 \right) \,
          \left( 1 + |{z_2}|^2 \right)  \right) \,
       \left( \Sigma_2\right) \,
       \left( \Sigma_1 \right) }{{\left( {{r_2}}^2\,
          |z_1^{18}+z_2^{18}-3\phi z_1^6z_2^6|^2\right) }^{\frac{4}{3}}} $

            $ -\frac{6\,3^{\frac{1}{9}}\,{{\bar z_2}}^5\,
       \left( -\left( \phi\,{{\bar z_1}}^6 \right)  + {{\bar z_2}}^{12}
         \right) \,\left( \left( {{z_1}}^{18} + \phi\,{{z_1}}^6\,{{z_2}}^6 - {{z_2}}^{18}
            \right) \,{\bar z_1} + 2\,{{z_1}}^5\,\left( {{z_1}}^{12} - \phi\,{{z_2}}^6 \right) \,
          \left( 1 + |{z_2}|^2 \right)  \right) \,
       {\left( \Sigma_1 \right) }^2}{{\left(
          {{r_2}}^2\,|z_1^{18}+z_2^{18}-3\phi z_1^6z_2^6|^2\right) }^{\frac{4}{3}}} $

       $- \frac{1}{
       \left( {{z_1}}^{18} - 3\,\phi\,{{z_1}}^6\,{{z_2}}^6 + {{z_2}}^{18} \right) \,
       {\left(\Sigma_1\right) }^2} \Biggl[3\,{r_1}\,\left( 3^{\frac{1}{9}}\,{z_2} +
         \frac{2\,{{r_2}}^3\,\left( {{z_1}}^{18} - 3\,\phi\,{{z_1}}^6\,{{z_2}}^6 +
              {{z_2}}^{18} \right) \,\left( \phi\,{{\bar z_1}}^6\,
               {{\bar z_2}}^5 - {{\bar z_2}}^{17} \right) }{{\left(
               {{r_2}}^2\,|z_1^{18}+z_2^{18}-3\phi z_1^6z_2^6|^2 \right) }^{\frac{8}{9}}} \right) \,$

               $\times
       \Biggl( 3^{\frac{1}{9}}\,\left( 7\,{{z_1}}^{18} - 5\,\phi\,{{z_1}}^6\,{{z_2}}^6 -
            {{z_2}}^{18} \right) \,{\bar z_1} +
         2\,{{z_1}}^5\,\left( {{z_1}}^{12} - \phi\,{{z_2}}^6 \right) \,$

         $
          \left( 4\,3^{\frac{1}{9}} + 4\,3^{\frac{1}{9}}\,|{z_2}|^2 -
            3\,{r_2}\,{\left( {{r_2}}^2\,
                 |z_1^{18}+z_2^{18}-3\phi z_1^6z_2^6|^2 \right) }^{\frac{1}{9}} \right)  \Biggr)\Biggr] \Biggr\}$

$\hskip-0.4cm \stackrel{LVS}{\sim}\frac{6\,{r_1}\,{{z_1}}^5\,\left( {{z_1}}^{12} - \phi\,{{z_2}}^6 \right) \,
       \left( 4\,3^{\frac{1}{9}}\,{z_2} + \frac{6\,{{r_2}}^3\,
            \left( {{z_1}}^{18} - 3\,\phi\,{{z_1}}^6\,{{z_2}}^6 + {{z_2}}^{18} \right) \,
            \left( \phi\,{{\bar z_1}}^6\,{{\bar z_2}}^5 -
              {{\bar z_2}}^{17} \right) }{{\left( {{r_2}}^2\,
               |z_1^{18}+z_2^{18}-3\phi z_1^6z_2^6|^2 \right) }^{\frac{8}{9}}} \right) }{
       \left( {{z_1}}^{18} - 3\,\phi\,{{z_1}}^6\,{{z_2}}^6 + {{z_2}}^{18} \right) \,
       \left( \Sigma_1\right) } -\frac{1}{
       \left( {{z_1}}^{18} - 3\,\phi\,{{z_1}}^6\,{{z_2}}^6 + {{z_2}}^{18} \right) \,
       {\left(\Sigma_1\right) }^2}$

       $\times\Biggl[3\,{r_1}\,\left( 3^{\frac{1}{9}}\,{z_2} +
         \frac{2\,{{r_2}}^3\,\left( {{z_1}}^{18} - 3\,\phi\,{{z_1}}^6\,{{z_2}}^6 +
              {{z_2}}^{18} \right) \,\left( \phi\,{{\bar z_1}}^6\,
               {{\bar z_2}}^5 - {{\bar z_2}}^{17} \right) }{{\left(
               {{r_2}}^2\,|z_1^{18}+z_2^{18}-3\phi z_1^6z_2^6|^2\right) }^{\frac{8}{9}}} \right) \,$

               $
       \Biggl( 3^{\frac{1}{9}}\,\left( 7\,{{z_1}}^{18} - 5\,\phi\,{{z_1}}^6\,{{z_2}}^6 -
            {{z_2}}^{18} \right) \,{\bar z_1} +
         2\,{{z_1}}^5\,\left( {{z_1}}^{12} - \phi\,{{z_2}}^6 \right) \,$

         $\times
          \left( 4\,3^{\frac{1}{9}} + 4\,3^{\frac{1}{9}}\,|{z_2}|^2 -
            3\,{r_2}\,{\left( {{r_2}}^2\,
                 |z_1^{18}+z_2^{18}-3\phi z_1^6z_2^6|^2\right) }^{\frac{1}{9}} \right)  \Biggr)\Biggr] \sim\sqrt{ln {\cal V}}{\cal V}^{-\frac{1}{18}}.$

\section{Intermediate Expansions Relevant to Evaluation of the Complete K\"{a}hler Potential as a Power Series in the Matter Fields}
\setcounter{equation}{0}
\seceqee

The following are relevant to the expansion of the geometric K\"{a}hler potential in $\delta z_i$:

\begin{itemize}
\item
\begin{eqnarray*}
& & \hskip-1in 3\phi_0z_1^6z_2^6 - z_1^{18} - z_2^{18}\sim\sqrt{\cal V}\left[1 - {\cal V}^{-\frac{1}{36}}(\delta z_1 + \delta z_2) - {\cal V}^{-\frac{1}{18}}((\delta z_1)^2 + (\delta z_2)^2) + ...\right],\nonumber\\
\end{eqnarray*}
\begin{eqnarray*}
& & \hskip-0.4in r_2 - \left(\frac{\zeta}{r_1}\right)^{\frac{1}{6}}\left(1 + |z_1|^2 + |z_2|^2\right)\frac{1}{\left(\left|3\phi_0z_1^6z_2^6 - z_1^{18} - z_2^{18}\right|^2\right)^{\frac{1}{18}}}\sim r_2 - \left(\frac{\zeta}{r_1}\right)^{\frac{1}{6}}{\cal V}^{-\frac{1}{18}}\Bigl[{\cal V}^{\frac{1}{18}}+ {\cal V}^{\frac{1}{36}}\nonumber\\
& & \hskip-0.4in  (\delta z_1 + \delta z_2 + c.c.) + |\delta z_1|^2 + |\delta z_2|^2 + (\delta z_1)^2 + (\delta z_2)^2 + \delta z_1 \delta z_2 + c.c. + |\delta z_1 + \delta z_2|^2 + ..\Bigr],\nonumber\\
\end{eqnarray*}
\begin{eqnarray*}
& & \hskip-0.4in -r_2ln\left(\frac{\zeta}{r_1\left|3\phi_0z_1^6z_2^6 - z_1^{18} - z_2^{18}\right|^2}\right)^{\frac{1}{6}}\sim -r_2ln\left\{\left(\frac{\zeta}{r_1}\right)^{\frac{1}{6}}{\cal V}^{-\frac{1}{18}}\right\} + r_2\frac{\delta z_1 + \delta z_2 + c.c.}{{\cal V}^{\frac{1}{36}}}\nonumber\\
& & \hskip-0.4in + r_2\frac{(\delta z_1)^2 + (\delta z_2)^2 + \delta z_1\delta z_2 + c.c. + |\delta z_1 + \delta z_2|^2}{{\cal V}^{\frac{1}{18}}}
- r_2\frac{(\delta z_1 + \delta z_2 + c.c.)^2}{{\cal V}^{\frac{1}{18}}} + ...,\nonumber\\
\end{eqnarray*}
\begin{eqnarray*}
& & \hskip-0.4in \frac{\zeta}{9r_1}\left[r_2 - \left(\frac{\zeta}{r_1}\right)^{\frac{1}{6}}\left(1 + |z_1|^2 + |z_2|^2\right)\frac{1}{\left(\left|3\phi_0z_1^6z_2^6 - z_1^{18} - z_2^{18}\right|^2\right)^{\frac{1}{18}}}\right]^2\nonumber\\
& & \hskip-0.4in \sim \frac{\zeta r_2^2}{9r_1}\biggl[1 - \frac{1}{r_2}\left(\frac{\zeta}{r_1}\right)^{\frac{1}{6}}\frac{\delta z_1 + \delta z_2}{{\cal V}^{\frac{1}{36}}} + \frac{1}{r_2}\frac{(\delta z_1)^2 + (\delta z_2)^2 + \delta z_1\delta z_2 + c.c. + |\delta z_1 + \delta z_2|^2}{{\cal V}^{\frac{1}{18}}}\nonumber\\
 & & \hskip-0.4in + \frac{1}{r_2^2}\left(\frac{\zeta}{r_1}\right)^{\frac{1}{3}}\frac{(\delta z_1 + \delta z_2 + c.c.)^2}{{\cal V}^{\frac{1}{18}}} + ...\biggr],\nonumber\\
\end{eqnarray*}
\begin{eqnarray*}
& & \hskip-0.4in -r_2ln\left[\frac{1}{3}\left(\frac{\zeta}{r_1}\right)^{\frac{1}{2}}\frac{r_2 - \left(\frac{\zeta}{r_1}\right)^{\frac{1}{6}}\left(1 + |z_1|^2 + |z_2|^2\right)\frac{1}{\left(\left|3\phi_0z_1^6z_2^6 - z_1^{18} - z_2^{18}\right|^2\right)^{\frac{1}{18}}}}{\sqrt{\left|3\phi_0z_1^6z_2^6 - z_1^{18} - z_2^{18}\right|^2}}\right] \sim - r_2ln\left(r_2\sqrt{\frac{\zeta}{r_1}}\right)\nonumber\\
 & & \hskip-0.4in - r_2\Biggl[\sqrt{\frac{\zeta}{r_1}}\frac{(|\delta z_1|^2 + |\delta z_2|^2 + (\delta z_1)^2 + (\delta z_2)^2 + \delta z_1 \delta z_2 + c.c. + |\delta z_1 + \delta z_2|^2)}{r_2{\cal V}^{\frac{1}{18}}}\nonumber\\
 & & \hskip-0.4in - \frac{1}{r_2}\sqrt{\frac{\zeta}{r_1}}\frac{(\delta z_1 + \delta z_2 + c.c.}{{\cal V}^{\frac{1}{36}}} - \left(\frac{\zeta}{r_1}\right)^{\frac{1}{3}}\frac{((\delta z_1 + \delta z_2 + c.c.)^2}{{\cal V}^{\frac{1}{18}}}\Biggr]+ r_1\Bigl[-\frac{(\delta z_1 + \delta z_2 + c.c.)}{{\cal V}^{\frac{1}{36}}}-\nonumber\\
& & \hskip-0.5in \frac{(|\delta z_1|^2 + |\delta z_2|^2 + (\delta z_1)^2 + (\delta z_2)^2 + \delta z_1 \delta z_2 + c.c. + |\delta z_1 + \delta z_2|^2)}{r_2{\cal V}^{\frac{1}{18}}} - \frac{((\delta z_1 + \delta z_2 + c.c.)^2}{{\cal V}^{\frac{1}{18}}}\Bigr] + ...\nonumber\\
  \end{eqnarray*}
\begin{eqnarray}
\label{eq:K_geom_fluc}
& & \hskip-0.4in \sim  - r_2ln\left(r_2\sqrt{\frac{\zeta}{r_1}}\right) + r_1\Bigl[-\frac{(\delta z_1 + \delta z_2 + c.c.)}{{\cal V}^{\frac{1}{36}}} - \frac{((\delta z_1 + \delta z_2 + c.c.)^2}{{\cal V}^{\frac{1}{18}}}\nonumber\\
& & \hskip-0.4in - \frac{(|\delta z_1|^2 + |\delta z_2|^2 + (\delta z_1)^2 + (\delta z_2)^2 + \delta z_1 \delta z_2 + c.c. + |\delta z_1 + \delta z_2|^2)}{r_2{\cal V}^{\frac{1}{18}}} \Bigr] + ...\nonumber\\
& &
\end{eqnarray}

Using (\ref{eq:K_geom_fluc}), one obtains:
\begin{eqnarray}
\label{eq:Kexp}
& & K_{\rm geom}|_{D_5}\sim\frac{r_2^2\zeta}{r_1} + \frac{r_2(\delta z_1 + \delta z_2 + c.c.)}{{\cal V}^{\frac{1}{36}}} + r_2\frac{((\delta z_1)^2 + (\delta z_2)^2 + \delta z_1 \delta z_2 + c.c.)}{{\cal V}^{\frac{1}{18}}}\nonumber\\
 & & + r_2\frac{(|\delta z_1|^2 + |\delta z_2|^2 + \delta z_1 {\bar\delta z_2} + \delta z_2{\bar\delta z_1}}{{\cal V}^{\frac{1}{18}}} + ...
\end{eqnarray}

\item

\begin{eqnarray*}
& & \hskip-0.4in -2ln\Biggl[\Biggl({\cal T}_B(\sigma^B,{\bar\sigma ^B};{\cal G}^a,{\bar{\cal G}^a};\tau,{\bar\tau}) + \mu_3\Biggl\{{\cal V}^{\frac{1}{18}} + {\cal V}^{\frac{1}{36}}(\delta z_1 + \delta z_2) + (\delta z_1)^2 + (\delta z_2)^2 + \delta z_1\delta z_2 \nonumber\\
& & \hskip-0.4in + {\cal V}^{\frac{1}{36}}(\delta z_1 + \delta z_2 + c.c.) + |\delta z_1|^2 + |\delta z_2|^2 + \delta z_1{\bar\delta z_2} + \delta z_2{\bar\delta z_1})\Biggr\}+ i\kappa_4^2\mu_7C_{1{\bar 1}}\nonumber\\
& & \hskip-0.4in \times \left[{\cal V}^{-\frac{1}{2}} + {\cal V}^{-\frac{1}{4}}(\delta a_1 + {\bar\delta a_1}) + |\delta a_1|^2\right] -\gamma\Biggl[\frac{\zeta r_2^2}{r_1} + r_2\frac{(\delta z_1 + \delta z_2 + c.c.)}{{\cal V}^{\frac{1}{36}}} \nonumber\\
& & \hskip-0.4in + r_2\frac{(|\delta z_1|^2 + |\delta z_2|^2 + (\delta z_1)^2 + (\delta z_2)^2 + \delta z_1 \delta z_2 + c.c. + \delta z_1{\bar\delta z_2} + \delta z_2{\bar\delta z_1} }{{\cal V}^{\frac{1}{18}}}\Biggr]\Biggr)^{\frac{3}{2}}\nonumber\\
& & \hskip-0.4in - \Biggl({\cal T}_S(\sigma^S,{\bar\sigma^S};{\cal G}^a,{\bar{\cal G}^a};\tau,{\bar\tau})+ \mu_3\Biggl\{{\cal V}^{\frac{1}{18}} + {\cal V}^{\frac{1}{36}}(\delta z_1 + \delta z_2) + (\delta z_1)^2 + (\delta z_2)^2 + \delta z_1\delta z_2 \nonumber\\
& & \hskip-0.4in + {\cal V}^{\frac{1}{36}}(\delta z_1 + \delta z_2 + c.c.) + |\delta z_1|^2 + |\delta z_2|^2 + \delta z_1{\bar\delta z_2} + \delta z_2{\bar\delta z_1})\Biggr\} \nonumber\\
& & \hskip-0.4in -\gamma\Biggl[ r_2\frac{(|\delta z_1|^2 + |\delta z_2|^2 + (\delta z_1)^2 + (\delta z_2)^2 + \delta z_1 \delta z_2 + c.c. + \delta z_1{\bar\delta z_2} + \delta z_2{\bar\delta z_1} }{{\cal V}^{\frac{1}{18}}}\nonumber\\
& & \hskip-0.4in +\frac{\zeta r_2^2}{r_1} + r_2\frac{(\delta z_1 + \delta z_2 + c.c.)}{{\cal V}^{\frac{1}{36}}} +\Biggr]\Biggr)^{\frac{3}{2}} +\sum_\beta n^0_\beta f({\cal G}^a,{\bar{\cal G}^a};\tau,{\bar\tau})\Biggr]\nonumber\\
\end{eqnarray*}

\begin{eqnarray*}
& & \hskip-0.4in \sim-2ln\Biggl[\sum_\beta n^0_\beta f({\cal G}^a,{\bar{\cal G}^a};\tau,{\bar\tau}) + \Biggl\{{\cal T}_B(\sigma^B,{\bar\sigma ^B};{\cal G}^a,{\bar{\cal G}^a};\tau,{\bar\tau}) + \mu_3{\cal V}^{\frac{1}{18}} + i\kappa_4^2\mu_7C_{1{\bar 1}}{\cal V}^{-\frac{1}{2}} \nonumber\\
& & \hskip-0.4in - \gamma\left(r_2 + \frac{r_2^2\zeta}{r_1}\right)\Biggr\}^{\frac{3}{2}}- \Biggl\{{\cal T}_S(\sigma^S,{\bar\sigma ^S};{\cal G}^a,{\bar{\cal G}^a};\tau,{\bar\tau}) + \mu_3{\cal V}^{\frac{1}{18}}  - \gamma\left(r_2 + \frac{r_2^2\zeta}{r_1}\right)\Biggr\}^{\frac{3}{2}}\Biggr] + (\delta z_1 + \delta z_2 \nonumber\\
& & \hskip-0.4in + c.c.) \Biggl(\mu_3{\cal V}^{\frac{1}{36}}\Biggl[\sqrt{{\cal T}_B(\sigma^B,{\bar\sigma ^B};{\cal G}^a,{\bar{\cal G}^a};\tau,{\bar\tau}) + \mu_3{\cal V}^{\frac{1}{18}} + i\kappa_4^2\mu_7C_{1{\bar 1}}{\cal V}^{-\frac{1}{2}} - \gamma\left(r_2 + \frac{r_2^2\zeta}{r_1}\right)}\nonumber\\
 & & \hskip-0.4in - \sqrt{{\cal T}_S(\sigma^S,{\bar\sigma ^S};{\cal G}^a,{\bar{\cal G}^a};\tau,{\bar\tau}) + \mu_3{\cal V}^{\frac{1}{18}}  - \gamma\left(r_2 + \frac{r_2^2\zeta}{r_1}\right)}\Biggr]\nonumber\\
  & & \hskip-0.4in + \gamma r_2{\cal V}^{-\frac{1}{36}}\Biggl[\sqrt{{\cal T}_B(\sigma^B,{\bar\sigma ^B};{\cal G}^a,{\bar{\cal G}^a};\tau,{\bar\tau}) + \mu_3{\cal V}^{\frac{1}{18}} + i\kappa_4^2\mu_7C_{1{\bar 1}}{\cal V}^{-\frac{1}{2}} - \gamma\left(r_2 + \frac{r_2^2\zeta}{r_1}\right)}\nonumber\\
 & & \hskip-0.4in - \sqrt{{\cal T}_S(\sigma^S,{\bar\sigma ^S};{\cal G}^a,{\bar{\cal G}^a};\tau,{\bar\tau}) + \mu_3{\cal V}^{\frac{1}{18}}  - \gamma\left(r_2 + \frac{r_2^2\zeta}{r_1}\right)}\Biggr]\Biggr) + \left(\delta {\cal A}_1 + \delta{\bar{\cal A}_1}\right)\nonumber\\
& & \hskip-0.4in \Biggl(i\kappa^2_4\mu_7C_{1{\bar 1}}{\cal V}^{-\frac{1}{2}}\sqrt{{\cal T}_B(\sigma^B,{\bar\sigma ^B};{\cal G}^a,{\bar{\cal G}^a};\tau,{\bar\tau}) + \mu_3{\cal V}^{\frac{1}{18}} + i\kappa_4^2\mu_7C_{1{\bar 1}}{\cal V}^{-\frac{1}{2}} - \gamma\left(r_2 + \frac{r_2^2\zeta}{r_1}\right)}\Biggr)\nonumber\\
& & \hskip-0.4in \left(|\delta z_1|^2 + |\delta z_2|^2 + \delta z_1{\bar\delta z_2} + \delta z_2{\bar\delta z_1}\right)\Biggl(\Biggl[\frac{-1}{\sqrt{{\cal T}_S(\sigma^S,{\bar\sigma ^S};{\cal G}^a,{\bar{\cal G}^a};\tau,{\bar\tau}) + \mu_3{\cal V}^{\frac{1}{18}}  - \gamma\left(r_2 + \frac{r_2^2\zeta}{r_1}\right)}}\nonumber\\
& & \hskip-0.4in +\frac{1}{\sqrt{{\cal T}_B(\sigma^B,{\bar\sigma ^B};{\cal G}^a,{\bar{\cal G}^a};\tau,{\bar\tau}) + \mu_3{\cal V}^{\frac{1}{18}} + i\kappa_4^2\mu_7C_{1{\bar 1}}{\cal V}^{-\frac{1}{2}} - \gamma\left(r_2 + \frac{r_2^2\zeta}{r_1}\right)}}\Biggr](\mu_3{\cal V}^{\frac{1}{36}})^2 \nonumber\\
& & \hskip-0.4in+ (\gamma r_2{\cal V}^{-\frac{1}{36}})^2\Biggl[\sqrt{{\cal T}_B(\sigma^B,{\bar\sigma ^B};{\cal G}^a,{\bar{\cal G}^a};\tau,{\bar\tau}) + \mu_3{\cal V}^{\frac{1}{18}} + i\kappa_4^2\mu_7C_{1{\bar 1}}{\cal V}^{-\frac{1}{2}} - \gamma\left(r_2 + \frac{r_2^2\zeta}{r_1}\right)}\nonumber\\
 & & \hskip-0.4in - \sqrt{{\cal T}_S(\sigma^S,{\bar\sigma ^S};{\cal G}^a,{\bar{\cal G}^a};\tau,{\bar\tau}) + \mu_3{\cal V}^{\frac{1}{18}}  - \gamma\left(r_2 + \frac{r_2^2\zeta}{r_1}\right)}\Biggr]\nonumber\\
 & &\hskip-0.4in +\mu_3\gamma r_2\Biggl[\frac{1}{\sqrt{{\cal T}_B(\sigma^B,{\bar\sigma ^B};{\cal G}^a,{\bar{\cal G}^a};\tau,{\bar\tau}) + \mu_3{\cal V}^{\frac{1}{18}} + i\kappa_4^2\mu_7C_{1{\bar 1}}{\cal V}^{-\frac{1}{2}} - \gamma\left(r_2 + \frac{r_2^2\zeta}{r_1}\right)}}\nonumber\\
 & & \hskip-0.4in -\frac{1}{\sqrt{{\cal T}_S(\sigma^S,{\bar\sigma ^S};{\cal G}^a,{\bar{\cal G}^a};\tau,{\bar\tau}) + \mu_3{\cal V}^{\frac{1}{18}}  - \gamma\left(r_2 + \frac{r_2^2\zeta}{r_1}\right)}}\Biggr]\nonumber\\
 & & \hskip-0.4in+\mu_3\Biggl[\sqrt{{\cal T}_B(\sigma^B,{\bar\sigma ^B};{\cal G}^a,{\bar{\cal G}^a};\tau,{\bar\tau}) + \mu_3{\cal V}^{\frac{1}{18}} + i\kappa_4^2\mu_7C_{1{\bar 1}}{\cal V}^{-\frac{1}{2}} - \gamma\left(r_2 + \frac{r_2^2\zeta}{r_1}\right)}\nonumber\\
  \end{eqnarray*}

\begin{eqnarray}
\label{eq:K_fluc_1}
 & & \hskip-0.4in - \sqrt{{\cal T}_S(\sigma^S,{\bar\sigma ^S};{\cal G}^a,{\bar{\cal G}^a};\tau,{\bar\tau}) + \mu_3{\cal V}^{\frac{1}{18}}  - \gamma\left(r_2 + \frac{r_2^2\zeta}{r_1}\right)}\Biggr]\Biggr)\nonumber\\
 & & \hskip-0.4in +|{\cal A}_1|^2\Biggl\{\frac{i\kappa_4^2\mu_7C_{1{\bar 1}}{\cal V}^{-\frac{1}{2}}}{\sqrt{{\cal T}_B(\sigma^B,{\bar\sigma ^B};{\cal G}^a,{\bar{\cal G}^a};\tau,{\bar\tau}) + \mu_3{\cal V}^{\frac{1}{18}} + i\kappa_4^2\mu_7C_{1{\bar 1}}{\cal V}^{-\frac{1}{2}} - \gamma\left(r_2 + \frac{r_2^2\zeta}{r_1}\right)}} \nonumber\\
 & & \hskip-0.4in + i\kappa_4^2\mu_7C_{1{\bar 1}}\sqrt{{\cal T}_B(\sigma^B,{\bar\sigma ^B};{\cal G}^a,{\bar{\cal G}^a};\tau,{\bar\tau}) + \mu_3{\cal V}^{\frac{1}{18}} + i\kappa_4^2\mu_7C_{1{\bar 1}}{\cal V}^{-\frac{1}{2}} - \gamma\left(r_2 + \frac{r_2^2\zeta}{r_1}\right)}\Biggr\}
 \nonumber\\
 & & \hskip-0.4in + \left(\delta z_1{\delta\bar{\cal A}_1} + \delta z_2{\delta\bar{\cal A}_1} + \delta{\bar z_1}\delta{\cal A}_1 + \delta{\bar z_2}\delta{\cal A}_1\right)\nonumber\\
 & & \hskip-0.4in
 \Biggl\{\frac{i\kappa_4^2\mu_7{\cal V}^{-\frac{1}{4}}\mu_3{\cal V}^{\frac{1}{36}}}{\sqrt{{\cal T}_B(\sigma^B,{\bar\sigma ^B};{\cal G}^a,{\bar{\cal G}^a};\tau,{\bar\tau}) + \mu_3{\cal V}^{\frac{1}{18}} + i\kappa_4^2\mu_7C_{1{\bar 1}}{\cal V}^{-\frac{1}{2}} - \gamma\left(r_2 + \frac{r_2^2\zeta}{r_1}\right)}}\Biggr\}\Biggr]\nonumber\\
\end{eqnarray}

\item
Using (\ref{eq:K_geom_fluc}) and (\ref{eq:K_fluc_1}), one arrives at (\ref{eq:K2}), wherein:
\begin{eqnarray*}
& & \hskip-0.4in K_{z_i{\bar z}_j}\sim\Biggl[\Biggl(\sqrt{{\cal T}_B(\sigma^B,{\bar\sigma ^B};{\cal G}^a,{\bar{\cal G}^a};\tau,{\bar\tau}) + \mu_3{\cal V}^{\frac{1}{18}} + i\kappa_4^2\mu_7C_{1{\bar 1}}{\cal V}^{-\frac{1}{2}} - \gamma\left(r_2 + \frac{r_2^2\zeta}{r_1}\right)}\nonumber\\
& & \hskip-0.4in   -\sqrt{{\cal T}_S(\sigma^S,{\bar\sigma ^S};{\cal G}^a,{\bar{\cal G}^a};\tau,{\bar\tau}) + \mu_3{\cal V}^{\frac{1}{18}}  - \gamma\left(r_2 + \frac{r_2^2\zeta}{r_1}\right)}\Biggr)^2 \times \frac{(\mu_3{\cal V}^{\frac{1}{36}}+\gamma r_2{\cal V}^{-\frac{1}{36}})^2}{\Xi^2} \nonumber\\
   & & \hskip-0.4in + \Biggl\{\frac{1}{\sqrt{{\cal T}_B(\sigma^B,{\bar\sigma ^B};{\cal G}^a,{\bar{\cal G}^a};\tau,{\bar\tau}) + \mu_3{\cal V}^{\frac{1}{18}} + i\kappa_4^2\mu_7C_{1{\bar 1}}{\cal V}^{-\frac{1}{2}} - \gamma\left(r_2 + \frac{r_2^2\zeta}{r_1}\right)}}\nonumber\\
 & & \hskip-0.4in -\frac{1}{\sqrt{{\cal T}_S(\sigma^S,{\bar\sigma ^S};{\cal G}^a,{\bar{\cal G}^a};\tau,{\bar\tau}) + \mu_3{\cal V}^{\frac{1}{18}}  - \gamma\left(r_2 + \frac{r_2^2\zeta}{r_1}\right)}}\Biggr\}\times\left(\frac{(\mu_3{\cal V}^{\frac{1}{36}})^2 + \mu_3\gamma r_2}{\Xi}  \right)\nonumber\\
 & & \hskip-0.4in +\frac{(\gamma r_2{\cal V}^{-\frac{1}{36}})^2 + \mu_3}{\Xi}\Biggl\{\sqrt{{\cal T}_B(\sigma^B,{\bar\sigma ^B};{\cal G}^a,{\bar{\cal G}^a};\tau,{\bar\tau}) + \mu_3{\cal V}^{\frac{1}{18}} + i\kappa_4^2\mu_7C_{1{\bar 1}}{\cal V}^{-\frac{1}{2}} - \gamma\left(r_2 + \frac{r_2^2\zeta}{r_1}\right)}\nonumber\\
 & & \hskip-0.4in - \sqrt{{\cal T}_S(\sigma^S,{\bar\sigma ^S};{\cal G}^a,{\bar{\cal G}^a};\tau,{\bar\tau}) + \mu_3{\cal V}^{\frac{1}{18}}  - \gamma\left(r_2 + \frac{r_2^2\zeta}{r_1}\right)}\Biggr\}\Biggr];\nonumber\\
  \end{eqnarray*}

\begin{eqnarray}
\label{eq:K2defs}
   & & \hskip-0.4in K_{{\cal A}_1\bar{\cal A}_1}\sim\Biggl\{\frac{(i\kappa_4^2\mu_7C_{1{\bar 1}}{\cal V}^{-\frac{1}{4}})^2}{\Xi\sqrt{{\cal T}_B(\sigma^B,{\bar\sigma ^B};{\cal G}^a,{\bar{\cal G}^a};\tau,{\bar\tau}) + \mu_3{\cal V}^{\frac{1}{18}} + i\kappa_4^2\mu_7C_{1{\bar 1}}{\cal V}^{-\frac{1}{2}} - \gamma\left(r_2 + \frac{r_2^2\zeta}{r_1}\right)}} \nonumber\\
   & & \hskip-0.4in + \frac{i\kappa_4^2\mu_7C_{1{\bar 1}}}{\Xi}\sqrt{{\cal T}_B(\sigma^B,{\bar\sigma ^B};{\cal G}^a,{\bar{\cal G}^a};\tau,{\bar\tau}) + \mu_3{\cal V}^{\frac{1}{18}} + i\kappa_4^2\mu_7C_{1{\bar 1}}{\cal V}^{-\frac{1}{2}} - \gamma\left(r_2 + \frac{r_2^2\zeta}{r_1}\right)}\nonumber\\
   & & \hskip-0.4in + \frac{\left(i\kappa_4^2\mu_7C_{1{\bar 1}}{\cal V}^{-\frac{1}{4}}\sqrt{{\cal T}_B(\sigma^B,{\bar\sigma ^B};{\cal G}^a,{\bar{\cal G}^a};\tau,{\bar\tau}) + \mu_3{\cal V}^{\frac{1}{18}} + i\kappa_4^2\mu_7C_{1{\bar 1}}{\cal V}^{-\frac{1}{2}} - \gamma\left(r_2 + \frac{r_2^2\zeta}{r_1}\right)}\right)^2}{\Xi^2}\Biggr\};\nonumber\\
   & & \hskip-0.4in K_{z_i\bar{\cal A}_1}\sim
\Biggl\{\frac{(i\kappa_4^2\mu_7C_{1{\bar 1}}{\cal V}^{-\frac{1}{4}})(\mu_3{\cal V}^{\frac{1}{36}})}{\Xi\sqrt{{\cal T}_B(\sigma^B,{\bar\sigma ^B};{\cal G}^a,{\bar{\cal G}^a};\tau,{\bar\tau}) + \mu_3{\cal V}^{\frac{1}{18}} + i\kappa_4^2\mu_7C_{1{\bar 1}}{\cal V}^{-\frac{1}{2}} - \gamma\left(r_2 + \frac{r_2^2\zeta}{r_1}\right)}} \nonumber\\
   & & \hskip-0.4in + \frac{\sqrt{{\cal T}_B(\sigma^B,{\bar\sigma ^B};{\cal G}^a,{\bar{\cal G}^a};\tau,{\bar\tau}) + \mu_3{\cal V}^{\frac{1}{18}} + i\kappa_4^2\mu_7C_{1{\bar 1}}{\cal V}^{-\frac{1}{2}} - \gamma\left(r_2 + \frac{r_2^2\zeta}{r_1}\right)}}{\Xi^2}\nonumber\\
   & & \hskip-0.4in \times \Biggl(\sqrt{{\cal T}_B(\sigma^B,{\bar\sigma ^B};{\cal G}^a,{\bar{\cal G}^a};\tau,{\bar\tau}) + \mu_3{\cal V}^{\frac{1}{18}} + i\kappa_4^2\mu_7C_{1{\bar 1}}{\cal V}^{-\frac{1}{2}} - \gamma\left(r_2 + \frac{r_2^2\zeta}{r_1}\right)}\nonumber\\
   & & \hskip-0.4in    -\sqrt{{\cal T}_S(\sigma^S,{\bar\sigma ^S};{\cal G}^a,{\bar{\cal G}^a};\tau,{\bar\tau}) + \mu_3{\cal V}^{\frac{1}{18}}  - \gamma\left(r_2 + \frac{r_2^2\zeta}{r_1}\right)}\Biggr)i\kappa_4^2\mu_7C_{1{\bar 1}}{\cal V}^{-\frac{1}{4}}(\mu_3{\cal V}^{\frac{1}{36}}+\gamma r_2{\cal V}^{-\frac{1}{36}})\Biggr\};\nonumber\\
   & &
\end{eqnarray}
and
\begin{eqnarray}
\label{eq:Xi_def}
& &
\Xi\equiv \left({\cal T}_B(\sigma^B,{\bar\sigma ^B};{\cal G}^a,{\bar{\cal G}^a};\tau,{\bar\tau}) + \mu_3{\cal V}^{\frac{1}{18}} + i\kappa_4^2\mu_7C_{1{\bar 1}}{\cal V}^{-\frac{1}{2}} - \gamma\left(r_2 + \frac{r_2^2\zeta}{r_1}\right)\right)^\frac{3}{2}\nonumber\\
& &
-\left({\cal T}_S(\sigma^S,{\bar\sigma ^S};{\cal G}^a,{\bar{\cal G}^a};\tau,{\bar\tau}) + \mu_3{\cal V}^{\frac{1}{18}}  - \gamma\left(r_2 + \frac{r_2^2\zeta}{r_1}\right)\right)^{\frac{3}{2}}
+\sum_\beta n^0_\beta f({\cal G}^a,{\bar{\cal G}^a};\tau,{\bar\tau}).\nonumber\\
\end{eqnarray}

\end{itemize}

\section{First and Second Derivatives of $\hat{K}_{{\cal Z}_i{\bar{\cal Z}}_i}$ and $\hat{K}_{\tilde{{\cal A}_1}{\bar{\tilde{{\cal A}_1}}}}$ and First Derivatives of det$\hat{K}_{i{\bar j}}$ with respect to Closed String Moduli $\sigma^\alpha,{\cal G}^a$}
\setcounter{equation}{0}
\seceqff

The first and second derivatives of $\hat{K}_{{\cal Z}_i{\bar{\cal Z}}_i}$ and $\hat{K}_{\tilde{{\cal A}_1}{\bar{\tilde{\cal A}_1}}}$ are relevant to the calculation of the soft SUSY breaking parameters in section {\bf 5}. We can show that:
\begin{eqnarray*}
& & \partial_{\sigma^\alpha}\hat{K}_{{\cal Z}_i{\bar{\cal Z}}_i}\sim\frac{\sqrt{{\cal T}_\alpha(\sigma^\alpha,{\bar\sigma ^\alpha};{\cal G}^a,{\bar{\cal G}^a};\tau,{\bar\tau}) + \mu_3{\cal V}^{\frac{1}{18}} + i\kappa_4^2\mu_7C_{1{\bar 1}}{\cal V}^{-\frac{1}{2}} - \gamma\left(r_2 + \frac{r_2^2\zeta}{r_1}\right)}}{\Xi^2}\nonumber\\
& & \times\Biggl\{\frac{1}{\sqrt{{\cal T}_B(\sigma^B,{\bar\sigma ^B};{\cal G}^a,{\bar{\cal G}^a};\tau,{\bar\tau}) + \mu_3{\cal V}^{\frac{1}{18}} + i\kappa_4^2\mu_7C_{1{\bar 1}}{\cal V}^{-\frac{1}{2}} - \gamma\left(r_2 + \frac{r_2^2\zeta}{r_1}\right)}}\nonumber\\
 & & -\frac{1}{\sqrt{{\cal T}_S(\sigma^S,{\bar\sigma ^S};{\cal G}^a,{\bar{\cal G}^a};\tau,{\bar\tau}) + \mu_3{\cal V}^{\frac{1}{18}}  - \gamma\left(r_2 + \frac{r_2^2\zeta}{r_1}\right)}}\Biggr\}(\mu_3{\cal V}^{\frac{1}{36}}+\gamma r_2{\cal V}^{-\frac{1}{36}})\nonumber\\
 & & + \frac{(\mu_3{\cal V}^{\frac{1}{36}}+\gamma r_2{\cal V}^{-\frac{1}{36}})}{\left({\cal T}_\alpha(\sigma^\alpha,{\bar\sigma ^\alpha};{\cal G}^a,{\bar{\cal G}^a};\tau,{\bar\tau}) + \mu_3{\cal V}^{\frac{1}{18}} + i\kappa_4^2\mu_7C_{1{\bar 1}}{\cal V}^{-\frac{1}{2}} - \gamma\left(r_2 + \frac{r_2^2\zeta}{r_1}\right)\right)^{\frac{3}{2}}\Xi}\nonumber\\
 & & + \frac{\sqrt{{\cal T}_\alpha(\sigma^\alpha,{\bar\sigma ^\alpha};{\cal G}^a,{\bar{\cal G}^a};\tau,{\bar\tau}) + \mu_3{\cal V}^{\frac{1}{18}} + i\kappa_4^2\mu_7C_{1{\bar 1}}{\cal V}^{-\frac{1}{2}} - \gamma\left(r_2 + \frac{r_2^2\zeta}{r_1}\right)}\left((\gamma r_2{\cal V}^{-\frac{1}{36}})^2 + \mu_3\right)}{\Xi^2}\nonumber\\
 & & \frac{\left((\gamma r_2{\cal V}^{-\frac{1}{36}})^2 + \mu_3\right)}{\Xi\sqrt{{\cal T}_\alpha(\sigma^\alpha,{\bar\sigma ^\alpha};{\cal G}^a,{\bar{\cal G}^a};\tau,{\bar\tau}) + \mu_3{\cal V}^{\frac{1}{18}} + i\kappa_4^2\mu_7C_{1{\bar 1}}{\cal V}^{-\frac{1}{2}} - \gamma\left(r_2 + \frac{r_2^2\zeta}{r_1}\right)}}
 \sim{\cal V}^{-\frac{35}{18}};\nonumber\\
 & & \partial_{{\cal G}^a}\hat{K}_{{\cal Z}_i{\bar{\cal Z}}_i}\sim\frac{(\mu_3{\cal V}^{\frac{1}{36}}+\gamma r_2{\cal V}^{-\frac{1}{36}})}{\Xi^2} \Biggl[\sum_\beta k^a n^0_\beta sin(...) + ({\cal G}^a,{\bar{\cal G}^a})\nonumber\\
 & & \times \Biggl\{\sqrt{{\cal T}_B(\sigma^B,{\bar\sigma ^B};{\cal G}^a,{\bar{\cal G}^a};\tau,{\bar\tau}) + \mu_3{\cal V}^{\frac{1}{18}} + i\kappa_4^2\mu_7C_{1{\bar 1}}{\cal V}^{-\frac{1}{2}} - \gamma\left(r_2 + \frac{r_2^2\zeta}{r_1}\right)}\nonumber\\
       & & - \sqrt{{\cal T}_S(\sigma^S,{\bar\sigma ^S};{\cal G}^a,{\bar{\cal G}^a};\tau,{\bar\tau}) + \mu_3{\cal V}^{\frac{1}{18}}  - \gamma\left(r_2 + \frac{r_2^2\zeta}{r_1}\right)}\Biggr\}\Biggr]\nonumber\\
       \end{eqnarray*}

\begin{eqnarray}
\label{eq:dKhat_z}
 & & \times\Biggl\{\frac{1}{\sqrt{{\cal T}_B(\sigma^B,{\bar\sigma ^B};{\cal G}^a,{\bar{\cal G}^a};\tau,{\bar\tau}) + \mu_3{\cal V}^{\frac{1}{18}} + i\kappa_4^2\mu_7C_{1{\bar 1}}{\cal V}^{-\frac{1}{2}} - \gamma\left(r_2 + \frac{r_2^2\zeta}{r_1}\right)}}\nonumber\\
 & & -\frac{1}{\sqrt{{\cal T}_S(\sigma^S,{\bar\sigma ^S};{\cal G}^a,{\bar{\cal G}^a};\tau,{\bar\tau}) + \mu_3{\cal V}^{\frac{1}{18}}  - \gamma\left(r_2 + \frac{r_2^2\zeta}{r_1}\right)}}\Biggr\}\frac{\left((\gamma r_2{\cal V}^{-\frac{1}{36}})^2 + \mu_3\right)}{\Xi^2}\nonumber\\
 & & \times\Biggl[({\cal G}^a,{\bar{\cal G}^a})\Biggl\{\sqrt{{\cal T}_B(\sigma^B,{\bar\sigma ^B};{\cal G}^a,{\bar{\cal G}^a};\tau,{\bar\tau}) + \mu_3{\cal V}^{\frac{1}{18}} + i\kappa_4^2\mu_7C_{1{\bar 1}}{\cal V}^{-\frac{1}{2}} - \gamma\left(r_2 + \frac{r_2^2\zeta}{r_1}\right)} \nonumber\\
 & & -\sqrt{{\cal T}_S(\sigma^S,{\bar\sigma ^S};{\cal G}^a,{\bar{\cal G}^a};\tau,{\bar\tau}) + \mu_3{\cal V}^{\frac{1}{18}}  - \gamma\left(r_2 + \frac{r_2^2\zeta}{r_1}\right)}\Biggr\}+\sum_\beta k^a n^0_\beta sin(...) \Biggr]\nonumber\\
 & & \times \Biggl\{\sqrt{{\cal T}_B(\sigma^B,{\bar\sigma ^B};{\cal G}^a,{\bar{\cal G}^a};\tau,{\bar\tau}) + \mu_3{\cal V}^{\frac{1}{18}} + i\kappa_4^2\mu_7C_{1{\bar 1}}{\cal V}^{-\frac{1}{2}} - \gamma\left(r_2 + \frac{r_2^2\zeta}{r_1}\right)}\nonumber\\
 & &    -\sqrt{{\cal T}_S(\sigma^S,{\bar\sigma ^S};{\cal G}^a,{\bar{\cal G}^a};\tau,{\bar\tau}) + \mu_3{\cal V}^{\frac{1}{18}}  - \gamma\left(r_2 + \frac{r_2^2\zeta}{r_1}\right)}\Biggr\} \sim{\cal V}^{\frac{41}{36}}.\nonumber\\
   & &
\end{eqnarray}
Hence,
\begin{eqnarray*}
& & \partial_{\sigma^B{\bar\sigma}^B}\hat{K}_{{\cal Z}_i{\bar{\cal Z}}_i}\sim\frac{\sqrt{{\cal T}_B(\sigma^B,{\bar\sigma ^B};{\cal G}^a,{\bar{\cal G}^a};\tau,{\bar\tau}) + \mu_3{\cal V}^{\frac{1}{18}} + i\kappa_4^2\mu_7C_{1{\bar 1}}{\cal V}^{-\frac{1}{2}} - \gamma\left(r_2 + \frac{r_2^2\zeta}{r_1}\right)}}{\Xi^3}\nonumber\\
& & \times\Biggl\{\frac{1}{\sqrt{{\cal T}_B(\sigma^B,{\bar\sigma ^B};{\cal G}^a,{\bar{\cal G}^a};\tau,{\bar\tau}) + \mu_3{\cal V}^{\frac{1}{18}} + i\kappa_4^2\mu_7C_{1{\bar 1}}{\cal V}^{-\frac{1}{2}} - \gamma\left(r_2 + \frac{r_2^2\zeta}{r_1}\right)}}\nonumber\\
 & & -\frac{1}{\sqrt{{\cal T}_S(\sigma^S,{\bar\sigma ^S};{\cal G}^a,{\bar{\cal G}^a};\tau,{\bar\tau}) + \mu_3{\cal V}^{\frac{1}{18}}  - \gamma\left(r_2 + \frac{r_2^2\zeta}{r_1}\right)}}\Biggr\}\times (\mu_3{\cal V}^{\frac{1}{36}}+\gamma r_2{\cal V}^{-\frac{1}{36}})\nonumber\\
 & & + \frac{(\mu_3{\cal V}^{\frac{1}{36}}+\gamma r_2{\cal V}^{-\frac{1}{36}})}{\Xi\left({\cal T}_B(\sigma^B,{\bar\sigma ^B};{\cal G}^a,{\bar{\cal G}^a};\tau,{\bar\tau}) + \mu_3{\cal V}^{\frac{1}{18}} + i\kappa_4^2\mu_7C_{1{\bar 1}}{\cal V}^{-\frac{1}{2}} - \gamma\left(r_2 + \frac{r_2^2\zeta}{r_1}\right)\right)}\nonumber\\
   \end{eqnarray*}

\begin{eqnarray*}
 & & \times\Biggl\{\frac{1}{\sqrt{{\cal T}_B(\sigma^B,{\bar\sigma ^B};{\cal G}^a,{\bar{\cal G}^a};\tau,{\bar\tau}) + \mu_3{\cal V}^{\frac{1}{18}} + i\kappa_4^2\mu_7C_{1{\bar 1}}{\cal V}^{-\frac{1}{2}} - \gamma\left(r_2 + \frac{r_2^2\zeta}{r_1}\right)}}\nonumber\\
 & & -\frac{1}{\sqrt{{\cal T}_S(\sigma^S,{\bar\sigma ^S};{\cal G}^a,{\bar{\cal G}^a};\tau,{\bar\tau}) + \mu_3{\cal V}^{\frac{1}{18}}  - \gamma\left(r_2 + \frac{r_2^2\zeta}{r_1}\right)}}\Biggr\}\nonumber\\
 & & \frac{(\mu_3{\cal V}^{\frac{1}{36}}+\gamma r_2{\cal V}^{-\frac{1}{36}})}{\Xi^2\left({\cal T}_B(\sigma^B,{\bar\sigma ^B};{\cal G}^a,{\bar{\cal G}^a};\tau,{\bar\tau}) + \mu_3{\cal V}^{\frac{1}{18}} + i\kappa_4^2\mu_7C_{1{\bar 1}}{\cal V}^{-\frac{1}{2}} - \gamma\left(r_2 + \frac{r_2^2\zeta}{r_1}\right)\right)}\nonumber\\
 & & \frac{(\mu_3{\cal V}^{\frac{1}{36}}+\gamma r_2{\cal V}^{-\frac{1}{36}})}{\Xi\left({\cal T}_B(\sigma^B,{\bar\sigma ^B};{\cal G}^a,{\bar{\cal G}^a};\tau,{\bar\tau}) + \mu_3{\cal V}^{\frac{1}{18}} + i\kappa_4^2\mu_7C_{1{\bar 1}}{\cal V}^{-\frac{1}{2}} - \gamma\left(r_2 + \frac{r_2^2\zeta}{r_1}\right)\right)^{\frac{5}{2}}}\nonumber\\
 & & + \frac{\left(\mu_3+\left\{\gamma r_2{\cal V}^{-\frac{1}{36}}\right\}\right)}{\Xi^2}+\left(\mu_3+\left\{\gamma r_2{\cal V}^{-\frac{1}{36}}\right\}\right)\nonumber\\
 & & \times\frac{\left[{\cal T}_B(\sigma^B,{\bar\sigma ^B};{\cal G}^a,{\bar{\cal G}^a};\tau,{\bar\tau}) + \mu_3{\cal V}^{\frac{1}{18}} + i\kappa_4^2\mu_7C_{1{\bar 1}}{\cal V}^{-\frac{1}{2}} - \gamma\left(r_2 + \frac{r_2^2\zeta}{r_1}\right)\right]^{\frac{3}{2}}}{\Xi^3}\nonumber\\
 & &  + \frac{\left(\mu_3+\left\{\gamma r_2{\cal V}^{-\frac{1}{36}}\right\}\right)}{\Xi\left[{\cal T}_B(\sigma^B,{\bar\sigma ^B};{\cal G}^a,{\bar{\cal G}^a};\tau,{\bar\tau}) + \mu_3{\cal V}^{\frac{1}{18}} + i\kappa_4^2\mu_7C_{1{\bar 1}}{\cal V}^{-\frac{1}{2}} - \gamma\left(r_2 + \frac{r_2^2\zeta}{r_1}\right)\right]^{\frac{3}{2}}}
 \nonumber\\
 & & \sim {\cal V}^{-\frac{19}{18}}\sim\partial_{\sigma^S{\bar\sigma}^S}\hat{K}_{{\cal Z}_i{\bar{\cal Z}}_i};\nonumber\\
& & \partial_{\sigma^B}{\bar\partial}_{{\bar\sigma}^S}\hat{K}_{{\cal Z}_i{\bar{\cal Z}}_i}\sim \frac{(\mu_3{\cal V}^{\frac{1}{36}}+\gamma r_2{\cal V}^{-\frac{1}{36}})}{\Xi^3}\nonumber\\
& & \times \sqrt{{\cal T}_B(\sigma^B,{\bar\sigma ^B};{\cal G}^a,{\bar{\cal G}^a};\tau,{\bar\tau}) + \mu_3{\cal V}^{\frac{1}{18}} + i\kappa_4^2\mu_7C_{1{\bar 1}}{\cal V}^{-\frac{1}{2}} - \gamma\left(r_2 + \frac{r_2^2\zeta}{r_1}\right)}\nonumber\\
& & \times \sqrt{{\cal T}_S(\sigma^S,{\bar\sigma ^S};{\cal G}^a,{\bar{\cal G}^a};\tau,{\bar\tau}) + \mu_3{\cal V}^{\frac{1}{18}}  - \gamma\left(r_2 + \frac{r_2^2\zeta}{r_1}\right)}\nonumber\\
& & \times\Biggl\{\frac{1}{\sqrt{{\cal T}_B(\sigma^B,{\bar\sigma ^B};{\cal G}^a,{\bar{\cal G}^a};\tau,{\bar\tau}) + \mu_3{\cal V}^{\frac{1}{18}} + i\kappa_4^2\mu_7C_{1{\bar 1}}{\cal V}^{-\frac{1}{2}} - \gamma\left(r_2 + \frac{r_2^2\zeta}{r_1}\right)}}\nonumber\\
 \end{eqnarray*}

\begin{eqnarray*}
 & & -\frac{1}{\sqrt{{\cal T}_S(\sigma^S,{\bar\sigma ^S};{\cal G}^a,{\bar{\cal G}^a};\tau,{\bar\tau}) + \mu_3{\cal V}^{\frac{1}{18}}  - \gamma\left(r_2 + \frac{r_2^2\zeta}{r_1}\right)}}\Biggr\}\nonumber\\
 & & + \frac{(\mu_3{\cal V}^{\frac{1}{36}}+\gamma r_2{\cal V}^{-\frac{1}{36}})}{\Xi^2\left[{\cal T}_S(\sigma^S,{\bar\sigma ^S};{\cal G}^a,{\bar{\cal G}^a};\tau,{\bar\tau}) + \mu_3{\cal V}^{\frac{1}{18}}  - \gamma\left(r_2 + \frac{r_2^2\zeta}{r_1}\right)\right]}\nonumber\\
 & & \times \sqrt{\frac{{\cal T}_B(\sigma^B,{\bar\sigma ^B};{\cal G}^a,{\bar{\cal G}^a};\tau,{\bar\tau}) + \mu_3{\cal V}^{\frac{1}{18}} + i\kappa_4^2\mu_7C_{1{\bar 1}}{\cal V}^{-\frac{1}{2}} - \gamma\left(r_2 + \frac{r_2^2\zeta}{r_1}\right)}{{\cal T}_S(\sigma^S,{\bar\sigma ^S};{\cal G}^a,{\bar{\cal G}^a};\tau,{\bar\tau}) + \mu_3{\cal V}^{\frac{1}{18}}  - \gamma\left(r_2 + \frac{r_2^2\zeta}{r_1}\right)}}\nonumber\\
 & & + \frac{(\mu_3{\cal V}^{\frac{1}{36}}+\gamma r_2{\cal V}^{-\frac{1}{36}})}{\Xi^2\left[{\cal T}_B(\sigma^B,{\bar\sigma^B};{\cal G}^a,{\bar{\cal G}^a};\tau,{\bar\tau}) + \mu_3{\cal V}^{\frac{1}{18}}  - \gamma\left(r_2 + \frac{r_2^2\zeta}{r_1}\right)\right]}\nonumber\\
 & & \times \sqrt{\frac{{\cal T}_S(\sigma^S,{\bar\sigma ^S};{\cal G}^a,{\bar{\cal G}^a};\tau,{\bar\tau}) + \mu_3{\cal V}^{\frac{1}{18}} + i\kappa_4^2\mu_7C_{1{\bar 1}}{\cal V}^{-\frac{1}{2}} - \gamma\left(r_2 + \frac{r_2^2\zeta}{r_1}\right)}{{\cal T}_B(\sigma^B,{\bar\sigma ^B};{\cal G}^a,{\bar{\cal G}^a};\tau,{\bar\tau}) + \mu_3{\cal V}^{\frac{1}{18}}  - \gamma\left(r_2 + \frac{r_2^2\zeta}{r_1}\right)}}\nonumber\\
 & & + \frac{(\mu_3{\cal V}^{\frac{1}{36}}+\gamma r_2{\cal V}^{-\frac{1}{36}})\sqrt{{\cal T}_S(\sigma^S,{\bar\sigma ^S};{\cal G}^a,{\bar{\cal G}^a};\tau,{\bar\tau}) + \mu_3{\cal V}^{\frac{1}{18}}  - \gamma\left(r_2 + \frac{r_2^2\zeta}{r_1}\right)}}{\Xi^2\left({\cal T}_B(\sigma^B,{\bar\sigma ^B};{\cal G}^a,{\bar{\cal G}^a};\tau,{\bar\tau}) + \mu_3{\cal V}^{\frac{1}{18}} + i\kappa_4^2\mu_7C_{1{\bar 1}}{\cal V}^{-\frac{1}{2}} - \gamma\left(r_2 + \frac{r_2^2\zeta}{r_1}\right)\right)^{\frac{3}{2}}}\nonumber\\
 & & + \frac{\left(\mu_3+\left\{\gamma r_2{\cal V}^{-\frac{1}{36}}\right\}\right)\sqrt{{\cal T}_S(\sigma^S,{\bar\sigma ^S};{\cal G}^a,{\bar{\cal G}^a};\tau,{\bar\tau}) + \mu_3{\cal V}^{\frac{1}{18}}  - \gamma\left(r_2 + \frac{r_2^2\zeta}{r_1}\right)}}{\Xi^3}\nonumber\\
  & & \times\left[{\cal T}_B(\sigma^B,{\bar\sigma ^B};{\cal G}^a,{\bar{\cal G}^a};\tau,{\bar\tau}) + \mu_3{\cal V}^{\frac{1}{18}} + i\kappa_4^2\mu_7C_{1{\bar 1}}{\cal V}^{-\frac{1}{2}} - \gamma\left(r_2 + \frac{r_2^2\zeta}{r_1}\right)\right]\nonumber\\
 & & + \sqrt{\frac{{\cal T}_B(\sigma^B,{\bar\sigma ^B};{\cal G}^a,{\bar{\cal G}^a};\tau,{\bar\tau}) + \mu_3{\cal V}^{\frac{1}{18}} + i\kappa_4^2\mu_7C_{1{\bar 1}}{\cal V}^{-\frac{1}{2}} - \gamma\left(r_2 + \frac{r_2^2\zeta}{r_1}\right)}{{\cal T}_S(\sigma^S,{\bar\sigma ^S};{\cal G}^a,{\bar{\cal G}^a};\tau,{\bar\tau}) + \mu_3{\cal V}^{\frac{1}{18}}  - \gamma\left(r_2 + \frac{r_2^2\zeta}{r_1}\right)}}\nonumber\\
 & & \times \left(\frac{\left(\mu_3+\left\{\gamma r_2{\cal V}^{-\frac{1}{36}}\right\}\right)}{\Xi^2}\right)\sim {\cal V}^{-2};\nonumber\\
 \end{eqnarray*}

\begin{eqnarray*}
 & & \partial_{\sigma^\alpha}{\bar\partial}_{{\bar{\cal G}}^a}\hat{K}_{{\cal Z}_i{\bar{\cal Z}}_i}
 \sim\frac{\left(\mu_3{\cal V}^{\frac{1}{36}}+\gamma r_2{\cal V}^{-\frac{1}{36}}\right)\sqrt{{\cal T}_\alpha(\sigma^\alpha,{\bar\sigma^\alpha};{\cal G}^a,{\bar{\cal G}^a};\tau,{\bar\tau}) + \mu_3{\cal V}^{\frac{1}{18}}  - \gamma\left(r_2 + \frac{r_2^2\zeta}{r_1}\right)}{\cal V}^{\frac{5}{6}}}{\Xi^3}\nonumber\\
 & & \times\Biggl\{\frac{1}{\sqrt{{\cal T}_B(\sigma^B,{\bar\sigma ^B};{\cal G}^a,{\bar{\cal G}^a};\tau,{\bar\tau}) + \mu_3{\cal V}^{\frac{1}{18}} + i\kappa_4^2\mu_7C_{1{\bar 1}}{\cal V}^{-\frac{1}{2}} - \gamma\left(r_2 + \frac{r_2^2\zeta}{r_1}\right)}}\nonumber\\
 & & -\frac{1}{\sqrt{{\cal T}_S(\sigma^S,{\bar\sigma ^S};{\cal G}^a,{\bar{\cal G}^a};\tau,{\bar\tau}) + \mu_3{\cal V}^{\frac{1}{18}}  - \gamma\left(r_2 + \frac{r_2^2\zeta}{r_1}\right)}}\Biggr\}\nonumber\\
 & & + \frac{\left(\mu_3{\cal V}^{\frac{1}{36}}+\gamma r_2{\cal V}^{-\frac{1}{36}}\right){\cal V}^{\frac{5}{6}}}{\Xi^2\left[{\cal T}_\alpha(\sigma^\alpha,{\bar\sigma^\alpha};{\cal G}^a,{\bar{\cal G}^a};\tau,{\bar\tau}) + \mu_3{\cal V}^{\frac{1}{18}}  - \gamma\left(r_2 + \frac{r_2^2\zeta}{r_1}\right)\right]^{\frac{3}{2}}}\nonumber\\
 & & + \frac{\left({\cal T}_\alpha(\sigma^\alpha,{\bar\sigma^\alpha};{\cal G}^a,{\bar{\cal G}^a};\tau,{\bar\tau}) + \mu_3{\cal V}^{\frac{1}{18}}  - \gamma\left(r_2 + \frac{r_2^2\zeta}{r_1}\right)\right)\left(\mu_3+\left\{\gamma r_2{\cal V}^{-\frac{1}{36}}\right\}\right){\cal V}^{\frac{5}{6}}}{\Xi^3}\nonumber\\
 & & + \frac{\left(\mu_3{\cal V}^{\frac{1}{36}}+\gamma r_2{\cal V}^{-\frac{1}{36}}\right){\cal V}^{\frac{5}{6}}}{\Xi^2\sqrt{{\cal T}_\alpha(\sigma^\alpha,{\bar\sigma^\alpha};{\cal G}^a,{\bar{\cal G}^a};\tau,{\bar\tau}) + \mu_3{\cal V}^{\frac{1}{18}}  - \gamma\left(r_2 + \frac{r_2^2\zeta}{r_1}\right)}}\sim {\cal V}^{-\frac{43}{36}};\nonumber\\
 & & \partial_{{\cal G}^a}{\bar\partial}_{{\bar{\cal G}}^a}\hat{K}_{{\cal Z}_i{\bar{\cal Z}}_i}\sim\frac{\left((\mu_3{\cal V}^{\frac{1}{36}}+\gamma r_2{\cal V}^{-\frac{1}{36}})\right)}{\Xi^3}\times\Biggl[\sum_\beta k^a n^0_\beta sin(...)\nonumber\\
 & &  + ({\cal G}^a,{\bar{\cal G}^a})\Biggl\{\sqrt{{\cal T}_B(\sigma^B,{\bar\sigma ^B};{\cal G}^a,{\bar{\cal G}^a};\tau,{\bar\tau}) + \mu_3{\cal V}^{\frac{1}{18}} + i\kappa_4^2\mu_7C_{1{\bar 1}}{\cal V}^{-\frac{1}{2}} - \gamma\left(r_2 + \frac{r_2^2\zeta}{r_1}\right)} \nonumber\\
  & & - \sqrt{{\cal T}_S(\sigma^S,{\bar\sigma ^S};{\cal G}^a,{\bar{\cal G}^a};\tau,{\bar\tau}) + \mu_3{\cal V}^{\frac{1}{18}}  - \gamma\left(r_2 + \frac{r_2^2\zeta}{r_1}\right)}\Biggr\}\Biggr]^2\nonumber\\
 & & \times\Biggl\{\frac{1}{\sqrt{{\cal T}_B(\sigma^B,{\bar\sigma ^B};{\cal G}^a,{\bar{\cal G}^a};\tau,{\bar\tau}) + \mu_3{\cal V}^{\frac{1}{18}} + i\kappa_4^2\mu_7C_{1{\bar 1}}{\cal V}^{-\frac{1}{2}} - \gamma\left(r_2 + \frac{r_2^2\zeta}{r_1}\right)}}\nonumber\\
  & & -\frac{1}{\sqrt{{\cal T}_S(\sigma^S,{\bar\sigma ^S};{\cal G}^a,{\bar{\cal G}^a};\tau,{\bar\tau}) + \mu_3{\cal V}^{\frac{1}{18}}  - \gamma\left(r_2 + \frac{r_2^2\zeta}{r_1}\right)}}\Biggr\} + \frac{\left(\mu_3{\cal V}^{\frac{1}{36}}+\gamma r_2{\cal V}^{-\frac{1}{36}}\right)}{\Xi^2}\nonumber\\
  \end{eqnarray*}

\begin{eqnarray*}
 & & \times \Biggl[ \delta_{ab}\Biggl\{\sqrt{{\cal T}_B(\sigma^B,{\bar\sigma ^B};{\cal G}^a,{\bar{\cal G}^a};\tau,{\bar\tau}) + \mu_3{\cal V}^{\frac{1}{18}} + i\kappa_4^2\mu_7C_{1{\bar 1}}{\cal V}^{-\frac{1}{2}} - \gamma\left(r_2 + \frac{r_2^2\zeta}{r_1}\right)}\nonumber\\
  & & - \sqrt{{\cal T}_S(\sigma^S,{\bar\sigma ^S};{\cal G}^a,{\bar{\cal G}^a};\tau,{\bar\tau}) + \mu_3{\cal V}^{\frac{1}{18}}  - \gamma\left(r_2 + \frac{r_2^2\zeta}{r_1}\right)}\Biggr\}+\left(\sum_\beta n^0_\beta cos(...)k^ak^b +\right)\Biggr]\nonumber\\
 & & \times\Biggl\{\frac{1}{\sqrt{{\cal T}_B(\sigma^B,{\bar\sigma ^B};{\cal G}^a,{\bar{\cal G}^a};\tau,{\bar\tau}) + \mu_3{\cal V}^{\frac{1}{18}} + i\kappa_4^2\mu_7C_{1{\bar 1}}{\cal V}^{-\frac{1}{2}} - \gamma\left(r_2 + \frac{r_2^2\zeta}{r_1}\right)}}\nonumber\\
  & & -\frac{1}{\sqrt{{\cal T}_S(\sigma^S,{\bar\sigma ^S};{\cal G}^a,{\bar{\cal G}^a};\tau,{\bar\tau}) + \mu_3{\cal V}^{\frac{1}{18}}  - \gamma\left(r_2 + \frac{r_2^2\zeta}{r_1}\right)}}\Biggr\} + \frac{\left(\mu_3+\left\{\gamma r_2{\cal V}^{-\frac{1}{36}}\right\}\right)}{\Xi^3}\nonumber\\
 & & \times\Biggl[ ({\cal G}^a,{\bar{\cal G}^a})\Biggl\{\sqrt{{\cal T}_B(\sigma^B,{\bar\sigma ^B};{\cal G}^a,{\bar{\cal G}^a};\tau,{\bar\tau}) + \mu_3{\cal V}^{\frac{1}{18}} + i\kappa_4^2\mu_7C_{1{\bar 1}}{\cal V}^{-\frac{1}{2}} - \gamma\left(r_2 + \frac{r_2^2\zeta}{r_1}\right)} \nonumber\\
  & & - \sqrt{{\cal T}_S(\sigma^S,{\bar\sigma ^S};{\cal G}^a,{\bar{\cal G}^a};\tau,{\bar\tau}) + \mu_3{\cal V}^{\frac{1}{18}}  - \gamma\left(r_2 + \frac{r_2^2\zeta}{r_1}\right)}\Biggr\}+\left(\sum_\beta k^a n^0_\beta sin(...) +\right)\Biggr]^2\nonumber\\
   & & \times\Biggl\{\sqrt{{\cal T}_B(\sigma^B,{\bar\sigma ^B};{\cal G}^a,{\bar{\cal G}^a};\tau,{\bar\tau}) + \mu_3{\cal V}^{\frac{1}{18}} + i\kappa_4^2\mu_7C_{1{\bar 1}}{\cal V}^{-\frac{1}{2}} - \gamma\left(r_2 + \frac{r_2^2\zeta}{r_1}\right)}\nonumber\\
 & & -\sqrt{{\cal T}_S(\sigma^S,{\bar\sigma ^S};{\cal G}^a,{\bar{\cal G}^a};\tau,{\bar\tau}) + \mu_3{\cal V}^{\frac{1}{18}}  - \gamma\left(r_2 + \frac{r_2^2\zeta}{r_1}\right)}\Biggr\}\nonumber\\
   & & + \frac{\left(\mu_3+\left\{\gamma r_2{\cal V}^{-\frac{1}{36}}\right\}\right)}{\Xi^2}\Biggl[\sum_\beta n^0_\beta cos(...)k^ak^b \nonumber\\
   & &  + \delta_{ab}\Biggl\{\sqrt{{\cal T}_B(\sigma^B,{\bar\sigma ^B};{\cal G}^a,{\bar{\cal G}^a};\tau,{\bar\tau}) + \mu_3{\cal V}^{\frac{1}{18}} + i\kappa_4^2\mu_7C_{1{\bar 1}}{\cal V}^{-\frac{1}{2}} - \gamma\left(r_2 + \frac{r_2^2\zeta}{r_1}\right)}\nonumber\\
    & & - \sqrt{{\cal T}_S(\sigma^S,{\bar\sigma ^S};{\cal G}^a,{\bar{\cal G}^a};\tau,{\bar\tau}) + \mu_3{\cal V}^{\frac{1}{18}}  - \gamma\left(r_2 + \frac{r_2^2\zeta}{r_1}\right)}\Biggr\}\Biggr]\nonumber\\
    \end{eqnarray*}

\begin{eqnarray}
\label{eq:ddKhat_z}
 & & \times\Biggl\{\sqrt{{\cal T}_B(\sigma^B,{\bar\sigma ^B};{\cal G}^a,{\bar{\cal G}^a};\tau,{\bar\tau}) + \mu_3{\cal V}^{\frac{1}{18}} + i\kappa_4^2\mu_7C_{1{\bar 1}}{\cal V}^{-\frac{1}{2}} - \gamma\left(r_2 + \frac{r_2^2\zeta}{r_1}\right)}\nonumber\\
 & & -\sqrt{{\cal T}_S(\sigma^S,{\bar\sigma ^S};{\cal G}^a,{\bar{\cal G}^a};\tau,{\bar\tau}) + \mu_3{\cal V}^{\frac{1}{18}}  - \gamma\left(r_2 + \frac{r_2^2\zeta}{r_1}\right)}\Biggr\}\nonumber\\
   & & \sim {\cal V}^{-\frac{35}{36}}.
\end{eqnarray}
Similarly,
\begin{eqnarray}
\label{eq:dKhat_a}
& & \partial_{\sigma^B}\hat{K}_{{\tilde{{\cal A}_1}{\bar{\tilde{\cal A}_1}}}}\sim\frac{{\cal V}^{\frac{11}{6}}}{\Xi\left({\cal T}_B(\sigma^B,{\bar\sigma ^B};{\cal G}^a,{\bar{\cal G}^a};\tau,{\bar\tau}) + \mu_3{\cal V}^{\frac{1}{18}} + i\kappa_4^2\mu_7C_{1{\bar 1}}{\cal V}^{-\frac{1}{2}} - \gamma\left(r_2 + \frac{r_2^2\zeta}{r_1}\right)\right)^{\frac{3}{2}}}\nonumber\\
 & & \hskip 1in +\frac{{\cal V}^{\frac{11}{6}}}{\Xi^2}\sim{\cal V}^{\frac{3}{4}},\nonumber\\
& & \partial_{\sigma^S}\hat{K}_{{\tilde{{\cal A}_1}{\bar{\tilde{\cal A}_1}}}}\sim\frac{{\cal V}^{\frac{11}{6}}}{\Xi^2}\sqrt{\frac{{\cal T}_B(\sigma^B,{\bar\sigma ^B};{\cal G}^a,{\bar{\cal G}^a};\tau,{\bar\tau}) + \mu_3{\cal V}^{\frac{1}{18}} + i\kappa_4^2\mu_7C_{1{\bar 1}}{\cal V}^{-\frac{1}{2}} - \gamma\left(r_2 + \frac{r_2^2\zeta}{r_1}\right)}{{\cal T}_S(\sigma^S,{\bar\sigma ^S};{\cal G}^a,{\bar{\cal G}^a};\tau,{\bar\tau}) + \mu_3{\cal V}^{\frac{1}{18}}  - \gamma\left(r_2 + \frac{r_2^2\zeta}{r_1}\right)}}\nonumber\\
& & \hskip 1in \sim{\cal V}^{-\frac{1}{6}};\nonumber\\
& & \partial_{{\cal G}^a}\hat{K}_{{\tilde{{\cal A}_1}{\bar{\tilde{\cal A}_1}}}}\sim \frac{{\cal V}^{\frac{11}{6}}}{\Xi^2\sqrt{{\cal T}_B(\sigma^B,{\bar\sigma ^B};{\cal G}^a,{\bar{\cal G}^a};\tau,{\bar\tau}) + \mu_3{\cal V}^{\frac{1}{18}} + i\kappa_4^2\mu_7C_{1{\bar 1}}{\cal V}^{-\frac{1}{2}} - \gamma\left(r_2 + \frac{r_2^2\zeta}{r_1}\right)}}\nonumber\\
& & \times\Biggl[ ({\cal G}^a,{\bar{\cal G}^a})\Biggl\{\sqrt{{\cal T}_B(\sigma^B,{\bar\sigma ^B};{\cal G}^a,{\bar{\cal G}^a};\tau,{\bar\tau}) + \mu_3{\cal V}^{\frac{1}{18}} + i\kappa_4^2\mu_7C_{1{\bar 1}}{\cal V}^{-\frac{1}{2}} - \gamma\left(r_2 + \frac{r_2^2\zeta}{r_1}\right)} \nonumber\\
 & & -\sqrt{{\cal T}_S(\sigma^S,{\bar\sigma ^S};{\cal G}^a,{\bar{\cal G}^a};\tau,{\bar\tau}) + \mu_3{\cal V}^{\frac{1}{18}}  - \gamma\left(r_2 + \frac{r_2^2\zeta}{r_1}\right)}\Biggr\}+\sum_\beta k^a n^0_\beta sin(...) \Biggr]\nonumber\\
 & & \sim{\cal V}^{\frac{23}{36}},
\end{eqnarray}
from where one concludes:
\begin{eqnarray*}
& & \partial_{\sigma^B}{\bar\partial}_{{\bar\sigma}^B}\hat{K}_{{\tilde{{\cal A}_1}{\bar{\tilde{\cal A}_1}}}}\sim
\frac{{\cal V}^{\frac{11}{6}}}{\Xi^2\left({\cal T}_B(\sigma^B,{\bar\sigma ^B};{\cal G}^a,{\bar{\cal G}^a};\tau,{\bar\tau}) + \mu_3{\cal V}^{\frac{1}{18}} + i\kappa_4^2\mu_7C_{1{\bar 1}}{\cal V}^{-\frac{1}{2}} - \gamma\left(r_2 + \frac{r_2^2\zeta}{r_1}\right)\right)}\nonumber\\
\end{eqnarray*}

\begin{eqnarray}
\label{eq:ddKhat_a}
& & + \frac{{\cal V}^{\frac{11}{6}}}{\Xi\left({\cal T}_B(\sigma^B,{\bar\sigma ^B};{\cal G}^a,{\bar{\cal G}^a};\tau,{\bar\tau}) + \mu_3{\cal V}^{\frac{1}{18}} + i\kappa_4^2\mu_7C_{1{\bar 1}}{\cal V}^{-\frac{1}{2}} - \gamma\left(r_2 + \frac{r_2^2\zeta}{r_1}\right)\right)^{\frac{5}{2}}}
\nonumber\\
& & + \frac{{\cal V}^{\frac{11}{6}}\sqrt{{\cal T}_B(\sigma^B,{\bar\sigma ^B};{\cal G}^a,{\bar{\cal G}^a};\tau,{\bar\tau}) + \mu_3{\cal V}^{\frac{1}{18}} + i\kappa_4^2\mu_7C_{1{\bar 1}}{\cal V}^{-\frac{1}{2}} - \gamma\left(r_2 + \frac{r_2^2\zeta}{r_1}\right)}}{\Xi^3}
\sim {\cal V}^{\frac{25}{36}};\nonumber\\
& & \partial_{\sigma^S}{\bar\partial}_{{\bar\sigma}^S}\hat{K}_{{\tilde{{\cal A}_1}{\bar{\tilde{\cal A}_1}}}}\sim
\frac{{\cal V}^{\frac{11}{6}}\left({\cal T}_S(\sigma^S,{\bar\sigma ^S};{\cal G}^a,{\bar{\cal G}^a};\tau,{\bar\tau}) + \mu_3{\cal V}^{\frac{1}{18}}  - \gamma\left(r_2 + \frac{r_2^2\zeta}{r_1}\right)\right)}{\sqrt{{\cal T}_B(\sigma^B,{\bar\sigma ^B};{\cal G}^a,{\bar{\cal G}^a};\tau,{\bar\tau}) + \mu_3{\cal V}^{\frac{1}{18}} + i\kappa_4^2\mu_7C_{1{\bar 1}}{\cal V}^{-\frac{1}{2}} - \gamma\left(r_2 + \frac{r_2^2\zeta}{r_1}\right)}}
\nonumber\\
& & + \frac{{\cal V}^{\frac{11}{6}}}{\Xi\sqrt{\left({\cal T}_B(\sigma^B,{\bar\sigma ^B};{\cal G}^a,{\bar{\cal G}^a};\tau,{\bar\tau}) + \mu_3{\cal V}^{\frac{1}{18}} + i\kappa_4^2\mu_7C_{1{\bar 1}}{\cal V}^{-\frac{1}{2}} - \gamma\left(r_2 + \frac{r_2^2\zeta}{r_1}\right)\right)}}\nonumber\\
& & \times \frac{1}{\sqrt{\left({\cal T}_S(\sigma^S,{\bar\sigma ^S};{\cal G}^a,{\bar{\cal G}^a};\tau,{\bar\tau}) + \mu_3{\cal V}^{\frac{1}{18}}  - \gamma\left(r_2 + \frac{r_2^2\zeta}{r_1}\right)\right)}} \sim {\cal V}^{\frac{7}{9}};\nonumber\\
& & \partial_{\sigma^B}{\bar\partial}_{{\bar\sigma}^S}\hat{K}_{{\tilde{{\cal A}_1}{\bar{\tilde{\cal A}_1}}}}\sim
\frac{{\cal V}^{\frac{11}{6}}\sqrt{{\cal T}_S(\sigma^S,{\bar\sigma ^S};{\cal G}^a,{\bar{\cal G}^a};\tau,{\bar\tau}) + \mu_3{\cal V}^{\frac{1}{18}}  - \gamma\left(r_2 + \frac{r_2^2\zeta}{r_1}\right)}}{\Xi^2\left({\cal T}_B(\sigma^B,{\bar\sigma ^B};{\cal G}^a,{\bar{\cal G}^a};\tau,{\bar\tau}) + \mu_3{\cal V}^{\frac{1}{18}} + i\kappa_4^2\mu_7C_{1{\bar 1}}{\cal V}^{-\frac{1}{2}} - \gamma\left(r_2 + \frac{r_2^2\zeta}{r_1}\right)\right)^{\frac{3}{2}}}\nonumber\\
& & + \frac{{\cal V}^{\frac{11}{6}}\sqrt{{\cal T}_S(\sigma^S,{\bar\sigma ^S};{\cal G}^a,{\bar{\cal G}^a};\tau,{\bar\tau}) + \mu_3{\cal V}^{\frac{1}{18}}  - \gamma\left(r_2 + \frac{r_2^2\zeta}{r_1}\right)}}{\Xi^3}\sim {\cal V}^{-\frac{2}{9}};\nonumber\\
& & \partial_{\sigma^B}{\bar\partial}_{{\bar{\cal G}}^a}\hat{K}_{{\tilde{{\cal A}_1}{\bar{\tilde{\cal A}_1}}}}\sim
\frac{{\cal V}^{\frac{11}{6}+\frac{5}{6}}}{\Xi^2\left({\cal T}_B(\sigma^B,{\bar\sigma ^B};{\cal G}^a,{\bar{\cal G}^a};\tau,{\bar\tau}) + \mu_3{\cal V}^{\frac{1}{18}} + i\kappa_4^2\mu_7C_{1{\bar 1}}{\cal V}^{-\frac{1}{2}} - \gamma\left(r_2 + \frac{r_2^2\zeta}{r_1}\right)\right)^{\frac{3}{2}}}\nonumber\\
& &  + \frac{{\cal V}^{\frac{11}{6}+\frac{5}{6}}}{\Xi^3}\sim{\cal V}^{\frac{7}{12}}; \, \, \, \, \partial_{\sigma^S}{\bar\partial}_{{\bar{\cal G}}^a}\hat{K}_{{\tilde{{\cal A}_1}{\bar{\tilde{\cal A}_1}}}}\sim
\frac{{\cal V}^{\frac{11}{6}+\frac{5}{6}}}{\Xi^3}\nonumber\\
& & \times\sqrt{\frac{{\cal T}_B(\sigma^B,{\bar\sigma ^B};{\cal G}^a,{\bar{\cal G}^a};\tau,{\bar\tau}) + \mu_3{\cal V}^{\frac{1}{18}} + i\kappa_4^2\mu_7C_{1{\bar 1}}{\cal V}^{-\frac{1}{2}} - \gamma\left(r_2 + \frac{r_2^2\zeta}{r_1}\right)}{{\cal T}_S(\sigma^S,{\bar\sigma ^S};{\cal G}^a,{\bar{\cal G}^a};\tau,{\bar\tau}) + \mu_3{\cal V}^{\frac{1}{18}}  - \gamma\left(r_2 + \frac{r_2^2\zeta}{r_1}\right)}}\sim{\cal V}^{-\frac{1}{3}};
\nonumber\\
& & \partial_{{\cal G}^a}{\bar\partial}_{{\bar{\cal G}}^a}\hat{K}_{{\tilde{{\cal A}_1}{\bar{\tilde{\cal A}_1}}}}\sim{\cal V}^{\frac{29}{36}}.
\end{eqnarray}
In the above $\left(i\kappa_4^2\mu_7C_{1{\bar 1}}{\cal V}^{-\frac{1}{4}}\right)^2\sim{\cal V}^{\frac{11}{6}}$.

Now,
\begin{eqnarray}
\label{eq:detKhat}
& & {\rm det}\ \left(\hat{K}_{i{\bar j}}\right)=\left(\hat{K}_{{\cal Z}_1{\bar{\cal Z}}_1}\right)^2\left(\hat{K}_{{\tilde{{\cal A}_1}{\bar{\tilde{\cal A}_1}}}}\right) \sim\Biggl[\frac{\left(\left(\mu_3{\cal V}^{\frac{1}{36}}\right)^2+\mu_3\gamma r_2\right)}{\Xi}\nonumber\\
& & \times\Biggl\{\frac{1}{\sqrt{{\cal T}_B(\sigma^B,{\bar\sigma ^B};{\cal G}^a,{\bar{\cal G}^a};\tau,{\bar\tau}) + \mu_3{\cal V}^{\frac{1}{18}} + i\kappa_4^2\mu_7C_{1{\bar 1}}{\cal V}^{-\frac{1}{2}} - \gamma\left(r_2 + \frac{r_2^2\zeta}{r_1}\right)}}\nonumber\\
 & & -\frac{1}{\sqrt{{\cal T}_S(\sigma^S,{\bar\sigma ^S};{\cal G}^a,{\bar{\cal G}^a};\tau,{\bar\tau}) + \mu_3{\cal V}^{\frac{1}{18}}  - \gamma\left(r_2 + \frac{r_2^2\zeta}{r_1}\right)}}\Biggr\} +\frac{\left(\left(\gamma r_2{\cal V}^{-\frac{1}{36}}\right)^2\right)}{\Xi}\nonumber\\
 & & \times\Biggl\{\sqrt{{\cal T}_B(\sigma^B,{\bar\sigma ^B};{\cal G}^a,{\bar{\cal G}^a};\tau,{\bar\tau}) + \mu_3{\cal V}^{\frac{1}{18}} + i\kappa_4^2\mu_7C_{1{\bar 1}}{\cal V}^{-\frac{1}{2}} - \gamma\left(r_2 + \frac{r_2^2\zeta}{r_1}\right)} \nonumber\\
 & & -\sqrt{{\cal T}_S(\sigma^S,{\bar\sigma ^S};{\cal G}^a,{\bar{\cal G}^a};\tau,{\bar\tau}) + \mu_3{\cal V}^{\frac{1}{18}}  - \gamma\left(r_2 + \frac{r_2^2\zeta}{r_1}\right)}\Biggr\}\Biggr]^2\nonumber\\
 & & \times\frac{\left(i\kappa_4^2\mu_7C_{1{\bar 1}}{\cal V}^{-\frac{1}{4}}\right)^2}{\Xi\sqrt{{\cal T}_B(\sigma^B,{\bar\sigma ^B};{\cal G}^a,{\bar{\cal G}^a};\tau,{\bar\tau}) + \mu_3{\cal V}^{\frac{1}{18}} + i\kappa_4^2\mu_7C_{1{\bar 1}}{\cal V}^{-\frac{1}{2}} - \gamma\left(r_2 + \frac{r_2^2\zeta}{r_1}\right)}}\sim{\cal V}^{-\frac{41}{36}}.\nonumber\\
\end{eqnarray}
Hence, we see that:
\begin{eqnarray*}
& & \partial_{\sigma^B}{\rm det}\ \left(\hat{K}_{i{\bar j}}\right)\sim\frac{{\cal V}^{\frac{11}{6}}}{\Xi^4}\nonumber\\
& & \times\Biggl[{\cal V}^{\frac{1}{18}}\Biggl\{\frac{1}{\sqrt{{\cal T}_B(\sigma^B,{\bar\sigma ^B};{\cal G}^a,{\bar{\cal G}^a};\tau,{\bar\tau}) + \mu_3{\cal V}^{\frac{1}{18}} + i\kappa_4^2\mu_7C_{1{\bar 1}}{\cal V}^{-\frac{1}{2}} - \gamma\left(r_2 + \frac{r_2^2\zeta}{r_1}\right)}}\nonumber\\
 & & -\frac{1}{\sqrt{{\cal T}_S(\sigma^S,{\bar\sigma ^S};{\cal G}^a,{\bar{\cal G}^a};\tau,{\bar\tau}) + \mu_3{\cal V}^{\frac{1}{18}}  - \gamma\left(r_2 + \frac{r_2^2\zeta}{r_1}\right)}}\Biggr\}\nonumber\\
 \end{eqnarray*}

\begin{eqnarray*}
  & & +\mu_3\Biggl\{\sqrt{{\cal T}_B(\sigma^B,{\bar\sigma ^B};{\cal G}^a,{\bar{\cal G}^a};\tau,{\bar\tau}) + \mu_3{\cal V}^{\frac{1}{18}} + i\kappa_4^2\mu_7C_{1{\bar 1}}{\cal V}^{-\frac{1}{2}} - \gamma\left(r_2 + \frac{r_2^2\zeta}{r_1}\right)} \nonumber\\
 & & - \sqrt{{\cal T}_S(\sigma^S,{\bar\sigma ^S};{\cal G}^a,{\bar{\cal G}^a};\tau,{\bar\tau}) + \mu_3{\cal V}^{\frac{1}{18}}  - \gamma\left(r_2 + \frac{r_2^2\zeta}{r_1}\right)}\Biggr\}\Biggr]^2\nonumber\\
 & & + \frac{{\cal V}^{\frac{11}{6}}}{\Xi^3\Biggl[{\cal T}_B(\sigma^B,{\bar\sigma ^B};{\cal G}^a,{\bar{\cal G}^a};\tau,{\bar\tau}) + \mu_3{\cal V}^{\frac{1}{18}} + i\kappa_4^2\mu_7C_{1{\bar 1}}{\cal V}^{-\frac{1}{2}} - \gamma\left(r_2 + \frac{r_2^2\zeta}{r_1}\right)\Biggr]^{\frac{3}{2}}}\nonumber\\
 & & \times\Biggl[{\cal V}^{\frac{1}{18}}\Biggl\{\frac{1}{\sqrt{{\cal T}_B(\sigma^B,{\bar\sigma ^B};{\cal G}^a,{\bar{\cal G}^a};\tau,{\bar\tau}) + \mu_3{\cal V}^{\frac{1}{18}} + i\kappa_4^2\mu_7C_{1{\bar 1}}{\cal V}^{-\frac{1}{2}} - \gamma\left(r_2 + \frac{r_2^2\zeta}{r_1}\right)}}\nonumber\\
 & & -\frac{1}{\sqrt{{\cal T}_S(\sigma^S,{\bar\sigma ^S};{\cal G}^a,{\bar{\cal G}^a};\tau,{\bar\tau}) + \mu_3{\cal V}^{\frac{1}{18}}  - \gamma\left(r_2 + \frac{r_2^2\zeta}{r_1}\right)}}\Biggr\}\nonumber\\
 & & +\mu_3\Biggl\{\sqrt{{\cal T}_B(\sigma^B,{\bar\sigma ^B};{\cal G}^a,{\bar{\cal G}^a};\tau,{\bar\tau}) + \mu_3{\cal V}^{\frac{1}{18}} + i\kappa_4^2\mu_7C_{1{\bar 1}}{\cal V}^{-\frac{1}{2}} - \gamma\left(r_2 + \frac{r_2^2\zeta}{r_1}\right)} - \nonumber\\
 & & \sqrt{{\cal T}_S(\sigma^S,{\bar\sigma ^S};{\cal G}^a,{\bar{\cal G}^a};\tau,{\bar\tau}) + \mu_3{\cal V}^{\frac{1}{18}}  - \gamma\left(r_2 + \frac{r_2^2\zeta}{r_1}\right)}\Biggr\}\Biggr]^2\nonumber\\
 & & + \frac{{\cal V}^{\frac{11}{6}}}{\Xi^3\sqrt{{\cal T}_B(\sigma^B,{\bar\sigma ^B};{\cal G}^a,{\bar{\cal G}^a};\tau,{\bar\tau}) + \mu_3{\cal V}^{\frac{1}{18}} + i\kappa_4^2\mu_7C_{1{\bar 1}}{\cal V}^{-\frac{1}{2}} - \gamma\left(r_2 + \frac{r_2^2\zeta}{r_1}\right)}}\nonumber\\
  & & \times\Biggl[{\cal V}^{\frac{1}{18}}\Biggl\{\frac{1}{\sqrt{{\cal T}_B(\sigma^B,{\bar\sigma ^B};{\cal G}^a,{\bar{\cal G}^a};\tau,{\bar\tau}) + \mu_3{\cal V}^{\frac{1}{18}} + i\kappa_4^2\mu_7C_{1{\bar 1}}{\cal V}^{-\frac{1}{2}} - \gamma\left(r_2 + \frac{r_2^2\zeta}{r_1}\right)}}\nonumber\\
 & & -\frac{1}{\sqrt{{\cal T}_S(\sigma^S,{\bar\sigma ^S};{\cal G}^a,{\bar{\cal G}^a};\tau,{\bar\tau}) + \mu_3{\cal V}^{\frac{1}{18}}  - \gamma\left(r_2 + \frac{r_2^2\zeta}{r_1}\right)}}\Biggr\}\nonumber\\
 & & +\mu_3\Biggl\{\sqrt{{\cal T}_B(\sigma^B,{\bar\sigma ^B};{\cal G}^a,{\bar{\cal G}^a};\tau,{\bar\tau}) + \mu_3{\cal V}^{\frac{1}{18}} + i\kappa_4^2\mu_7C_{1{\bar 1}}{\cal V}^{-\frac{1}{2}} - \gamma\left(r_2 + \frac{r_2^2\zeta}{r_1}\right)}  \nonumber\\
   \end{eqnarray*}

\begin{eqnarray*}
 & & -\sqrt{{\cal T}_S(\sigma^S,{\bar\sigma ^S};{\cal G}^a,{\bar{\cal G}^a};\tau,{\bar\tau}) + \mu_3{\cal V}^{\frac{1}{18}}  - \gamma\left(r_2 + \frac{r_2^2\zeta}{r_1}\right)}\Biggr\}\Biggr]\nonumber\\
 & & \times\Biggl[\frac{{\cal V}^{\frac{1}{18}}}{\left({\cal T}_B(\sigma^B,{\bar\sigma ^B};{\cal G}^a,{\bar{\cal G}^a};\tau,{\bar\tau}) + \mu_3{\cal V}^{\frac{1}{18}} + i\kappa_4^2\mu_7C_{1{\bar 1}}{\cal V}^{-\frac{1}{2}} - \gamma\left(r_2 + \frac{r_2^2\zeta}{r_1}\right)\right)^{\frac{3}{2}}}\nonumber\\
 & & + \frac{\mu_3}{\sqrt{{\cal T}_B(\sigma^B,{\bar\sigma ^B};{\cal G}^a,{\bar{\cal G}^a};\tau,{\bar\tau}) + \mu_3{\cal V}^{\frac{1}{18}} + i\kappa_4^2\mu_7C_{1{\bar 1}}{\cal V}^{-\frac{1}{2}} - \gamma\left(r_2 + \frac{r_2^2\zeta}{r_1}\right)}}\Biggr]\sim {\cal V}^{-\frac{43}{36}};
 \nonumber\\
 & & \partial_{\sigma^S}{\rm det}\ \left(\hat{K}_{i{\bar j}}\right)\sim\frac{{\cal V}^{\frac{11}{6}}\sqrt{{\cal T}_B(\sigma^B,{\bar\sigma ^B};{\cal G}^a,{\bar{\cal G}^a};\tau,{\bar\tau}) + \mu_3{\cal V}^{\frac{1}{18}} + i\kappa_4^2\mu_7C_{1{\bar 1}}{\cal V}^{-\frac{1}{2}} - \gamma\left(r_2 + \frac{r_2^2\zeta}{r_1}\right)}}{\Xi^4\sqrt{{\cal T}_S(\sigma^S,{\bar\sigma ^S};{\cal G}^a,{\bar{\cal G}^a};\tau,{\bar\tau}) + \mu_3{\cal V}^{\frac{1}{18}}  - \gamma\left(r_2 + \frac{r_2^2\zeta}{r_1}\right)}}\nonumber\\
& & \times\Biggl[{\cal V}^{\frac{1}{18}}\Biggl\{\frac{1}{\sqrt{{\cal T}_B(\sigma^B,{\bar\sigma ^B};{\cal G}^a,{\bar{\cal G}^a};\tau,{\bar\tau}) + \mu_3{\cal V}^{\frac{1}{18}} + i\kappa_4^2\mu_7C_{1{\bar 1}}{\cal V}^{-\frac{1}{2}} - \gamma\left(r_2 + \frac{r_2^2\zeta}{r_1}\right)}}\nonumber\\
 & & -\frac{1}{\sqrt{{\cal T}_S(\sigma^S,{\bar\sigma ^S};{\cal G}^a,{\bar{\cal G}^a};\tau,{\bar\tau}) + \mu_3{\cal V}^{\frac{1}{18}}  - \gamma\left(r_2 + \frac{r_2^2\zeta}{r_1}\right)}}\Biggr\}\nonumber\\
 & & +\mu_3\Biggl\{\sqrt{{\cal T}_B(\sigma^B,{\bar\sigma ^B};{\cal G}^a,{\bar{\cal G}^a};\tau,{\bar\tau}) + \mu_3{\cal V}^{\frac{1}{18}} + i\kappa_4^2\mu_7C_{1{\bar 1}}{\cal V}^{-\frac{1}{2}} - \gamma\left(r_2 + \frac{r_2^2\zeta}{r_1}\right)} - \nonumber\\
 & & \sqrt{{\cal T}_S(\sigma^S,{\bar\sigma ^S};{\cal G}^a,{\bar{\cal G}^a};\tau,{\bar\tau}) + \mu_3{\cal V}^{\frac{1}{18}}  - \gamma\left(r_2 + \frac{r_2^2\zeta}{r_1}\right)}\Biggr\}\Biggr]^2\nonumber\\
 & & + \frac{{\cal V}^{\frac{11}{6}}}{\Xi^3\sqrt{{\cal T}_B(\sigma^B,{\bar\sigma ^B};{\cal G}^a,{\bar{\cal G}^a};\tau,{\bar\tau}) + \mu_3{\cal V}^{\frac{1}{18}} + i\kappa_4^2\mu_7C_{1{\bar 1}}{\cal V}^{-\frac{1}{2}} - \gamma\left(r_2 + \frac{r_2^2\zeta}{r_1}\right)}}\nonumber\\
 & & \times\Biggl[{\cal V}^{\frac{1}{18}}\Biggl\{\frac{1}{\sqrt{{\cal T}_B(\sigma^B,{\bar\sigma ^B};{\cal G}^a,{\bar{\cal G}^a};\tau,{\bar\tau}) + \mu_3{\cal V}^{\frac{1}{18}} + i\kappa_4^2\mu_7C_{1{\bar 1}}{\cal V}^{-\frac{1}{2}} - \gamma\left(r_2 + \frac{r_2^2\zeta}{r_1}\right)}}\nonumber\\
 & & -\frac{1}{\sqrt{{\cal T}_S(\sigma^S,{\bar\sigma ^S};{\cal G}^a,{\bar{\cal G}^a};\tau,{\bar\tau}) + \mu_3{\cal V}^{\frac{1}{18}}  - \gamma\left(r_2 + \frac{r_2^2\zeta}{r_1}\right)}}\Biggr\}\nonumber\\
 & & +\mu_3\Biggl\{\sqrt{{\cal T}_B(\sigma^B,{\bar\sigma ^B};{\cal G}^a,{\bar{\cal G}^a};\tau,{\bar\tau}) + \mu_3{\cal V}^{\frac{1}{18}} + i\kappa_4^2\mu_7C_{1{\bar 1}}{\cal V}^{-\frac{1}{2}} - \gamma\left(r_2 + \frac{r_2^2\zeta}{r_1}\right)}  \nonumber\\
 \end{eqnarray*}

\begin{eqnarray*}
 & & -\sqrt{{\cal T}_S(\sigma^S,{\bar\sigma ^S};{\cal G}^a,{\bar{\cal G}^a};\tau,{\bar\tau}) + \mu_3{\cal V}^{\frac{1}{18}}  - \gamma\left(r_2 + \frac{r_2^2\zeta}{r_1}\right)}\Biggr\}\Biggr]\nonumber\\
 & & \times\Biggl[\frac{{\cal V}^{\frac{1}{18}}}{\left({\cal T}_S(\sigma^S,{\bar\sigma ^S};{\cal G}^a,{\bar{\cal G}^a};\tau,{\bar\tau}) + \mu_3{\cal V}^{\frac{1}{18}} + i\kappa_4^2\mu_7C_{1{\bar 1}}{\cal V}^{-\frac{1}{2}} - \gamma\left(r_2 + \frac{r_2^2\zeta}{r_1}\right)\right)^{\frac{3}{2}}}\nonumber\\
 & & + \frac{\mu_3}{\sqrt{{\cal T}_S(\sigma^S,{\bar\sigma ^S};{\cal G}^a,{\bar{\cal G}^a};\tau,{\bar\tau}) + \mu_3{\cal V}^{\frac{1}{18}} + i\kappa_4^2\mu_7C_{1{\bar 1}}{\cal V}^{-\frac{1}{2}} - \gamma\left(r_2 + \frac{r_2^2\zeta}{r_1}\right)}}\Biggr]\sim {\cal V}^{-\frac{43}{36}};
 \nonumber\\
    & & \partial_{{\cal G}^a}{\rm det}\ \left(\hat{K}_{i{\bar j}}\right)\sim\frac{{\cal V}^{\frac{11}{6}}}{\Xi^4} \times\Biggl[\sum_\beta n^0_\beta k^a sin(...)+({\cal G}^a,{\bar{\cal G}}^a)\times\nonumber\\
   & & \Biggl(\sqrt{{\cal T}_B(\sigma^B,{\bar\sigma ^B};{\cal G}^a,{\bar{\cal G}^a};\tau,{\bar\tau}) + \mu_3{\cal V}^{\frac{1}{18}} + i\kappa_4^2\mu_7C_{1{\bar 1}}{\cal V}^{-\frac{1}{2}} - \gamma\left(r_2 + \frac{r_2^2\zeta}{r_1}\right)}\nonumber\\
   & &    -\sqrt{{\cal T}_S(\sigma^S,{\bar\sigma ^S};{\cal G}^a,{\bar{\cal G}^a};\tau,{\bar\tau}) + \mu_3{\cal V}^{\frac{1}{18}}  - \gamma\left(r_2 + \frac{r_2^2\zeta}{r_1}\right)}\Biggr)\Biggr]\nonumber\\
 & &  \times\Biggl[{\cal V}^{\frac{1}{18}}\Biggl\{\frac{1}{\sqrt{{\cal T}_B(\sigma^B,{\bar\sigma ^B};{\cal G}^a,{\bar{\cal G}^a};\tau,{\bar\tau}) + \mu_3{\cal V}^{\frac{1}{18}} + i\kappa_4^2\mu_7C_{1{\bar 1}}{\cal V}^{-\frac{1}{2}} - \gamma\left(r_2 + \frac{r_2^2\zeta}{r_1}\right)}}\nonumber\\
 & & -\frac{1}{\sqrt{{\cal T}_S(\sigma^S,{\bar\sigma ^S};{\cal G}^a,{\bar{\cal G}^a};\tau,{\bar\tau}) + \mu_3{\cal V}^{\frac{1}{18}}  - \gamma\left(r_2 + \frac{r_2^2\zeta}{r_1}\right)}}\Biggr\}\nonumber\\
 & & +\mu_3\Biggl\{\sqrt{{\cal T}_B(\sigma^B,{\bar\sigma ^B};{\cal G}^a,{\bar{\cal G}^a};\tau,{\bar\tau}) + \mu_3{\cal V}^{\frac{1}{18}} + i\kappa_4^2\mu_7C_{1{\bar 1}}{\cal V}^{-\frac{1}{2}} - \gamma\left(r_2 + \frac{r_2^2\zeta}{r_1}\right)} - \nonumber\\
 & & \sqrt{{\cal T}_S(\sigma^S,{\bar\sigma ^S};{\cal G}^a,{\bar{\cal G}^a};\tau,{\bar\tau}) + \mu_3{\cal V}^{\frac{1}{18}}  - \gamma\left(r_2 + \frac{r_2^2\zeta}{r_1}\right)}\Biggr\}\Biggr]^2\nonumber\\
 & & + \frac{{\cal V}^{\frac{11}{6}}\left({\cal G}^a,{\bar{\cal G}}^a\right)}{\Xi^3\left({\cal T}_\alpha(\sigma^\alpha,{\bar\sigma ^\alpha};{\cal G}^a,{\bar{\cal G}^a};\tau,{\bar\tau}) + \mu_3{\cal V}^{\frac{1}{18}}  - \gamma\left(r_2 + \frac{r_2^2\zeta}{r_1}\right)\right)^{\frac{3}{2}}}
 \nonumber\\
    & &  \times\Biggl[{\cal V}^{\frac{1}{18}}\Biggl\{\frac{1}{\sqrt{{\cal T}_B(\sigma^B,{\bar\sigma ^B};{\cal G}^a,{\bar{\cal G}^a};\tau,{\bar\tau}) + \mu_3{\cal V}^{\frac{1}{18}} + i\kappa_4^2\mu_7C_{1{\bar 1}}{\cal V}^{-\frac{1}{2}} - \gamma\left(r_2 + \frac{r_2^2\zeta}{r_1}\right)}}\nonumber\\
    \end{eqnarray*}

\begin{eqnarray}
\label{eq:ddetKhat}
 & & -\frac{1}{\sqrt{{\cal T}_S(\sigma^S,{\bar\sigma ^S};{\cal G}^a,{\bar{\cal G}^a};\tau,{\bar\tau}) + \mu_3{\cal V}^{\frac{1}{18}}  - \gamma\left(r_2 + \frac{r_2^2\zeta}{r_1}\right)}}\Biggr\}\nonumber\\
 & & +\mu_3\Biggl\{\sqrt{{\cal T}_B(\sigma^B,{\bar\sigma ^B};{\cal G}^a,{\bar{\cal G}^a};\tau,{\bar\tau}) + \mu_3{\cal V}^{\frac{1}{18}} + i\kappa_4^2\mu_7C_{1{\bar 1}}{\cal V}^{-\frac{1}{2}} - \gamma\left(r_2 + \frac{r_2^2\zeta}{r_1}\right)} - \nonumber\\
 & & \sqrt{{\cal T}_S(\sigma^S,{\bar\sigma ^S};{\cal G}^a,{\bar{\cal G}^a};\tau,{\bar\tau}) + \mu_3{\cal V}^{\frac{1}{18}}  - \gamma\left(r_2 + \frac{r_2^2\zeta}{r_1}\right)}\Biggr\}\Biggr]^2\nonumber\\
 & & + \frac{{\cal V}^{\frac{11}{6}}}{\Xi^3\sqrt{{\cal T}_B(\sigma^B,{\bar\sigma ^B};{\cal G}^a,{\bar{\cal G}^a};\tau,{\bar\tau}) + \mu_3{\cal V}^{\frac{1}{18}} + i\kappa_4^2\mu_7C_{1{\bar 1}}{\cal V}^{-\frac{1}{2}} - \gamma\left(r_2 + \frac{r_2^2\zeta}{r_1}\right)}}\nonumber\\
 & &
 \times\Biggl[{\cal V}^{\frac{1}{18}}\Biggl\{\frac{1}{\sqrt{{\cal T}_B(\sigma^B,{\bar\sigma ^B};{\cal G}^a,{\bar{\cal G}^a};\tau,{\bar\tau}) + \mu_3{\cal V}^{\frac{1}{18}} + i\kappa_4^2\mu_7C_{1{\bar 1}}{\cal V}^{-\frac{1}{2}} - \gamma\left(r_2 + \frac{r_2^2\zeta}{r_1}\right)}}\nonumber\\
 & & -\frac{1}{\sqrt{{\cal T}_S(\sigma^S,{\bar\sigma ^S};{\cal G}^a,{\bar{\cal G}^a};\tau,{\bar\tau}) + \mu_3{\cal V}^{\frac{1}{18}}  - \gamma\left(r_2 + \frac{r_2^2\zeta}{r_1}\right)}}\Biggr\}\nonumber\\
 & & +\mu_3\Biggl\{\sqrt{{\cal T}_B(\sigma^B,{\bar\sigma ^B};{\cal G}^a,{\bar{\cal G}^a};\tau,{\bar\tau}) + \mu_3{\cal V}^{\frac{1}{18}} + i\kappa_4^2\mu_7C_{1{\bar 1}}{\cal V}^{-\frac{1}{2}} - \gamma\left(r_2 + \frac{r_2^2\zeta}{r_1}\right)} - \nonumber\\
 & & \sqrt{{\cal T}_S(\sigma^S,{\bar\sigma ^S};{\cal G}^a,{\bar{\cal G}^a};\tau,{\bar\tau}) + \mu_3{\cal V}^{\frac{1}{18}}  - \gamma\left(r_2 + \frac{r_2^2\zeta}{r_1}\right)}\Biggr\}\Biggr]\left({\cal G}^a,{\bar{\cal G}}^a\right)\nonumber\\
& & \times\Biggl[\frac{{\cal V}^{\frac{1}{18}}}{\left({\cal T}_B(\sigma^B,{\bar\sigma ^B};{\cal G}^a,{\bar{\cal G}^a};\tau,{\bar\tau}) + \mu_3{\cal V}^{\frac{1}{18}} + i\kappa_4^2\mu_7C_{1{\bar 1}}{\cal V}^{-\frac{1}{2}} - \gamma\left(r_2 + \frac{r_2^2\zeta}{r_1}\right)\right)^{\frac{3}{2}}}\nonumber\\
 & & + \frac{\mu_3}{\sqrt{{\cal T}_B(\sigma^B,{\bar\sigma ^B};{\cal G}^a,{\bar{\cal G}^a};\tau,{\bar\tau}) + \mu_3{\cal V}^{\frac{1}{18}} + i\kappa_4^2\mu_7C_{1{\bar 1}}{\cal V}^{-\frac{1}{2}} - \gamma\left(r_2 + \frac{r_2^2\zeta}{r_1}\right)}}\Biggr]\sim{\cal V}^{-\frac{109}{36} - g_a}.\nonumber\\
 \end{eqnarray}
Using (\ref{eq:ddetKhat}) and (\ref{eq:Fs}), one obtains:
\begin{equation}
\label{eq:F.dlndetKhat}
F^m\partial_m\ ln\ {\rm det}\left(\hat{K}_{i{\bar j}}\right)\sim m_{\frac{3}{2}}.
\end{equation}

\section{First and Second Derivatives of $Z$ with respect to Closed String Moduli $\sigma^\alpha,{\cal G}^a$}
\setcounter{equation}{0}
\seceqgg

From (\ref{eq:K2}), one sees:
\begin{eqnarray*}
& & Z_{z_1z_2}\sim \Biggl[\sqrt{{\cal T}_B(\sigma^B,{\bar\sigma ^B};{\cal G}^a,{\bar{\cal G}^a};\tau,{\bar\tau}) + \mu_3{\cal V}^{\frac{1}{18}} + i\kappa_4^2\mu_7C_{1{\bar 1}}{\cal V}^{-\frac{1}{2}} - \gamma\left(r_2 + \frac{r_2^2\zeta}{r_1}\right)}\nonumber\\
& &    -\sqrt{{\cal T}_S(\sigma^S,{\bar\sigma ^S};{\cal G}^a,{\bar{\cal G}^a};\tau,{\bar\tau}) + \mu_3{\cal V}^{\frac{1}{18}}  - \gamma\left(r_2 + \frac{r_2^2\zeta}{r_1}\right)}\ \ \Biggr]^2\times \frac{(\mu_3{\cal V}^{\frac{1}{36}}+\gamma r_2{\cal V}^{-\frac{1}{36}})^2}{\Xi^2} ;\nonumber\\
\end{eqnarray*}
\begin{eqnarray*}
& & Z_{{\cal A}_1{\cal A}_1}\nonumber\\
& & \sim\frac{\left(i\kappa_4^2\mu_7C_{1{\bar 1}}{\cal V}^{-\frac{1}{4}}\sqrt{{\cal T}_B(\sigma^B,{\bar\sigma ^B};{\cal G}^a,{\bar{\cal G}^a};\tau,{\bar\tau}) + \mu_3{\cal V}^{\frac{1}{18}} + i\kappa_4^2\mu_7C_{1{\bar 1}}{\cal V}^{-\frac{1}{2}} - \gamma\left(r_2 + \frac{r_2^2\zeta}{r_1}\right)}\right)^2}{\Xi^2};\nonumber\\
   \end{eqnarray*}
\begin{eqnarray}
\label{eq:Zcoeffs}
& & Z_{z_i{\cal A}_1}\sim\Biggl\{\frac{1}{\Xi^2}{i\kappa_4^2\mu_7C_{1{\bar 1}}{\cal V}^{-\frac{1}{4}}(\mu_3{\cal V}^{\frac{1}{36}}+\gamma r_2{\cal V}^{-\frac{1}{36}})}\nonumber\\
& & \sqrt{{\cal T}_B(\sigma^B,{\bar\sigma ^B};{\cal G}^a,{\bar{\cal G}^a};\tau,{\bar\tau}) + \mu_3{\cal V}^{\frac{1}{18}} + i\kappa_4^2\mu_7C_{1{\bar 1}}{\cal V}^{-\frac{1}{2}} - \gamma\left(r_2 + \frac{r_2^2\zeta}{r_1}\right)}\nonumber\\
& & \Biggl(\sqrt{{\cal T}_B(\sigma^B,{\bar\sigma ^B};{\cal G}^a,{\bar{\cal G}^a};\tau,{\bar\tau}) + \mu_3{\cal V}^{\frac{1}{18}} + i\kappa_4^2\mu_7C_{1{\bar 1}}{\cal V}^{-\frac{1}{2}} - \gamma\left(r_2 + \frac{r_2^2\zeta}{r_1}\right)} \nonumber\\
& & -\sqrt{{\cal T}_S(\sigma^S,{\bar\sigma ^S};{\cal G}^a,{\bar{\cal G}^a};\tau,{\bar\tau}) + \mu_3{\cal V}^{\frac{1}{18}}  - \gamma\left(r_2 + \frac{r_2^2\zeta}{r_1}\right)}\Biggr)\Biggr\}.\nonumber\\
  \end{eqnarray}
The first and second derivatives of $Z$ are also relevant to the evaluation of the soft SUSY breaking parameters in section {\bf 5}. The same are given as under:

\begin{eqnarray}
\label{eq:dZ_z}
& & \partial_{\sigma^\alpha}Z_{{\cal Z}_i{\cal Z}_i}\sim\frac{{\cal V}^{\frac{1}{18}}}{\Xi^2}\nonumber\\
& & \times \Biggl(\sqrt{{\cal T}_B(\sigma^B,{\bar\sigma ^B};{\cal G}^a,{\bar{\cal G}^a};\tau,{\bar\tau}) + \mu_3{\cal V}^{\frac{1}{18}} + i\kappa_4^2\mu_7C_{1{\bar 1}}{\cal V}^{-\frac{1}{2}} - \gamma\left(r_2 + \frac{r_2^2\zeta}{r_1}\right)}\nonumber\\
& & -\sqrt{{\cal T}_S(\sigma^S,{\bar\sigma ^S};{\cal G}^a,{\bar{\cal G}^a};\tau,{\bar\tau}) + \mu_3{\cal V}^{\frac{1}{18}}  - \gamma\left(r_2 + \frac{r_2^2\zeta}{r_1}\right)}\Biggr)+\frac{{\cal V}^{\frac{1}{18}}}{\Xi^3} \nonumber\\
      & & \times \Biggl(\sqrt{{\cal T}_B(\sigma^B,{\bar\sigma ^B};{\cal G}^a,{\bar{\cal G}^a};\tau,{\bar\tau}) + \mu_3{\cal V}^{\frac{1}{18}} + i\kappa_4^2\mu_7C_{1{\bar 1}}{\cal V}^{-\frac{1}{2}} - \gamma\left(r_2 + \frac{r_2^2\zeta}{r_1}\right)}\nonumber\\
      & &    -\sqrt{{\cal T}_S(\sigma^S,{\bar\sigma ^S};{\cal G}^a,{\bar{\cal G}^a};\tau,{\bar\tau}) + \mu_3{\cal V}^{\frac{1}{18}}  - \gamma\left(r_2 + \frac{r_2^2\zeta}{r_1}\right)}\Biggr)^2\nonumber\\
   & & \times\sqrt{{\cal T}_\alpha(\sigma^S,{\bar\sigma ^S};{\cal G}^a,{\bar{\cal G}^a};\tau,{\bar\tau}) + \mu_3{\cal V}^{\frac{1}{18}}  - \gamma\left(r_2 + \frac{r_2^2\zeta}{r_1}\right)}\sim{\cal V}^{-\frac{23}{12}};\nonumber\\
   & & \partial_{{\cal G}^a}Z_{{\cal Z}_i{\cal Z}_i}\sim \frac{{\cal V}^{\frac{1}{18}}}{\Xi^3}\nonumber\\
   & & \times \Biggl(\sqrt{{\cal T}_B(\sigma^B,{\bar\sigma ^B};{\cal G}^a,{\bar{\cal G}^a};\tau,{\bar\tau}) + \mu_3{\cal V}^{\frac{1}{18}} + i\kappa_4^2\mu_7C_{1{\bar 1}}{\cal V}^{-\frac{1}{2}} - \gamma\left(r_2 + \frac{r_2^2\zeta}{r_1}\right)}\nonumber\\
     & & -\sqrt{{\cal T}_S(\sigma^S,{\bar\sigma ^S};{\cal G}^a,{\bar{\cal G}^a};\tau,{\bar\tau}) + \mu_3{\cal V}^{\frac{1}{18}}  - \gamma\left(r_2 + \frac{r_2^2\zeta}{r_1}\right)}\Biggr)^2\nonumber\\
   & & \times\Biggl[\sum_\beta n^0_\beta k^a sin(...)+({\cal G}^a,{\bar{\cal G}}^a)\nonumber\\
   & & \times\Biggl(\sqrt{{\cal T}_B(\sigma^B,{\bar\sigma ^B};{\cal G}^a,{\bar{\cal G}^a};\tau,{\bar\tau}) + \mu_3{\cal V}^{\frac{1}{18}} + i\kappa_4^2\mu_7C_{1{\bar 1}}{\cal V}^{-\frac{1}{2}} - \gamma\left(r_2 + \frac{r_2^2\zeta}{r_1}\right)}\nonumber\\
   & &  -\sqrt{{\cal T}_S(\sigma^S,{\bar\sigma ^S};{\cal G}^a,{\bar{\cal G}^a};\tau,{\bar\tau}) + \mu_3{\cal V}^{\frac{1}{18}}  - \gamma\left(r_2 + \frac{r_2^2\zeta}{r_1}\right)}\Biggr)\Biggr]\nonumber\\
   & & \sim{\cal V}^{-\frac{7}{54}}.
\end{eqnarray}

\begin{eqnarray}
\label{eq:ddZ_z}
& & \partial_{\sigma^\alpha}{\bar\partial}_{{\bar\sigma}^{\bar\beta}}Z_{{\cal Z}_i{\cal Z}_i}\sim\frac{{\cal V}^{\frac{1}{18}}}{\Xi^2\left({\cal T}_\alpha(\sigma^\alpha,{\bar\sigma ^\alpha};{\cal G}^a,{\bar{\cal G}^a};\tau,{\bar\tau}) + \mu_3{\cal V}^{\frac{1}{18}} + i\kappa_4^2\mu_7C_{1{\bar 1}}{\cal V}^{-\frac{1}{2}} - \gamma\left(r_2 + \frac{r_2^2\zeta}{r_1}\right)\right)}\nonumber\\
& &  + \frac{{\cal V}^{\frac{1}{18}}}{\Xi^3}\Biggl(\sqrt{{\cal T}_B(\sigma^B,{\bar\sigma ^B};{\cal G}^a,{\bar{\cal G}^a};\tau,{\bar\tau}) + \mu_3{\cal V}^{\frac{1}{18}} + i\kappa_4^2\mu_7C_{1{\bar 1}}{\cal V}^{-\frac{1}{2}} - \gamma\left(r_2 + \frac{r_2^2\zeta}{r_1}\right)}\nonumber\\
& &    -\sqrt{{\cal T}_S(\sigma^S,{\bar\sigma ^S};{\cal G}^a,{\bar{\cal G}^a};\tau,{\bar\tau}) + \mu_3{\cal V}^{\frac{1}{18}}  - \gamma\left(r_2 + \frac{r_2^2\zeta}{r_1}\right)}\Biggr)\nonumber\\
   & & \times \sqrt{{\cal T}_\alpha(\sigma^\alpha,{\bar\sigma ^\alpha};{\cal G}^a,{\bar{\cal G}^a};\tau,{\bar\tau}) + \mu_3{\cal V}^{\frac{1}{18}} + i\kappa_4^2\mu_7C_{1{\bar 1}}{\cal V}^{-\frac{1}{2}} - \gamma\left(r_2 + \frac{r_2^2\zeta}{r_1}\right)}\sim {\cal V}^{-\frac{26}{27}};\nonumber\\
   & & \partial_{{\cal G}^a}{\bar\partial}_{{\bar\sigma}^\alpha}Z_{{\cal Z}_i{\cal Z}_i}\sim\frac{{\cal V}^{\frac{1}{18}}}{\Xi^2}\Biggl(\frac{\left({\cal G}^a,{\bar{\cal G}^a}\right)}{\sqrt{{\cal T}_B(\sigma^B,{\bar\sigma ^B};{\cal G}^a,{\bar{\cal G}^a};\tau,{\bar\tau}) + \mu_3{\cal V}^{\frac{1}{18}} + i\kappa_4^2\mu_7C_{1{\bar 1}}{\cal V}^{-\frac{1}{2}} - \gamma\left(r_2 + \frac{r_2^2\zeta}{r_1}\right)}}\nonumber\\
   & &    - \frac{\left({\cal G}^a,{\bar{\cal G}^a}\right)}{\sqrt{{\cal T}_S(\sigma^S,{\bar\sigma ^S};{\cal G}^a,{\bar{\cal G}^a};\tau,{\bar\tau}) + \mu_3{\cal V}^{\frac{1}{18}}  - \gamma\left(r_2 + \frac{r_2^2\zeta}{r_1}\right)}}\Biggr)+\frac{{\cal V}^{\frac{1}{18}}}{\Xi^3}\nonumber\\
   & & \times\Biggl(\sqrt{{\cal T}_B(\sigma^B,{\bar\sigma ^B};{\cal G}^a,{\bar{\cal G}^a};\tau,{\bar\tau}) + \mu_3{\cal V}^{\frac{1}{18}} + i\kappa_4^2\mu_7C_{1{\bar 1}}{\cal V}^{-\frac{1}{2}} - \gamma\left(r_2 + \frac{r_2^2\zeta}{r_1}\right)}\nonumber\\
   & &   -\sqrt{{\cal T}_S(\sigma^S,{\bar\sigma ^S};{\cal G}^a,{\bar{\cal G}^a};\tau,{\bar\tau}) + \mu_3{\cal V}^{\frac{1}{18}}  - \gamma\left(r_2 + \frac{r_2^2\zeta}{r_1}\right)}\Biggr)\times{\cal V}^{\frac{5}{6}}k_a\sim{\cal V}^{-\frac{25}{36}}k_a;\nonumber\\
   & & \partial_{{\cal G}^a}{\bar\partial}_{{\bar{\cal G}^a}}Z_{{\cal Z}_i{\cal Z}_i}\sim \frac{{\cal V}^{\frac{1}{18}}}{\Xi^3}\Biggl(\sqrt{{\cal T}_B(\sigma^B,{\bar\sigma ^B};{\cal G}^a,{\bar{\cal G}^a};\tau,{\bar\tau}) + \mu_3{\cal V}^{\frac{1}{18}} + i\kappa_4^2\mu_7C_{1{\bar 1}}{\cal V}^{-\frac{1}{2}} - \gamma\left(r_2 + \frac{r_2^2\zeta}{r_1}\right)}\nonumber\\
   & &    -\sqrt{{\cal T}_S(\sigma^S,{\bar\sigma ^S};{\cal G}^a,{\bar{\cal G}^a};\tau,{\bar\tau}) + \mu_3{\cal V}^{\frac{1}{18}}  - \gamma\left(r_2 + \frac{r_2^2\zeta}{r_1}\right)}\Biggr)\times \left({\cal G}^a,{\cal G}^b\right)k_a{\cal V}^{\frac{5}{6}}\nonumber\\
   & & +\frac{{\cal V}^{\frac{1}{18}}}{\Xi^3}\Biggl(\sqrt{{\cal T}_B(\sigma^B,{\bar\sigma ^B};{\cal G}^a,{\bar{\cal G}^a};\tau,{\bar\tau}) + \mu_3{\cal V}^{\frac{1}{18}} + i\kappa_4^2\mu_7C_{1{\bar 1}}{\cal V}^{-\frac{1}{2}} - \gamma\left(r_2 + \frac{r_2^2\zeta}{r_1}\right)}\nonumber\\
   & & -\sqrt{{\cal T}_S(\sigma^S,{\bar\sigma ^S};{\cal G}^a,{\bar{\cal G}^a};\tau,{\bar\tau}) + \mu_3{\cal V}^{\frac{1}{18}}  - \gamma\left(r_2 + \frac{r_2^2\zeta}{r_1}\right)}\Biggr)^2\times {\cal V}k_ak_b\nonumber\\
    & & + \frac{{\cal V}^{\frac{1}{18}}}{\Xi^4}\Biggl(\sqrt{{\cal T}_B(\sigma^B,{\bar\sigma ^B};{\cal G}^a,{\bar{\cal G}^a};\tau,{\bar\tau}) + \mu_3{\cal V}^{\frac{1}{18}} + i\kappa_4^2\mu_7C_{1{\bar 1}}{\cal V}^{-\frac{1}{2}} - \gamma\left(r_2 + \frac{r_2^2\zeta}{r_1}\right)}\nonumber\\
     & & -\sqrt{{\cal T}_S(\sigma^S,{\bar\sigma ^S};{\cal G}^a,{\bar{\cal G}^a};\tau,{\bar\tau}) + \mu_3{\cal V}^{\frac{1}{18}}  - \gamma\left(r_2 + \frac{r_2^2\zeta}{r_1}\right)}\Biggr)^2\nonumber\\
   & & \times k_ak_b{\cal V}^{\frac{5}{6}}\sim{\cal V}^{-\frac{17}{9}}.
    \end{eqnarray}
Similarly,
\begin{eqnarray}
\label{eq:dZ_a}
& & \partial_{\sigma^B}Z_{{\cal A}_1{\cal A}_1}\sim\frac{{\cal V}^{\frac{11}{6}}}{\Xi^2} + \frac{\left({\cal T}_B(\sigma^B,{\bar\sigma ^B};{\cal G}^a,{\bar{\cal G}^a};\tau,{\bar\tau}) + \mu_3{\cal V}^{\frac{1}{18}} + i\kappa_4^2\mu_7C_{1{\bar 1}}{\cal V}^{-\frac{1}{2}} - \gamma\left(r_2 + \frac{r_2^2\zeta}{r_1}\right)\right)^{\frac{3}{2}}}{\Xi^3}\nonumber\\
& & \times {\cal V}^{\frac{11}{6}} \sim{\cal V}^{-\frac{1}{6}};\nonumber\\
& & \partial_{\sigma^S}Z_{{\cal A}_1{\cal A}_1}\sim\frac{{\cal V}^{\frac{11}{6}}\left({\cal T}_B(\sigma^B,{\bar\sigma ^B};{\cal G}^a,{\bar{\cal G}^a};\tau,{\bar\tau}) + \mu_3{\cal V}^{\frac{1}{18}} + i\kappa_4^2\mu_7C_{1{\bar 1}}{\cal V}^{-\frac{1}{2}} - \gamma\left(r_2 + \frac{r_2^2\zeta}{r_1}\right)\right)}{\Xi^3}\nonumber\\
& & \times\sqrt{\left({\cal T}_S(\sigma^B,{\bar\sigma ^B};{\cal G}^a,{\bar{\cal G}^a};\tau,{\bar\tau}) + \mu_3{\cal V}^{\frac{1}{18}} + i\kappa_4^2\mu_7C_{1{\bar 1}}{\cal V}^{-\frac{1}{2}} - \gamma\left(r_2 + \frac{r_2^2\zeta}{r_1}\right)\right)}\sim {\cal V}^{-\frac{13}{12}};
\nonumber\\
& & \partial_{{\cal G}^a}Z_{{\cal A}_1{\cal A}_1}\sim\frac{{\cal V}^{\frac{11}{6}}({\cal G}^a,{\bar{\cal G}^a})}{\Xi^2}\nonumber\\
& & +\frac{{\cal V}^{\frac{11}{6}}\left({\cal T}_B(\sigma^B,{\bar\sigma ^B};{\cal G}^a,{\bar{\cal G}^a};\tau,{\bar\tau}) + \mu_3{\cal V}^{\frac{1}{18}} + i\kappa_4^2\mu_7C_{1{\bar 1}}{\cal V}^{-\frac{1}{2}} - \gamma\left(r_2 + \frac{r_2^2\zeta}{r_1}\right)\right)}{\Xi^3}\nonumber\\
& & \times\Biggl[\sum_\beta n^0_\beta k^a sin(...)+({\cal G}^a,{\bar{\cal G}}^a)\nonumber\\
   & & \times \Biggl(\sqrt{{\cal T}_B(\sigma^B,{\bar\sigma ^B};{\cal G}^a,{\bar{\cal G}^a};\tau,{\bar\tau}) + \mu_3{\cal V}^{\frac{1}{18}} + i\kappa_4^2\mu_7C_{1{\bar 1}}{\cal V}^{-\frac{1}{2}} - \gamma\left(r_2 + \frac{r_2^2\zeta}{r_1}\right)}\nonumber\\
   & & -\sqrt{{\cal T}_S(\sigma^S,{\bar\sigma ^S};{\cal G}^a,{\bar{\cal G}^a};\tau,{\bar\tau}) + \mu_3{\cal V}^{\frac{1}{18}}  - \gamma\left(r_2 + \frac{r_2^2\zeta}{r_1}\right)}\Biggr)\Biggr]\nonumber\\
   & & \sim{\cal V}^{-\frac{1}{6}-g_a},
\end{eqnarray}
where ${\cal G}^a\sim{\cal V}^{-g_a},\ 0<g_a<\frac{1}{9}$, and

\begin{eqnarray*}
& & \partial_{\sigma^B}{\bar\partial}_{{\bar\sigma}^B}Z_{\tilde{\cal A}_1\tilde{\cal A}_1}\sim\frac{{\cal V}^{\frac{11}{6}}\sqrt{{\cal T}_B(\sigma^B,{\bar\sigma ^B};{\cal G}^a,{\bar{\cal G}^a};\tau,{\bar\tau}) + \mu_3{\cal V}^{\frac{1}{18}} + i\kappa_4^2\mu_7C_{1{\bar 1}}{\cal V}^{-\frac{1}{2}} - \gamma\left(r_2 + \frac{r_2^2\zeta}{r_1}\right)}}{\Xi^3}\nonumber\\
& & +\frac{{\cal V}^{\frac{11}{6}}\left({\cal T}_B(\sigma^B,{\bar\sigma ^B};{\cal G}^a,{\bar{\cal G}^a};\tau,{\bar\tau}) + \mu_3{\cal V}^{\frac{1}{18}} + i\kappa_4^2\mu_7C_{1{\bar 1}}{\cal V}^{-\frac{1}{2}} - \gamma\left(r_2 + \frac{r_2^2\zeta}{r_1}\right)\right)^2}{\Xi^4} \sim {\cal V}^{\frac{-41}{36}};\nonumber\\
& & \nonumber\\
& & \partial_{\sigma^S}{\bar\partial}_{{\bar\sigma}^S}Z_{\tilde{\cal A}_1\tilde{\cal A}_1}\sim\frac{{\cal V}^{\frac{11}{6}}\left({\cal T}_B(\sigma^B,{\bar\sigma ^B};{\cal G}^a,{\bar{\cal G}^a};\tau,{\bar\tau}) + \mu_3{\cal V}^{\frac{1}{18}} + i\kappa_4^2\mu_7C_{1{\bar 1}}{\cal V}^{-\frac{1}{2}} - \gamma\left(r_2 + \frac{r_2^2\zeta}{r_1}\right)\right)}{\Xi^3\sqrt{{\cal T}_S(\sigma^S,{\bar\sigma ^S};{\cal G}^a,{\bar{\cal G}^a};\tau,{\bar\tau}) + \mu_3{\cal V}^{\frac{1}{18}}  - \gamma\left(r_2 + \frac{r_2^2\zeta}{r_1}\right)}}\nonumber\\
\end{eqnarray*}
\begin{eqnarray}
\label{eq:ddZ_a}
& & + \frac{{\cal V}^{\frac{11}{6}}\left({\cal T}_B(\sigma^B,{\bar\sigma ^B};{\cal G}^a,{\bar{\cal G}^a};\tau,{\bar\tau}) + \mu_3{\cal V}^{\frac{1}{18}} + i\kappa_4^2\mu_7C_{1{\bar 1}}{\cal V}^{-\frac{1}{2}} - \gamma\left(r_2 + \frac{r_2^2\zeta}{r_1}\right)\right)}{\Xi^3\sqrt{{\cal T}_S(\sigma^S,{\bar\sigma ^S};{\cal G}^a,{\bar{\cal G}^a};\tau,{\bar\tau}) + \mu_3{\cal V}^{\frac{1}{18}}  - \gamma\left(r_2 + \frac{r_2^2\zeta}{r_1}\right)}}\nonumber\\
& & + \frac{{\cal V}^{\frac{11}{6}}\left({\cal T}_B(\sigma^B,{\bar\sigma ^B};{\cal G}^a,{\bar{\cal G}^a};\tau,{\bar\tau}) + \mu_3{\cal V}^{\frac{1}{18}} + i\kappa_4^2\mu_7C_{1{\bar 1}}{\cal V}^{-\frac{1}{2}} - \gamma\left(r_2 + \frac{r_2^2\zeta}{r_1}\right)\right)}{\Xi^4}\nonumber\\
& & \times \left({\cal T}_S(\sigma^S,{\bar\sigma ^S};{\cal G}^a,{\bar{\cal G}^a};\tau,{\bar\tau}) + \mu_3{\cal V}^{\frac{1}{18}}  - \gamma\left(r_2 + \frac{r_2^2\zeta}{r_1}\right)\right)\sim {\cal V}^{-\frac{41}{36}};\nonumber\\
& & \nonumber\\
& & \partial_{\sigma^B}{\bar\partial}_{{\bar\sigma}^S}Z_{\tilde{\cal A}_1\tilde{\cal A}_1}\sim\frac{{\cal V}^{\frac{11}{6}}\sqrt{{\cal T}_S(\sigma^S,{\bar\sigma ^S};{\cal G}^a,{\bar{\cal G}^a};\tau,{\bar\tau}) + \mu_3{\cal V}^{\frac{1}{18}}  - \gamma\left(r_2 + \frac{r_2^2\zeta}{r_1}\right)}}{\Xi^3}\nonumber\\
& & +\frac{{\cal V}^{\frac{11}{6}}\left({\cal T}_B(\sigma^B,{\bar\sigma ^B};{\cal G}^a,{\bar{\cal G}^a};\tau,{\bar\tau}) + \mu_3{\cal V}^{\frac{1}{18}} + i\kappa_4^2\mu_7C_{1{\bar 1}}{\cal V}^{-\frac{1}{2}} - \gamma\left(r_2 + \frac{r_2^2\zeta}{r_1}\right)\right)^{\frac{3}{2}}}{\Xi^4} \nonumber\\
& & \times \sqrt{{\cal T}_S(\sigma^S,{\bar\sigma ^S};{\cal G}^a,{\bar{\cal G}^a};\tau,{\bar\tau}) + \mu_3{\cal V}^{\frac{1}{18}}  - \gamma\left(r_2 + \frac{r_2^2\zeta}{r_1}\right)} \sim {\cal V}^{-\frac{41}{36}};\nonumber\\
& & \partial_{\sigma^B}{\bar\partial}_{{\bar{\cal G}^a}}Z_{\tilde{\cal A}_1\tilde{\cal A}_1}\nonumber\\
& & \sim\frac{{\cal V}^{\frac{11}{6}}\left({\cal G}^a,{\bar{\cal G}^a}\right)\sqrt{{\cal T}_B(\sigma^B,{\bar\sigma ^B};{\cal G}^a,{\bar{\cal G}^a};\tau,{\bar\tau}) + \mu_3{\cal V}^{\frac{1}{18}} + i\kappa_4^2\mu_7C_{1{\bar 1}}{\cal V}^{-\frac{1}{2}} - \gamma\left(r_2 + \frac{r_2^2\zeta}{r_1}\right)}}{\Xi^3}\nonumber\\
& & +\frac{{\cal V}^{\frac{11}{6}+\frac{5}{6}}k_a\left({\cal T}_B(\sigma^B,{\bar\sigma ^B};{\cal G}^a,{\bar{\cal G}^a};\tau,{\bar\tau}) + \mu_3{\cal V}^{\frac{1}{18}} + i\kappa_4^2\mu_7C_{1{\bar 1}}{\cal V}^{-\frac{1}{2}} - \gamma\left(r_2 + \frac{r_2^2\zeta}{r_1}\right)\right)^{\frac{3}{2}}}{\Xi^4}\nonumber\\
& & + \frac{{\cal V}^{\frac{11}{6}+\frac{5}{6}}k_a}{\Xi^3} \sim{\cal V}^{-\frac{13}{6}};\nonumber\\
& & {\rm Similarly,}\ \partial_{\sigma^S}{\bar\partial}_{{\bar{\cal G}^a}}Z_{\tilde{\cal A}_1\tilde{\cal A}_1}\sim{\cal V}^{-\frac{5}{4}};
\nonumber\\
& & \partial_{{\cal G}^a}{\bar\partial}_{{\bar{\cal G}^b}}Z_{\tilde{\cal A}_1\tilde{\cal A}_1}\sim\frac{{\cal V}^{\frac{11}{6}}\delta_{ab}}{\Xi^2}
+ \frac{{\cal V}^{\frac{11}{6}+\frac{5}{6}}\left({\cal G}^a,{\bar{\cal G}^a}\right)k_b}{\Xi^3}\nonumber\\
& & + \frac{{\cal V}^{\frac{11}{6}+1}\left({\cal T}_B(\sigma^B,{\bar\sigma ^B};{\cal G}^a,{\bar{\cal G}^a};\tau,{\bar\tau}) + \mu_3{\cal V}^{\frac{1}{18}} + i\kappa_4^2\mu_7C_{1{\bar 1}}{\cal V}^{-\frac{1}{2}} - \gamma\left(r_2 + \frac{r_2^2\zeta}{r_1}\right)\right)}{\Xi^3}
\nonumber\\
& & + \frac{{\cal V}^{\frac{11}{6}+\frac{10}{6}}k_ak_b\left({\cal T}_B(\sigma^B,{\bar\sigma ^B};{\cal G}^a,{\bar{\cal G}^a};\tau,{\bar\tau}) + \mu_3{\cal V}^{\frac{1}{18}} + i\kappa_4^2\mu_7C_{1{\bar 1}}{\cal V}^{-\frac{1}{2}} - \gamma\left(r_2 + \frac{r_2^2\zeta}{r_1}\right)\right)}{\Xi^4}\nonumber\\
& & \sim{\cal V}^{-\frac{4}{9}}.
\end{eqnarray}

\section{Periods for Swiss-Cheese Calabi Yau}
\setcounter{equation}{0}
\seceqhh

Here, we fill in the details relevant to evaluation of periods in different portions of the complex structure moduli space of section ${\bf 6.2}$.

$\underline{|\phi^3|<1,\ {\rm large}\ \psi}$: The expressions for $P_{1,2,3}$ relevant to (\ref{eq:smallphilargepsi2}) are:
\begin{equation}
\label{eq:smallphilargepsi3}
P_1\equiv\left(\begin{array}{c}
\sum_{m=0}^\infty\sum_{n=0}^\infty A_{m,n}\frac{\phi_0^m}{\rho_0^{18n + 6m}}\\
\sum_{m=0}^\infty\sum_{n=0}^\infty e^{\frac{i\pi(-35m + 128 n)}{9}}
A_{m,n}\frac{\phi_0^m}{\rho_0^{18n + 6m}}\\
\sum_{m=0}^\infty\sum_{n=0}^\infty e^{\frac{2i\pi(-35m + 128 n)}{9}}
A_{m,n}\frac{\phi_0^m}{\rho_0^{18n + 6m}}\\
\sum_{m=0}^\infty\sum_{n=0}^\infty e^{\frac{i\pi(-35m + 128 n)}{3}}
A_{m,n}\frac{\phi_0^m}{\rho_0^{18n + 6m}}\\
\sum_{m=0}^\infty\sum_{n=0}^\infty e^{\frac{4i\pi(-35m + 128 n)}{9}}
A_{m,n}\frac{\phi_0^m}{\rho_0^{18n + 6m}}\\
\sum_{m=0}^\infty\sum_{n=0}^\infty e^{\frac{5i\pi(-35m + 128 n)}{9}}
A_{m,n}\frac{\phi_0^m}{\rho_0^{18n + 6m}}\\
\end{array}\right),
\end{equation}
\begin{equation}
\label{eq:smallphilargepsi4}
P_2\equiv\left(\begin{array}{c}
\sum_{m=0}^\infty\sum_{n=0}^\infty A_{m,n}\frac{m\phi_0^{m-1}}{\rho_0^{18n + 6m}}\\
\sum_{m=0}^\infty\sum_{n=0}^\infty A_{m,n}\frac{m\phi_0^{m-1}}{\rho_0^{18n + 6m}}
e^{\frac{i\pi(-35m + 128 n)}{9}}\\
\sum_{m=0}^\infty\sum_{n=0}^\infty A_{m,n}\frac{m\phi_0^{m-1}}{\rho_0^{18n + 6m}}
e^{2\frac{i\pi(-35m + 128 n)}{9}}\\
\sum_{m=0}^\infty\sum_{n=0}^\infty A_{m,n}\frac{m\phi_0^{m-1}}{\rho_0^{18n + 6m}}
e^{\frac{i\pi(-35m + 128 n)}{3}}\\
\sum_{m=0}^\infty\sum_{n=0}^\infty A_{m,n}\frac{m\phi_0^{m-1}}{\rho_0^{18n + 6m}}
e^{\frac{4i\pi(-35m + 128 n)}{9}}\\
\sum_{m=0}^\infty\sum_{n=0}^\infty A_{m,n}\frac{m\phi_0^{m-1}}{\rho_0^{18n + 6m}}
e^{\frac{5i\pi(-35m + 128 n)}{9}}\\
\end{array}\right)
\end{equation}
\begin{equation}
\label{eq:smallphilargepsi5}
P_3\equiv\left(\begin{array}{c}
-\sum_{m=0}^\infty\sum_{n=0}^\infty A_{m,n}\frac{(18n + 6m)\phi_0^m}{\rho_0^{18n + 6m + 1}}\\
-\sum_{m=0}^\infty\sum_{n=0}^\infty A_{m,n}\frac{(18n + 6m)\phi_0^m}{\rho_0^{18n + 6m + 1}}e^{\frac{i\pi(-35m + 128 n)}{9}}\\
-\sum_{m=0}^\infty\sum_{n=0}^\infty A_{m,n}\frac{(18n + 6m)\phi_0^m}{\rho_0^{18n + 6m + 1}}e^{\frac{2i\pi(-35m + 128 n)}{9}}\\
-\sum_{m=0}^\infty\sum_{n=0}^\infty A_{m,n}\frac{(18n + 6m)\phi_0^m}{\rho_0^{18n + 6m + 1}}e^{\frac{i\pi(-35m + 128 n)}{3}}\\
-\sum_{m=0}^\infty\sum_{n=0}^\infty A_{m,n}\frac{(18n + 6m)\phi_0^m}{\rho_0^{18n + 6m + 1}}e^{\frac{4i\pi(-35m + 128 n)}{9}}\\
-\sum_{m=0}^\infty\sum_{n=0}^\infty A_{m,n}\frac{(18n + 6m)\phi_0^m}{\rho_0^{18n + 6m + 1}}e^{\frac{5i\pi(-35m + 128 n)}{9}}\\
\end{array}\right).
\end{equation}
The coefficients $A_{m,n}$ appearing in (\ref{eq:smallphilargepsi3})-(\ref{eq:smallphilargepsi5}) are given by:
$$A_{m,n}\equiv \frac{(18n + 6m)!(-3\phi)^m(3^4.2)^{18n + 6m}}{(9n + 3m)!(6n + 2m)!(n!)^3m!18^{18n + 6m}}.$$

$\underline{|\frac{\rho^6}{\phi - \omega^{0,-1,-2}}|<1}$: The expressions for $M_{1,2,3}$ relevant to (\ref{eq:awaycl12}) are:
\begin{equation}
\label{eq:awaycl13}
M_1\equiv\left(\begin{array}{c}
\sum_{r=1,5}\sum_{k=0}^\infty\sum_{m=0}^\infty A_{k,m,r}\rho_0^{6k+r}\phi_0^m
\\
\sum_{r=1,5}\sum_{k=0}^\infty\sum_{m=0}^\infty A_{k,m,r}\rho_0^{6k+r}\phi_0^m
e^{\frac{2i\pi(k+\frac{r}{6})}{3} + \frac{2i\pi m}{3}}
\\
\sum_{r=1,5}\sum_{k=0}^\infty\sum_{m=0}^\infty A_{k,m,r}\rho_0^{6k+r}\phi_0^m
 e^{\frac{4i\pi(k+\frac{r}{6})}{3} + \frac{4i\pi m}{3}}
 \\
\sum_{r=1,5}\sum_{k=0}^\infty\sum_{m=0}^\infty A_{k,m,r}\rho_0^{6k+r}\phi_0^m e^{\frac{i\pi r}{3}}
\\
\sum_{r=1,5}\sum_{k=0}^\infty\sum_{m=0}^\infty A_{k,m,r}\rho_0^{6k+r}\phi_0^m
e^{\frac{2i\pi(k+\frac{r}{6})}{3} + \frac{2i\pi m}{3} + \frac{i\pi r}{3}}
\\
\sum_{r=1,5}\sum_{k=0}^\infty\sum_{m=0}^\infty A_{k,m,r}\rho_0^{6k+r}\phi_0^m
e^{\frac{4i\pi(k+\frac{r}{6})}{3} + \frac{4i\pi m}{3} + \frac{i\pi r}{3}}
\end{array}\right),
\end{equation}
\begin{equation}
\label{eq:awaycfl4}
M_2\equiv\left(\begin{array}{c}
\sum_{r=1,5}\sum_{k=0}^\infty\sum_{m=0}^\infty A_{k,m,r}(6k + r)\rho_0^{6k + r - 1}\phi_0^m
\\
\sum_{r=1,5}\sum_{k=0}^\infty\sum_{m=0}^\infty A_{k,m,r}(6k + r)\rho_0^{6k + r - 1}\phi_0^m
e^{\frac{2i\pi(k+\frac{r}{6})}{3} + \frac{2i\pi m}{3}}
\\
\sum_{r=1,5}\sum_{k=0}^\infty\sum_{m=0}^\infty A_{k,m,r}(6k + r)\rho_0^{6k + r - 1}\phi_0^m
 e^{\frac{4i\pi(k+\frac{r}{6})}{3} + \frac{4i\pi m}{3}}
 \\
\sum_{r=1,5}\sum_{k=0}^\infty\sum_{m=0}^\infty A_{k,m,r}(6k + r)\rho_0^{6k + r - 1}\phi_0^{m - 1} e^{\frac{i\pi r}{3}}\\
\sum_{r=1,5}\sum_{k=0}^\infty\sum_{m=0}^\infty A_{k,m,r}(6k + r)\rho_0^{6k + r - 1}\phi_0^m e^{\frac{2i\pi(k+\frac{r}{6})}{3} + \frac{2i\pi m}{3} + \frac{i\pi r}{3}}
\\
\sum_{r=1,5}\sum_{k=0}^\infty\sum_{m=0}^\infty A_{k,m,r}(6k + r)\rho_0^{6k + r - 1}\phi_0^m
e^{\frac{4i\pi(k+\frac{r}{6})}{3} + \frac{4i\pi m}{3} + \frac{i\pi r}{3}}
\end{array}\right)
\end{equation}
\begin{equation}
\label{eq:awaycl15}
M_3\equiv\left(\begin{array}{c}
\sum_{r=1,5}\sum_{k=0}^\infty\sum_{m=0}^\infty A_{k,m,r}m\rho_0^{6k + r}\phi_0^{m - 1}\\
\sum_{r=1,5}\sum_{k=0}^\infty\sum_{m=0}^\infty A_{k,m,r}m\rho_0^{6k + r}\phi_0^{m - 1}
e^{\frac{2i\pi(k+\frac{r}{6})}{3} + \frac{2i\pi m}{3}}\\
\sum_{r=1,5}\sum_{k=0}^\infty\sum_{m=0}^\infty A_{k,m,r}m\rho_0^{6k + r}\phi_0^{m - 1}
e^{\frac{4i\pi(k+\frac{r}{6})}{3} + \frac{4i\pi m}{3}}\\
\sum_{r=1,5}\sum_{k=0}^\infty\sum_{m=0}^\infty A_{k,m,r}m
\rho_0^{6k + r}\phi_0^{m-1} e^{\frac{i\pi r}{3}}\\
\sum_{r=1,5}\sum_{k=0}^\infty\sum_{m=0}^\infty A_{k,m,r}m
\rho_0^{6k + r}\phi_0^{m - 1} e^{\frac{2i\pi(k+\frac{r}{6})}{3} + \frac{2i\pi m}{3} + \frac{i\pi r}{3}}\\
\sum_{r=1,5}\sum_{k=0}^\infty\sum_{m=0}^\infty A_{k,m,r}m\rho_0^{6k + r}\phi_0^{m - 1}
e^{\frac{4i\pi(k+\frac{r}{6})}{3} + \frac{4i\pi m}{3} + \frac{i\pi r}{3}}
\end{array}\right).
\end{equation}
In equations (\ref{eq:awaycl13})-(\ref{eq:awaycl15}), the coefficients $A_{k,m,r}$ are given by:
$$A_{k,m,r}\equiv e^{\frac{i\pi ar}{3}}sin\left(\frac{\pi r}{3}\right)\frac{(-)^k3^{-1 + k+\frac{r}{6} + m}e^{-\frac{i\pi(k + \frac{r}{6})}{3} + \frac{2im\pi}{3}}\Gamma(\frac{m + k + \frac{r}{6}}{3})(\Gamma(k+\frac{r}{6}))^2}{(\Gamma(1 - \frac{m + k + \frac{r}{6}}{3}))^2m!}.$$

$\underline{{\rm Near\ the\ conifold\ locus:}\ \rho^6 + \phi = 1}$: The expressions for $N_{1,2,3}$ relevant for evaluation of (\ref{eq:confl14}) are:
\begin{equation}
\label{eq:confl15}
N_1\equiv\left(\begin{array}{c}
\sum_{r=1,5}\sum_{k=0}^\infty A_{k,0,r}e^{-\frac{i\pi(k + \frac{r}{6})}{3}}\\
\sum_{r=1,5}\sum_{k=0}^\infty A_{k,0,r}e^{\frac{i\pi(k + \frac{r}{6})}{3}}\\
\sum_{r=1,5}\sum_{k=0}^\infty A_{k,0,r}e^{i\pi(k + \frac{r}{6}}\\
\sum_{r=1,5}\sum_{k=0}^\infty A_{k,0,r}e^{\frac{i\pi(-k + 5\frac{r}{6})}{3}}\\
\sum_{r=1,5}\sum_{k=0}^\infty A_{k,0,r}e^{-\frac{i\pi(k + 7\frac{r}{6})}{3}}\\
\sum_{r=1,5}\sum_{k=0}^\infty A_{k,0,r}e^{-\frac{i\pi(3k + 3\frac{r}{6})}{3}}\end{array}\right),
\end{equation}
\begin{equation}
\label{eq:confl16}
N_2\equiv\left(\begin{array}{c}
\sum_{r=1,5}\sum_{k=0}^\infty (k + \frac{r}{6}) A_{k,0,r}e^{-\frac{i\pi(k + \frac{r}{6})}{3}}\\
\sum_{r=1,5}\sum_{k=0}^\infty (k + \frac{r}{6}) A_{k,0,r}e^{\frac{i\pi(k + \frac{r}{6})}{3}}\\
\sum_{r=1,5}\sum_{k=0}^\infty (k + \frac{r}{6}) A_{k,0,r}e^{i\pi(k + \frac{r}{6}}\\
\sum_{r=1,5}\sum_{k=0}^\infty (k + \frac{r}{6}) A_{k,0,r}e^{\frac{i\pi(-k + 5\frac{r}{6})}{3}}\\
\sum_{r=1,5}\sum_{k=0}^\infty (k + \frac{r}{6}) A_{k,0,r}e^{-\frac{i\pi(k + 7\frac{r}{6})}{3}}\\
\sum_{r=1,5}\sum_{k=0}^\infty (k + \frac{r}{6}) A_{k,0,r}e^{-\frac{i\pi(3k + 3\frac{r}{6})}{3}}\end{array}\right)
\end{equation}
\begin{equation}
\label{eq:confl17}
N_3\equiv\left(\begin{array}{c}
\sum_{r=1,5}\sum_{k=0}^\infty (A_{k,1,r} e^{\frac{2i\pi}{3}} - A_{k,0,r}(k + \frac{r}{6}))
e^{-\frac{i\pi(k + \frac{r}{6})}{3}}\\
\sum_{r=1,5}\sum_{k=0}^\infty (A_{k,1,r} e^{\frac{4i\pi}{3}} - A_{k,0,r}(k + \frac{r}{6}))
e^{\frac{i\pi(k + \frac{r}{6})}{3}}\\
\sum_{r=1,5}\sum_{k=0}^\infty (A_{k,1,r}  - A_{k,0,r}(k + \frac{r}{6}))
e^{i\pi(k + \frac{r}{6}}\\
\sum_{r=1,5}\sum_{k=0}^\infty (A_{k,1,r} e^{\frac{2i\pi}{3}} - A_{k,0,r}(k + \frac{r}{6}))
e^{\frac{i\pi(-k + 5\frac{r}{6})}{3}}\\
\sum_{r=1,5}\sum_{k=0}^\infty (A_{k,1,r} e^{\frac{4i\pi}{3}} - A_{k,0,r}(k + \frac{r}{6}))
e^{-\frac{i\pi(k + 7\frac{r}{6})}{3}}\\
\sum_{r=1,5}\sum_{k=0}^\infty (A_{k,1,r}  - A_{k,0,r}(k + \frac{r}{6}))
e^{-\frac{i\pi(3k + 3\frac{r}{6})}{3}}
\end{array}\right).
\end{equation}
The coefficients $A_{k,m,r}$ figuring in (\ref{eq:confl15})-(\ref{eq:confl17}) are given by:
$$A_{k,m,r}\equiv \frac{3^{-1 + k + \frac{r}{6} + m}e^{\frac{2i\pi(\sigma + 1)}{3} + \frac{-i\pi(k + \frac{r}{6})}{3}}(-)^k(\Gamma(k + \frac{r}{6}))^2}{\Gamma(k + 1)\Gamma(k + \frac{r}{6})\Gamma(k + \frac{r}{6})(\Gamma(1 - \frac{m + k + \frac{r}{6}}{3}))^2m!}sin\left(\frac{\pi r}{3}\right).$$

$\underline{{\rm Near}\ \phi^3=1,\ {\rm Large}\ \rho}$: The expressions for $\varpi_{0,..,5}$ relevant for evaluation of (\ref{eq:confl210}):
\begin{eqnarray}
\label{eq:confl24}
& & (i) \varpi_0 \sim -\int_{\Gamma^\prime}\frac{d\mu}{4\pi^2i}\frac{\Gamma(-\mu)\Gamma(\mu+\frac{1}{6})}{\Gamma(1+\mu)}\rho^{-6\mu}
\frac{\sqrt{3}\sum_{\tau=0}^2\gamma^{0,\tau}_\mu\omega^\tau(\phi-\omega^{-\tau})^{\mu+1}}{2\pi(\mu+1)}\nonumber\\
& & \sim A_{0,0}\left[(\phi-1) - 2\omega(\phi - \omega^{-1}) + \omega^2(\phi - \omega^2)\right] + \frac{A_{0,1}}{\rho^6}\Bigl[(\phi - 1)^2 - 2\omega(\phi - \omega^{-1})^2 \nonumber\\
& & + \omega^2(\phi - \omega^{-2})^2\Bigr],
\end{eqnarray}
where $A_{0,0}=-\frac{\sqrt{3}}{16\pi^4i}\Gamma(\frac{1}{6})\Gamma(\frac{5}{6})$ and $A_{0,1}=\frac{\sqrt{3}}{16\pi^4i}\Gamma(\frac{7}{6})\Gamma(\frac{11}{6})$.
\begin{eqnarray}
\label{eq:confl25}
& & (ii) \varpi_1\sim
\int_{\Gamma^\prime}\frac{d\mu}{8\pi^3}(\Gamma(-\mu))^2\Gamma(\mu+\frac{1}{6})\Gamma(\mu+\frac{5}{6})
\rho^{-6\mu}\nonumber\\
& & \times\frac{\sqrt{3}}{2\pi(\mu+1)}\biggl[2i sin(\pi\mu)\Bigl(e^{-2i\pi\mu}(\phi - 1)^{\mu+1} + \omega(\phi - \omega^{-1})^{\mu+1} - 2\omega^2(\phi - \omega^{-2})^{\mu+1}\nonumber\\
& & + (\phi - 1)^{\mu+1}\Bigr)  + e^{-3i\pi\mu}(\phi - 1)^{\mu + 1}\biggr]
\nonumber\\
& & \sim A_{0,1}\Bigl[(\phi - 1)(A_0 + ln(\rho^{-6})) + i\pi((\phi - 1) + 2\omega(\phi - \omega^{-1} - 4\omega^2(\phi - \omega^{-2}) \nonumber\\
& &  + (\phi - 1)ln(\phi - 1)\Bigr]+ \frac{A_{0,1}}{\rho^6}\Biggl[(\phi - 1)^2(A_1 + ln(\rho^{-6}) + i\pi\Bigl((\phi - 1) + 2\omega(\phi - \omega^{-1}) \nonumber\\
& & - 4\omega^2(\phi - \omega^{-2})\Bigr) + (\phi - 1)^2ln(\phi - 1)\Biggr],
\end{eqnarray}
where $A_0\equiv -1 - 2\Psi(1) + \Psi(\frac{1}{6}) + \Psi(\frac{5}{6})$ and $A_1\equiv -\frac{1}{2} - 2\Psi(1) - 2 + \Psi(\frac{7}{6})+ \psi(\frac{11}{6})$.
\begin{eqnarray*}
& & (iii)\ \varpi_2\sim-\int_{\Gamma^\prime}\frac{d\mu}{8\pi^3}(\Gamma(-\mu))^2\Gamma(\mu+\frac{1}{6})\Gamma(\mu+\frac{5}{6})
\rho^{-6\mu}\nonumber\\
& & \times\frac{\sqrt{3}}{2\pi(\mu+1)}\biggl[2i sin(\pi\mu)\left(-2e^{-2i\pi\mu}(\phi - 1)^{\mu + 1} + e^{-2i\pi\mu}\omega(\phi - \omega^{-1})^{\mu+1} + \omega^2(\phi - \omega^{-2})^{\mu+1}\right)\nonumber\\
\end{eqnarray*}
 \begin{eqnarray}
\label{eq:confl26}
 & & - \left((\phi - 1)^{\mu+1} - \omega(\phi - \omega^{-1})^{\mu+1}\right) - (\phi - 1)^{\mu+1}\biggr]\nonumber\\
 & &
\sim A_{0,0}\Biggl[\left(\omega(\phi - \omega^{-1}) - 2(\phi - 1)\right)(A_0 + ln(\rho^{-6})) + 2i\pi\bigl(-2(\phi - 1) + \omega(\phi - \omega^{-1})\nonumber\\
 & &  + \omega^2(\phi - \omega^{-2})\bigr)\- 2(\phi - 1)ln(\phi - 1)\Biggr]
+\frac{A_{0,1}}{\rho^6}\biggl[\left(\omega(\phi - \omega^{-1}) - 2(\phi - 1)^2\right)(A_1 + ln(\rho^{-6}))\nonumber\\
& &  -2i\pi\biggl(-2(\phi - 1)^2 + \omega(\phi - \omega^{-1})^2
+ \omega^2(\phi - \omega^{-2})^2\biggr) - 2(\phi - 1)^2 ln(\phi - 1)\biggr]
\end{eqnarray}

\begin{eqnarray}
\label{eq:confl27}
& & (iv)\ \varpi_3\sim
\int_{\Gamma^\prime}\frac{d\mu}{8\pi^3}(\Gamma(-\mu))^2\Gamma(\mu+\frac{1}{6})\Gamma(\mu + \frac{5}{6})\rho^{-6\mu}e^{i\pi\mu}\nonumber\\
& & \times-\frac{\sqrt{3}}{2\pi(\mu+1)}\left[(\phi - 1)^{\mu+1} - 2\omega(\phi - \omega^{-1})^{\mu+1} + \omega^2(\phi - \omega^{-2})^{\mu+1}\right]\nonumber\\
& & \sim A_{0,0}\biggl[\left((\phi - 1) - 2\omega(\phi - \omega^{-1}) + \omega^2(\phi - \omega^{-2})\right)(A_0 + i\pi + ln(\rho^{-6}))\nonumber\\
& & +(\phi - 1)ln(\phi - 1) - 2\omega(\phi - \omega^{-1})ln(\phi - \omega^{-1}) + \omega^2(\phi - \omega^{-2})ln(\phi - \omega^{-2})\biggr] \nonumber\\
& & + \frac{A_{0,1}}{\rho^6}\biggl[\left((\phi - 1)^2
- 2\omega(\phi - \omega^{-1})^2 + \omega^2(\phi - \omega^{-2})^2\right)(A_1 + i\pi + ln(\rho^{-6}))\nonumber\\
& & +(\phi - 1)^2ln(\phi - 1) - 2\omega(\phi - \omega^{-1})^2ln(\phi - \omega^{-1}) + \omega^2(\phi - \omega^{-2})^2ln(\phi - \omega^{-2})\biggr]\nonumber\\
\end{eqnarray}

\begin{eqnarray*}
& & (v)\ \varpi_4\sim
\int_{\Gamma^\prime}\frac{d\mu}{8\pi^3}(\Gamma(-\mu))^2\Gamma(\mu+\frac{1}{6})\Gamma(\mu + \frac{5}{6})\rho^{-6\mu}e^{i\pi\mu}\nonumber\\
& & \times-\frac{\sqrt{3}}{2\pi(\mu+1)}\left[e^{-2i\pi\mu}(\phi - 1)^{\mu+1} + \omega\omega(\phi - \omega^{-1})^{\mu+1} -2\omega^2(\phi - \omega^{-2})^{\mu+1}\right]\nonumber\\
& & \sim A_{0,0}\biggl[\left((\phi - 1) - 2\omega(\phi - \omega^{-1}) + \omega^2(\phi - \omega^{-2})\right)(A_0 + i\pi + ln(\rho^{-6}))\nonumber\\
& & -2i\pi(\phi - 1) +(\phi - 1)ln(\phi - 1) + \omega(\phi - \omega^{-1})ln(\phi - \omega^{-1})  -2\omega^2(\phi - \omega^{-2})ln(\phi - \omega^{-2})\biggr] \nonumber\\
\end{eqnarray*}
\begin{eqnarray}
\label{eq:confl28}
& & + \frac{A_{0,1}}{\rho^6}\biggl[\left((\phi - 1)^2
- 2\omega(\phi - \omega^{-1})^2 + \omega^2(\phi - \omega^{-2})^2\right)(A_1 + i\pi + ln(\rho^{-6})) -2i\pi(\phi - 1)^2 \nonumber\\
& & +(\phi - 1)^2ln(\phi - 1) + \omega(\phi - \omega^{-1})^2ln(\phi - \omega^{-1}) -  2 \omega^2(\phi - \omega^{-2})^2ln(\phi - \omega^{-2})\biggr]
\end{eqnarray}
\begin{eqnarray}
\label{eq:confl29}
& & (vi)\ \varpi_5\sim
\int_{\Gamma^\prime}\frac{d\mu}{8\pi^3}(\Gamma(-\mu))^2\Gamma(\mu+\frac{1}{6})\Gamma(\mu + \frac{5}{6})\rho^{-6\mu}e^{i\pi\mu} \times-\frac{\sqrt{3}}{2\pi(\mu+1)}\Biggl[-2e^{-2i\pi\mu}(\phi - 1)^{\mu+1}\nonumber\\
& & + e^{-2i\pi\mu}\omega\omega(\phi - \omega^{-1})^{\mu+1} + \omega^2(\phi - \omega^{-2})^{\mu+1}\Biggr]\nonumber\\
& & \sim A_{0,0}\biggl[\left(-2(\phi - 1) + \omega(\phi - \omega^{-1}) + \omega^2(\phi - \omega^{-2})\right)(A_0 + i\pi + ln(\rho^{-6})) +4i\pi(\phi - 1) \nonumber\\
& & - 2(\phi - 1)ln(\phi - 1) -2i\pi\omega(\phi - \omega^{-1}) + \omega(\phi - \omega^{-1})ln(\phi - \omega^{-1}) + \omega^2(\phi - \omega^{-2})\nonumber\\
& & \times ln(\phi - \omega^{-2})\biggr]  + \frac{A_{0,1}}{\rho^6}\biggl[\left(-2(\phi - 1)^2 +
\omega(\phi - \omega^{-1})^2 + \omega^2(\phi - \omega^{-2})^2\right)(A_1 + i\pi + ln(\rho^{-6}))\nonumber\\
& & 4i\pi(\phi - 1)^2 -2(\phi - 1)^2ln(\phi - 1) - 2i\pi\omega(\phi - \omega^{-1})^2 + \omega(\phi - \omega^{-1})^2ln(\phi - \omega^{-1})\nonumber\\
 & & + \omega^2(\phi - \omega^{-2})^2ln(\phi - \omega^{-2})\biggr].
\end{eqnarray}

\section{Complex Structure Superpotential Extremization}
\setcounter{equation}{0}
\seceqjj
Here, the details pertaining to evaluation of the covariant derivative of the complex structure superpotential in (\ref{eq:Wconfl161}), are given.
\begin{equation}
\label{eq:Wconfl13}
\hskip - 3cm
\partial_x K\sim \frac{- ln x\left({\bar A_1} B_{42} - {\bar B_{12}} {\bar A_4} + {\bar A_2}B_{52} -
{\bar A_5}B_{22}\right)}{{\cal K}},
\end{equation}
\begin{eqnarray}
\label{eq:Wconfl14}
& & {\cal K}\equiv2i\ {\rm Im}\Biggl[ ({\bar A_0} A_3 + {\bar A_1} A_4 + {\bar A}_2 A_5) + ({\bar B_{01}} A_3 + {\bar A_0}B_{31} + {\bar A_1}B_{41} + {\bar B_{11}} A_4 + {\bar A_2} B_{51}\nonumber\\
& & + {\bar B_{21}} A_5)x
   + ({\bar A_1} B_{42} + {\bar B_{12}} A_4 + {\bar A_2} B_{52} + {\bar B_{22}} A_5)x ln x + ({\bar A_0} C_3 + {\bar C_0} A_3 + {\bar A_1} C_4 \nonumber\\
& & + {\bar C_1} A_4 + {\bar A_2} C_5 + {\bar C_2} A_5)(\rho - \rho_0)\Biggr].
\end{eqnarray}
At the extremum values of the complex structure moduli ($x,\rho - \rho_0$),
\begin{eqnarray}
\label{eq:Wconfl15}
& & \tau=\frac{f^T.{\bar\Pi}}{h^T.{\bar\Pi}}\approx\nonumber\\
& & \frac{1}{(\sum_{i=0}^5h_i{\bar A_i})}\Biggl[f_0\left({\bar A_0} + {\bar B_{01}}{\bar x} + {\bar C_0}({\bar\rho} - {\bar \rho_0})\right) + f_1\left({\bar A_1} + {\bar B_{11}}{\bar x} + {\bar B_{12}}{\bar x} ln{\bar x} + {\bar C_1}({\bar\rho} - {\bar \rho_0})\right)\nonumber\\
 & & + f_2\left({\bar A_2} + {\bar B_{21}}{\bar x} + {\bar B_{22}}{\bar x} ln{\bar x} + {\bar C_2}({\bar\rho} - {\bar \rho_0})\right)+ f_3\left(({\bar A_3} + {\bar B_{31}}{\bar x} + {\bar C_3}({\bar\rho} - {\bar \rho_0})\right)\nonumber\\
& & + f_4\left({\bar A_4} + {\bar B_{41}}{\bar x} + {\bar B_{42}}{\bar x} ln{\bar x} + {\bar C_4}({\bar\rho} - {\bar \rho_0})\right) + f_5\left({\bar A_5} + {\bar B_{51}}{\bar x} + {\bar B_{52}}{\bar x} ln{\bar x} + {\bar C_1}({\bar\rho} - {\bar \rho_0})\right)\Biggr]\nonumber\\
& & \times\Biggl( 1 - \frac{1}{(\sum_{i=0}^5h_i{\bar A_i})}\Biggl[h_0({\bar B_{01}}{\bar x} + {\bar C_0}({\bar\rho} - {\bar \rho_0})) + h_1({\bar B_{11}}{\bar x} + {\bar B_{12}}{\bar x} ln{\bar x} + {\bar C_1}({\bar\rho} - {\bar \rho_0}))\nonumber\\
 & & + h_2({\bar B_{21}}{\bar x} + {\bar B_{22}}{\bar x} ln{\bar x} + {\bar C_2}({\bar\rho} - {\bar \rho_0})) + h_3({\bar A_3} + {\bar B_{31}}{\bar x} + {\bar C_3}({\bar\rho} - {\bar \rho_0}))\nonumber\\
 & & + h_4({\bar A_4} + {\bar B_{41}}{\bar x} + {\bar B_{42}}{\bar x} ln{\bar x} + {\bar C_4}({\bar\rho} - {\bar \rho_0})) + h_5({\bar A_5} + {\bar B_{51}}{\bar x} + {\bar B_{52}}{\bar x} ln{\bar x} + {\bar C_1}({\bar\rho} - {\bar \rho_0}))\Biggr).\nonumber\\
& &
\end{eqnarray}
Hence,
\begin{eqnarray}
\label{eq:Wconfl16}
& & \partial_x W_{c.s.} \approx ln x\Biggl[ {\bar B_{12}}\Biggl(f_1 - \frac{h_1\Xi[f_i;{\bar x},({\bar\rho} - {\bar \rho_0})]}{\sum_{i=0}^5h_i{\bar A_i}} + \frac{h_1(\sum_{j=0}^5f_i{\bar A_i})\Xi[h_i;{\bar x},({\bar\rho} - {\bar \rho_0})]}{(\sum_{i=0}^5h_i{\bar A_i})^2}\Biggr)\nonumber\\
& & + {\bar B_{22}}\Biggl(f_2 - \frac{h_2\Xi[f_i;{\bar x},({\bar\rho} - {\bar \rho_0})]}{\sum_{i=0}^5h_i{\bar A_i}} + \frac{h_2(\sum_{j=0}^5f_i{\bar A_i})\Xi[h_i;{\bar x},({\bar\rho} - {\bar \rho_0})]}{(\sum_{i=0}^5h_i{\bar A_i})^2}\Biggr)\nonumber\\
& & + {\bar B_{42}}\Biggl(f_4 - \frac{h_2\Xi[f_i;{\bar x},({\bar\rho} - {\bar \rho_0})]}{\sum_{i=0}^5h_i{\bar A_i}} + \frac{h_2(\sum_{j=0}^5f_i{\bar A_i})\Xi[h_i;{\bar x},({\bar\rho} - {\bar \rho_0})]}{(\sum_{i=0}^5h_i{\bar A_i})^2}\Biggr)\nonumber\\
& & + {\bar B_{52}}\Biggl(f_5 - \frac{h_5\Xi[f_i;{\bar x},({\bar\rho} - {\bar \rho_0})]}{\sum_{i=0}^5h_i{\bar A_i}} + \frac{h_5(\sum_{j=0}^5f_i{\bar A_i})\Xi[h_i;{\bar x},({\bar\rho} - {\bar \rho_0})]}{(\sum_{i=0}^5h_i{\bar A_i})^2}\Biggr)\Biggr],
\end{eqnarray}
where
\begin{eqnarray}
\label{eq:Xisdefs}
& &
\Xi[f_i; {\bar x},({\bar\rho} - {\bar \rho_0})]\equiv f_0({\bar A_0} + {\bar B_{01}}{\bar x} + {\bar C_0}({\bar\rho} - {\bar \rho_0})) + f_1({\bar A_1} + {\bar B_{11}}{\bar x} + {\bar B_{12}}{\bar x} ln{\bar x} + {\bar C_1}({\bar\rho} - {\bar \rho_0}))\nonumber\\
 & & + f_2({\bar A_2} + {\bar B_{21}}{\bar x} + {\bar B_{22}}{\bar x} ln{\bar x} +
 {\bar C_2}({\bar\rho}-{\bar \rho_0})
 + f_3({\bar A_3}+{\bar B_{31}}{\bar x}+{\bar C_3}({\bar\rho}-{\bar \rho_0}))\nonumber\\
& & + f_4({\bar A_4}+{\bar B_{41}}{\bar x}+{\bar B_{42}}{\bar x} ln{\bar x}+{\bar C_4}({\bar\rho}- {\bar \rho_0}))+f_5({\bar A_5}+{\bar B_{51}}{\bar x}+{\bar B_{52}}{\bar x} ln{\bar x}+{\bar C_1}({\bar\rho}-{\bar \rho_0})),\nonumber\\
& & \equiv f^T{\bar\Pi}({\bar x},{\bar\rho} - {\bar\rho_0})\nonumber\\
& &
\Xi[h_i;{\bar x},({\bar\rho} - {\bar \rho_0})]= h_0({\bar B_{01}}{\bar x} + {\bar C_0}({\bar\rho} - {\bar \rho_0})) + h_1({\bar A_1} + {\bar B_{11}}{\bar x} + {\bar B_{12}}{\bar x} ln{\bar x} + {\bar C_1}({\bar\rho} - {\bar \rho_0}))\nonumber\\
 & & + h_2({\bar B_{21}}{\bar x} + {\bar B_{22}}{\bar x} ln{\bar x} + {\bar C_2}({\bar\rho} - {\bar \rho_0})) + h_3({\bar B_{31}}{\bar x} + {\bar C_3}({\bar\rho} - {\bar \rho_0}))\nonumber\\
& & + h_4({\bar B_{41}}{\bar x} + {\bar B_{42}}{\bar x} ln{\bar x} + {\bar C_4}({\bar\rho} - {\bar \rho_0})) + h_5({\bar A_5} + {\bar B_{51}}{\bar x} + {\bar B_{52}}{\bar x} ln{\bar x} + {\bar C_1}({\bar\rho} - {\bar \rho_0}))\nonumber\\
& & \equiv h^T\left[{\bar\Pi}({\bar x},{\bar\rho} - {\bar\rho_0}) - {\bar\Pi}(x=0,\rho=\rho_0)\right].
\end{eqnarray}

\section{Ingredients for Evaluation of ${\cal N}_{IJ}$}
\setcounter{equation}{0}
\seceqll
In this appendix we fill in the details relevant to evaluation of $X^I Im(F_{IJ}) X^J$ to arrive at (\ref{eq:BHinv15}). First, using (\ref{eq:BHinv12}), one arrives at;
\begin{eqnarray*}
& &  Im(F_{0I})X^I=-\frac{i}{2}\Biggl[\left(\frac{B_{01}}{B_3}+\frac{C_0}{C_3}-\frac{{\bar B_{01}}}{{\bar B_3}}-\frac{{\bar C_0}}{{\bar C_3}}\right)(A_3+B_{31}x+C_3(\rho-\rho_0)\nonumber\\
& & +\left(\frac{C_1}{C_3}-\frac{{\bar C_1}}{{\bar C_3}}\right)(A_4+B_{41}x+B_{42}x ln x+C_4(\rho-\rho_0))\Biggr] +\left(\frac{C_2}{C_3}-\frac{{\bar C_2}}{{\bar C_3}}\right)\nonumber\\
& & \times\left(A_5+B_{51}x+B_{52}x ln x+C_5(\rho-\rho_0)\right)\Biggr];\nonumber\\
& & Im(F_{1I})X^I=-\frac{i}{2}\Biggl[\left(\frac{C_1}{C_3}-\frac{{\bar C_1}}{{\bar C_3}}\right)(A_3+B_{31}x+C_3(\rho-\rho_0)\nonumber\\
& & +\left(\frac{C_1}{C_4}-\frac{{\bar C_1}}{{\bar C_4}}\right)(A_4+B_{41}x+B_{42}x ln x+C_4(\rho-\rho_0))\Biggr] +\left(\frac{C_2}{C_4}-\frac{{\bar C_2}}{{\bar C_4}}\right)\nonumber\\
\end{eqnarray*}
\begin{eqnarray}
\label{eq:BHinv13}
& & \times\left(A_5+B_{51}x+B_{52}x ln x+C_5(\rho-\rho_0)\right)\Biggr];\nonumber\\
& & Im(F_{2I})X^I=-\frac{i}{2}\Biggl[\left(\frac{C_2}{C_3}-\frac{{\bar C_2}}{{\bar C_3}}\right)(A_3+B_{31}x+C_3(\rho-\rho_0)\nonumber\\
& & +\left(\frac{C_2}{C_4}-\frac{{\bar C_2}}{{\bar C_4}}\right)(A_4+B_{41}x+B_{42}x ln x+C_4(\rho-\rho_0))\Biggr] +\left(\frac{C_2}{C_5}-\frac{{\bar C_2}}{{\bar C_5}}\right)\nonumber\\
& & \times\left(A_5+B_{51}x+B_{52}x ln x+C_5(\rho-\rho_0)\right)\Biggr].
\end{eqnarray}
This hence yields
\begin{eqnarray}
\label{eq:BHinv14}
& & X^I Im(F_{IJ} X^J = (X^0)^2 Im(F_{00}) + (X^1)^2 Im(F_{11}) + (X^2)^2 Im(F_{22})\nonumber\\
& & + 2x^0X1 Im(F_{01}) + 2X^0X^2 Im(F_{02}) + 2X^1X^2 Im(F_{12})\nonumber\\
& & \approx-\frac{i}{2}\Biggl[\Biggl(A_3^2 + 2A_3B_{31} x + 2A_3C_3 (\rho-\rho_0)\Biggr)\left(\frac{B_{01}}{B_3} + \frac{C_0}{C_3} - \frac{{\bar B_{01}}}{{\bar B_3}} - \frac{{\bar C_0}}{{\bar C_3}}\right)\nonumber\\
& & + \Biggl(A_4^2 + 2A_4B_{41} x + 2A_4B_{42} x ln x + 2A_4C_4 (\rho-\rho_0)\Biggr)\left(\frac{C_1}{C_4} - \frac{{\bar C_1}}{{\bar C_4}}\right)\nonumber\\
& & + \Biggl(A_5^2 + 2A_5B_{51} x + 2A_5B_{52} x ln x + 2A_5C_5 (\rho-\rho_0)\Biggr)
\left(\frac{C_2}{C_4} - \frac{{\bar C_2}}{{\bar C_5}}\right)\nonumber\\
& & + \Biggl(A_3A_4 + [A_3B_{41}+A_4B_{31}] x + A_3B_{42} x ln x + [A_3C_4 + A_4C_3](\rho - \rho_0)\Biggr)\left(\frac{C_1}{C_3} - \frac{{\bar C_1}}{{\bar C_3}}\right)\nonumber\\
& & + \Biggl(A_3A_5 + [A_3B_{51}+A_5B_{31}] x + A_3B_{52} x ln x + [A_3C_5 + A_5C_3](\rho - \rho_0)\Biggr)\left(\frac{C_2}{C_3} - \frac{{\bar C_2}}{{\bar C_3}}\right)\nonumber\\
& & + \Biggl(A_4A_5 + [A_4B_{51}+A_5B_{41}] x + A_4B_{52} x ln x + [A_4C_5 + A_5C_4](\rho - \rho_0)\Biggr)\left(\frac{C_2}{C_4} - \frac{{\bar C_2}}{{\bar C_4}}\right)\Biggr].\nonumber\\
\end{eqnarray}

\newpage
\section{Tables}
\setcounter{equation}{0}
\begin{table}[htbp]
\centering
\begin{tabular}{|c|c|}
\hline
&  \\
$(G^{-1})^{\rho_1{\bar\rho}_1}|\partial_{\rho_1}W_{np}|^2$ &
$\frac{{\cal V}\sqrt{ln {\cal V}}}{{\cal V}^{2n^1e^{-\phi}}}$ \\
&  \\
\hline
&  \\
$(G^{-1})^{\rho_2{\bar\rho}_2}|\partial_{\rho_2}W_{np}|^2$ &
${\cal V}^{\frac{4}{3}}e^{-2n^2e^{-\phi}{\cal V}^{\frac{2}{3}}}$ \\
&  \\
\hline
&  \\
$(G^{-1})^{\rho_1{\bar\rho_2}}\partial_{\rho_1}W_{np}{\bar\partial}_{\bar\rho_2}{\bar W}_{np} + c.c.$
& $\frac{{\cal V}^{\frac{2}{3}} ln {\cal V} e^{-n^2e^{-\phi}{\cal V}^{\frac{2}{3}}}}{{\cal V}^{n^1e^{-\phi}}}$\\
&  \\
\hline
&  \\
$(G^{-1})^{G^1{\bar G^1}}|\partial_{G^1}W|^2 + (G^{-1})^{G^2{\bar G^2}}|\partial_{G^2}W|^2 + $ &
${\cal V}^{1-(2n^1\ or\ 2n^2\ or\ n^1+n^2)e^{-\phi}}$\\
$(G^{-1})^{G^1{\bar G^2}}\partial_{G^1}W_{np}{\bar\partial}_{\bar G^2}{\bar W}_{np} + c.c.$ & \\
&  \\
\hline
&  \\
$(G^{-1})^{\rho_1{\bar G^1}}\partial_{\rho_1}W_{np}{\bar\partial}_{\bar G^1}{\bar W}_{np} + c.c.$
& $\frac{ln {\cal V} }{{\cal V}^{2n^1e^{-\phi}}}$ \\
&  \\
\hline
&  \\
$(G^{-1})^{\rho_2{\bar G^1}}\partial_{\rho_2}W_{np}{\bar\partial}_{\bar G^1}{\bar W}_{np} + c.c.$
& $\frac{{\cal V}^{\frac{2}{3}} e^{-n^1e^{-\phi}{\cal V}^{\frac{2}{3}}}}{{\cal V}^{n^1e^{-\phi}+\frac{1}{3}}}$ \\
&  \\
\hline
\end{tabular}
\caption{$(G^{-1})^{A{\bar B}}\partial_A W_{np}{\bar\partial}_{\bar B}{\bar W}_{np}$}
\end{table}

\begin{table}[htbp]
\centering
\begin{tabular}{|c|c|}
\hline
&  \\
$(G^{-1})^{\rho_1{\bar\rho}_1}\partial_{\rho_1}K{\bar\partial}_{\bar\rho_1}{\bar W}_{np}+c.c.$ &
$\frac{ln {\cal V}}{{\cal V}^{n^1e^{-\phi}}}$ \\
&  \\
\hline
&  \\
$(G^{-1})^{\rho_2{\bar\rho}_2}\partial_{\rho_2}K{\bar\partial}_{\bar\rho_1}{\bar W}_{np}+c.c.$ &
${\cal V}^{\frac{2}{3}}e^{-n^2e^{-\phi}{\cal V}^{\frac{2}{3}}}$ \\
&  \\
\hline
&  \\
$(G^{-1})^{\rho_1{\bar\rho}_2}\partial_{\rho_1}K{\bar\partial}_{\bar\rho_2}{\bar W}_{np}+c.c.$ &
$\frac{(ln {\cal V})^{\frac{3}{2}}e^{-n^2e^{-\phi}{\cal V}^{\frac{2}{3}}}}{{\cal V}^{\frac{1}{3}}}$
\\
&  \\
\hline
&  \\
$(G^{-1})^{G^1{\bar G^1}}\partial_{G^1}K{\bar\partial}_{\bar G^1}{\bar W}_{np} +
(G^{-1})^{G^1{\bar G^1}}\partial_{G^1}K{\bar\partial}_{\bar G^1}{\bar W}_{np} +$ & $\frac{1}{{\cal V}^{n^1e^{-\phi}}}$\\
$(G^{-1})^{G^1{\bar G^2}}\partial_{G^1}K{\bar\partial}_{\bar G^2}{\bar W}_{np} +
(G^{-1})^{\rho_2{\bar G^1}}\partial_{\rho_2}K{\bar\partial}_{\bar G^1}{\bar W}_{np} + c.c.$  & \\
 &  \\
 \hline
 &  \\
$(G^{-1})^{\rho_1{\bar G^1}}\partial_{\rho_1}K{\bar\partial}_{\bar G^1}{\bar W}_{np} + c.c.$
& $\frac{(ln {\cal V})^{\frac{3}{2}}}{{\cal V}^{1+n^1e^{-\phi}}}$ \\
&  \\
\hline
\end{tabular}
\caption{$(G^{-1})^{A{\bar B}}\partial_AK{\bar\partial}_{\bar B}{\bar W}_{np} + c.c.$}
\end{table}

\begin{table}[htbp]
\centering
\begin{tabular}{|c|c|}
\hline
&  \\
$(G^{-1})^{\rho_1{\bar\rho}_1}|\partial_{\rho_1}K|^2$ & $\frac{(ln {\cal V})^{\frac{3}{2}}}{{\cal V}}$ \\
&  \\
\hline
&  \\
$(G^{-1})^{\rho_2{\bar\rho}_2}|\partial_{\rho_2}K|^2$  & ${\cal O}(1)$ \\
&  \\
\hline
&  \\
$(G^{-1})^{\rho_1{\bar\rho}_2}\partial_{\rho_1}{\bar\partial}_{\bar\rho_2}K + c.c.$
& $\frac{(ln {\cal V})^{\frac{3}{2}}}{{\cal V}}$ \\
&  \\
\hline
&  \\
$(G^{-1})^{G^1{\bar G^1}}|\partial_{G^1}K|^2 + (G^{-1})^{G^2{\bar G^2}}|\partial_{G^2}K|^2 + $ &
$\frac{1}{\cal V}$\\
$(G^{-1})^{G^1{\bar G^2}}\partial_{G^1}K{\bar\partial}_{\bar G^2}K
+ (G^{-1})^{\rho_2{\bar G^1}}\partial_{\rho_2}K{\bar\partial}_{\bar G^1}K + c.c.$
& \\ &  \\
\hline
&  \\
$(G^{-1})^{\rho_1{\bar G^1}}\partial_{\rho_1}K{\bar\partial}_{\bar G^1}K + c.c.$
& $\frac{(ln {\cal V})^{\frac{3}{2}}}{{\cal V}^2}$\\
&  \\
\hline
\end{tabular}
\caption{$(G^{-1})^{A{\bar B}}\partial_AK{\bar\partial}_{\bar B}K$}
\end{table}

\begin{table}[htbp]
\centering
\begin{tabular}{|l|l|}
\hline
 &  \\
Gravitino mass &  $ m_{\frac{3}{2}}\sim{\cal V}^{-\frac{n^s}{2} - 1}$ \\
&  \\
Gaugino mass & $ M_{\tilde g}\sim m_{\frac{3}{2}}$\\
&  \\
\hline
&  \\
$D3$-brane position moduli mass & $ m_{{\cal Z}_i}\sim {\cal V}^{\frac{19}{36}}m_{\frac{3}{2}}$ \\
&  \\
Wilson line moduli mass & $ m_{\tilde{\cal A}_1}\sim {\cal V}^{\frac{73}{72}}m_{\frac{3}{2}}$\\
&  \\
\hline
&  \\
& $A_{{\cal Z}_i{\cal Z}_j{\cal Z}_k}\sim n^s{\cal V}^{\frac{37}{36}}m_{\frac{3}{2}}$\\
A-terms & $A_{{\tilde{\cal A}_1}{\tilde{\cal A}_1}{\tilde{\cal A}_1}}\sim n^s{\cal V}^{\frac{37}{36}}m_{\frac{3}{2}}$\\
& $A_{{{\tilde{\cal A}_1}^2}{{\cal Z}_i}}\sim n^s{\cal V}^{\frac{37}{36}}m_{\frac{3}{2}}$\\
& $A_{{\tilde{\cal A}_1}{{\cal Z}_i}{{\cal Z}_j}}\sim n^s{\cal V}^{\frac{37}{36}}m_{\frac{3}{2}}$\\
&  \\
\hline
&  \\
Physical $\mu$-terms & $\hat{\mu}_{{\cal Z}_i{\cal Z}_j}\sim{\cal V}^{\frac{37}{36}}m_{\frac{3}{2}}$\\
& $\hat{\mu}_{{\cal A}_1{\cal Z}_i}\sim{\cal V}^{-\frac{3}{4}}m_{\frac{3}{2}}$\\
& $\hat{\mu}_{{\cal A}_1{\cal A}_1}\sim{\cal V}^{-\frac{33}{36}}m_{\frac{3}{2}}$\\
&  \\
\hline
&  \\
& $\hat{Y}_{{\cal Z}_i{\cal Z}_i{\cal Z}_i}\sim {\cal V}^{\frac{19}{24}-\frac{n^s}{2}}$\\
& $\hat{Y}_{{\cal Z}_i^2{\cal Z}_j}\sim {\cal V}^{\frac{13}{24}-\frac{n^s}{2}}$\\
Physical Yukawa couplings&$\hat{Y}_{{\cal Z}_i^2\tilde{\cal A}_1}\sim {\cal V}^{-\frac{71}{72}-\frac{n^s}{2}}$\\
&$\hat{Y}_{{\cal Z}_1{\cal Z}_2\tilde{\cal A}_1}\sim {\cal V}^{-\frac{89}{72}-\frac{n^s}{2}}$\\
&$\hat{Y}_{\tilde{\cal A}_1^2{\cal Z}_i}\sim {\cal V}^{-\frac{199}{72}-\frac{n^s}{2}}$\\
&$\hat{Y}_{\tilde{\cal A}_1\tilde{\cal A}_1\tilde{\cal A}_1}\sim {\cal V}^{-\frac{109}{24}-\frac{n^s}{2}}$\\
&  \\
\hline
&  \\
&$\left(\hat{\mu}B\right)_{{\cal Z}_i{\cal Z}_i}\sim{\cal V}^{\frac{223}{108}}m_{\frac{3}{2}}^2$\\
$\hat{\mu}B$-terms & $\left(\hat{\mu}B\right)_{{\cal Z}_1{\cal Z}_2}\sim{\cal V}^{\frac{37}{18}}m_{\frac{3}{2}}^2$\\
 &$\left(\hat{\mu}B\right)_{\tilde{\cal A}_1\tilde{\cal A}_1}\sim{\cal V}^{\frac{5}{36}}m_{\frac{3}{2}}^2$\\
&$\left(\hat{\mu}B\right)_{{\cal Z}_i\tilde{{\cal A}_1}}\sim{\cal V}^{-\frac{13}{18}}m_{\frac{3}{2}}^2$\\
&  \\
\hline
\end{tabular}
\caption{Gravitino/gaugino masses and soft Susy breaking parameters.}
\end{table}